\documentclass[12pt]{article}
\usepackage{amsmath}
\usepackage{graphicx}
\usepackage{enumerate}

\usepackage{authblk}
\usepackage[]{natbib}

\usepackage{url} 

\usepackage{ amsbsy, epsfig, epsf, psfrag, booktabs, amssymb, amsthm}

\newcommand{\blind}{1}

\newcommand{\bm}[1]{{\boldsymbol {#1}}}

\usepackage{tikz}             
\usetikzlibrary{arrows}
\usetikzlibrary{calc}
\usetikzlibrary{positioning}

\newtheorem{proposition}{Proposition}
\newtheorem{corollary}{Corollary}
\newtheorem{lemma}{Lemma}
\newtheorem{theorem}{Theorem}

\newcommand{\beginsupplementS}{%
        \setcounter{table}{0}
        \renewcommand{\thetable}{S\arabic{table}}%
        \setcounter{figure}{0}
        \renewcommand{\thefigure}{S\arabic{figure}}%
     }

\begin{document}

\def\spacingset#1{\renewcommand{\baselinestretch}%
{#1}\small\normalsize} \spacingset{1}


\if1\blind
{
  \title{\bf Fast Bayesian inference of Block Nearest Neighbor Gaussian process for large data}
  \author{Zaida C. Quiroz
  \thanks{E-mail: zquiroz@pucp.edu.pe
    }\hspace{.2cm}\\
    Department of Sciences, Pontif\'{i}cia Universidad Cat\'{o}lica del Per\'{u},\\
    Marcos O. Prates \\
    Department of Statistics, Universidade Federal de Minas Gerais,\\
    Dipak K. Dey \\
    Department of Statistics, University of Connecticut
        and \\
    H{\aa}vard Rue \\
    Computer, Electrical and Mathematical Science and Engineering Division, King Abdullah University of Science and Technology,\\
}
  \maketitle
} \fi

\if0\blind
{
  \bigskip
  \bigskip
  \bigskip
  \begin{center}
    {\LARGE\bf Fast Bayesian inference of Block Nearest Neighbor Gaussian process for large data}
\end{center}
  \medskip
} \fi

\bigskip
\begin{abstract}
This paper presents the development of a spatial block-Nearest Neighbor Gaussian process (block-NNGP) for location-referenced large spatial data. The key idea behind this approach is to divide the spatial domain into several blocks which are dependent under some constraints. The cross-blocks capture the large-scale spatial dependence, while each block captures the small-scale spatial dependence. The resulting block-NNGP enjoys Markov properties reflected on its sparse precision matrix. It is embedded as a prior within the class of latent Gaussian models, thus Bayesian inference is obtained using the integrated nested Laplace approximation (INLA). The  performance of the block-NNGP is illustrated on simulated examples and massive real data for locations in the order of $10^4$.
\end{abstract}

\noindent%
\vfill

\newpage
\spacingset{1.5} 
\section{Introduction}
\label{Section intro}

New technologies such as GPS and remote sensing enable the collection of massive amounts of high-resolution geographically referenced observations over large areas. When georeferenced units depends on the distance between the locations, these data are quite often  analyzed through a spatial Gaussian process (GP) due to its good analytic properties \citep{cressie93}. Specifically, in a domain $\bm{D} \in \Re^2$ with coordinates $s \in \bm D$, $w(s)$ is a continuously indexed GP if all finite collections $\{w(s_i)\}$ are jointly Gaussian distributed, $\forall s_i \in \bm D$. The GP is usually specified by using a mean function $\bm \mu$ and a covariance function $\bm C$, so the covariance matrix is $\bm C_S$ is composed of $C(s_i, s_j)$ elements. Since many calculations depend on a dense $\bm C_S$ and possibly also dense precision matrix $\bm C_S^{-1}$, it is well-known that computations over this GP can be prohibitive for large data \citep{Banerjeeetal:2008, Lindgren:2011}.

A useful approach to deal with large geostatistical data 
relies on representing the GP as a Gaussian Markov random process (GMRP), inducing sparsity on $\bm {C_S}^{-1}$, given that the spatial correlation between pairs of distantly located observations is nearly zero \citep{Rue:2002}. In particular, the sparsity can be achieved either through stochastic partial differential equations (SPDE - \citealt{Lindgren:2011})  or Nearest Neighbor Gaussian processes (NNGP- \citealt{Dattaetal:2016}). In the SPDE approach, a GP with a Mat\'{e}rn covariance function is the solution of a specific SPDE \citep{Whittle:1954}, so it can be approximated to a GMRP through finite element methods. On the other hand, the NNGP works for any valid covariance function, it is built from lower-dimensional conditional distributions, based on  a collection of some ``past'' neighbors. A major drawback of relying on the NNGP is how to choose the number of nearest neighbors to achieve an accurate approximations.

Another approach  to deal with computationally intractable large matrices of GP is the spatial independent blocking, specifically, the partition of the spatial domain into independent disjoint blocks. This approach was often restricted to covariance matrices, ignoring the dependence between blocks. For instance, \citet{Stein:2013}  showed that this simple approach is better than the covariance tapering approach \citep{Kaufman:2008, BolinWallin:2016}, method that  strategically set some observations of the covariance matrix into zero to get its sparsity. Otherwise, \citet{Steinetal:2004}, \citet{Caragea:2007} and \citet{Eidsviketal:2014} used independent or conditionally independent blocks of observations to build composite-likelihood functions, but it was not clear how to obtain proper joint predictive distributions. All of these approaches were restricted to get a sparse covariance matrix. 

The main contribution of this paper is to present an accurate and computationally efficient block nearest neighbor Gaussian process (block-NNGP). It assumes that the blocks are conditionally independent given some specific nearest neighbor blocks. The motivation is that aggregation of small blocks plays an important role to explain small and large scale spatial dependencies, being able to capture the spatial dependence at all scales. Specifically, the cross-blocks are able to capture the large-scale spatial dependence while the small-scale spatial dependence is captured through the observations within each block. Thus, the block-NNGP enables a consistent way to combine parameter estimation and spatial prediction. 

Even though the framework allows both frequentist and Bayesian inference, we adopted a Bayesian framework due to the flexibility to fit non-Gaussian responses \citep{Diggle:1998}.  Because the block-NNGP hierarchical model belongs to the class of  latent Gaussian models, where the block-NNGP, represented as a Gaussian Markov random field (GMRF), belongs to the second stage, we used the Integrated Nested Laplace Approximation (INLA) algorithm for estimation \citep{Rue:2009}. Therefore, using this framework, a direct contribution of this paper is that we are able to scale the inference process for massive data. As will be shown, the proposed structure only requires to store small dense matrices to compute the sparse precision matrix of the block-NNGP which is done by fast parallel computation. Moreover, we introduce a full Markov Chain Monte Carlo (MCMC) algorithm that enables simulation of the block-NNGP  and demonstrates the full inferential capabilities in terms of estimation, prediction and goodness of fit. In particular, the parameters were also estimated through the collapsed MCMC method \citep{Finleyetal:2019} to improve convergence and running times. 

The paper is organized as follows. Section~\ref{s:blockNNGP} gives details of the new block-NNGP spatial process and inference through INLA and MCMC. In Section~\ref{s:sim}, simulations are assessed to evaluate the  performance of the proposed process under different block setting and different covariance functions. It also presented a predictive performance assessment. 
Section~\ref{s:app}  illustrates the benefits of the proposed process by analysing mining and precipitation data. Finally, some discussions and concluding remarks are given in Section~\ref{sec:conc}. 

\section{Block NNGP process}
\label{s:blockNNGP}

Let $w(s)$ be a spatial Gaussian process, $\mbox{GP}(0, C(\theta))$, defined for all $s \in \bm{D} \subset  \Re^2$, where  $C(\theta)$  is any valid covariance function. Let $\bm{S} = \{s_1, \dots, s_n\}$ be a fixed set of locations in $\bm{D}$ and consider a partition of $\bm{D}$ in $M$ blocks  $b_1,\dots ,b_M$, where $U_{k=1}^M b_k = \bm{D}, b_k \cap b_l = \phi$, for all pairs of blocks $b_k$ and $b_l$. The vector $\bm{w_{b_k}} = \{w (s_i ); s_i \in b_k \}$ and $\mbox{dim}(\bm{w_{b_k}}) = n_{k}$ such that $\sum_{k=1}^M n_{k} = n$. In this context, the joint density of $\bm{w_S} = (w(s_1), \dots, w(s_n))'$  can be written as follows:  
\begin{equation}
{\pi}(\bm{w_S}) =  \pi(w_{b1}) \prod_{k=2}^M \pi(w_{\bm{b_k}}|\bm{w_{b_1}}, \dots, \bm{w_{b_{k-1}}}).
 \label{eq:eq1}
\end{equation}
Following \cite{Steinetal:2004},  a subset $\bm{N(b_k)}$ is composed of  $nb$ or less neighbor blocks of a block
 $b_k$, from  $\bm{w_{b_1}}, \dots, \bm{w_{b_{k-1}}}$.  It is assumed that  $\bm{w_{b_k}}$ and $\bm{w_{b_j}}$, given a subset $\bm{N(b_k)}$, are conditionally independent, then the joint density of $\bm{w_S}$ in \eqref{eq:eq1} is approximated by 
\begin{equation}
\widetilde{\pi}(\bm{w_S}) =  \pi(\bm{w_{b1}}) \prod_{k=2}^M \pi(\bm{w_{b_k}}|\bm{w_{N(b_k)}}). 
\label{eq:eq2}  
\end{equation}
Here, $\widetilde{\pi}(.|.)$ is an approximated (conditional) density of its arguments. To ensure that this approximation is also a joint density we state the next proposition. 

\begin{proposition}
Let $\bm{G} = \{\bm{S}, \bm{\xi}\}$ be a chain graph, where $\bm{S} = \{s_1, \dots, s_n\}$ is a set of nodes,  and $\bm\xi$ is a set of edges, composed of: (i) directed edges from every node in the set $\bm{s_{b_k}} = \{ s_i \in b_k, \forall i =1, \dots, n\}$ to all nodes in $\bm N(b_k)$, $\forall  k= 1, \dots, M$, and
(ii)  undirected edges between every pair of nodes on each block $b_k$.
Let $\bm G^b$ be a subgraph of $\bm G$, composed of one node from each set $\bm s_{b_k}$. 
If $\bm G^b$ is a directed acyclic graph (DAG) and $\pi(\bm w_S)$ is a multivariate joint density, then $\widetilde{\pi}(\bm w_S)$ in (\ref{eq:eq2}) is also a multivariate joint  density.
\label{prop1}
\end{proposition}

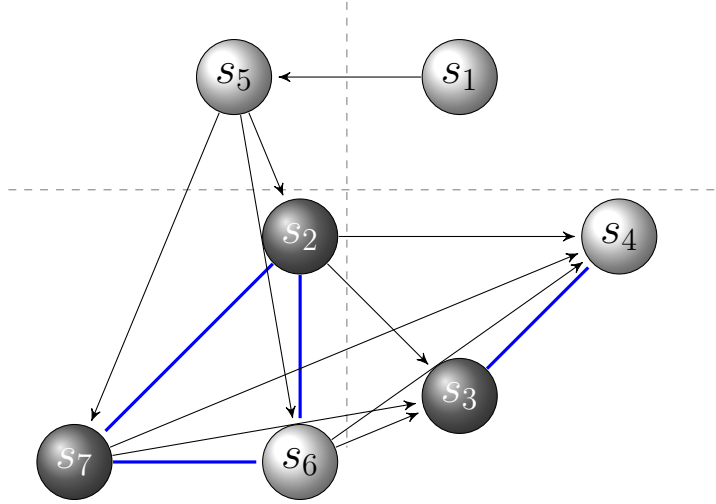
\begin{figure}
\begin{center}
\begin{tikzpicture}
\tikzset{>=stealth',shorten >=2pt,auto,node distance=3cm,thick}
\tikzset{main node/.style={ball color=white,text=black, circle,draw,font=\sffamily\Large\bfseries}}
  \node[main node] (1) {$s_1$};
\tikzset{main node/.style={ball color=gray,text=white, circle,draw,font=\sffamily\Large\bfseries}}
  \node[main node] (2) [below left of=1] {$s_2$};
  \node[main node] (3) [below right of=2] {$s_3$};
\tikzset{main node/.style={ball color=white,text=black, circle,draw,font=\sffamily\Large\bfseries}}
  \node[main node] (4) [below right of=1] {$s_4$};
  \node[main node] (5) [ left of=1] {$s_5$};
\node[main node] (6) [ below  of=2] {$s_6$};
\tikzset{main node/.style={ball color=gray,text=white, circle,draw,font=\sffamily\Large\bfseries}}
  \node[main node] (7) [  left of=6] {$s_7$}; 
\draw[gray,dashed] (-6,-1.5) -- (3.5,-1.5);
\draw[gray,dashed] (-1.5,1) -- (-1.5,-5);
  \draw[blue,very thick] (3) -- (4);
\draw[blue,very thick] (2) -- (6);
\draw[blue,very thick] (7) -- (6);
\draw[blue,very thick] (2) -- (7); 
\tikzset{->,>=stealth',shorten >=2pt,auto,node distance=3cm,thick}
\path[every node/.style={font=\sffamily\small}]
    (1) edge node [left] {} (5)
    (2) edge node {} (4)
        edge node[left] {} (3)
    (6) edge node {} (4)
        edge node[left] {} (3)
    (7) edge node {} (4)
        edge node[left] {} (3)
    (5) edge  node[right] {} (2)
        edge node[left] {} (6)
        edge node[left] {} (7);
\end{tikzpicture}

\end{center}
\caption{\small An example of a chain graph $\bm G$ with $n = 7$ nodes and directed edges (black arrows) and undirected edges (solid blue lines). It is considered  $M = 4$ blocks: $\bm{b_1} = \{s_1\}$, $\bm{b_2} = \{s_5\}$, $\bm{b_3} = \{s_2, s_6, s_7\}$, $\bm{b_4} = \{s_3,s_4\}$, and $nb=1$ neighbor block. $\bm{G^b}$ is a subgraph of $\bm G$, composed of $s_1, s_4, s_5, s_6 $ (light gray nodes).} 
  \label{fig:fig1}
\end{figure}

The proof of this proposition and subsequent proofs are found in the supplementary material A. Proposition 1 states that $\widetilde{\pi}(\bm w_S)$ is a joint density if  $\bm G$ is a chain graph and $\bm G^b$ is a directed acyclic subgraph of $\bm G$. A simple example of a chain graph is shown in Figure~\ref{fig:fig1}. The motivation for expressing the nodes in terms of a chain graph, is to ensure that each observation depends on any observation inside the same block, while it also depends on the neighbor blocks of $b_k$. In particular, if  $\bm N(b_k)$ is any subset of $\{ \bm N(b_1), \dots, \bm N(b_{k-1})\}$ then $\bm G^b$ is acyclic. 
Hence, it is assumed that each  $w (s_i )\in \bm w(b_k)$ depends on all $\bm w (s_{-i}) \in \bm w(b_k)$ and  $w (s_j) \in \bm w_{N(b_k)}$, where  $\bm w (s_{-i})$ represents the set  $\bm w(b_k)$ without  $w(s_i)$. Specifically, if $s_i \in b_k$, each $w (s_i )$ depends on $n_{bk} = n_k -1 + N_{b_k}$  neighbors, where $N_{b_k}$ is the number of locations in their neighbor blocks. This representation avoids loss of information at small scale while preserving information when the spatial process has a large range. In particular, Proposition~\ref{prop1} leads to the following corollary.  

\begin{corollary}
Let $\bm{w_S} = (w(s_1), \dots, w(s_n))'$ be a realization of a GP$(0, \bm{C(\theta_1)})$, with joint density  $p(w_S)$. If $\bm G$ is a chain graph and $\bm G^b$ is a DAG, then $\widetilde \pi(\bm {w_S})$ is a joint density.  
\end{corollary}
\noindent In particular, if $\bm{w_S}\sim N(0, \bm{C_S})$, the resulting approximated joint density  of $\bm{w_S}$ follows as
\begin{equation}
 \widetilde{\pi}(w_S) =  \prod_{k=1}^M f(w_{b_k}|B_{b_k} w_{N(b_k)}, F_{b_k}),
\label{eq:eq3}  
\end{equation}
where $f$ is the joint density of a normal distribution,
 $\bm{B}_{b_k} = \bm{C}_{b_k, \bm{N(b_k)}} \bm{C}^{-1}_{\bm{N(b_k)}}$ and $\bm{F}_{b_k} = \bm{C}_{b_k} - \bm{C}_{b_k, \bm{N(b_k)}} \bm{C}^{-1}_{b_k}\bm{C}_{ \bm{N(b_k)},b_k}$,  being $\bm{C}_{b_k}$, $\bm{C}^{-1}_{\bm{N(b_k)}}$ and $\bm{C}_{b_k, \bm{N(b_k)}}$ submatrices  of $\bm C_s$. 
This leads to the following  result concerning the  distribution associated to the joint density $\widetilde{\pi}(w_S)$.

\begin{proposition}
Let $\bm{w_S}$ be normally distributed random vector with zero mean,  covariance matrix $\bm{C_S}$ and joint density $\pi(\bm{w_S})$. Then  for a chain graph $\bm{G} = \{\bm{S}, \bm{\xi}\}$ and a directed acyclic subgraph $\bm{G^b}$, $\widetilde \pi(\bm{w_S})$ represents the joint density of a n-variate normal distribution with zero mean and positive definite precision matrix $\bm{\widetilde Q_S} = (\bm{B^T_S F_S^{-1} B_S)} $.
\label{prop2}
\end{proposition}
\noindent Some remarks concerning this result are as follows:
a) Since $\widetilde \pi(w_S)$ is a proper multivariate joint density, inference can be performed directly from a likelihood function not a composite or pseudo-likelihood.
b) The matrices $\bm B_S$ and $\bm F_S$  are easy to compute using parallel algorithms for block matrices. The elements of  $\bm B_S$ are non-zero only for neighbor blocks and each block itself. The elements of $\bm F_S$ are non-zero only for diagonal blocks. Explicit formulae are given in supplementary material A.
c) The covariance matrix $\bm{\widetilde{Q}_S^{-1}}$ is positive definite, but it is not guarantee that it is sparse.
d) The precision matrix $\bm{\widetilde{Q}_S}$ is sparse for $n_{bk} \ll n; \forall k=1,\dots,M$, hence we are able to perform related computations using sparse matrix algorithms. The nonzero pattern of $\bm{\widetilde{Q}_S}$ determines the chain graph $\bm G$. If we assume more blocks, the precision matrix will be more sparse. However, the sparsity pattern will depend on the range of the spatial process \citep{Lindgren:2011}. 
e) Doing the Cholesky decomposition, $\bm{\widetilde{Q}_S} = \bm{LL^T}$, the triangular matrix $\bm{L}$ will inherit the lower bandwidth of $\bm{\widetilde{Q}_S}$. 

Following the {idea of} the NNGP approach, we use $ \bm{\widetilde \pi(w_S)}$ to build a well-defined Gaussian process. We assume that $\bm S$ is a set of fixed locations and define $\bm U = \{u_1, \dots, u_l\}$ as any finite set of locations, such that $\bm S \cap \bm U = \emptyset$ and $\bm V = \bm S' \cup \bm U$ for $\bm S'\subset \bm S $.  
Since $\{\bm{w_U}, \bm{w_S}\}$ is a realization of the $GP(0, C(\theta_1))$, the conditional density of $\bm {w_U|w_S}$ is known and it can be approximated as follows
\begin{equation}
\widetilde{\pi}(\bm{w_U}|\bm{w_S}) = \prod_{i=1}^{l}\pi(w_{u_i}|\bm{w_{N(u_i)}})=
 \prod_{i=1}^l f(w_{u_i}|B_{u_i} \bm{w_{N(u_i)}}, F_{u_i}),
\label{eq:eq31}  
\end{equation}
where  $B_{u_i} = C_{u_i, N(u_i)} C^{-1}_{N(u_i)} $ and  $F_{u_i} = C_{u_i} - \bm{C_{u_i, N(u_i)}} C^{-1}_{u_i}\bm{C_{ N(u_i),u_i}}$, $C_{i,j}$ and $C_i$ are elements of $\bm{C_s}$. 
Specifically, to ensure that these conditionals are consistent so that this approximation is a joint density, we assume that the set of neighbors of $u_i$ is defined by $\bm{N(u_i)}= \{s_i; s_i\in \bm{S'}, s_i \subset b_k\},$  $\forall u_i\subset b_k$, such that $w_{u_i} \perp w_{u_j}| \bm{w_{N(u_i)}, w_{N(u_j)}}$.  This is the simplest (and cheapest) neighbor set using the structure of the blocks, where any set of neighbors of $u_i$ depends on the same fixed observations in $\bm{S'}$ for any reordering of $\bm{U}$. 

\begin{lemma} 
Let $\bm{S}$ be a fixed set and define the labelled graph $\mathcal{G}=\{\bm V,  \bm\xi'\}$, where $\bm V = \bm S' \cup \bm U$ for $\bm S'\subset \bm S$ and $\bm U \subset D$, $\bm\xi'= \bm\xi_{S'} \cup \bm\xi_U$, $\bm\xi_{S'}\subset \bm\xi$ and $\bm\xi_U$ is the set of undirected edges between $u_i$ and $\bm{N(u_i)}= \{s_i; s_i\in \bm{S}, s_i \subset b_k \}$, $\forall u_i\subset b_k$. Given the densities $\widetilde{\pi}(\bm{w_S})$
 and $\widetilde{\pi}(\bm{w_U}|\bm{w_S})$ defined as in  (\ref{eq:eq3}) and (\ref{eq:eq31}), then the finite-dimensional distributions with joint density 
\begin{equation}
 \widetilde{\pi}(\bm{w_V}) = \int \widetilde{\pi}(\bm{w_U}|\bm{w_S})\widetilde{\pi}(\bm{w_S}) \prod_{s_i \in (\bm{S'})^c} d(w(s_i)), 
\label{eq:eq4}  
\end{equation}
support a valid random process $\bm{w_V}$.
\end{lemma}
Here, $(\bm{S'})^c$ is the complement of $\bm{S'}$. If $\bm{S} = \bm{S'}$ the last term related  to $\bm{S'}$ in (\ref{eq:eq4}) is not necessary. 
This lemma states that the density $\widetilde{\pi}(\bm{w_V})$  is consistent with some well-defined spatial process, in the sense that the Kolmogorov's consistency conditions are verified, that is, the symmetry and compatibility conditions hold for the process defined through the finite-dimensional distributions with joint density of  (\ref{eq:eq4}). In summary, $\widetilde{\pi}(\bm{w_V})$ will be the same under reordering of the sites or for any new site $v_0\notin \bm V$ in $\bm D$. Hence, Lemma 1 defines a new valid spatial process and the next theorem proves that such spatial process derived from a GP is a Gaussian Markov random process. 
\begin{theorem}
 For any finite set $\bm{V} \subset \bm{D}$, $ \widetilde{\pi}(\bm{w_V})$  is the finite dimensional joint density of a Gaussian  process with respect to  $\mathcal{G}=\{\bm V,  \bm\xi'\}$,  with cross covariance funcion 
\[ \widetilde C _{v_i,v_j} =
  \begin{cases}
    \widetilde C _{s_i,s_j}      & \quad \text{if } (v_1= s_i , v_2 = s_j) \in \bm{S}  \\
    \bm{B_{v_1} \widetilde C_{N(v_1),s_j}}  & \quad \text{if }  v_1 \notin \bm{S} \mbox{ and } v_2 = s_j \in \bm{S}\\
  \delta_{(v_1=v_2)}F_{v_1} + \bm{B_{v_1} \widetilde C_{N(v_1),N(v_2)}B^T_{v_2}}  & \quad \text{if } (v_1, v_2) \notin \bm{S}, 
  \end{cases}
\] 
where $\bm{\widetilde C_{m,n}}$ is a covariance matrix of $\bm{\widetilde C_S}$.
 \end{theorem}
The covariance matrix can be written in terms of blocks as follows,
\[
 \bm{\widetilde C} =   \left[
    \begin{array}{ccccc}
     \bm{\widetilde C_{s,s}}   &  \bm{\widetilde C_{s,v}}  \\ 
     \bm{(\widetilde C_{s,v})}^T   &  \bm{\widetilde C_{v,v}}  \\ 
\end{array}\right]
,\]
where $\bm{\widetilde C_{s,s}}$ represents the covariance matrix for all $s  \in \bm S$,  $\bm{\widetilde C_{v,v}}$ represents the covariance matrix for all $v\notin \bm S$, and $\bm{\widetilde C_{s,v}}$ represents the cross-covariance matrix for all $s \in \bm S$ and $v\notin \bm S$. Then the precision matrix is found as
\begin{equation}
 \bm{\widetilde Q} =   \left[
    \begin{array}{ccccc}
     \bm{\widetilde Q_{s,s}}   &  \bm{\widetilde Q_{s,v}}  \\ 
     \bm{(\widetilde Q_{s,v})}^T   &  \bm{\widetilde Q_{v,v}}  \\ 
\end{array}\right],
\label{eq:eqQ}
\end{equation}
where  $ \bm {\widetilde Q_{s,s}} =   (I - \bm{\widetilde Q_{s,v} \widetilde C_{v,s}})\bm{\widetilde Q_S} $,  
$\bm{\widetilde Q_{s,v}} = -\bm{\widetilde Q_{S} \widetilde C_{s,v} \widetilde Q_{v,v}}$, and 
$\bm{\widetilde Q_{v,v}} = (\bm{\widetilde C_v} - \bm{\widetilde C_{v,s}}\widetilde Q_S \bm{\widetilde C_{s,v}})^{-1}$. 
The benefit of this reordering is that $\bm { \widetilde Q_{s,s}}$  will inherit the sparse property of  $\bm {\widetilde Q_S}$, then the corresponding precision matrix $\bm {\widetilde Q}$ is an sparse block matrix, and hence we obtain GMRF models which are called block-NNGP.

The block-NNGP also contains existing processes as special cases, giving rise to the next two corollaries.

\begin{corollary}
The NNGP  is a special case of the block-NNGP when the number of blocks is the same as the number of observations in $\bf{S}$, that is, $M=n$.
\end{corollary}

\begin{corollary}
The spatial independent blocking, where each block $w_{N(b_k)}$ is independent of any other block, is a special case of the block-NNGP when the number of neighbor blocks for every block is zero, that is, $nb=0$, or equivalently ${N(b_k)}=\emptyset$. 
\end{corollary}

\begin{figure}
 \hspace{1.5cm} \textbf{$\phi = 12$} \hspace{2.8cm} \textbf{$\phi = 6$}\hspace{2.8cm} \textbf{{$\phi = 3$}}\hspace{2.8cm} \textbf{Q}\\
\begin{center}
\includegraphics[scale=0.22]{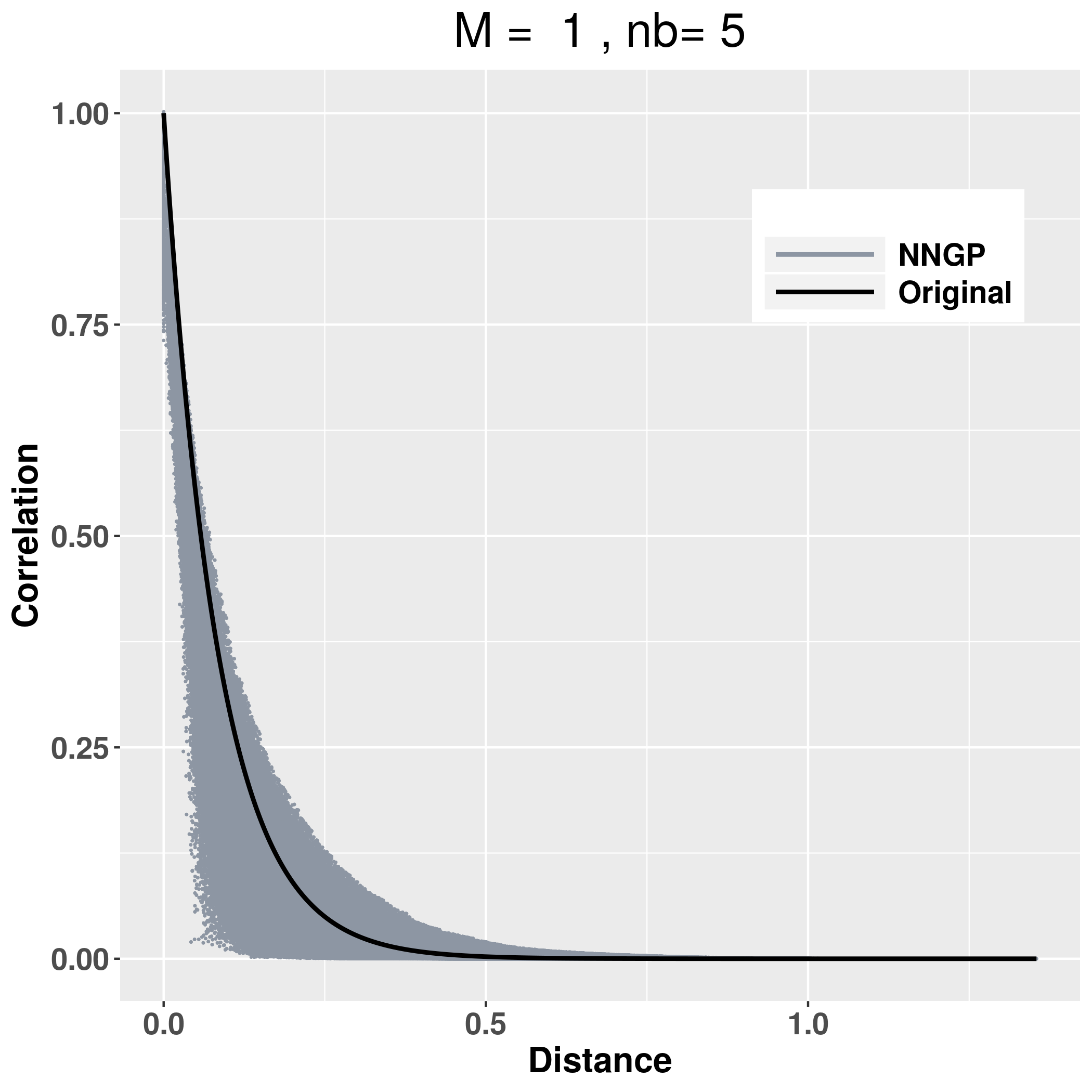} 
\includegraphics[scale=0.22]{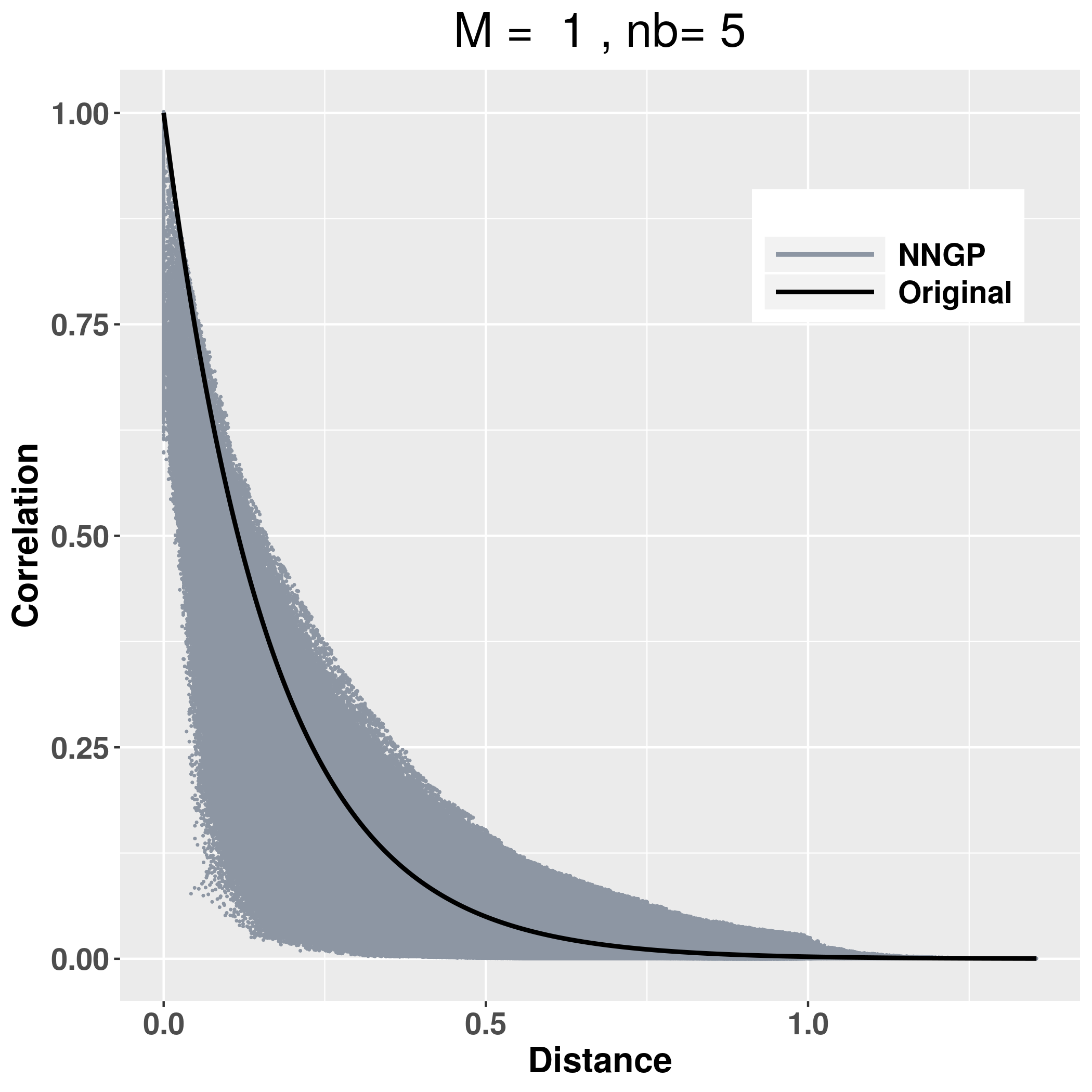} 
\includegraphics[scale=0.22]{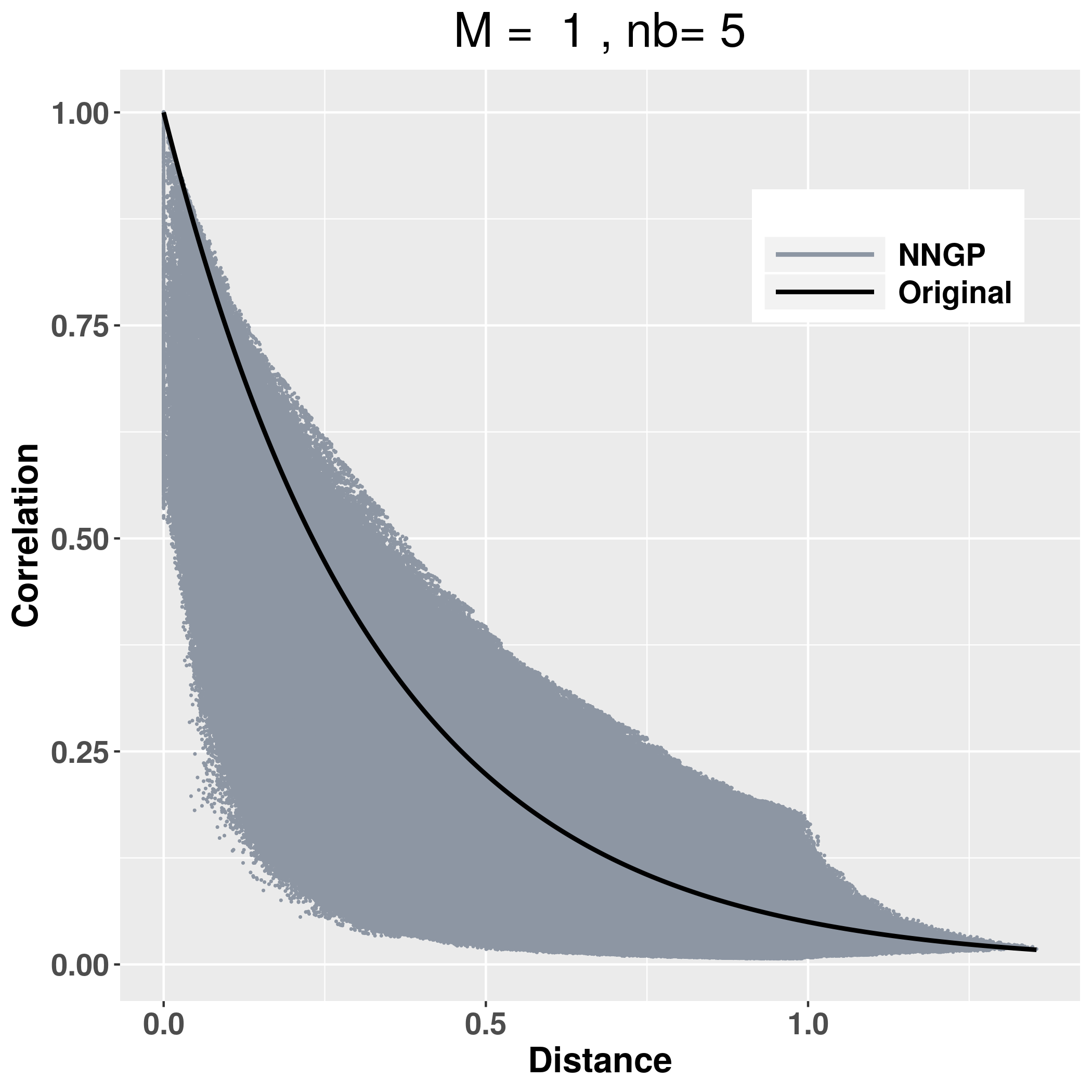} 
\includegraphics[scale=0.22]{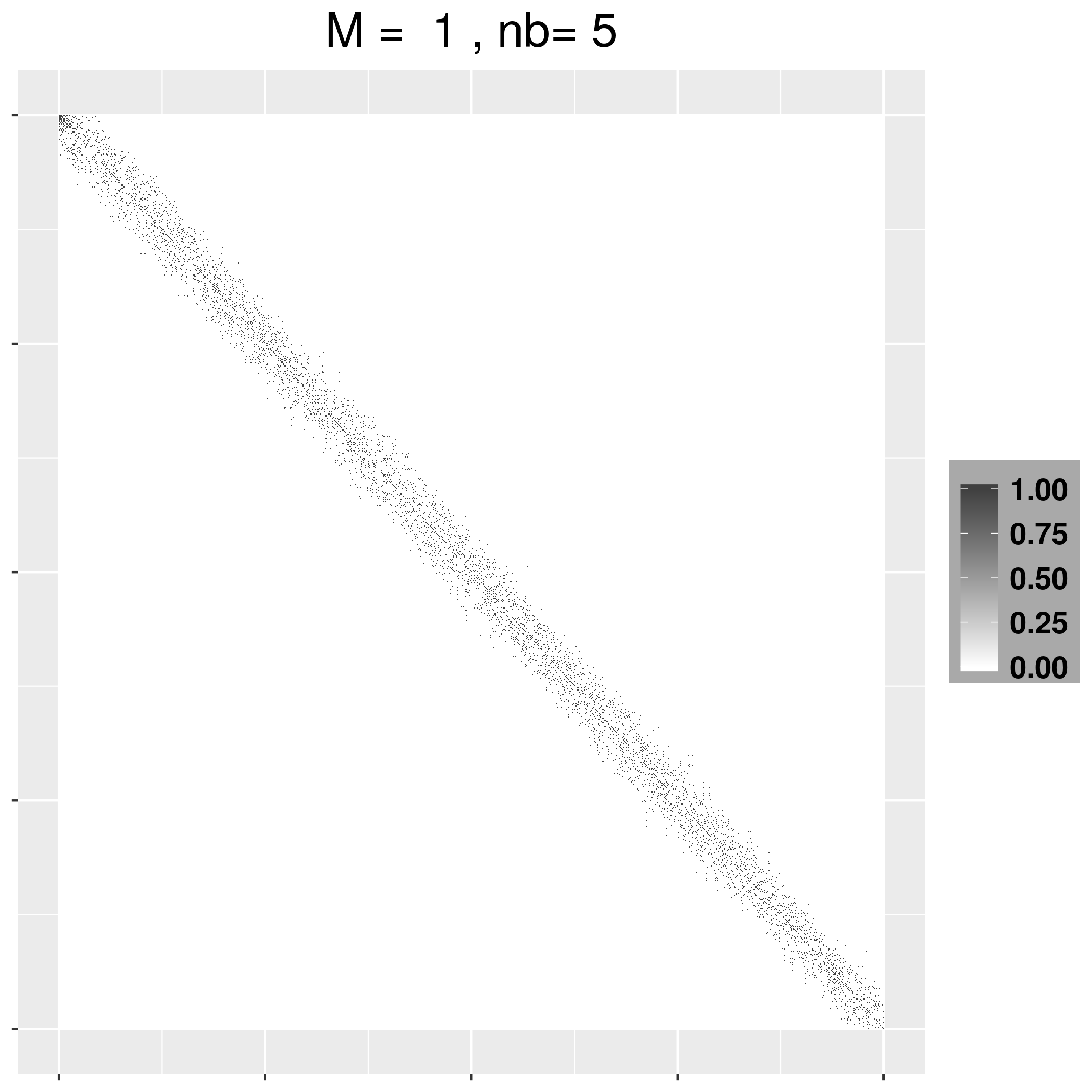} \\
\includegraphics[scale=0.22]{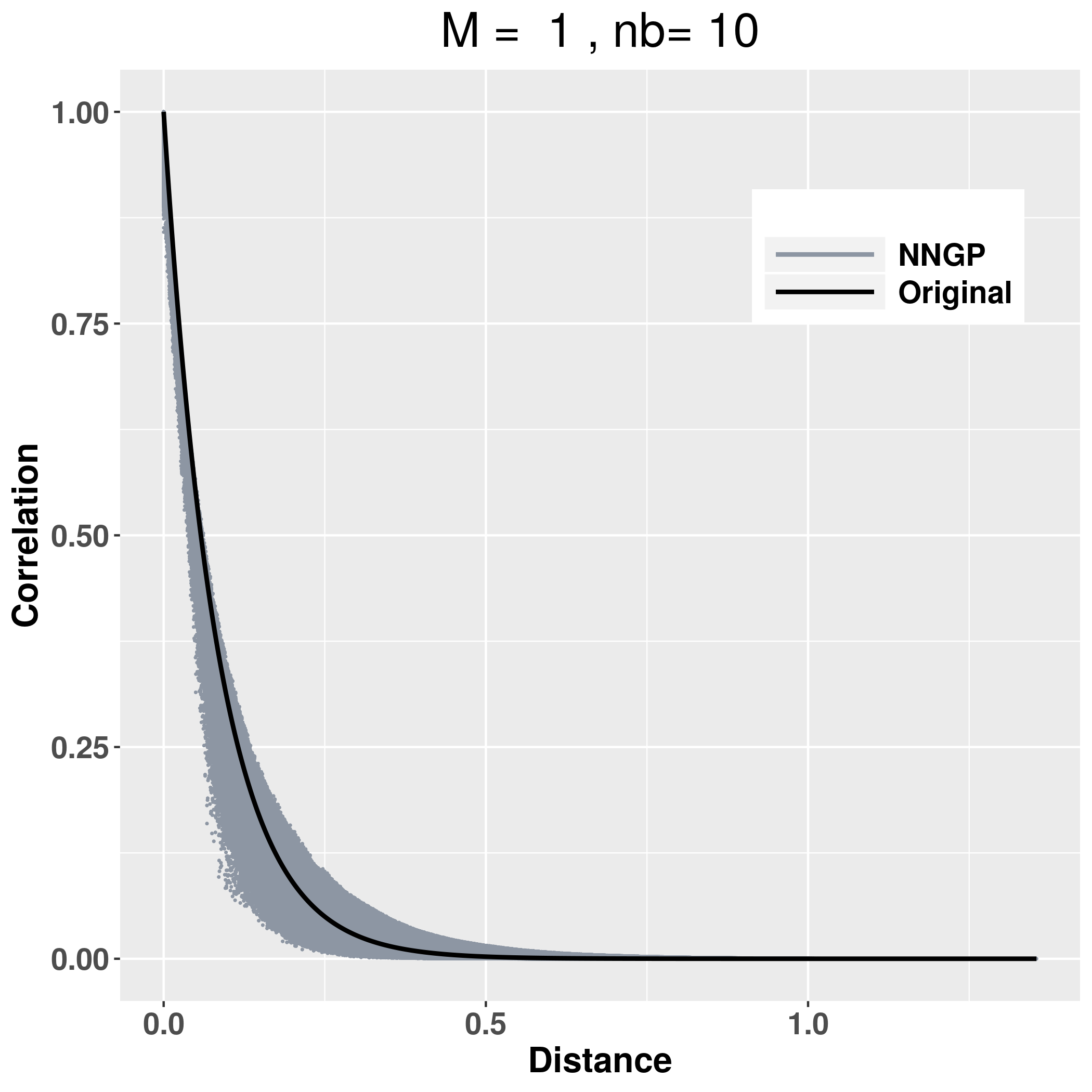} 
\includegraphics[scale=0.22]{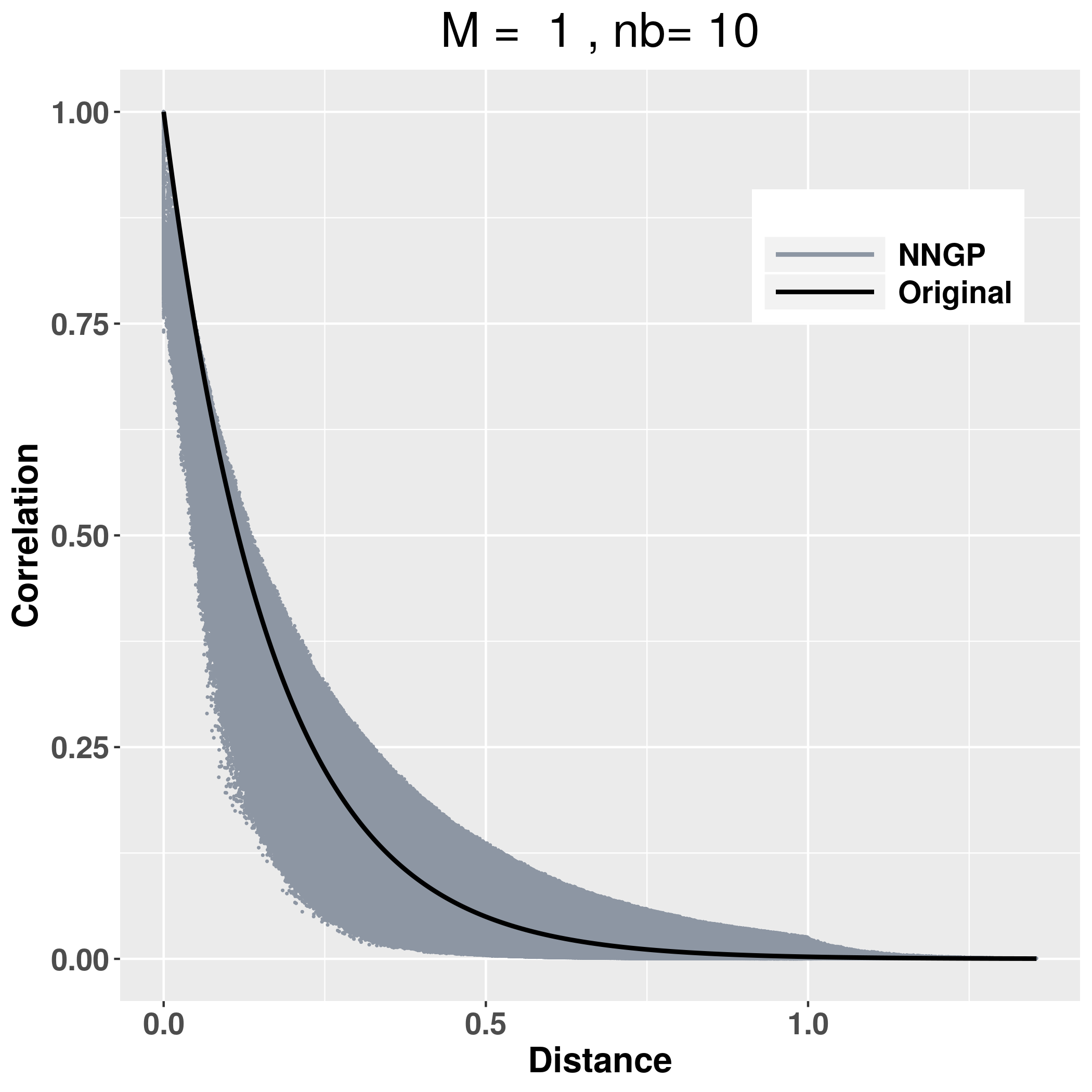} 
\includegraphics[scale=0.22]{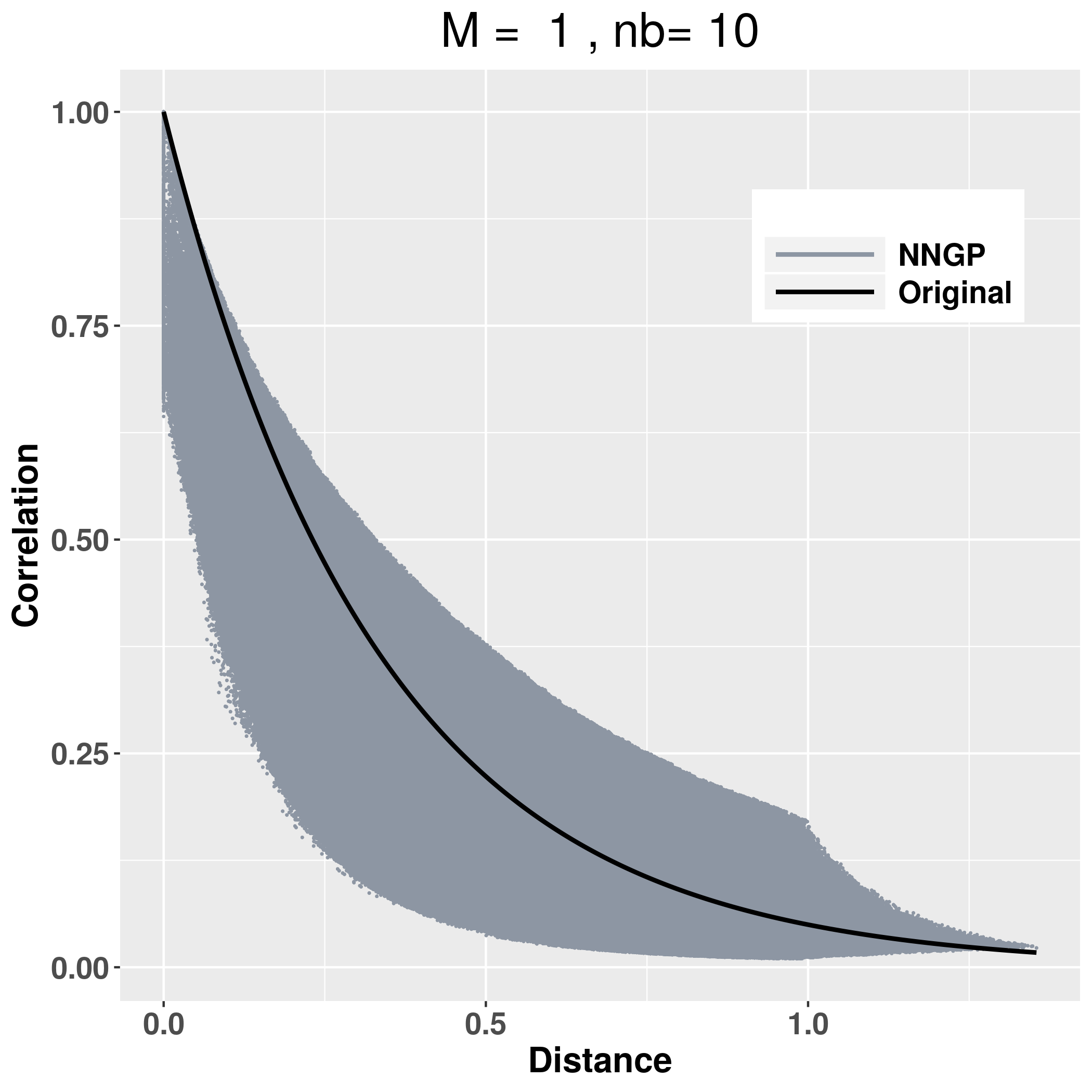} 
\includegraphics[scale=0.22]{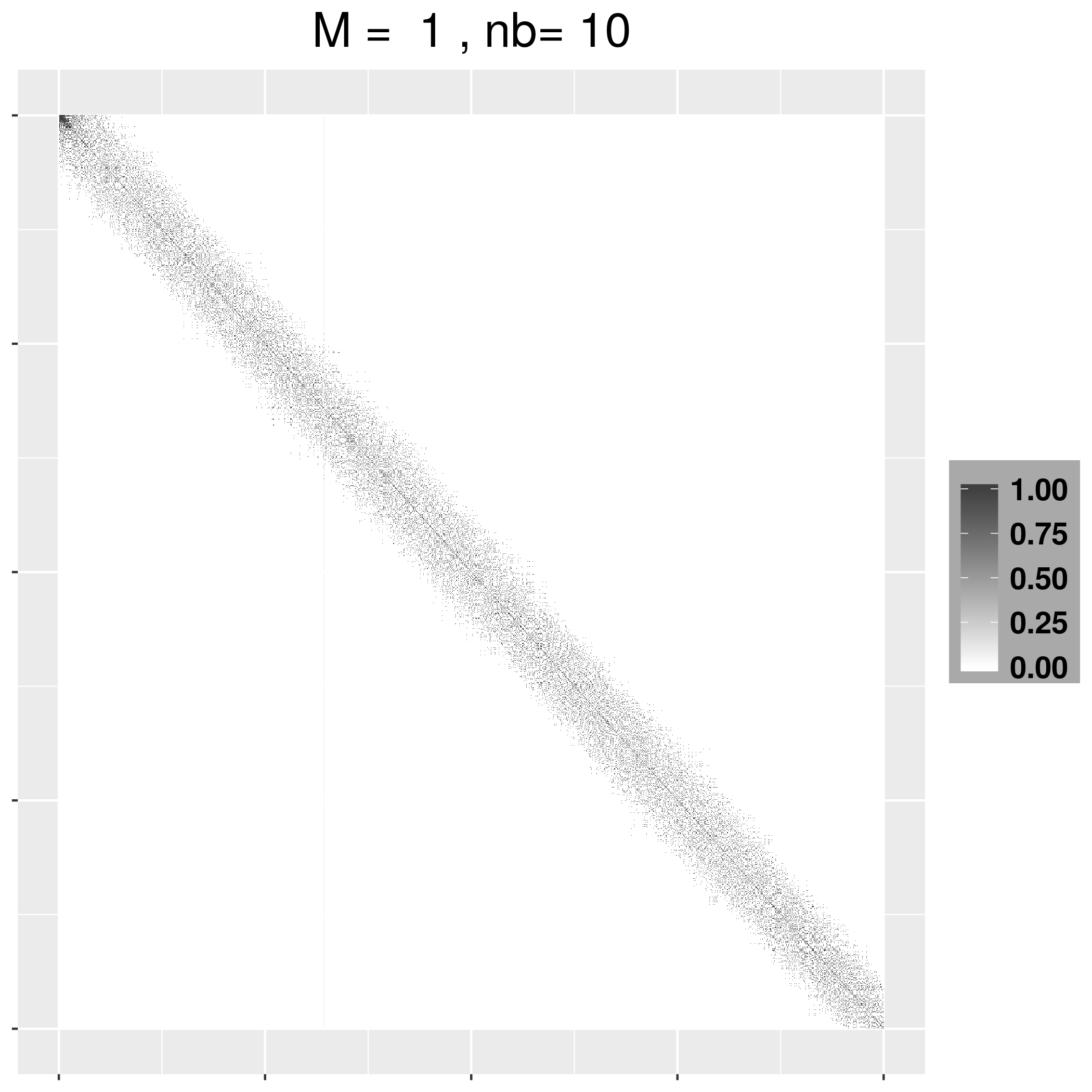} \\
\includegraphics[scale=0.22]{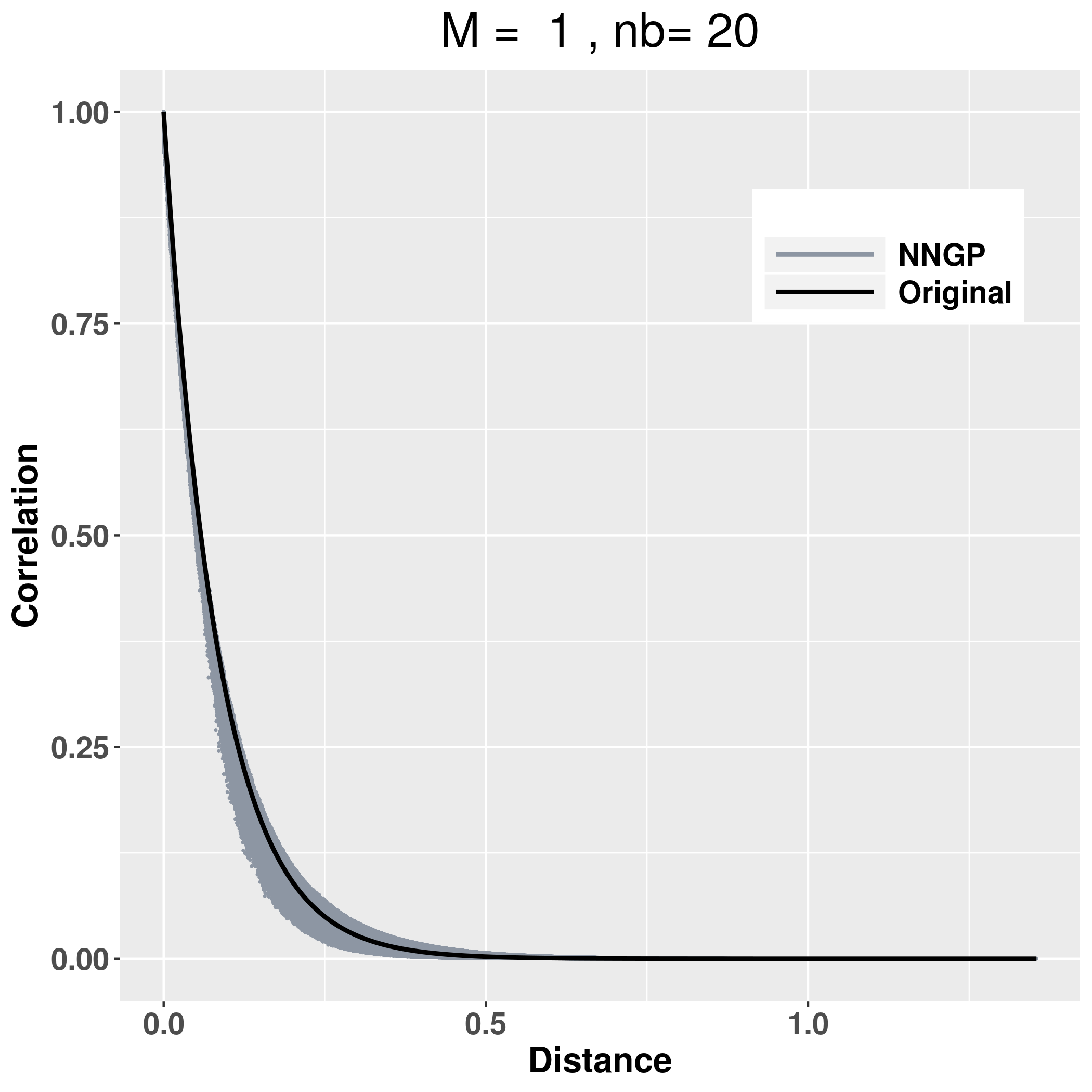} 
\includegraphics[scale=0.22]{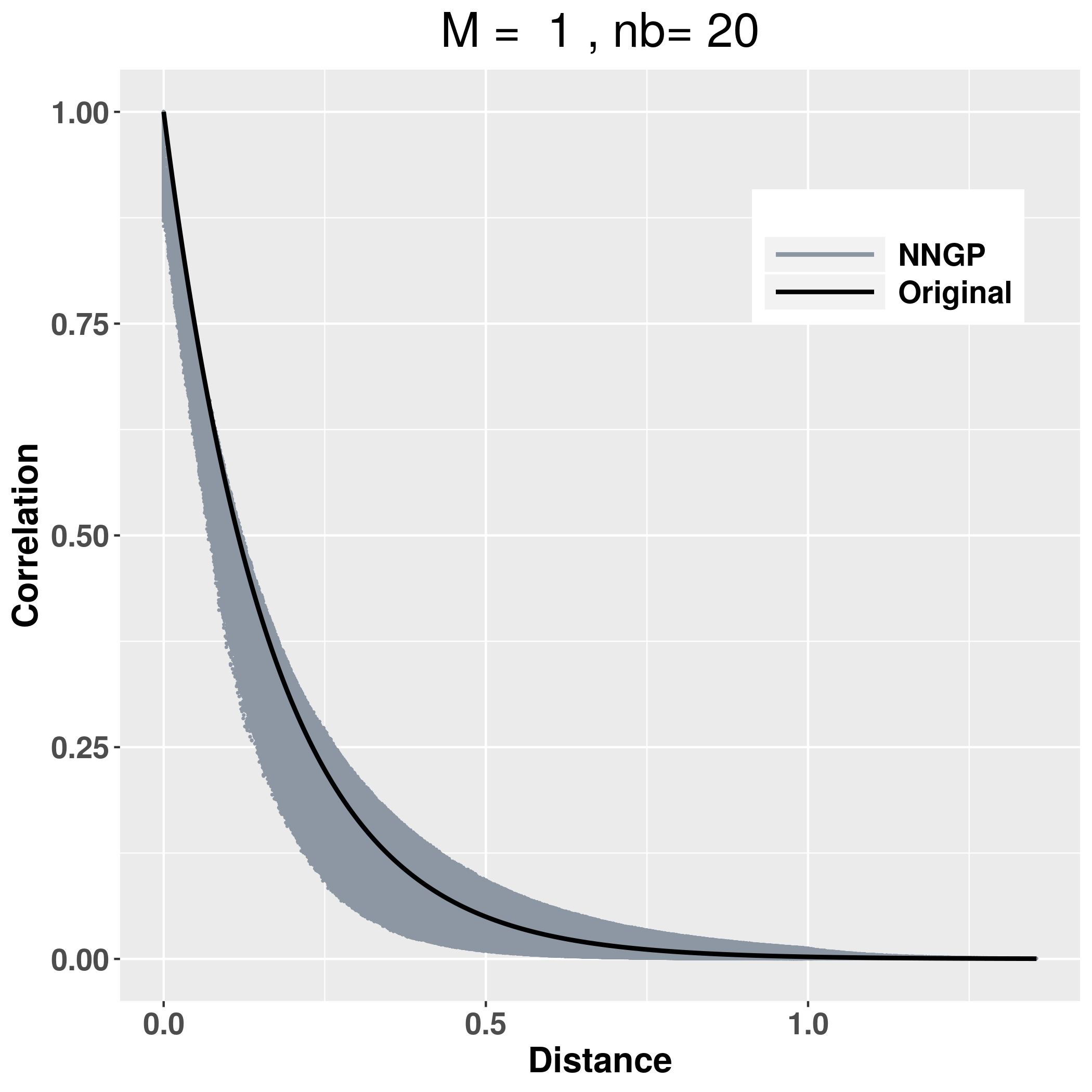} 
\includegraphics[scale=0.22]{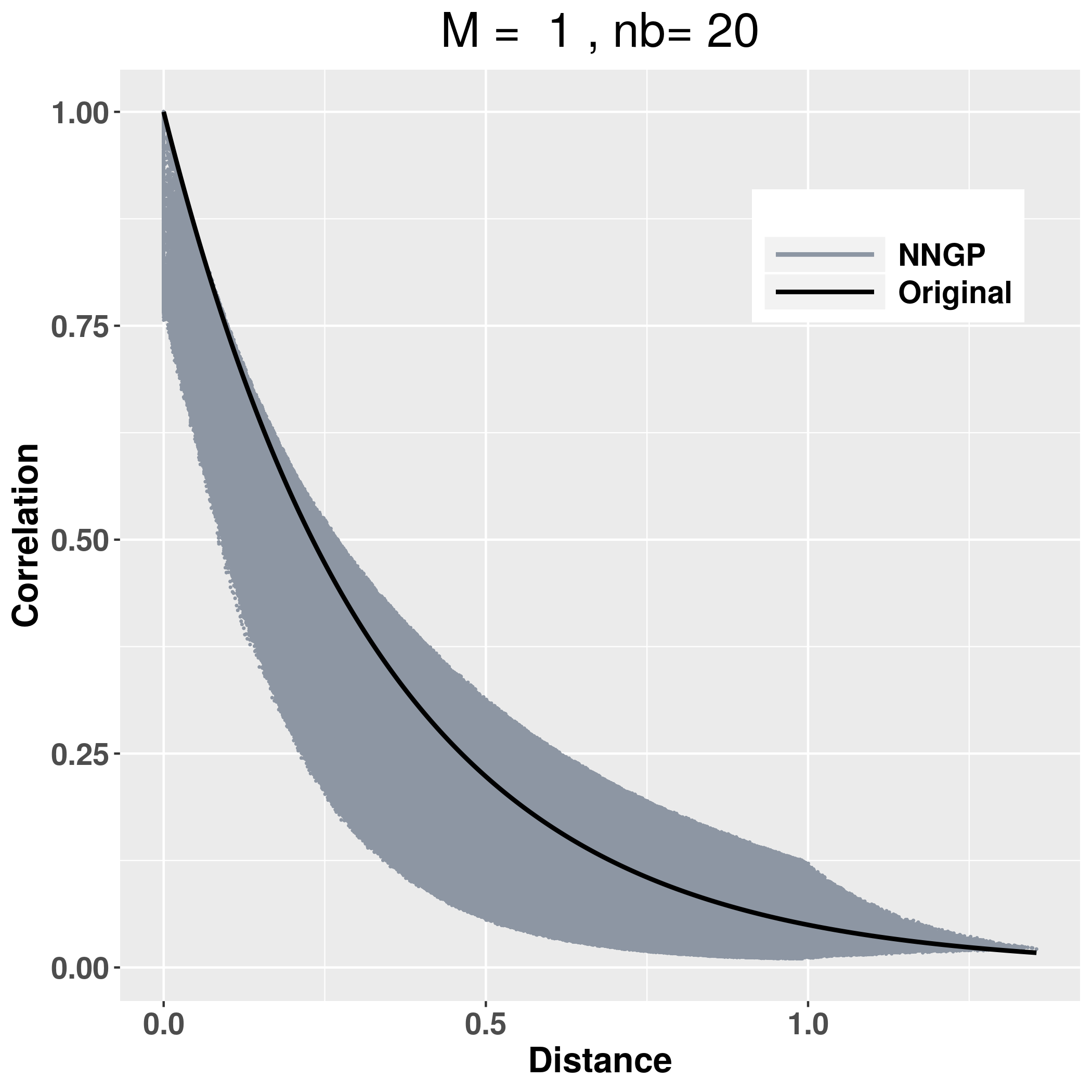} 
\includegraphics[scale=0.22]{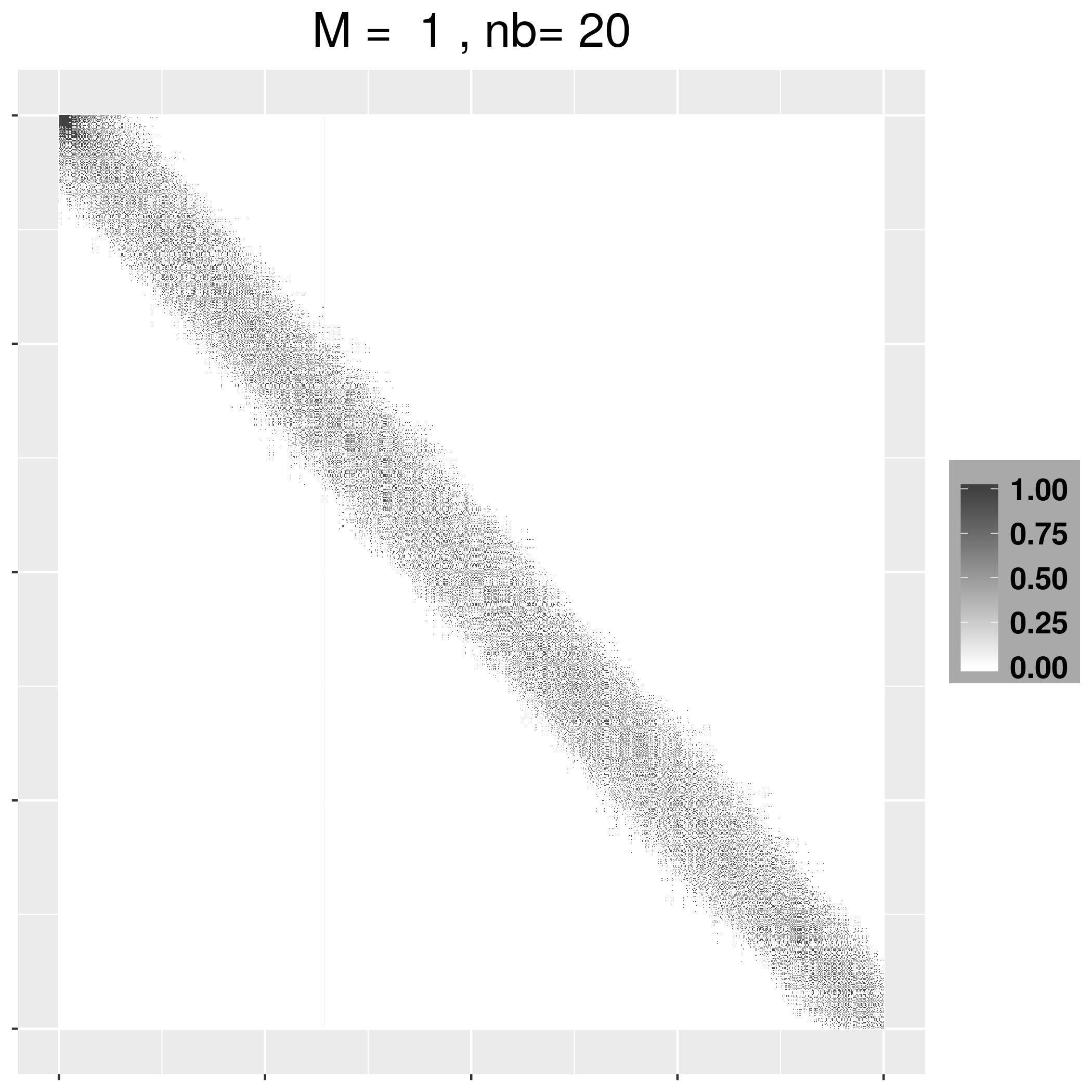} 
\end{center}
\caption{\small  True exponential correlation of GP  against distance (black lines) and empirical approximated correlations 
of NNGP aginst distance (blue dots) for:
short range($\phi$ = 12, first column), medium range ($\phi$ = 6, second column) and large range ($\phi$ = 3, third column), and $nb = 5$ neighbors (first row), $nb = 10$ neighbors (second row) and $nb = 20$ neighbors (third row). The fourth column presents Precision matrices for $nb = 5$ neighbors (first row), $nb = 10$ neighbors (second row) and $nb = 20$ neighbors (third row), gray squares represent non-zero values. 
}
  \label{fig:figk3}
\end{figure}

To see the capacity of the NNGP and block-NNGP to approximate the original spatial processes, we simulate spatial zero mean Gaussian processes with an exponential covariance function with elements  $C(s_i, s_j) = \sigma^2 \exp(-\phi \|s_i-s_j\|)$, where  $ \|s_i-s_j\|$ is the Euclidean distance between two locations; $\sigma^2$ is the marginal variance, and the spatial decay $\phi$ is related to the so-called effective range ($r$), the distance at which the correlation decays to 0.1, that is, $\phi = 2/r$. We set  $\sigma^2=1$ and approximate effective ranges 0.17 ($\phi=12$), 0.33 ($\phi=6$) and 0.67 ($\phi=3$).
Figure~\ref{fig:figk3} displays the theoretical exponential correlation function (black line), the empirical correlation function  for the NNGP (blue dots) and precision matrices for different number of neighbors. We note that when the range increases the match is far away from the true correlation function. As previous studies state, it is not surprising that when the true range parameter is small relative to the sampling domain, it can be well estimated from data \citep{KAUFMAN:2013}. The nonzero structure of the precision matrices for each scenario implies that the bigger the number of neighbors, less sparse the precision matrix. 
This implies that the number of neighbors should be as small as possible but still large enough to give a reasonably accurate approximation of the true spatial process. However, as we can see choosing an adequate number of neighbors is not trivial.

\begin{figure}
 \hspace{1.5cm} \textbf{$\phi = 12$} \hspace{2.8cm} \textbf{$\phi = 6$}\hspace{2.8cm} \textbf{{$\phi = 3$}}\hspace{2.8cm} \textbf{Q}\\
\begin{center}
\includegraphics[scale=0.22]{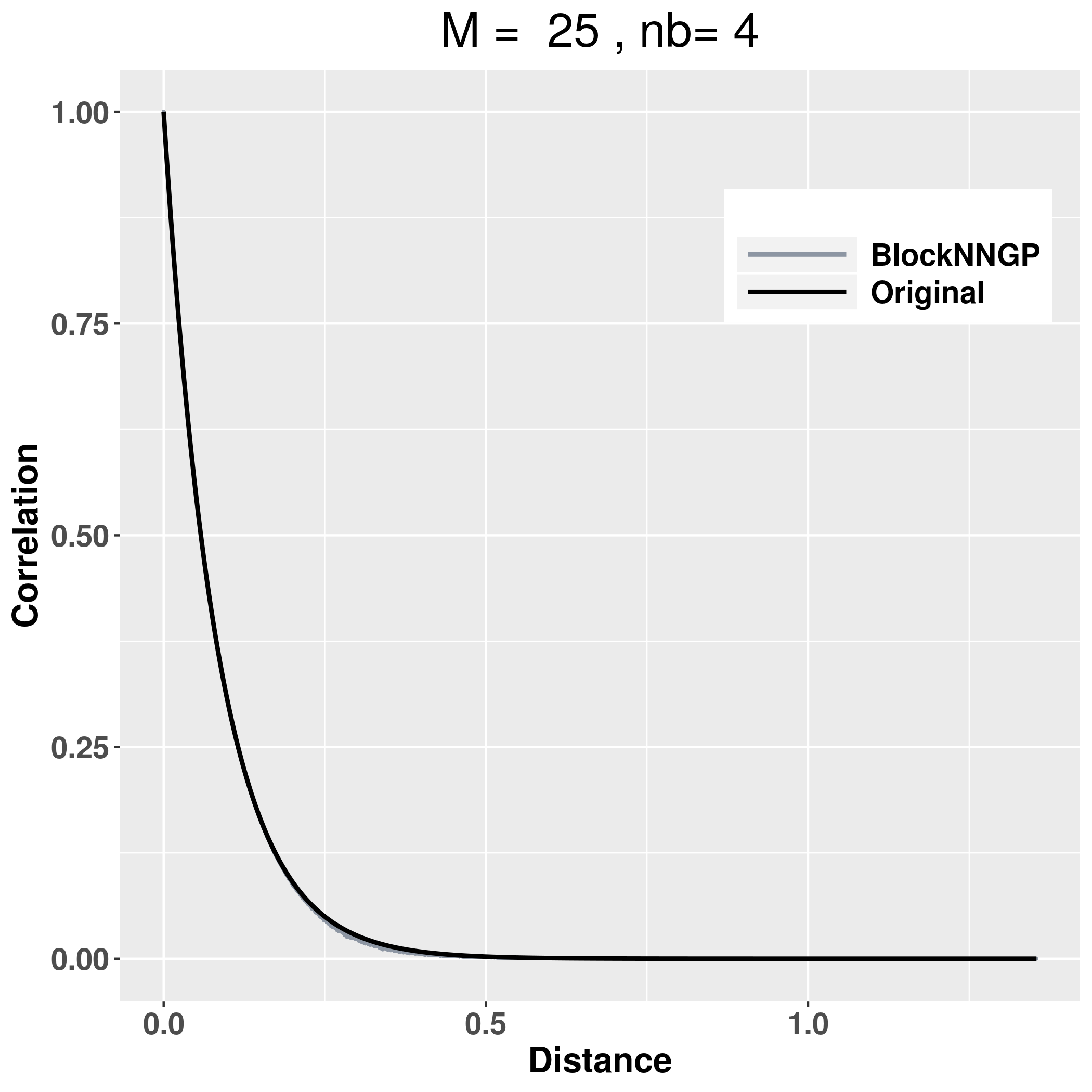} 
\includegraphics[scale=0.22]{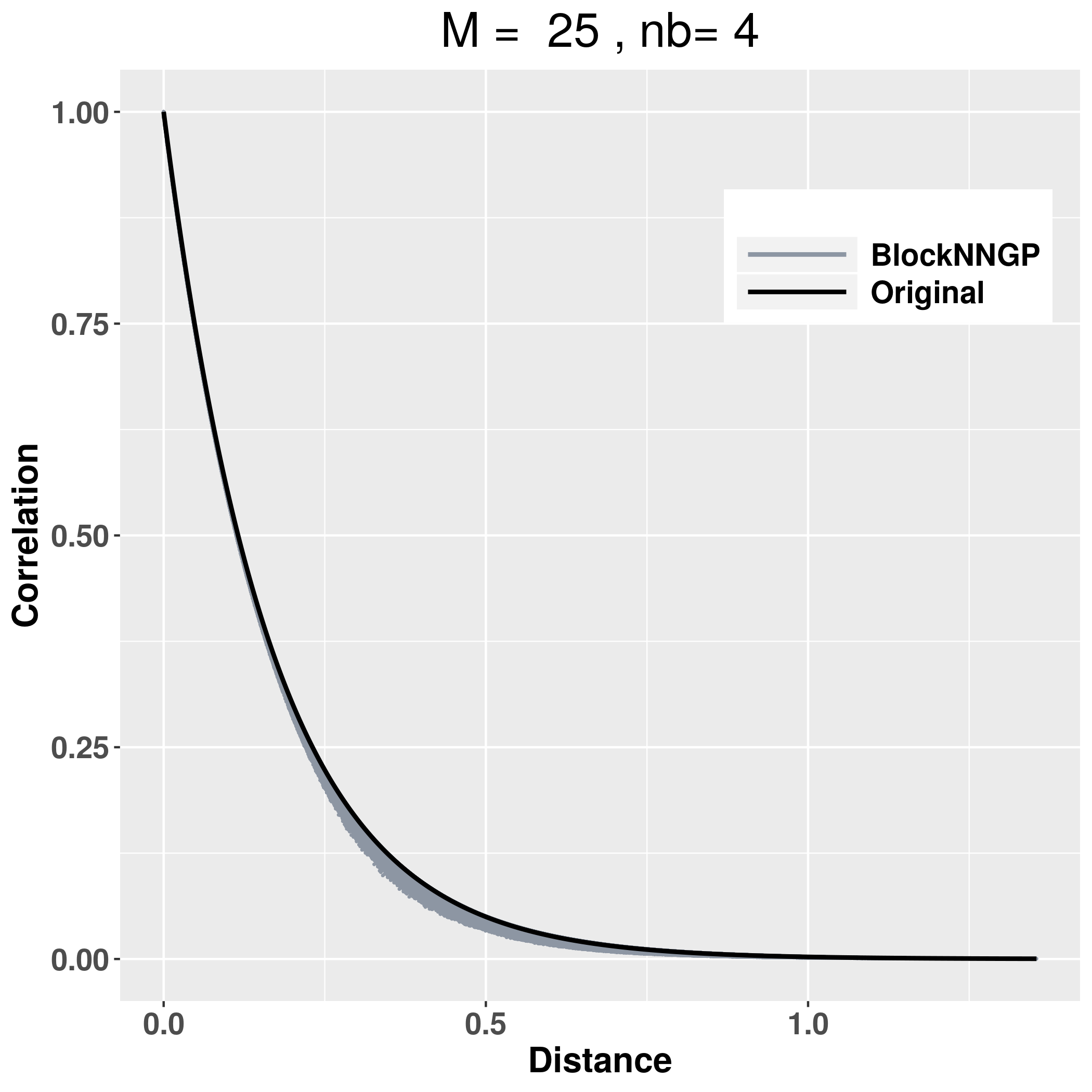} 
\includegraphics[scale=0.22]{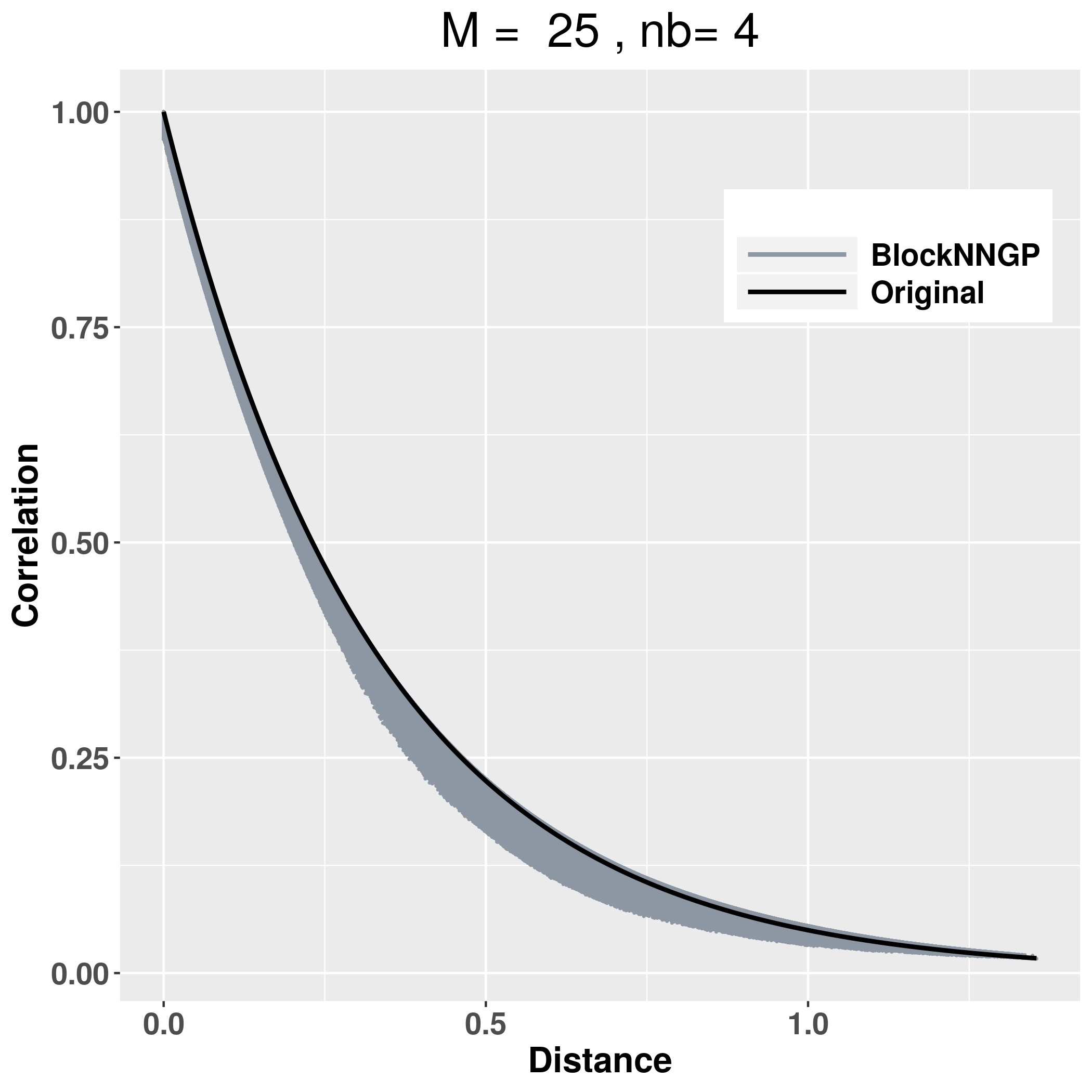} 
\includegraphics[scale=0.22]{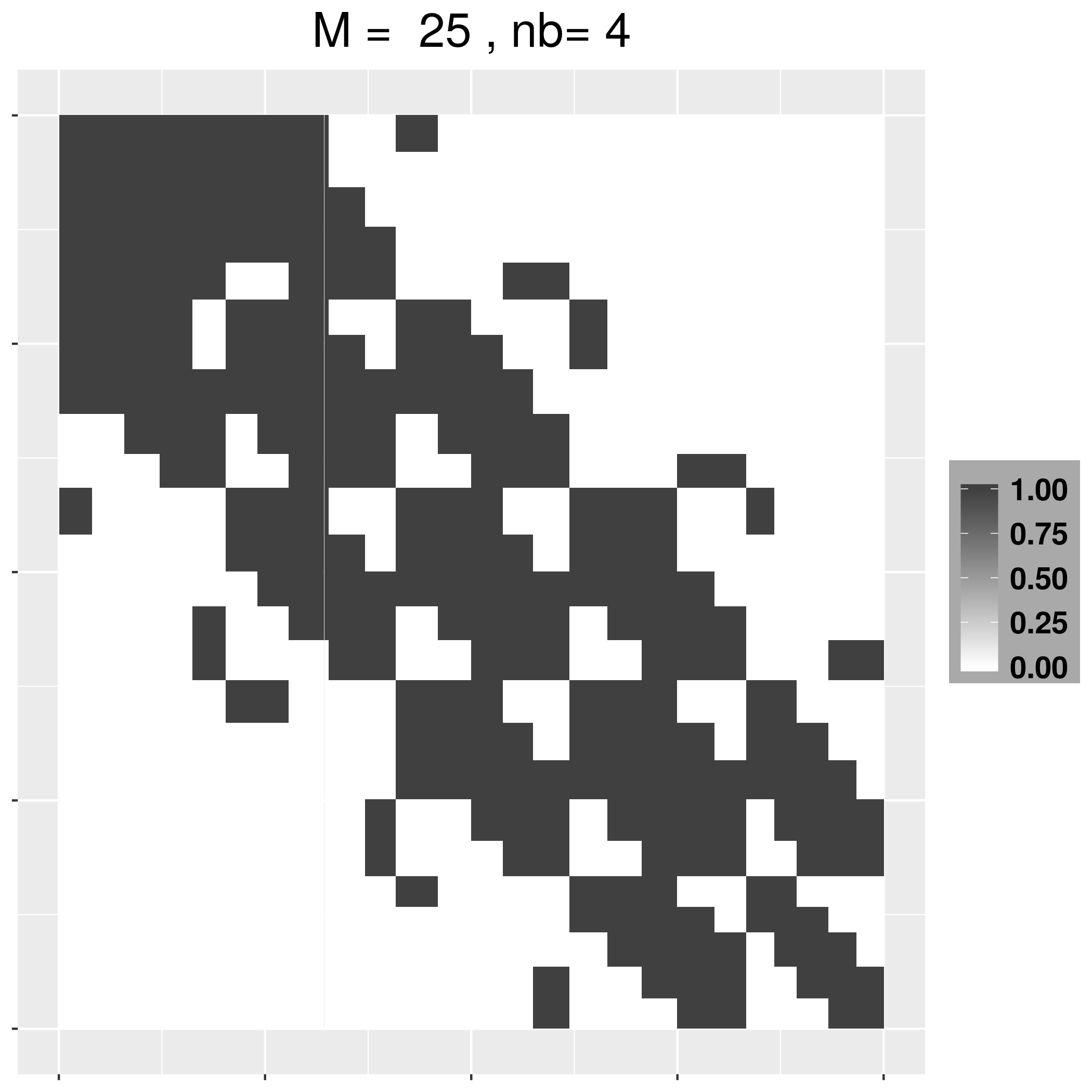} \\
\includegraphics[scale=0.22]{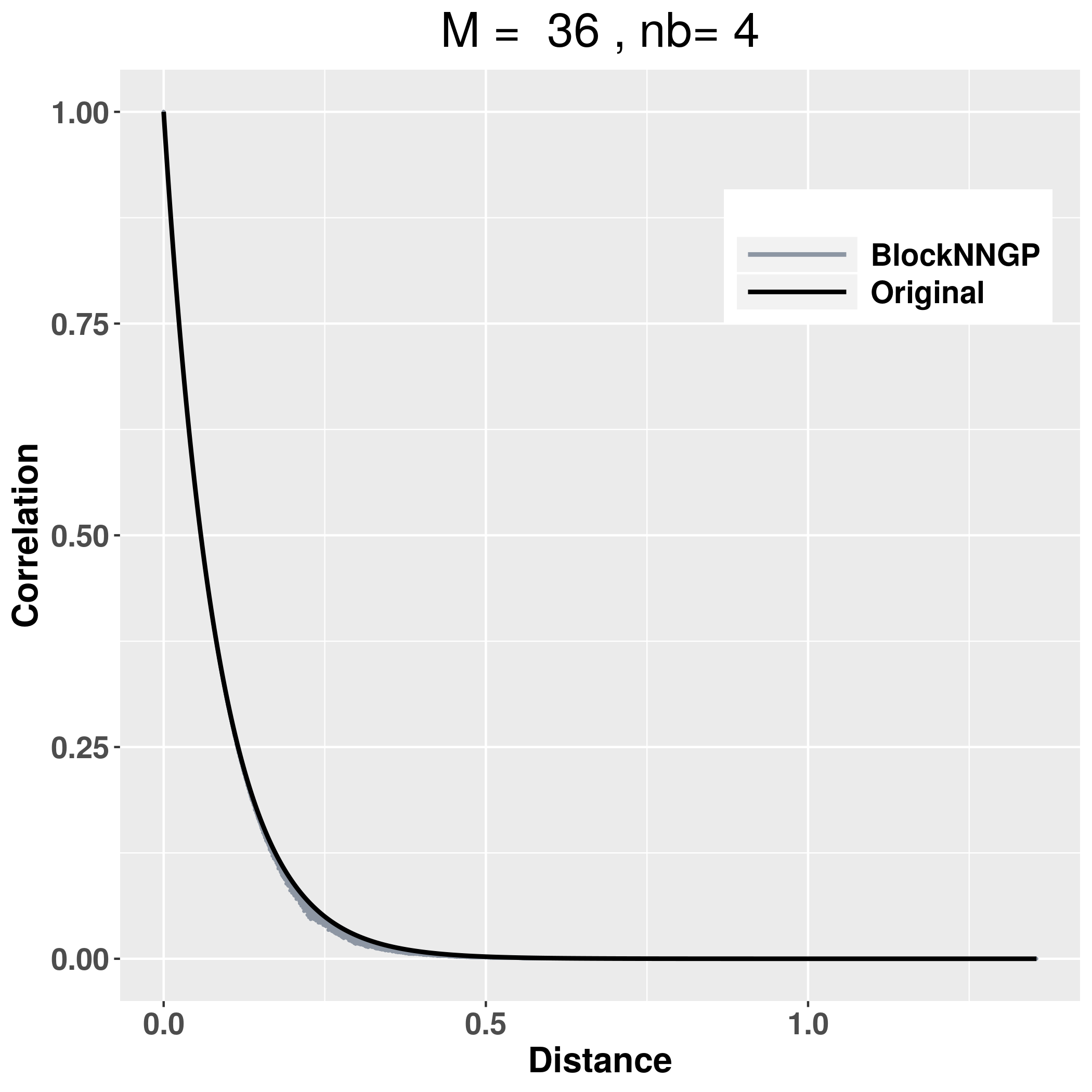} 
\includegraphics[scale=0.22]{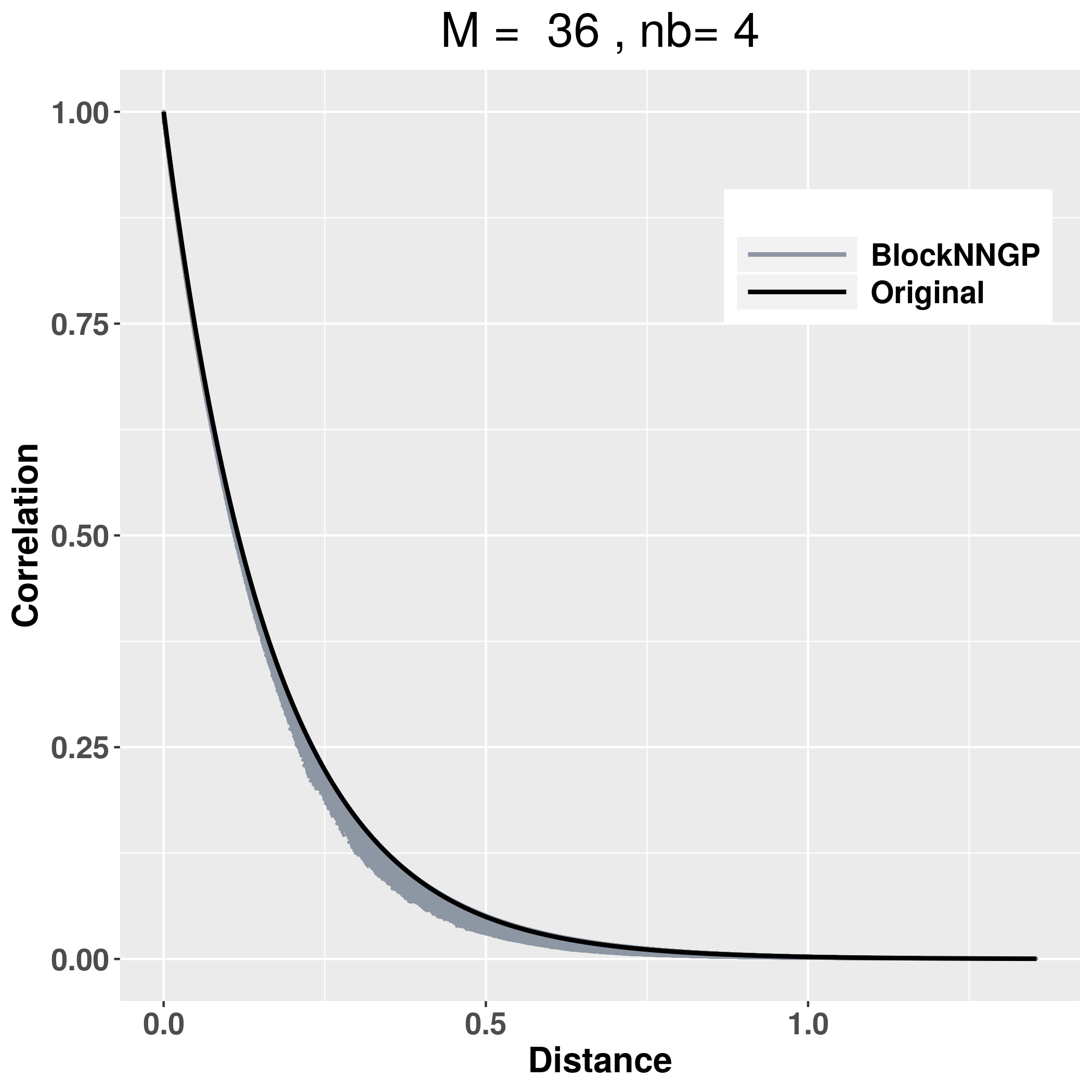} 
\includegraphics[scale=0.22]{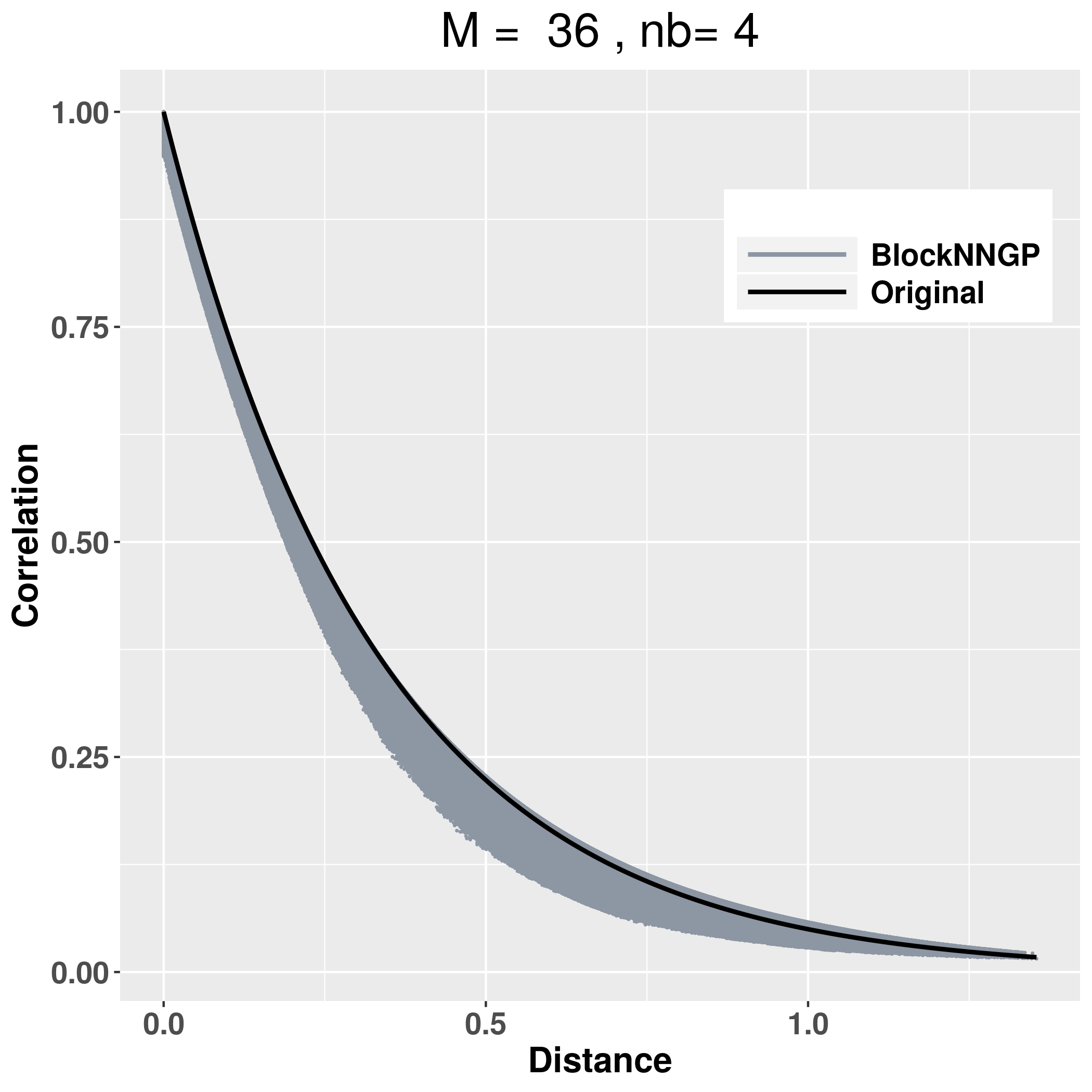} 
\includegraphics[scale=0.22]{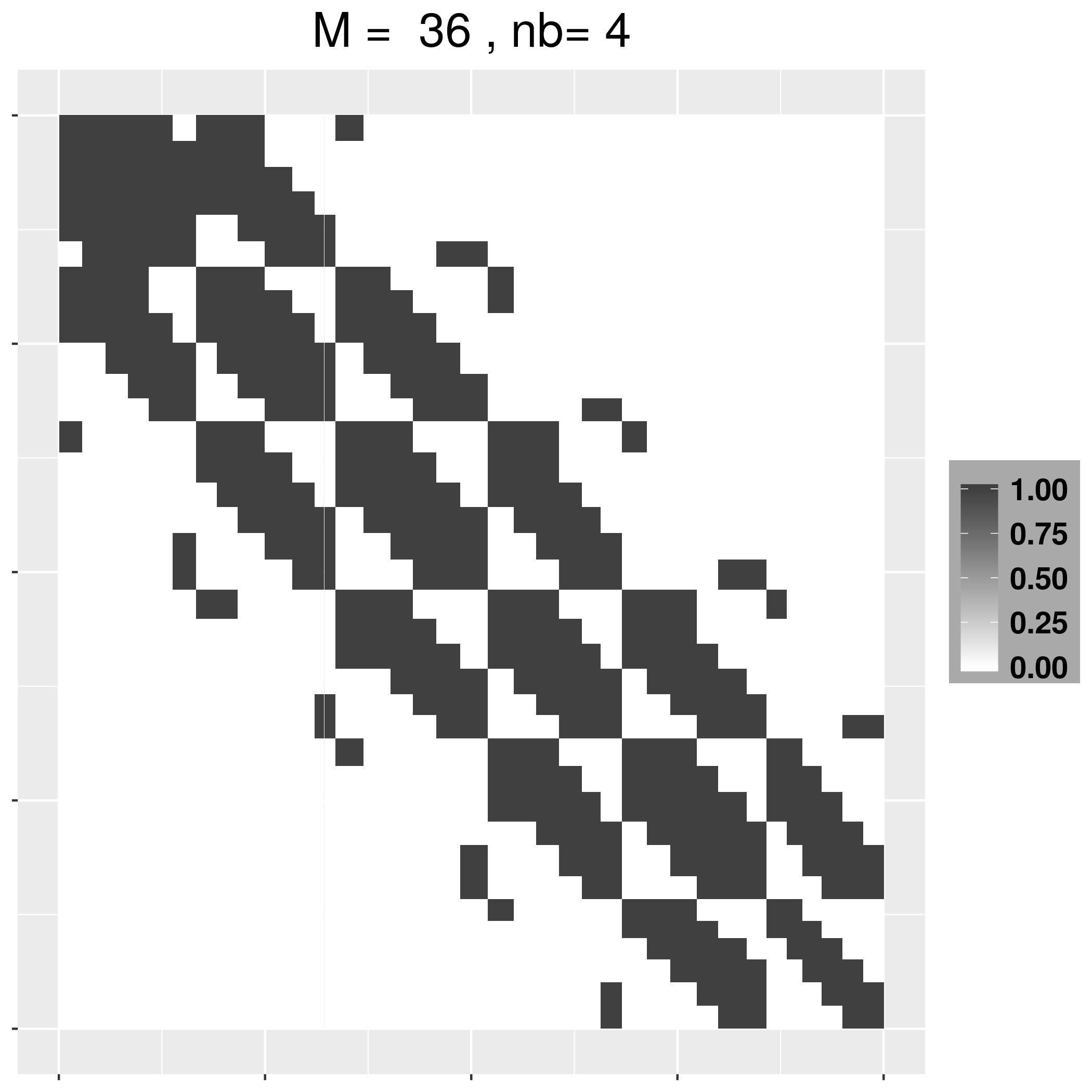} \\
\includegraphics[scale=0.22]{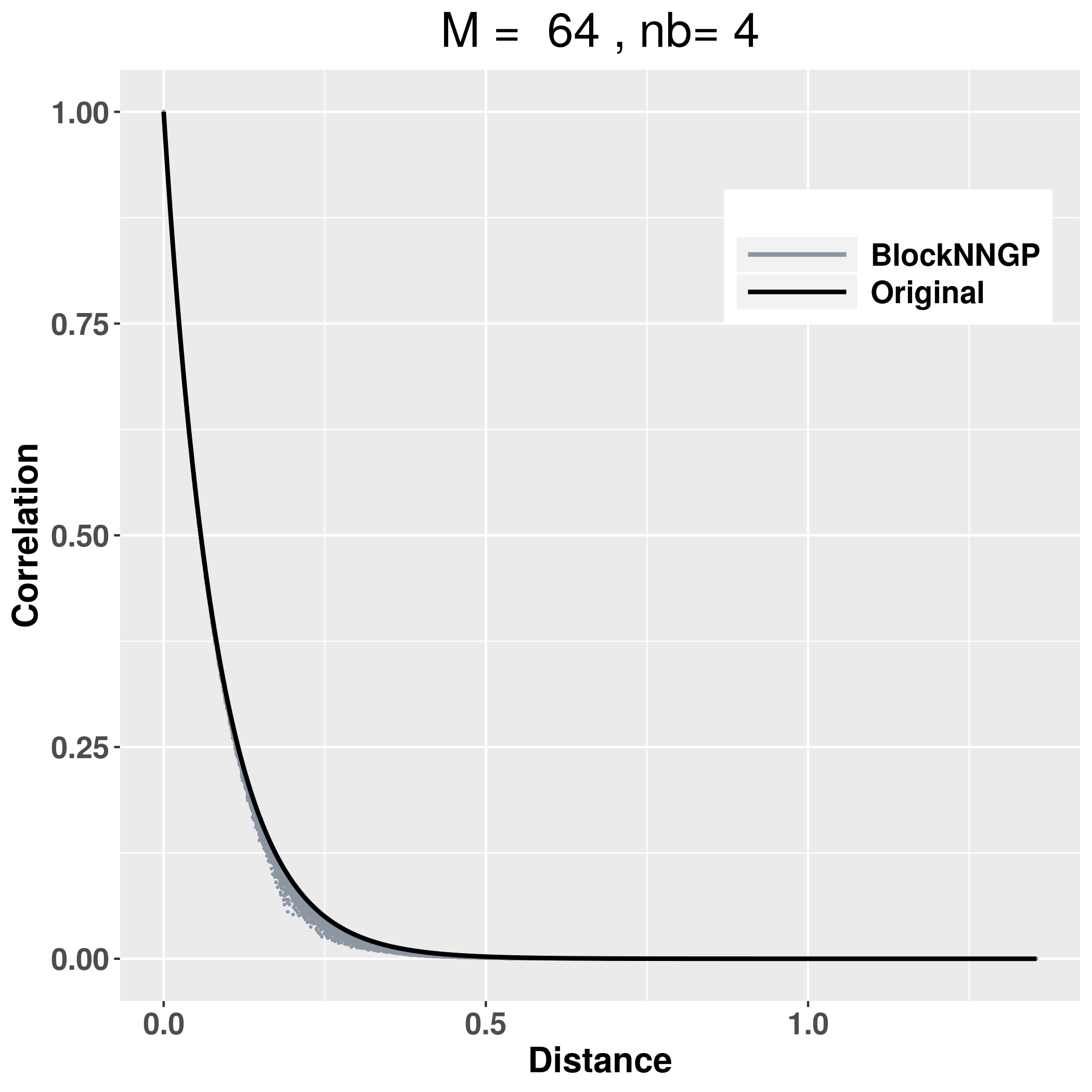} 
\includegraphics[scale=0.22]{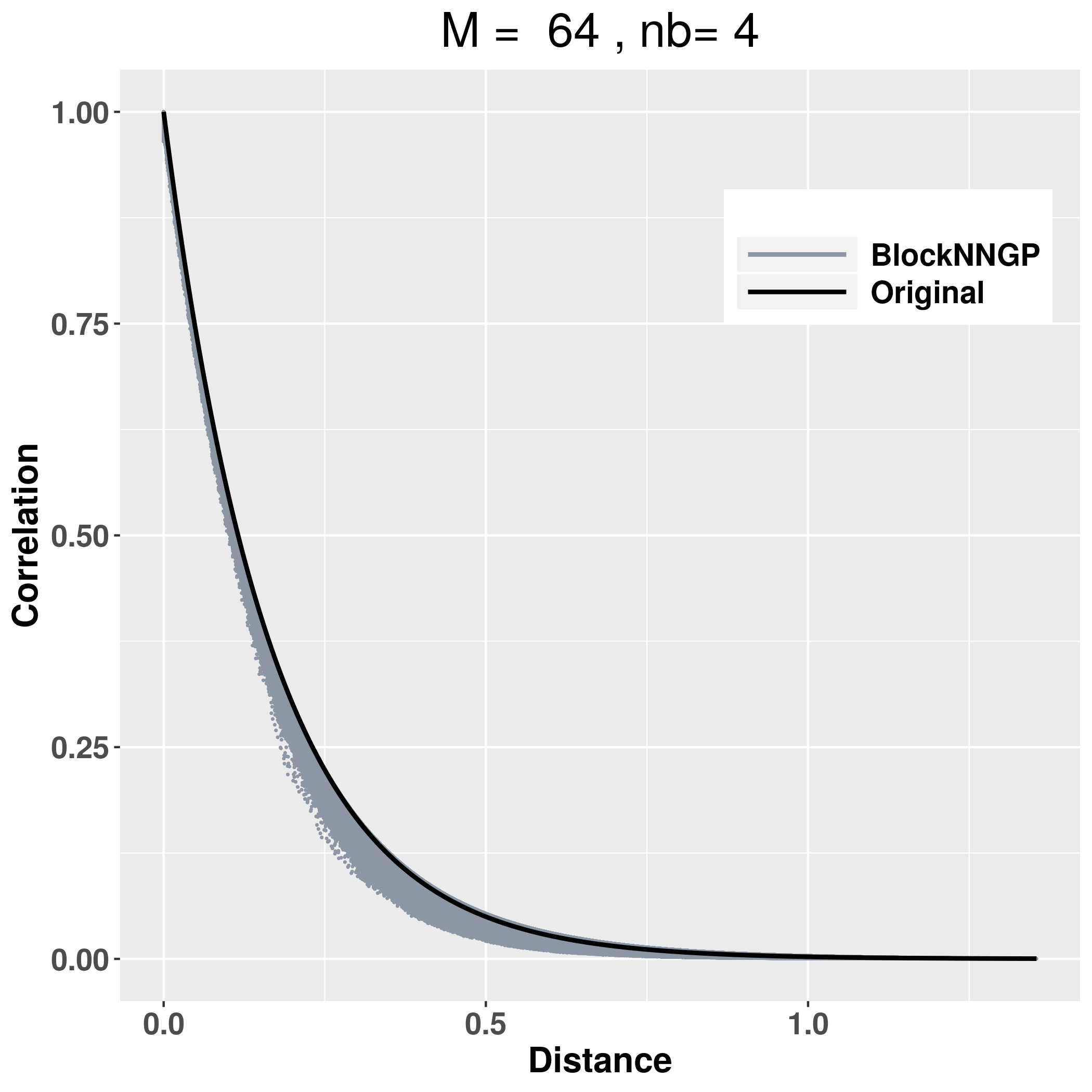} 
\includegraphics[scale=0.22]{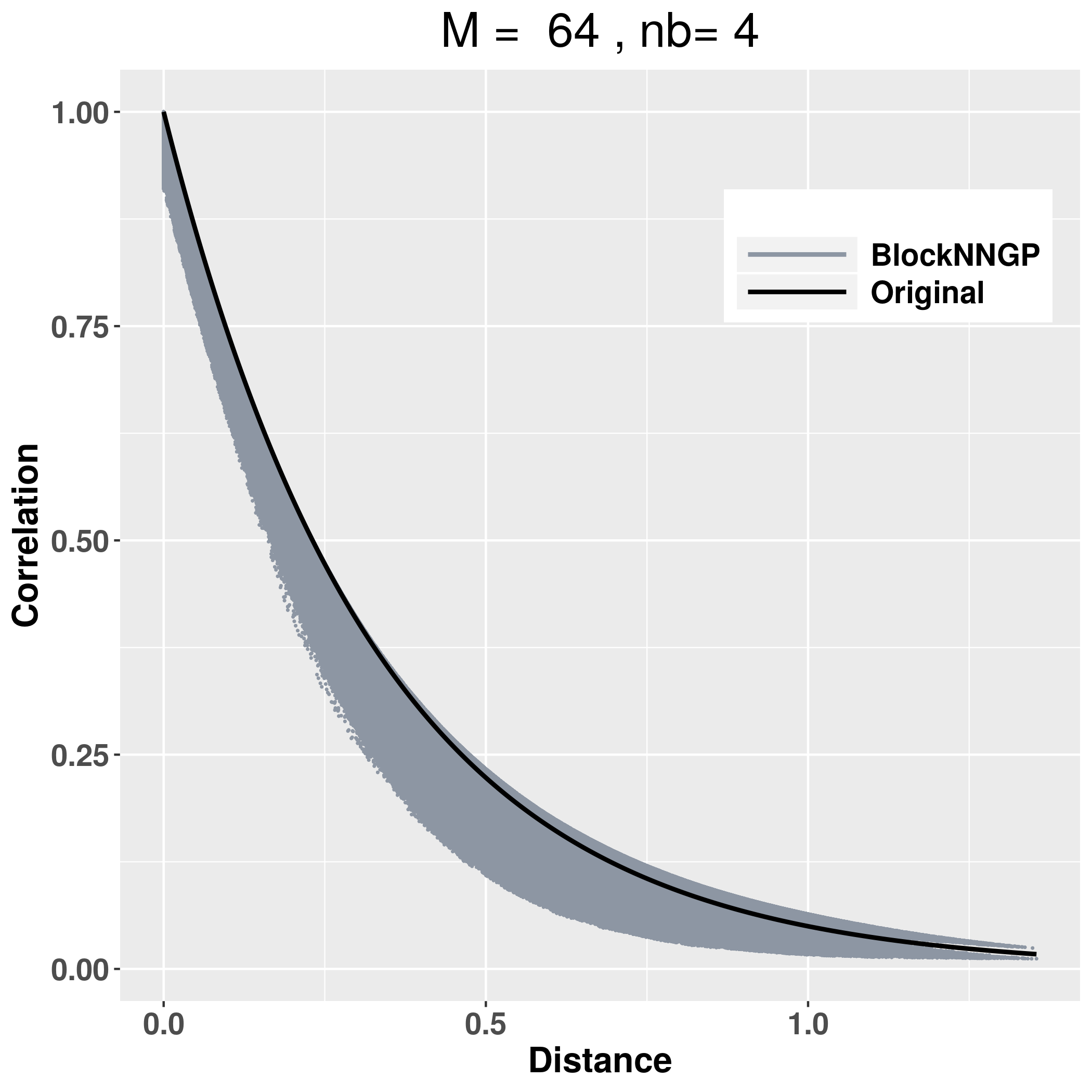} 
\includegraphics[scale=0.22]{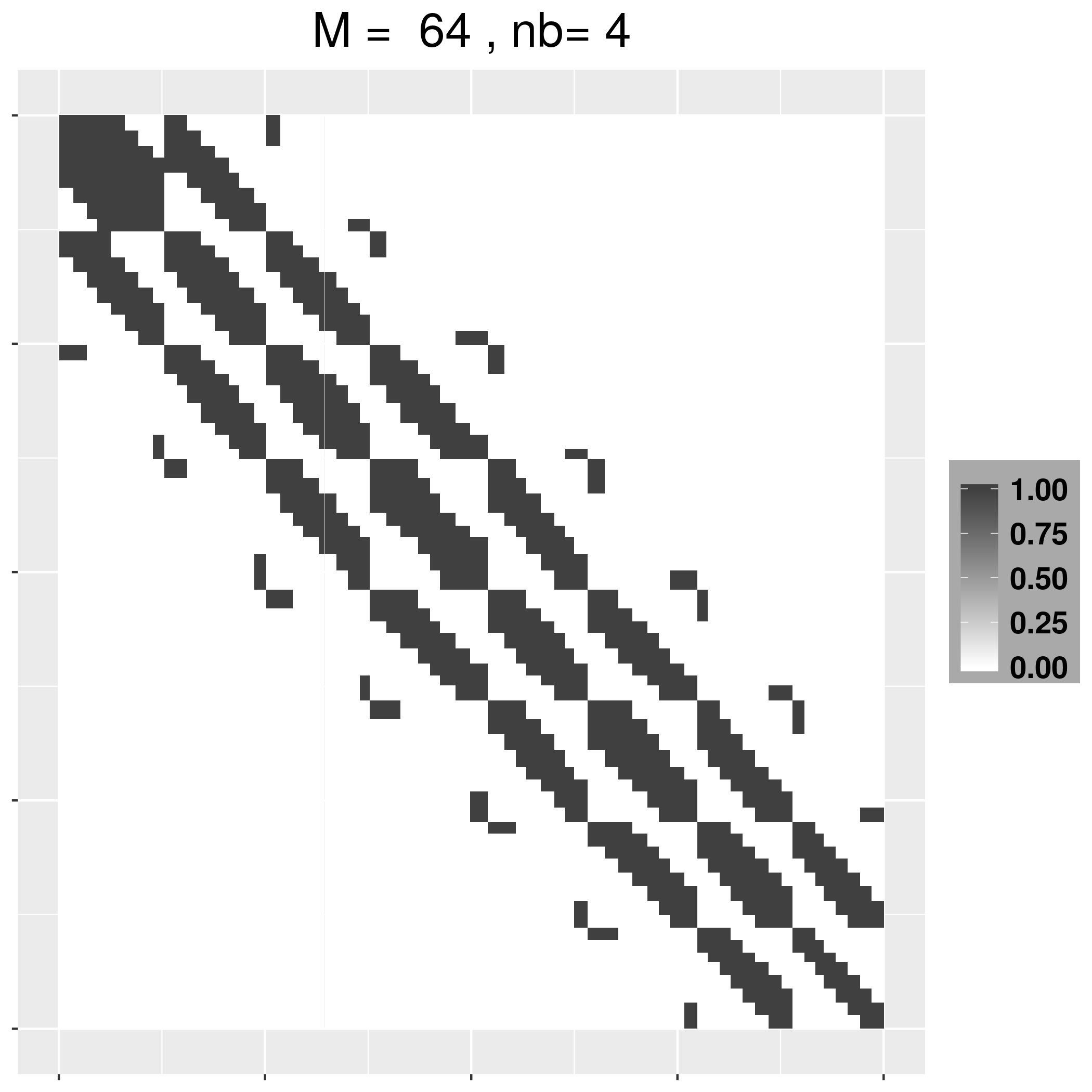} 
\end{center}
\caption{\small  Regular blocks. True exponential correlation of GP  against distance (black lines) and empirical approximated correlations of block-NNGP aginst distance (blue dots) for:
short range($\phi$ = 12, first column), medium range ($\phi$ = 6, second column) and large range ($\phi$ = 3, third column), and $M = 25$ blocks (first row), $M = 36$ blocks (second row) and $M = 64$ blocks (third row). The fourth column presents Precision matrices for $M = 16$ blocks (first row), $M = 36$ blocks (second row) and $M = 64$ blocks (third row), gray squares represent non-zero values. In all cases, each block has $nb=4$ neighbor blocks.
}
  \label{fig:figk2}
\end{figure}

Figure~\ref{fig:figk2} displays the theoretical exponential correlation function (black line), the empirical correlation function  and precision matrices for the block-NNGP, under different number of blocks (blue dots) varying the number of blocks ($M$) and neighbors ($nb$). In general, the match between the theoretical and empirical correlation is good. 

\noindent Nevertheless, some cases show a discrepancy from the true correlations, specially, when the range is large or when the number of blocks is large. The block-NNGP precision matrix is a block sparse matrix, a more optimal structure in terms of computational gains. In practice there is a trade-off between accuracy of the block-NNGP approximation and the number of blocks used.

\begin{figure}
 \hspace{1.8cm} \textbf{$\phi = 12$} \hspace{4.5cm} \textbf{$\phi = 6$}\hspace{4.5cm} \textbf{{$\phi = 3$}}\\
\includegraphics[scale=0.3]{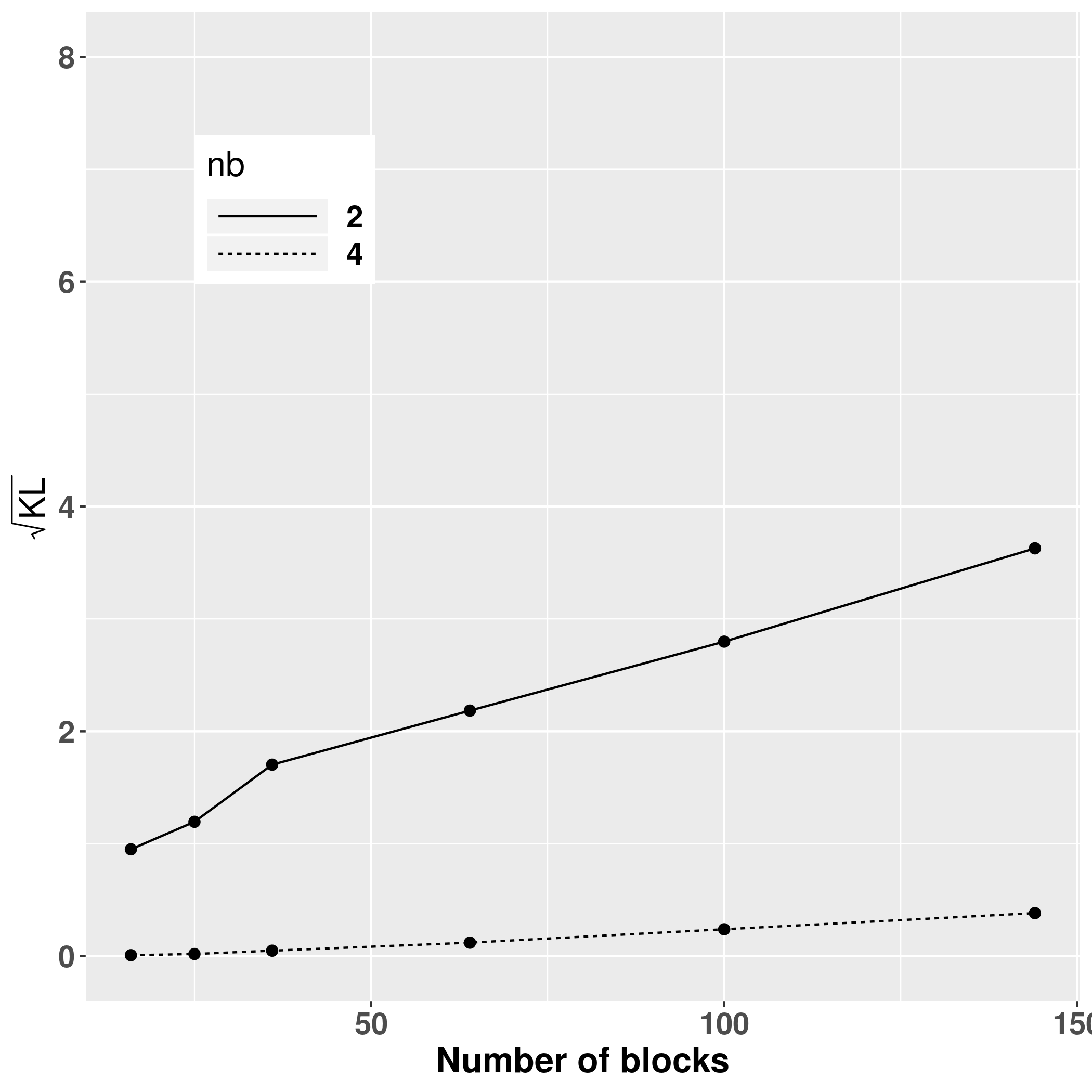} 
\includegraphics[scale=0.3]{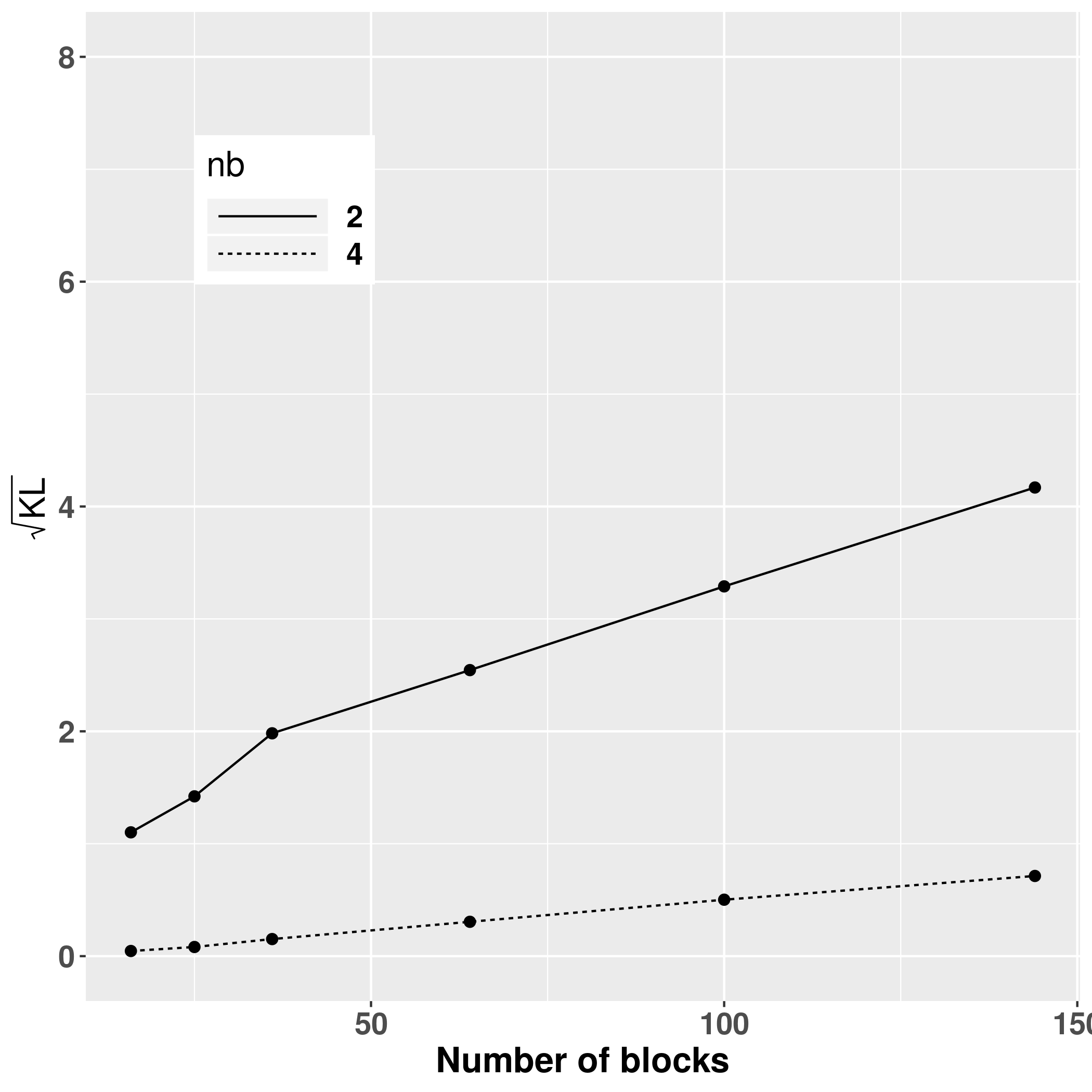} 
\includegraphics[scale=0.3]{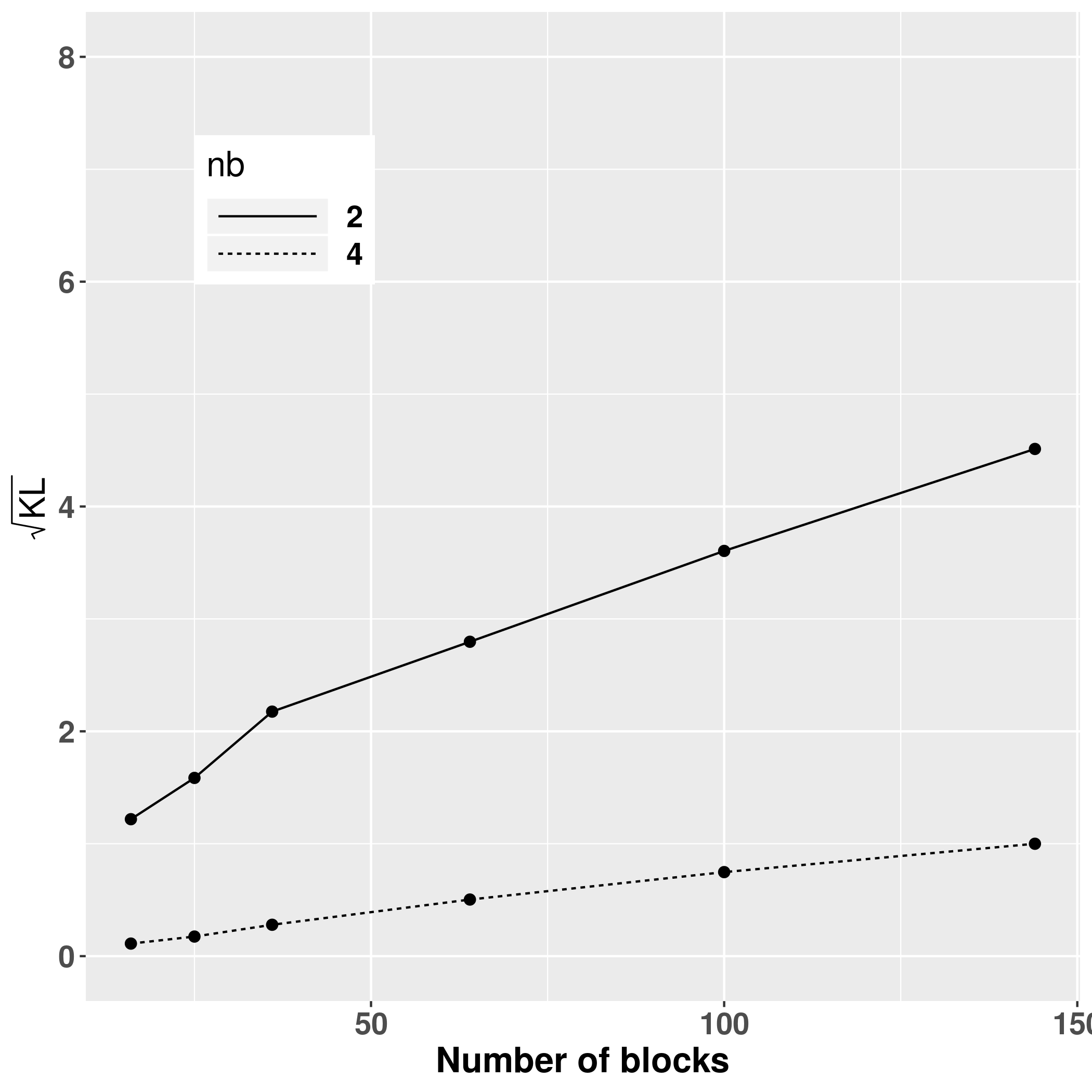} \\
\includegraphics[scale=0.3]{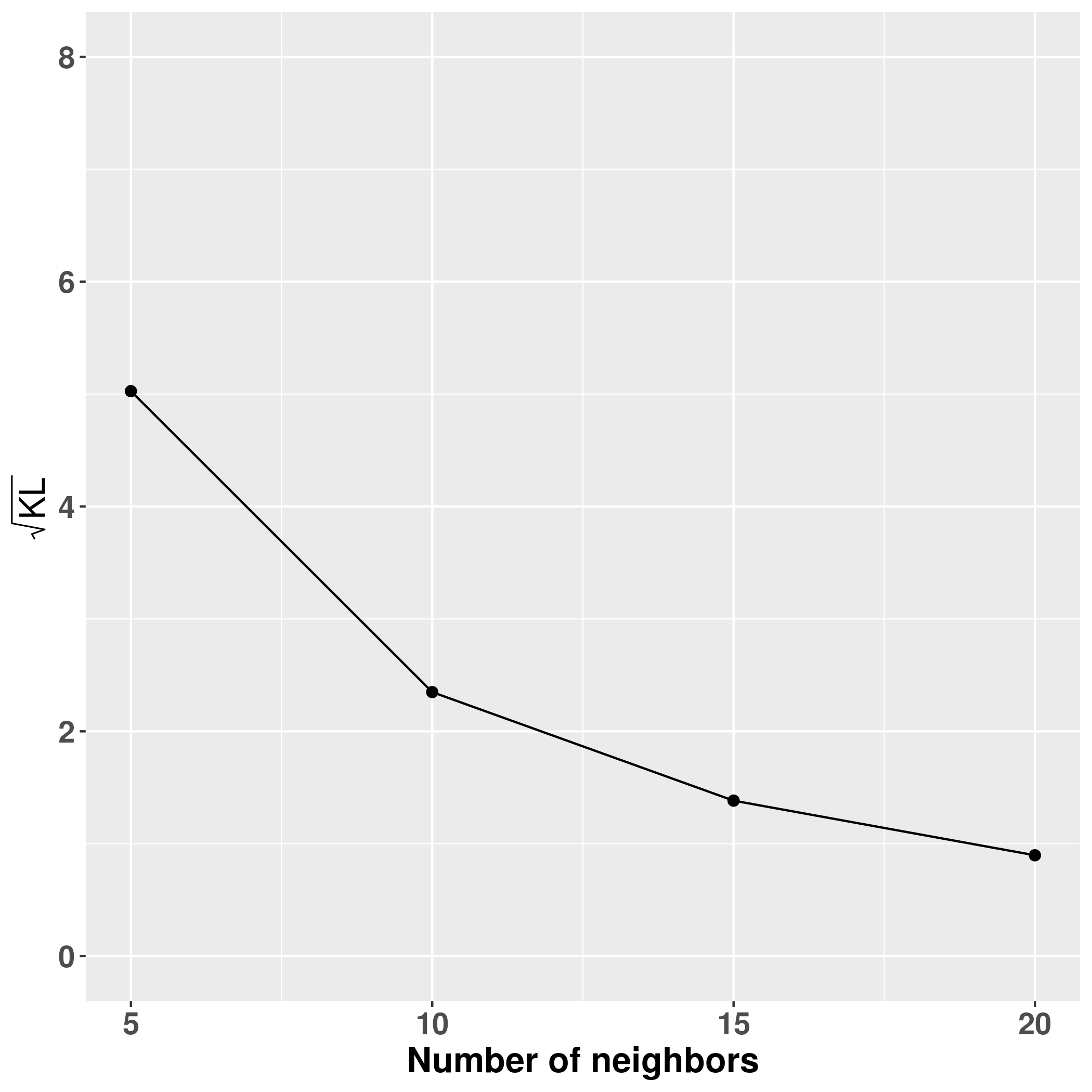} 
\includegraphics[scale=0.3]{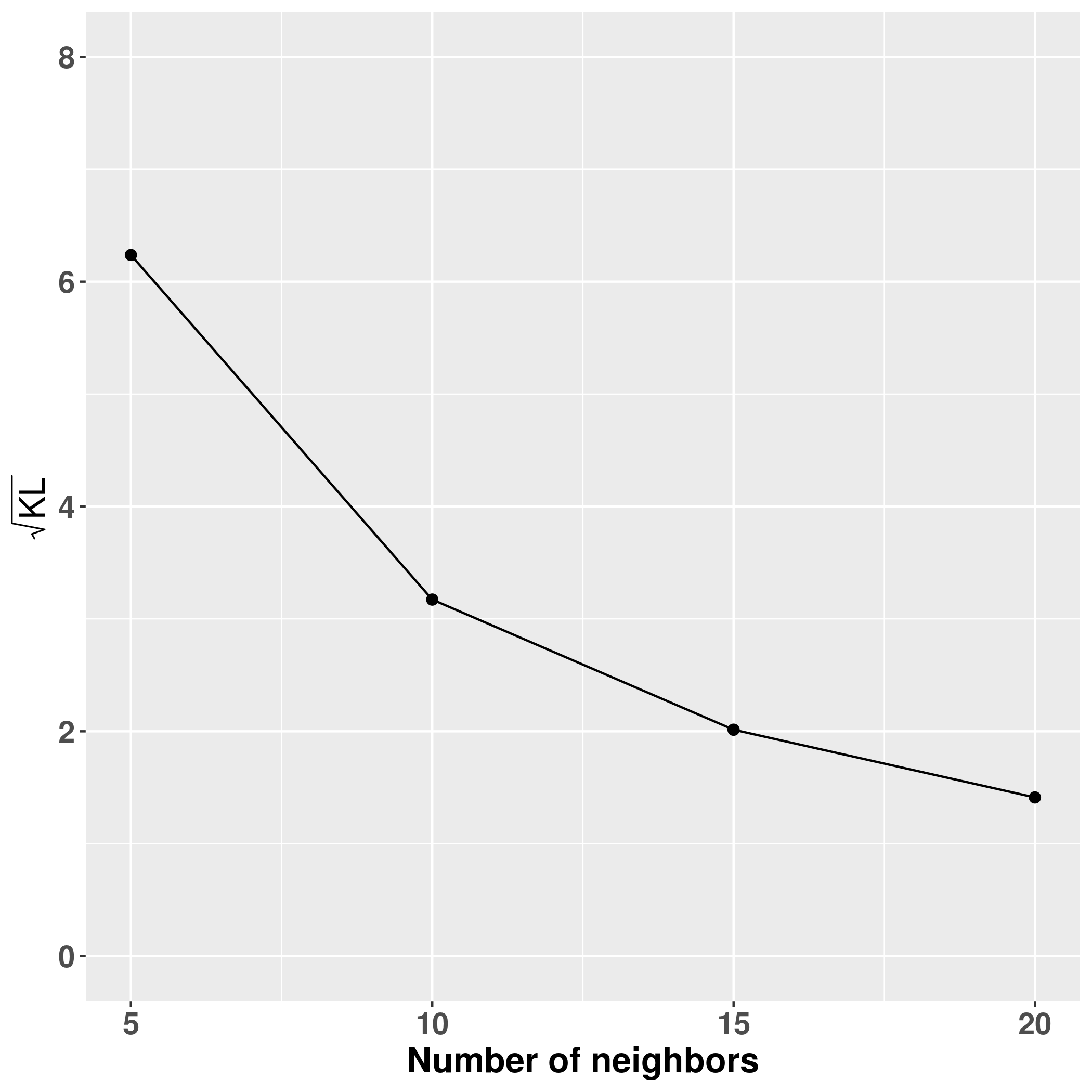} 
\includegraphics[scale=0.3]{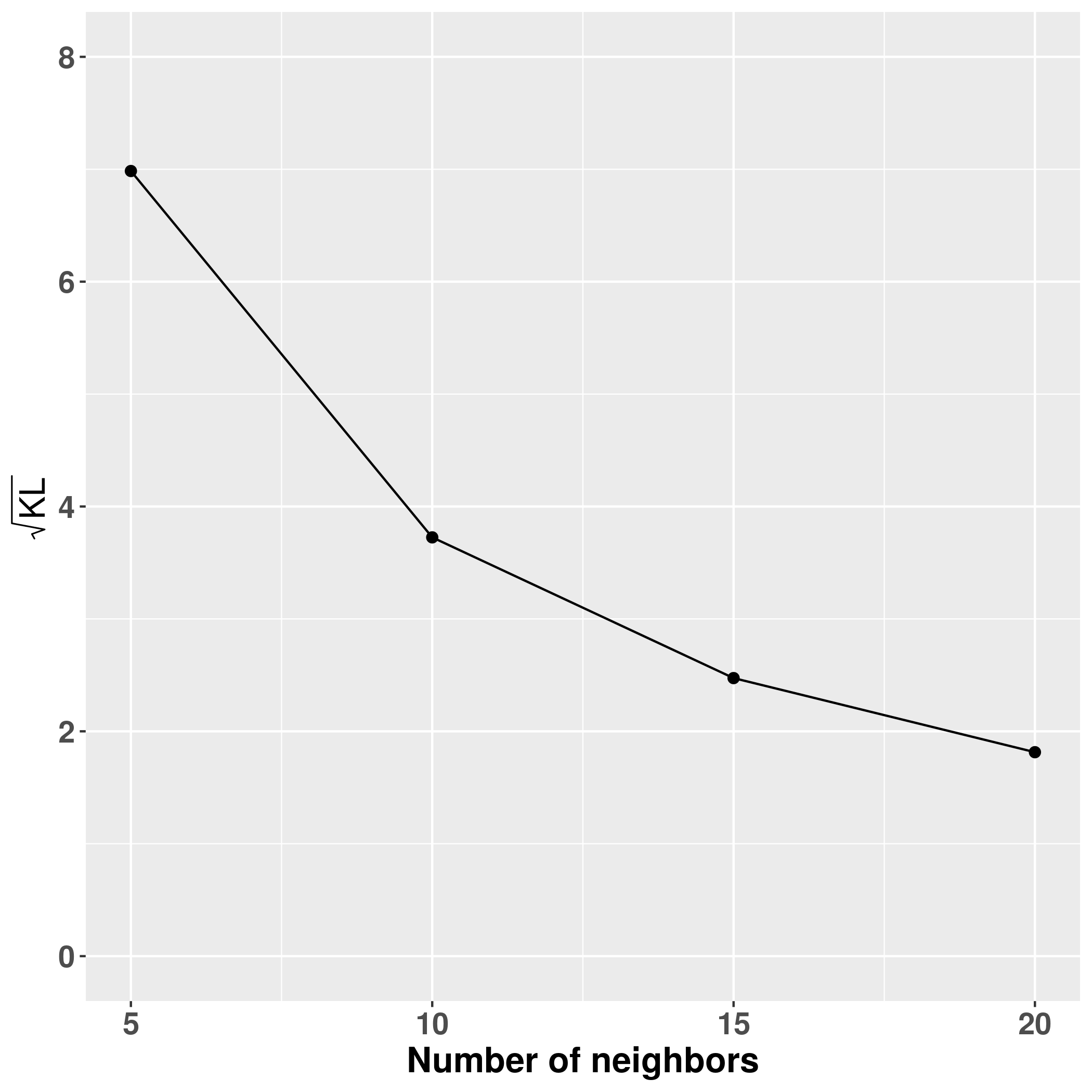} 
\caption{\small  Upper panel: The square root of the Kullback-Leibler divergence of block-NNGP as a function of
the number of blocks for 2000 locations and different ranges using the blockNNGP model with nb = 2  and nb = 4 
neighbor blocks.
Lower panel:   The square root of the Kullback-Leibler divergence of NNGP as a function of the number of observations
for 2000 locations and different ranges. 
First column: $\phi$ = 12 (short range), Second column: $\phi = 6$ (medium range) and Third column: $\phi$ = 3 (large range).
}
  \label{fig:figk4}
\end{figure}

Another way to illustrate the difference between the NNGP, the block-NNGP and the exact GP is in terms of the Kullback-Leibler divergence (KLD). Figure~\ref{fig:figk4} displays the {square root of the KLD (SR-KLD) of the block-NNGP (upper) for different number of blocks ($M$) and number of neighbor blocks ($nb$),  and the SR-KLD  of the NNGP (lower panel) for different number of neighbors. We notice that for NNGP as the number of neighbors increases it clearly gives an improvement, nevertheless the loss of information is higher when the range is larger.} While the block-NNGP with $nb = 4$ clearly gives an improvement over $nb = 2$, the results are quite similar for the three ranges, but the loss of information {increases when the number of blocks becomes larger.}
In general, since the block-NNGP is expected to have more neighbors than NNGP it will, necessarily, provide a better approximation to the full process. For a formal proof we refer to \citet{Guinness:2018, BANERJEE:2020}.

\subsection{Blocking strategies}

The spatial domain $\bm{D}$ can be partitioned into several regions, either using  a regular  or an irregular block design \citep{Eidsviketal:2014}. In general, in order to achieve fast inference it is useful that each block has the same number of locations, otherwise for some blocks it will be expensive to perform matrices operations {, instead of a balanced cost. With this in mind, the regular block design may not be practical}. In practice, when the observed locations are almost uniformly distributed over the domain $\bm{D}$, it is recommended to use the regular block design. But when the observed locations are far from uniform, it is recommended to have more complicated partitioning schemes like irregular block design.

In particular, for the regular block design, the spatial domain is simply partitioned into $M$ non-intersecting rectangular subregions of approximately the same area. For the irregular block design, we propose to split the spatial domain into $M$ non-intersecting rectangular subregions with approximately the same number of points through a 2-dimensional (2-d) tree \citep{Bentley1975}.  Given a set of coordinates $s_1, \dots, s_n$, the 2-d tree is constructed as follows: i) Compute the median ($Me$) of the coordinates, that is, at least 50\% of the coordinates are  equal or smaller than $Me$. ii) Then the spatial domain $\bm{D}$ is partitioned into rectangular regions $\bm{D_L}$ and $\bm{D_R}$, where $\bm{D_L}$ contains only the coordinates equal or smaller than $Me$. iii) Repeat i) and ii) recursively on both $\bm{D_L}$ and $\bm{D_R}$. iv) The recursion stops when the number of desired blocks $M$ is achieved. The last step was conditioned to fulfill our requirements.  

\begin{figure}[htb]
 \centering
\includegraphics[scale=0.47]{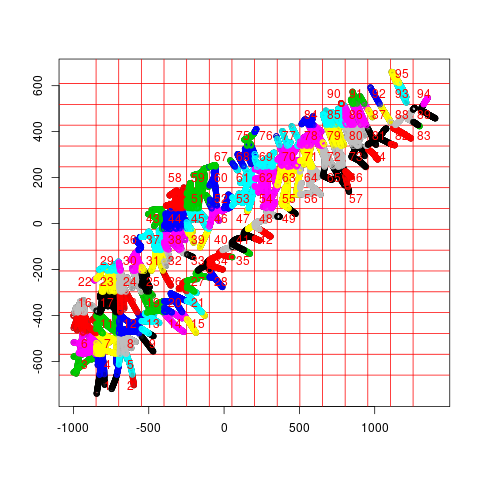}
\includegraphics[scale=0.47]{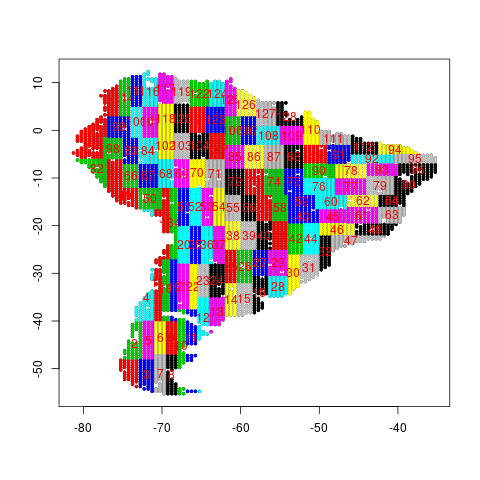}
  \caption{Left: Regular blocks for joint-frequency data,  $n = 10000$ locations. Right:  Irregular blocks for Precipitation data, $n=6000$ locations.} 
 \label{fig:fig2}
 \end{figure}

To illustrate the process of our two blocking strategies, Figure~\ref{fig:fig2} displays two examples. The first one shows the locations of $10000$ samples of joint frequency mining data in Norway, with a regular blocking, where $M = 95$. And the second one shows the locations of $6000$ samples of precipitation in South America, with an irregular blocking, here  $M = 128$ and the approximated number of samples in each block is $47$. 

\subsection{Bayesian inference for block-NNGP models}
\label{s:inference}

Let $\bf{Y} = (Y(s_1), \dots, Y(s_n))^T$ be a realization of a spatial process defined for all $s_i \in \bm{D} \subset \Re^2$,  $i =1, \dots, n$, where the random variable $Y_i = Y(s_i)$ follows a generic distribution with pdf $\pi(y_i|\mu_i, \theta_0)$, which depends on the mean $\mu(s_i)$ and some hyperparameters $\theta_0$. In  a  general geostatistical model the mean is linked to the linear predictor $\eta_i$ through  a suitable link function $g(.)$ and the structure of the linear predictor has the next general form:  
\begin{equation}
g(\mu(s_i)) = \eta_i = \bm{X}^T(s_i) \bm{\beta} +  w(s_i) + \epsilon(s_i),  
\label{eq:eqinf1} 
\end{equation}
where $\bm{\beta}$ is a coefficient vector (or regression parameter), $\bm{X}^T(s_i)$ is a  vector of covariates, $w(s_i)$ is a  spatial structured random effect, thus $\bm{w}=(w(s_1), \dots, w(s_n))^T$ captures the spatial association, and $\epsilon(s_i) \sim N(0, \tau^2)$ models the measurement error. 
The usual Gaussian process prior for $\bm{w} \sim N(0, \bm{C}(\theta_1))$, where $\bm{C}(.)$ is some specific covariance function which depends on $\boldsymbol{\theta_1}=(\phi, \sigma^2)$. Instead of this prior we assume that $\bm{w} \sim \mbox{block-NNGP}(0,\bm{\widetilde{C}})$. Further, we assign priors for $\bm{\beta} \sim N(\bm{\mu_{\beta}},  \bm{\Sigma_{\beta}})$ and for the parameters $ \theta_0, \phi, \sigma^2, \tau^2$. 

The model presented in (\ref{eq:eqinf1}) belongs to the latent Gaussian models (LGM) class. 
More specifically, we have that $Y_{i}$'s given the latent Gaussian process 
$\boldsymbol{x}$=$\{\bm{\beta}, \bm{w}\}$ and possible  hyperparameters 
$\boldsymbol{\theta}=(\theta_0, \sigma^2, \phi, \tau^2)$, are assumed conditionally independent. Hence, the posterior estimates of parameters are computed using INLA.

With regard to the distribution of the latent process, since  $\bm{w} \sim \mbox{block-NNGP}(0,\bm{\widetilde{Q}^{-1}})$, therefore the latent process $\boldsymbol{x}\sim N(0, \bm{Q})$, where  
\begin{equation}
\boldsymbol{Q} = \left[\begin{array}{cc}
 \boldsymbol{\widetilde Q} & \boldsymbol{0} \\
\mathbf{0} & \boldsymbol{Q}_{\beta}
\end{array}\right] \nonumber
\end{equation}
is sparse, and it can improve  computational performance dramatically. We also remark that the structure of this  precision matrix is particularly ideal, because  $\boldsymbol{\widetilde Q}$  is built from blocks (see  proposition 2 and (\ref{eq:eqQ})) and INLA can exploit fast parallel computation algorithms for blocks.  Moreover the whole algorithm does not need to store $n\times n$ dense distance matrices to compute $\boldsymbol{\widetilde Q}$, it only  stores $\bm M$ small dense matrices.

Then the joint posterior distribution of  the LGM can be computed using the likelihood distribution of $\boldsymbol{Y}$, the distribution of the latent
Gaussian field $\boldsymbol{x}$ and the distribution of hyperparameters $\boldsymbol{\theta}$ as
\begin{equation}
\pi(\boldsymbol{x},\boldsymbol{\theta}|\boldsymbol{y}) \propto \pi(\boldsymbol{\theta}) |\boldsymbol{Q}|^{1/2}\exp\left[-\frac{1}{2}x^{\top}Qx+
\sum_{i=1}^{n} \log\{\pi(y_{i}|\boldsymbol{x},\boldsymbol{\theta})\}\right], \nonumber
\end{equation}
where the determinant of $\bm{Q}$ is easily computed from the Cholesky decomposition of $\boldsymbol{\widetilde Q}$ which inherits the sparsity property of  $\boldsymbol{\widetilde Q_S}$ in Proposition~\ref{prop2}. 

The posterior marginals of the latent variables $\pi(x_{i}|y)$ and the 
posterior marginal of hyperparameters $\pi(\theta_s|y)$ are defined by
\begin{equation}
\pi(x_{i}|y)=\int \pi(x_{i}|\theta,y) \pi(\theta|y) d\theta \mbox{ and }
\pi(\theta_j|y)= \int \pi(\theta|y)d\theta_{-j}, \nonumber
\end{equation}
where $i=1,\ldots,n$.  INLA provides approximations for the posterior marginals of the latent variables and hyperparameters, as given by
\begin{equation}
   \widetilde{\pi}(x_{i}|y)=\int \widetilde{\pi}(x_{i}|\theta,y) \widetilde{\pi}(\theta|y) d\theta \mbox{ and }
   \widetilde{\pi}(\theta_j|y)= \int \widetilde{\pi}(\theta|y)d\theta_{-j},
\label{eq:dust}
\end{equation}
which are both very accurate and extremely fast to compute. 
Here, $\widetilde{\pi}$ denotes an approximation to a pdf.
In summary, the main idea of INLA is divided into the following tasks: First, it provides a Gaussian approximation of $\widetilde{\pi}(\theta|y)$ to the joint posterior of hyperparameters given the data $\pi(\theta|y)$. Then, it provides an approximation of $\widetilde{\pi}(x_{i}|\theta,y)$ for the marginals of the conditional distribution of the latent field given the data and the  hyperparameters $\pi(x_{i}|\theta,y)$. And finally, it explores $\widetilde{\pi}(\theta|y)$ on a grid and uses it to 
integrate out $\theta$ and $\theta_{-j}$ in (\ref{eq:dust}). For more details on INLA calculations, we refer to \citet*{Rue:2009}.

Following \cite{Finleyetal:2019}, spatial prediction can be carried out after parameter inference.  Conditioning on a particular estimated value of the parameters $(\bm{\theta}, \bm{\beta})$, spatial prediction amounts to finding the posterior predictive distribution on a set of prediction locations $u_i$, that is, $\pi(y(u_i)|\bm{y})$. 
Without loss of generality, if we consider all observed data for estimation, thus $\bm{S}$ comprises the observed locations, while the new location points for predictions belong to the finite set $\bm{U}$. Furthermore, since the components of $\bm{w_U}|\bm{w_S}$ are independent, we can update $w(u_i)$ for each $i=1, \dots, l$, from 
$\pi(w(u_i)|\bm{w_S}, \bm{\beta}, \bm{\theta}, \bm{y}) \sim N(m,v)$, where $m =  \bm{C}^T_{u_i,\bm{N(u_i)}}  \bm{C} ^{-1}_{\bm{N(u_i),N(u_i)}} \bm{w(N(s_0))}$ and $v = \sigma^2 -  \bm{C}^T_{u_i,\bm{N(u_i)}}  \bm{C} ^{-1}_{\bm{N(u_i),N(u_i)}} \bm{C}_{u_i, \bm{N(u_i)}}$. 
{Therefore, the block-NNGP is especially} useful here because the posterior sampling for $\bm{w_U}$ is cheap, given that their components are independent, each $w(u_i)$ is only based on the observations that lie in the block that it belongs. 
Now using the posterior samples of $w(u_i)$, the posterior predictive sampling is performed through $y(u_i)|\bm{w_U}, \bm{w_S}, \bm{\beta}, \bm{\theta}, \bm{y} \sim N(X^T(u_i) \bm{\beta} + w(u_i), \tau^2)$. 

In the supplementary material B we present in details a full-MCMC approach using a collapsed MCMC strategy. This strategy was shown to improve convergence and reduce running times in comparison to other MCMC strategies \citep{Finleyetal:2019}. 
It is also important to emphasize that INLA avoids the complex updating schemes, long running times, and diagnostic convergence checks of MCMC, and uses directly the sparse precision matrix of the block-NNGP which is also computed using fast parallel algorithms for blocks and Cholesky decomposition. For this reason we focus our results and discussions mainly using 
INLA. 

\section{Simulation results}
\label{s:sim}

In order to assess the performance of the block-NNGP models, we present the next simulation experiments. We generate $s_i$ sites on a spatial domain $[0, 1] \times [0, 1]$  where $i=1,\dots,2500$ and the random spatial effects $w(s_i)$, under a zero mean Gaussian process model  $\bm{w} \sim GP(0, \bm{C}(\phi, \sigma^2))$, where $\bm{C}(.)$ is an exponential covariance function with elements  $C(s_i, s_j) = \sigma^2 \exp(-\phi \|s_i-s_j\|)$. We set $\sigma^2 = 1$ and the effective range ($r$) is studied for a variety of settings chosen to mimic the range of behaviour we might observe in practice: 
(i) SIM I: $r$ = 0.16 ($\phi$ = 12);  
(ii) SIM II: $r$ = 0.33 ($\phi$ = 6);  
(iii) SIM III: $r$ = 0.67 ($\phi$ = 3),
which represent small, medium and large effective ranges, respectively. 
We also generate the covariates $x_i \sim N(0,1)$, then $\bm{X}(s_i) = (1 , x_i)^T$. The values of the regression parameters are given by $\bm{\beta}= (1, 5)^T$ and the value of the nugget effect is $\tau^2 = 0.1$. Finally we draw $n=2500$ samples from 
$Y(s_i)\mid \beta, \phi, \sigma^2, \tau^2 \sim N(\bm{X}^T(s_i) \bm{\beta} +  w(s_i) , \tau^2 ).$  
We let $\bm{S}$ be a set of $2000$ locations and $\bm{U}$ be the set of the remaining observations used to assess predictive performance.  

First we implemented  the full-MCMC (see supplementary material B) for full Gaussian process (full GP), NNGP models ($nb=10$ and $nb=20$), a few  regular (R) block-NNGP models and  irregular (I) block-NNGP models, under scenario SIM I. 
We used flat prior distributions for $\beta$, for the spatial decay $\phi$ we assigned a uniform prior $U(1, 30)$, which is equivalent to a range between approximately 0.067 and 2 units, and  for $\sigma^2$ w and  $\tau^2$ we assigned inverse gamma priors $IG(2, 1)$. Posterior inference was based upon three chains of 25000 iterations (with a burn-in of 5000 iterations). 
The results are are shown in Table~\ref{tab:tabs1}. The parameter estimates  were quite accurate but very expensive; in the best scenario the running time for a block-NNGP model was about 17000 seconds, showing that the computational time using MCMC is too long. We also point out that the computational cost for block-NNGP models was the lower one. 
\setlength{\tabcolsep}{0.3em}
\renewcommand{\arraystretch}{0.6}
\begin{table}[htb]
  \caption{ SIM I ($\phi$ = 12). Full MCMC results. Summary of mean posterior parameter estimates and parameter posterior summary credible intervals (2.5, 97.5). 
}
  \centering
\scriptsize
  \begin{tabular}{llllllllll}
    \toprule 
& 	&  \textbf{Full-GP} & \textbf{NNGP}  &  \textbf{NNGP}    & (R)\textbf{M=64} & (R)\textbf{M=100}   &  (I)\textbf{M=64} & (I)\textbf{M=128}\\ 
&True 		& &    \textbf{nb=10}  &  \textbf{nb=20}    &  \textbf{nb=2}  &  \textbf{nb=1}   & \textbf{nb=1}  & \textbf{nb=2} \\ 
   \midrule
\textbf{$\beta_0$}	& 1& 0.92	    &1.03	     &1.07  	    &1.05	        &1.05          &1.1         &1.13\\
                    & &(0.4,1.31)	&(0.69,1.46) &(0.65, 1.39)	&(0.72,1.33)	&(0.85,1.3)    &(0.89, 1.34)&(0.82,1.37) \\
\textbf{$\beta_1$}	& 5& 5.02	    &5.02        &5.02	        &5.02		    &5.02 	       &5.02        &5.02\\
                    & &(5, 5.05) 	&(5,5.04)	 &(5,5.05)	    &(5,5.05)   	&(5,5.05)      &(5,5.04)    &(5, 5.04)\\
\textbf{$\sigma^2$ }& 1& 1.09       &1.08        &1.06  	    &1.02	        &1.06          &1.01        &1.03  \\
                    & &	(0.85,1.54) &(0.86,1.52) &(0.84,1.54)	&(0.82,1.42)	& (0.87,1.32)  &(0.83,1.32) &(0.84,1.34)\\
\textbf{$\phi$ }	& 12& 11.08	    &11.23       &11.5 	        &12.15		    &11.62         &12.16       &11.38\\
                    & &	(7.28, 14.8)&(7.77,14.75)&(7.48,15.24)	&(8.31,15.62)	&(8.74,14.54)  &(8.83,15.2) &(0.08,0.12)\\
\textbf{$\tau^2$}	& 0.1&  0.1	    &0.1         &0.1 	        & 0.1 	        &0.1           &0.1         &0.1 \\
                    & &(0.08, 0.12)	&(0.08,0.12) &(0.08,0.12)	&(0.08,0.12) 	&(0.09,0.12 )  &(0.08,0.12) &(8.49,14.5)\\
   \midrule
\textbf{Time (sec)} && 21620.86 &  19514.39 & 22775.03 &	22738.14 &	16122.3&  17067.59 & 17583.05  \\
   \bottomrule    
    \end{tabular}
\label{tab:tabs1}    
\end{table}

We were able to fit all the next models throughout INLA: (i) NNGP models ($nb = 5, 10, 15, 20, 30, 50$), (ii) regular (R) block-NNGP models and (iii) irregular (I) block-NNGP models, under all the scenarios proposed, in few minutes or seconds. For the block-NNGP models we test different number of  blocks and neighbor blocks to investigate how different blocking schemes influence the estimation and prediction capabilities. The number of regular blocks are $M = 9,  16, 25,  36, 49, 81, 100.$  The number of irregular blocks are $M = 8, 16, 32, 64, 128.$ Moreover, the number of neighbor blocks are $nb = 2, 4, 6$. 
We use similar priors as in the full-MCMC, but for $\tau^2$ we assigned gamma priors $G(1,  5\times10^{-05})$. Our implementation in \texttt{R-INLA} (\verb www.r-inla.org ) and subsequent analysis were run on a Linux workstation with 32 GB of RAM and Intel quad-Xeon processor {with 12 cores}.

\begin{figure}
\begin{center}
\includegraphics[scale=0.3]{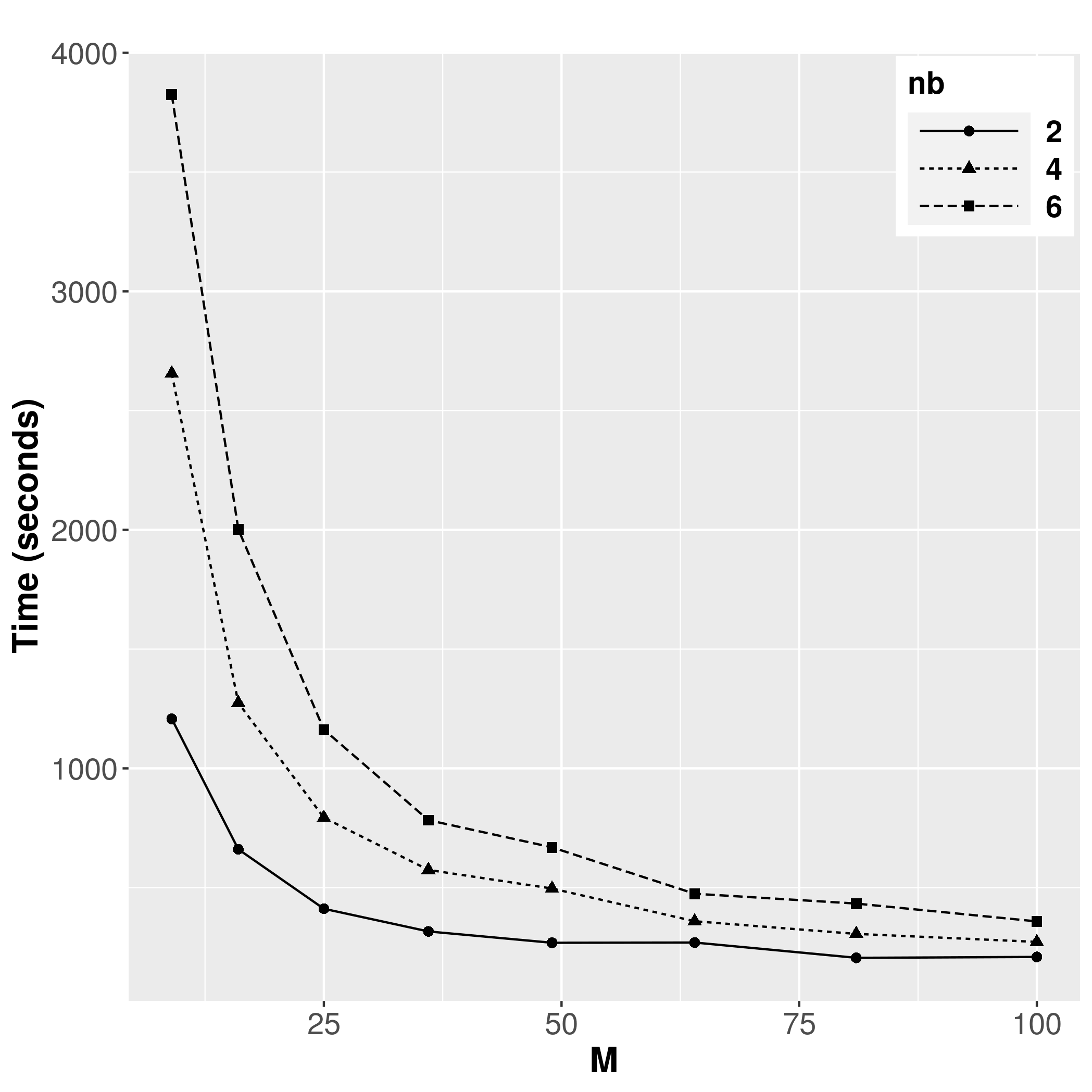} 
\includegraphics[scale=0.3]{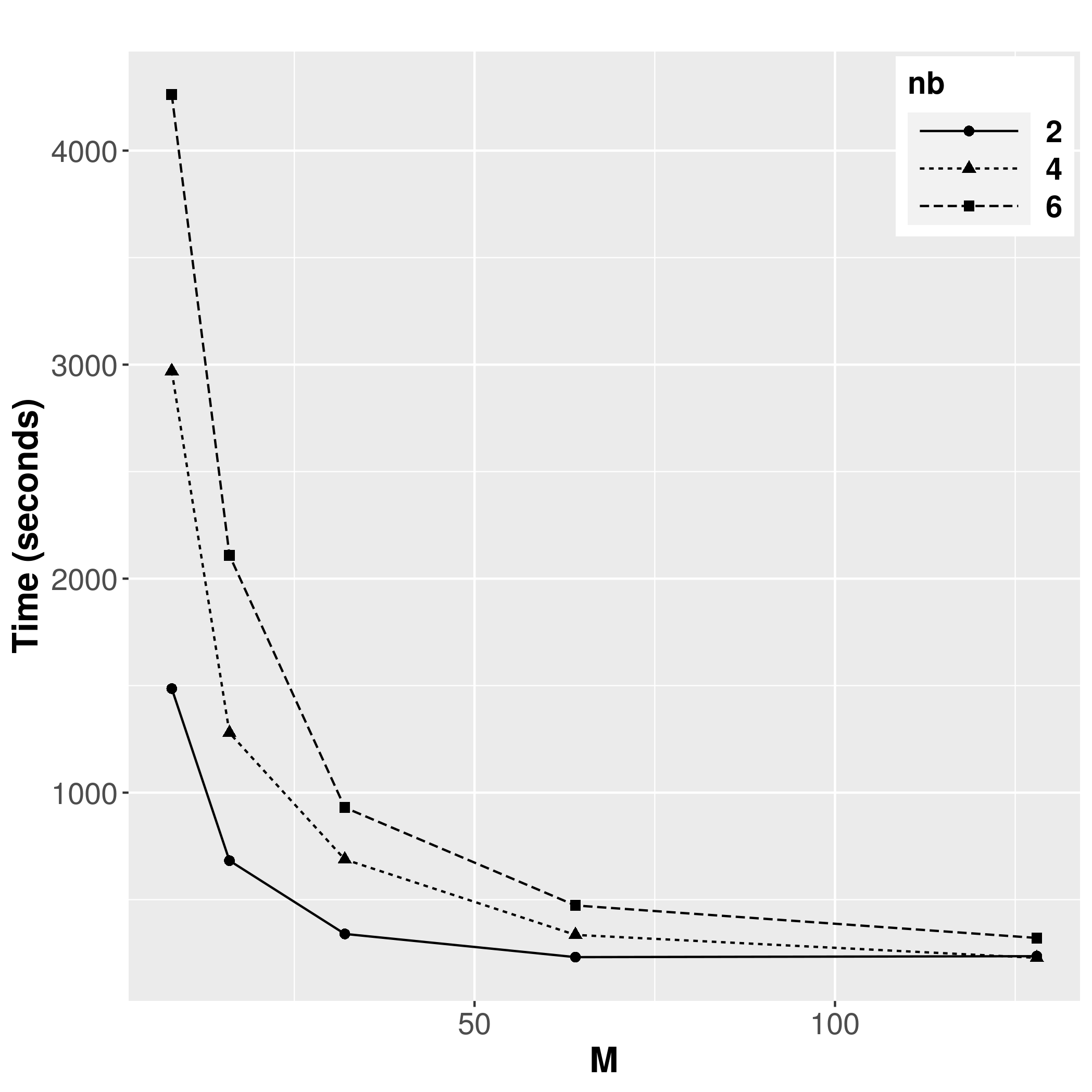} 
\includegraphics[scale=0.3]{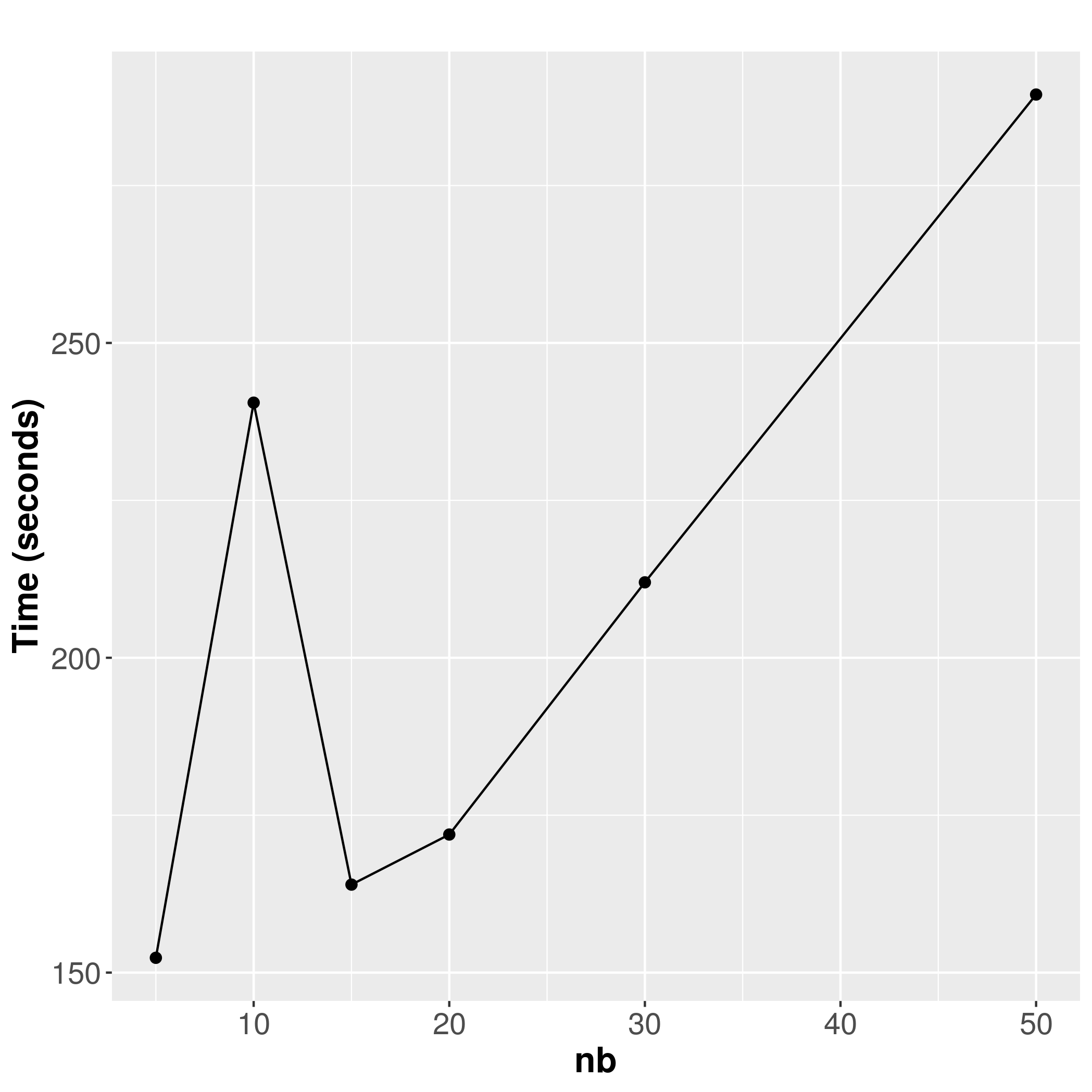} \\
\includegraphics[scale=0.3]{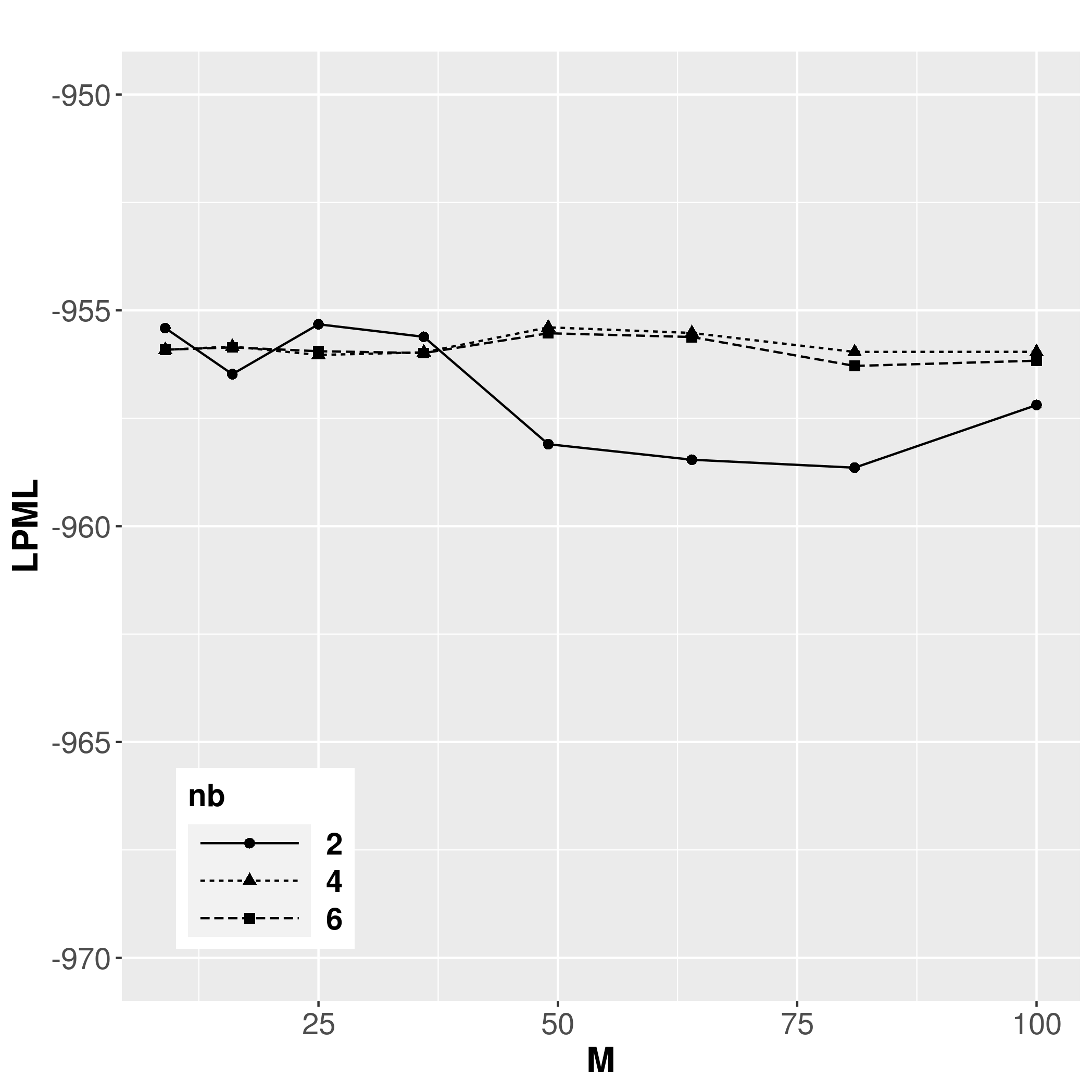} 
\includegraphics[scale=0.3]{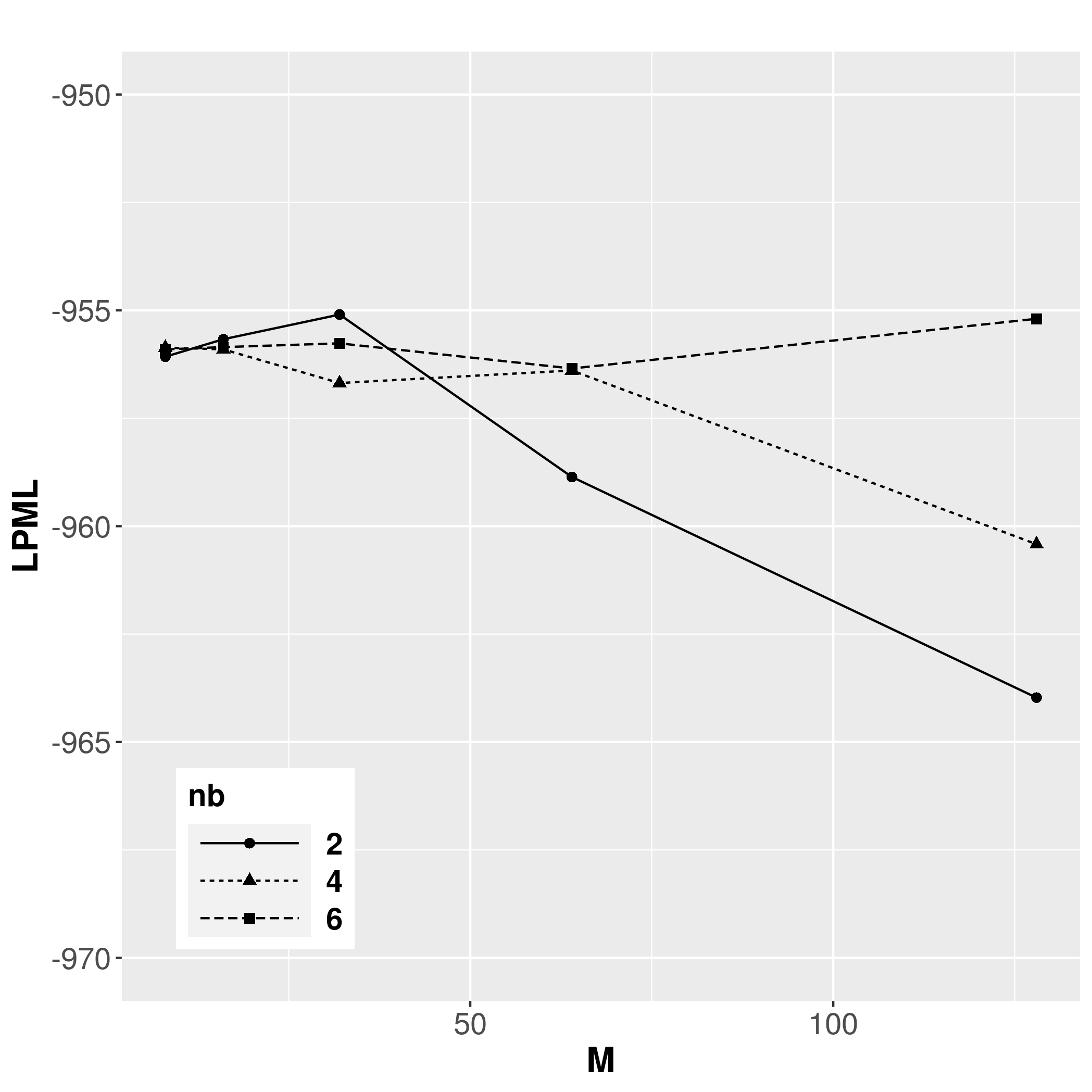} 
\includegraphics[scale=0.3]{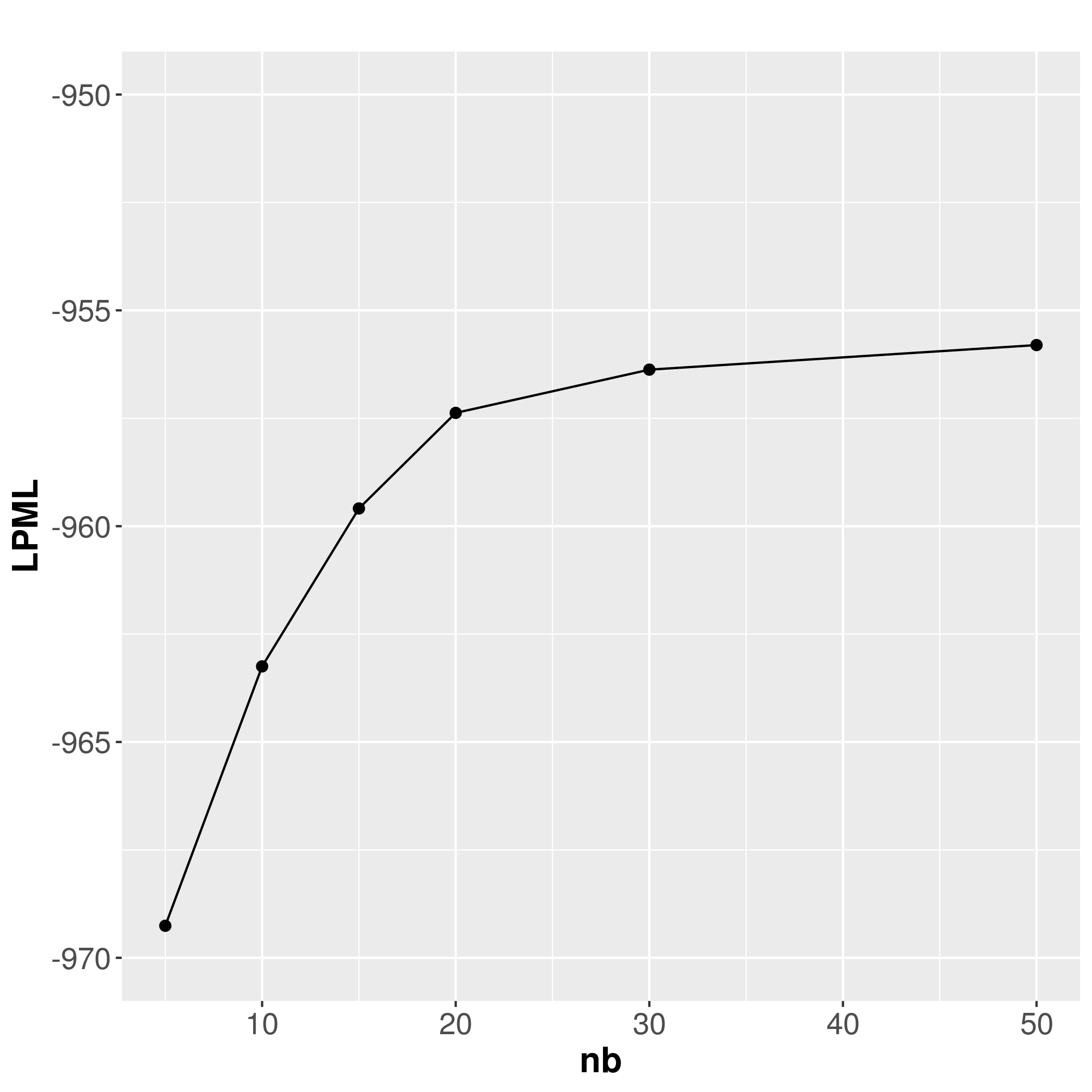} \\
\includegraphics[scale=0.3]{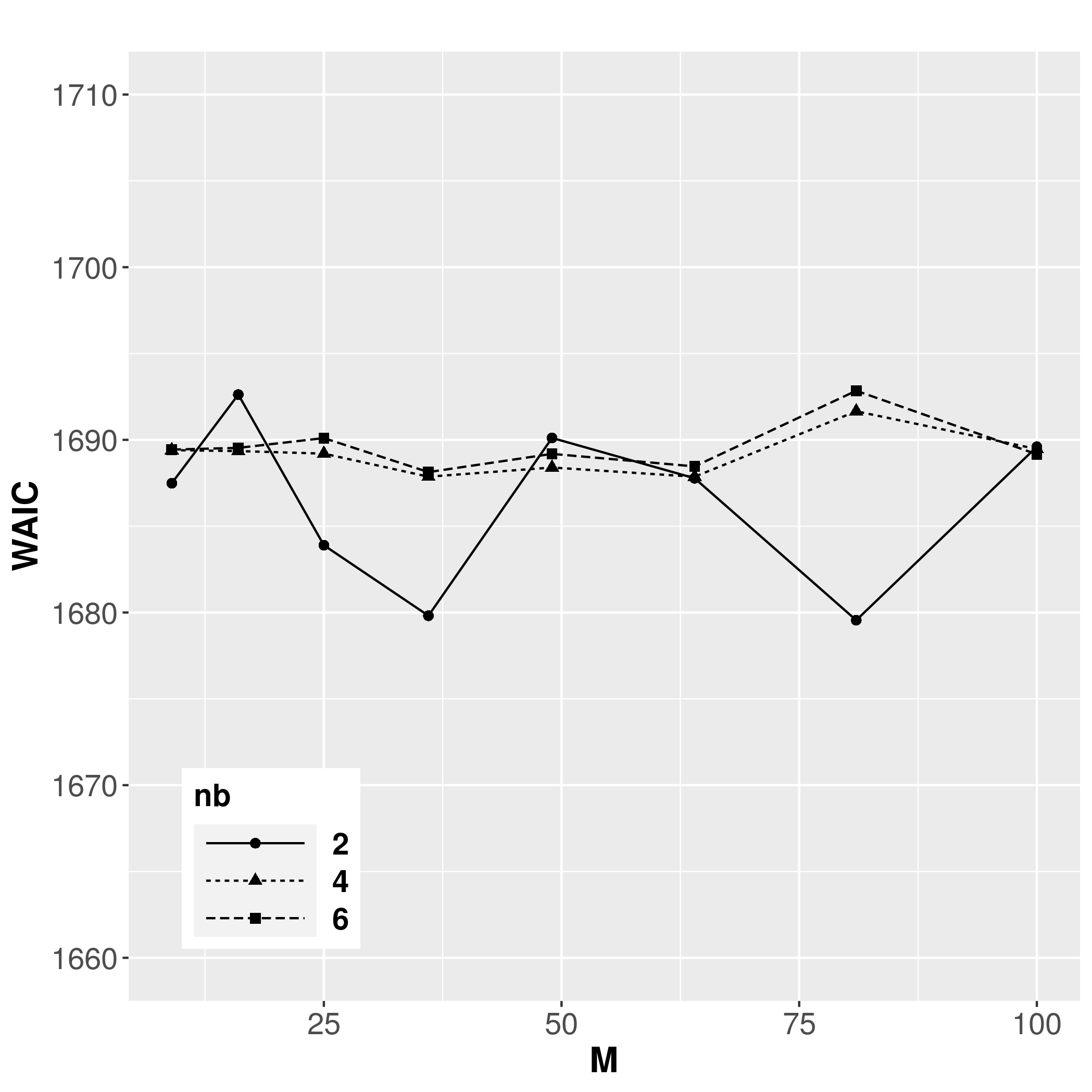} 
\includegraphics[scale=0.3]{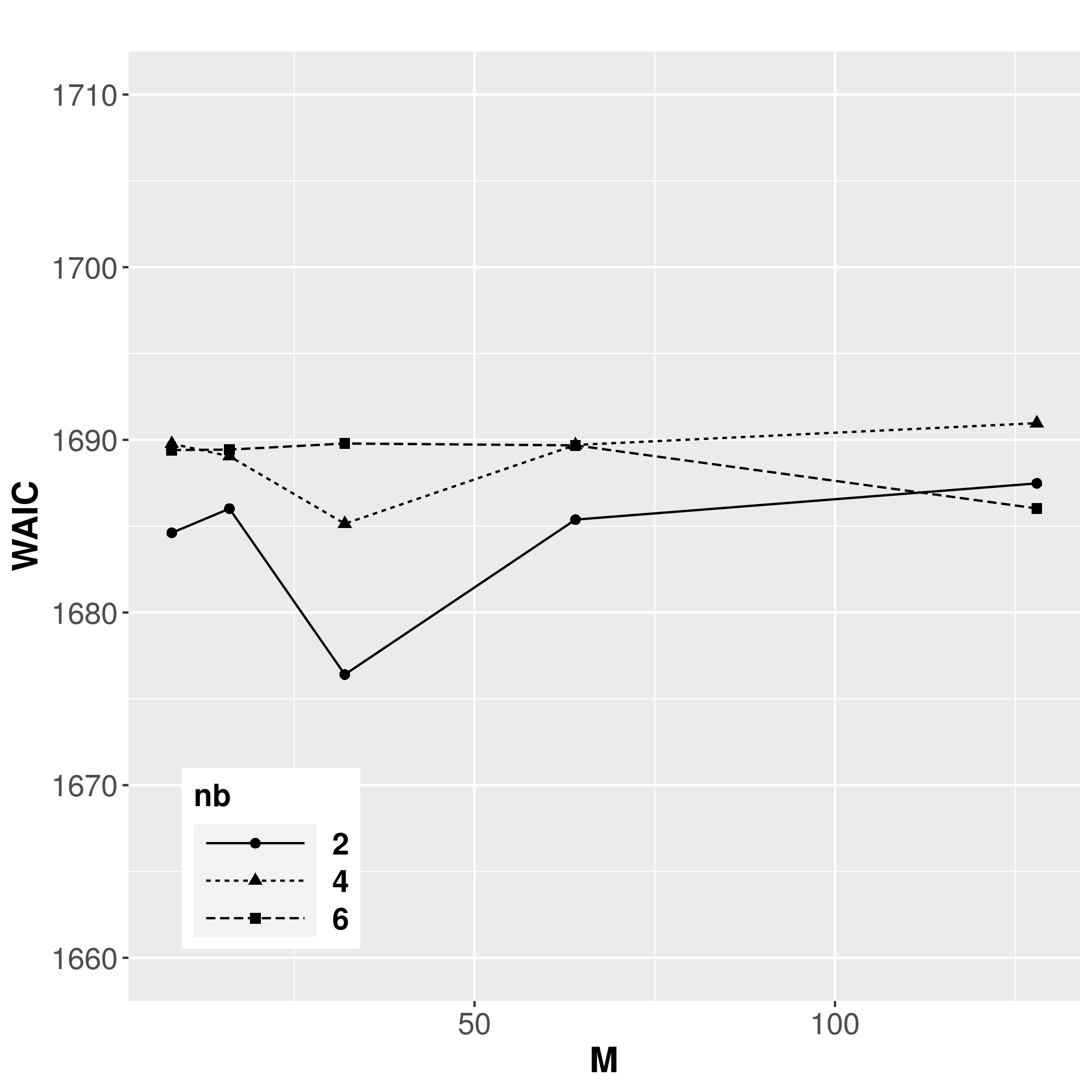} 
\includegraphics[scale=0.3]{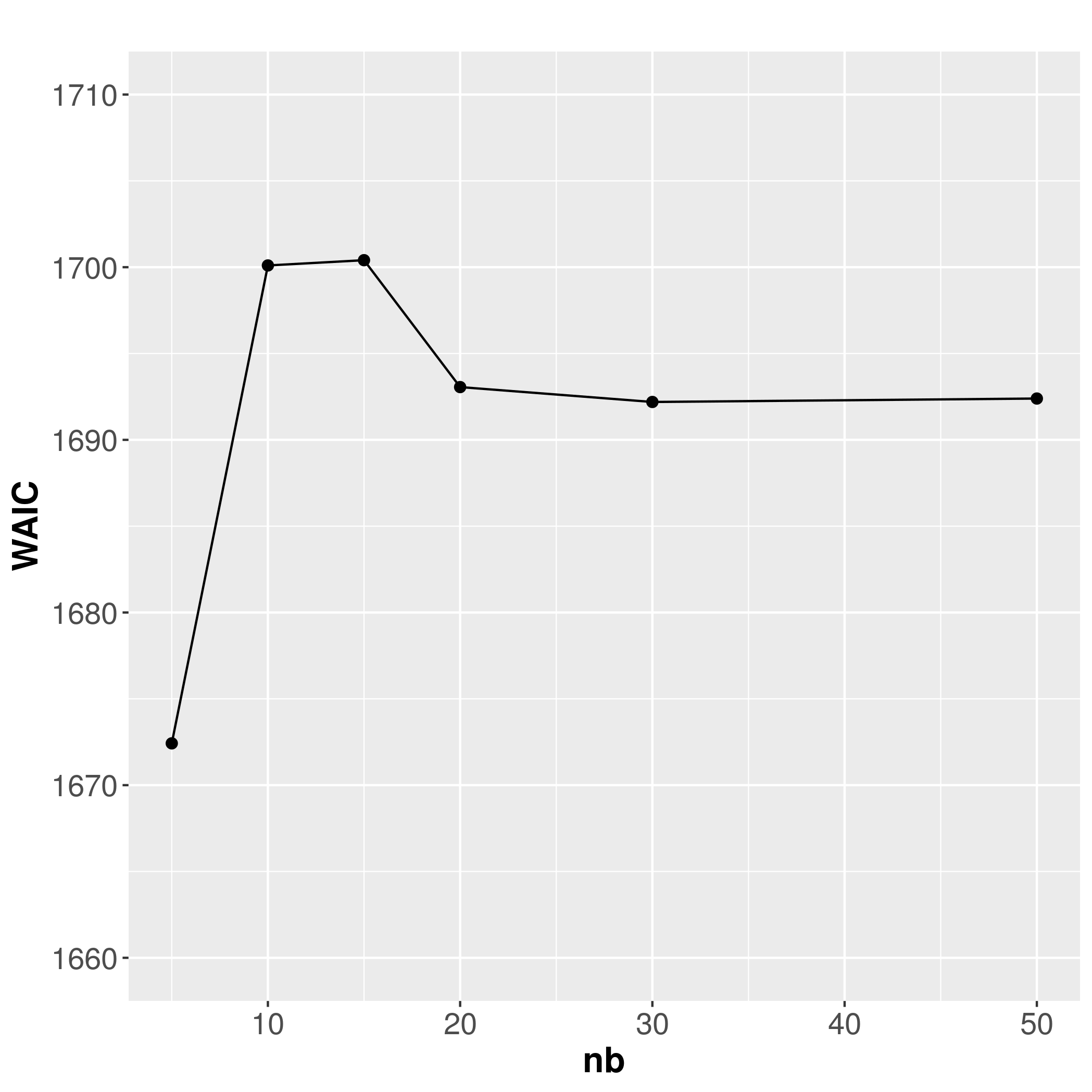} \\
\end{center}
\caption{\small  SIM III ($\phi$ = 3). INLA results. Criteria assessment:  Running times (first row), LPML (second row)  and WAIC (third row), under block-NNGP models using regular blocks (left column), irregular blocks (middle column) and NNGP models (right column). }
  \label{fig:fig3}
\end{figure}

The criteria assessment and time requirements for scenarios SIM I ($\phi = 12$) and SIM II ($\phi = 6$) are reported in the supplementary material C. 
The results obtained confirm that the NNGP and block-NNGP models show a very good performance when the range is not too large.  
Here we present the results for the scenario SIM III, corresponding to a process that has a large effective range.  Figure~\ref{fig:fig3} shows the running time, LPML and WAIC for block-NNGP models using regular or irregular blocks and NNGP models. We observe that the running time for all the block-NNGP models is fast, even for block-NNGP models with a high number of neighbors, for instance using $\ge36$ blocks, it takes less than $800$ seconds.   
We also remark a slightly time reduction for block-NNGP models with irregular blocks. In particular, the time requirements decreases when the number of blocks $M$ increases or the number of neighbor $nb$  decreases. Computing times requirements for the NNGP are lower due to the lower number of neighbors, but the running time for NNGP models increases as the number of neighbor increases too.

The LPML and WAIC seems to be quite similar for block-NNGP models with $nb = 4$ and $nb=6$ neighbor blocks, and the LPML  shows a better goodness of fit, as well as the number of neighbors is increased. We can also observe that in general the NNGP model needs $\ge30$ neighbors to achieve a similar goodness of fit than the block-NNGP models with $nb = 4$ and $nb=6$ neighbor blocks. The  WAIC of the NNGP models seems to decrease as the number of neighbors decreases, but we need to be careful because following this criteria the best model should be the NNGP model with $nb=5$ neighbors. 

Parameter estimates and the root of mean square prediction error (RMSP) for specific models under SIM III ($\phi=3$) are provided in Table~\ref{tab:tab2}.
 In general, the mean posterior estimates for all the models fitted are close to the true parameter values which are included within credible intervals, but the mean posterior estimates  of  $\beta_0$ for the NNGP are far away from the true values. 
Further the RMSP increases when the number of blocks is increased, being lower for a higher number of neighbor blocks. This result is a little bit more evident for irregular blocks. We also found that the RSMPE for NNGP model with $10$ neighbors is lower than the RSMPE with $20$ neighbors.

\setlength{\tabcolsep}{0.3em}
\renewcommand{\arraystretch}{0.6}
\begin{table}[htb]
  \caption{\small  SIM III ($\phi$ = 3). INLA results.  Summary of mean posterior parameter estimates, parameter posterior summary credible intervals (2.5, 97.5) and criteria assessment.
}
  \centering
\scriptsize
  \begin{tabular}{llllllllll}
    \toprule 
& 	&   \textbf{NNGP}  &  \textbf{NNGP}    & (R)\textbf{M=49} & (R)\textbf{M=64}  & (I)\textbf{M=32}  & (I)\textbf{M=64}   \\ 
&	&    \textbf{(10)}  &  \textbf{(20)}    &  \textbf{nb=4}  &  \textbf{nb=4}   &  \textbf{nb=4} & \textbf{nb=4}   \\ 
   \midrule
\textbf{$\beta_0$}	& 1&0.283 	        &0.412	        &0.594  	    &0.585	        & 0.592           &0.622  \\
                    & &(-0.653,1.053)	&(-0.500,1.203)	&(-0.321,1.501)	&(-0.300,1.459)	& (-0.237,1.504)   &(-0.249,1.523)  \\
\textbf{$\beta_1$}	& 5&4.994	        &4.994          &4.994	        &4.994		    & 4.994	           &4.994 \\
                    & & (4.977,5.011)	&(4.977,5.011)	&(4.977,5.011)	&(4.977,5.011)	& (4.977,5.011)    &(4.977,5.011) \\
\textbf{$\sigma^2$ }& 1&1.081          &1.044          &1.053  	    &1.045	        & 1.019            &1.036   \\
                    & &(0.845,1.250)	&(0.813,1.208)	&(0.829,1.216)	&(0.815,1.209)	& (0.775,1.183)    &(0.804,1.200) \\
\textbf{$\phi$ }	& 3&3.168	        & 3.278         & 3.240	        &3.286		    & 3.437            &3.322   \\
                    & &(2.473,3.717)	&(2.560,3.851)	&(2.559,3.780)	&(2.578,3.852)	& (2.652,4.084)    &(2.597,3.907) \\
\textbf{$\tau^2$}	& 0.1&0.092 	        &0.092          &0.092 	        &0.092 	        & 0.091            &0.092  \\
                    & &(0.088,0.096)	&(0.088,0.096)	&(0.088,0.095)	&(0.088,0.095) 	& (0.087,0.095)    &(0.088,0.095) \\
   \midrule
\textbf{RMSP}	    & &0.444     &0.457		&0.479 	    &0.451	    & \textbf{0.426}     & 0.453 \\
\textbf{Time (sec)} & &240.515   & \textbf{171.927}	&496.271	&359.056	&  687.980  & 335.161\\
   \bottomrule    
    \end{tabular}
\label{tab:tab2}    
\end{table}

\begin{figure}[htb]
\begin{center}
\includegraphics[scale=0.28]{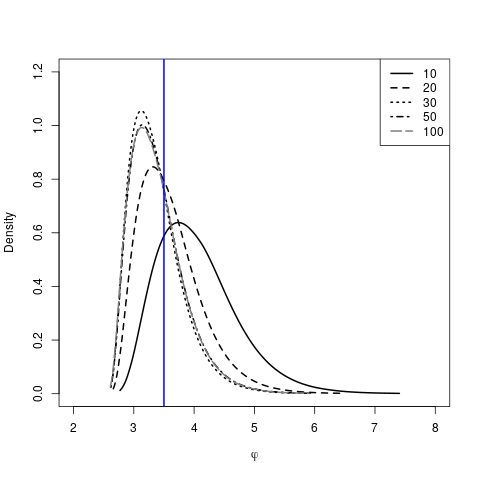}
\includegraphics[scale=0.28]{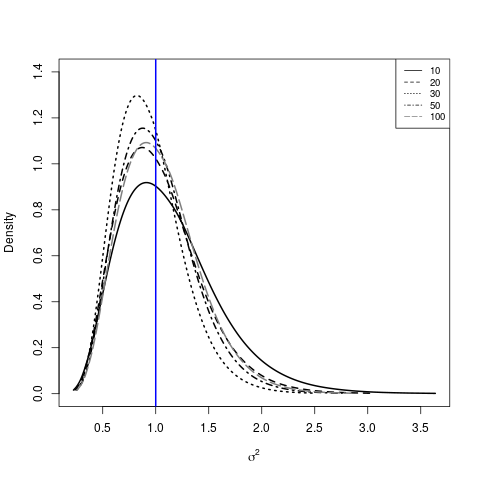}
\includegraphics[scale=0.28]{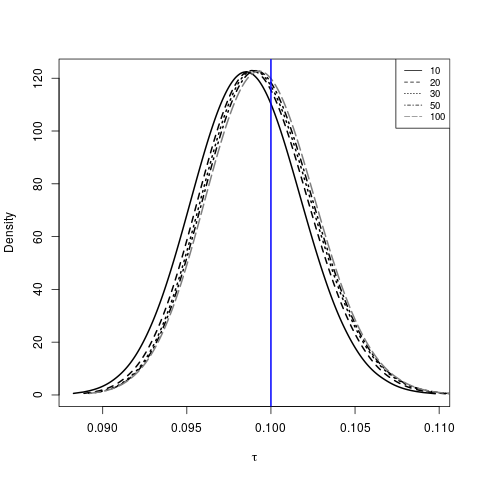}
\includegraphics[scale=0.28]{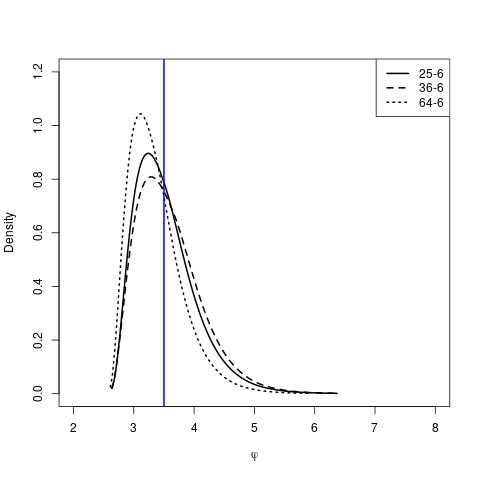}
\includegraphics[scale=0.28]{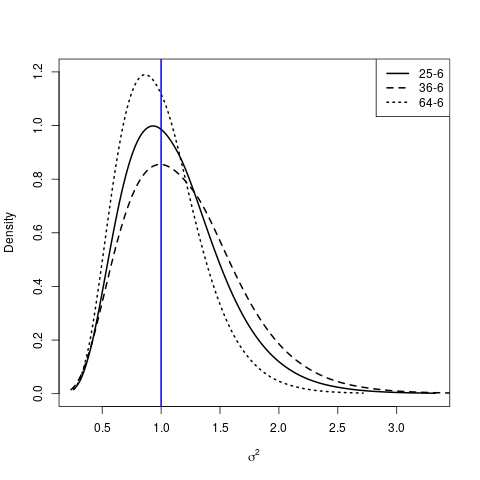}
\includegraphics[scale=0.28]{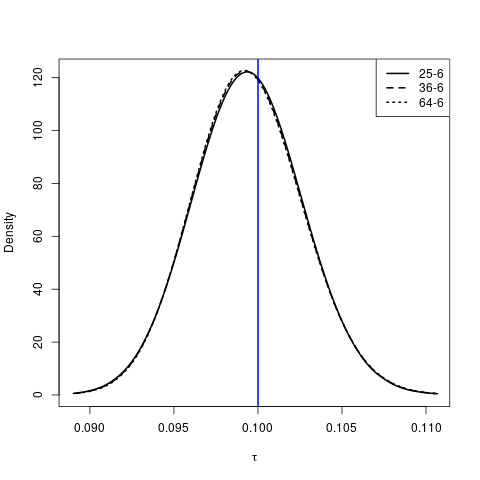}
\end{center}
\caption{\small Posterior marginal densities of the spatial decay parameter ($\phi$), the marginal variance ($\sigma^2$)  and nugget effect ($\tau$) for NNGP models (upper panel) with $nb = 10, 20, 30, 50, 100$ neighbors and block-NNGP models (lower panel) with $M = 25, 36, 64$ regular blocks and $nb=6$ neighbor blocks. The solid vertical blue lines represent the true parameter values. }
  \label{fig:figs7}
\end{figure}

The previous results for large effective range are magnified when the spatial process is smoother. To illustrate this statement, we simulate data from the same model, but 
we set $\nu=1.5$, $\phi=3.5$, $\sigma^2=1$, $\bm{\beta}= (1, 5)^T$ and  $\tau^2 = 0.1$. 
We fit the next models using \texttt{R-INLA}: NNGP models ($nb = 5, 10, 15, 20, 30, 50$) and regular (R) block-NNGP models where  $M = 9,  16, 25,  36, 49, 81, 100$ and $nb = 2, 4, 6.$ 
The approximate posterior marginals for the hyperparameters  
shows that the block-NNGP models give more accurate results than the NNGP models (Figure~\ref{fig:figs7}). For instance, the posterior marginals for $\tau^2$ with different number of blocks are overlaid, 
while the posterior marginals for $\sigma^2$ and $\phi$ with different number of blocks do not significantly change. A different result is obtained for NNGP models, where the posterior marginals for $\phi$ with different number of neighbors dramatically change.

\begin{figure}[htb]
\begin{center}
\includegraphics[scale=0.22]{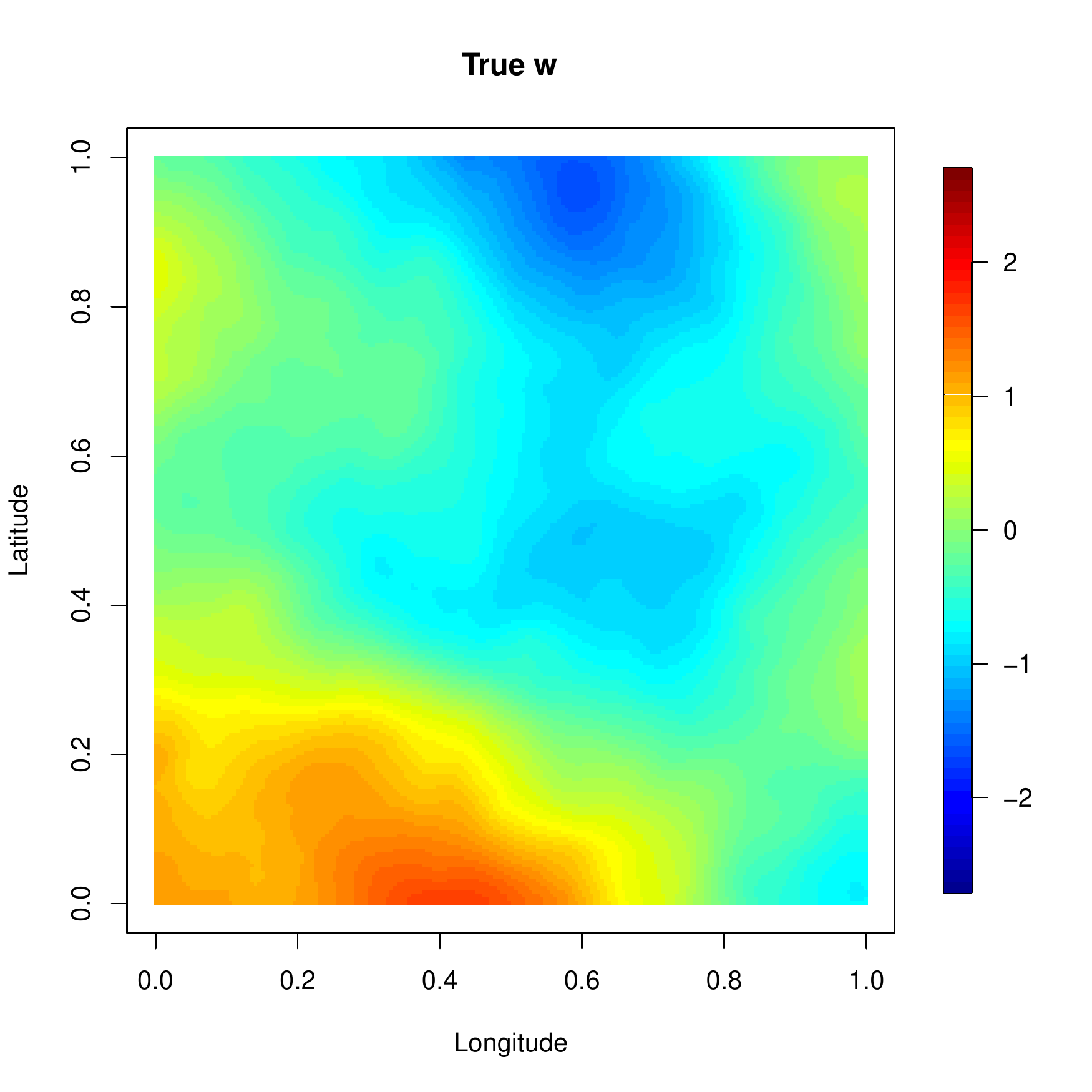}
\includegraphics[scale=0.22]{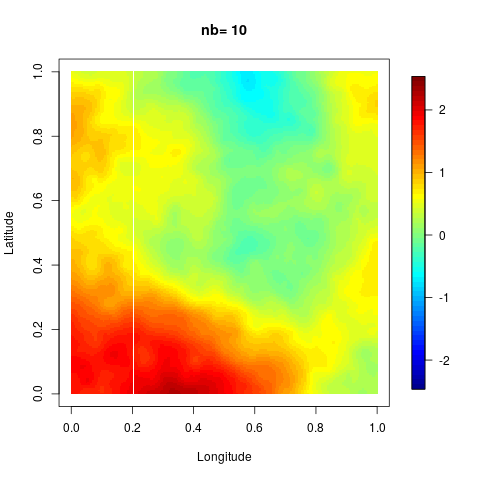}
\includegraphics[scale=0.22]{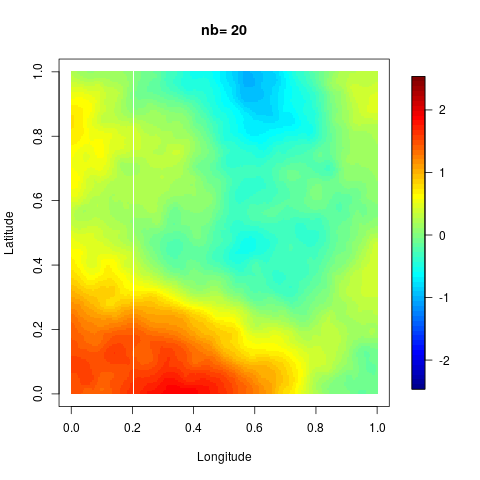}
\includegraphics[scale=0.22]{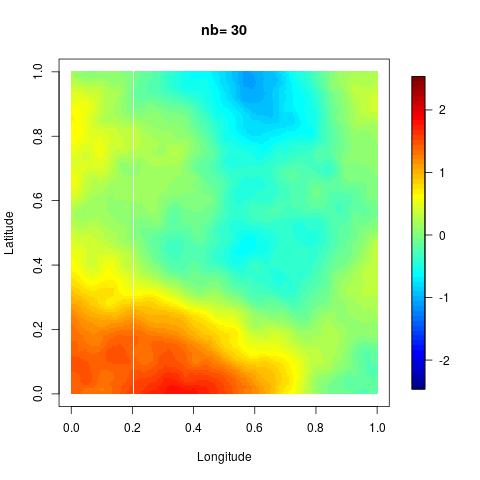}
\includegraphics[scale=0.22]{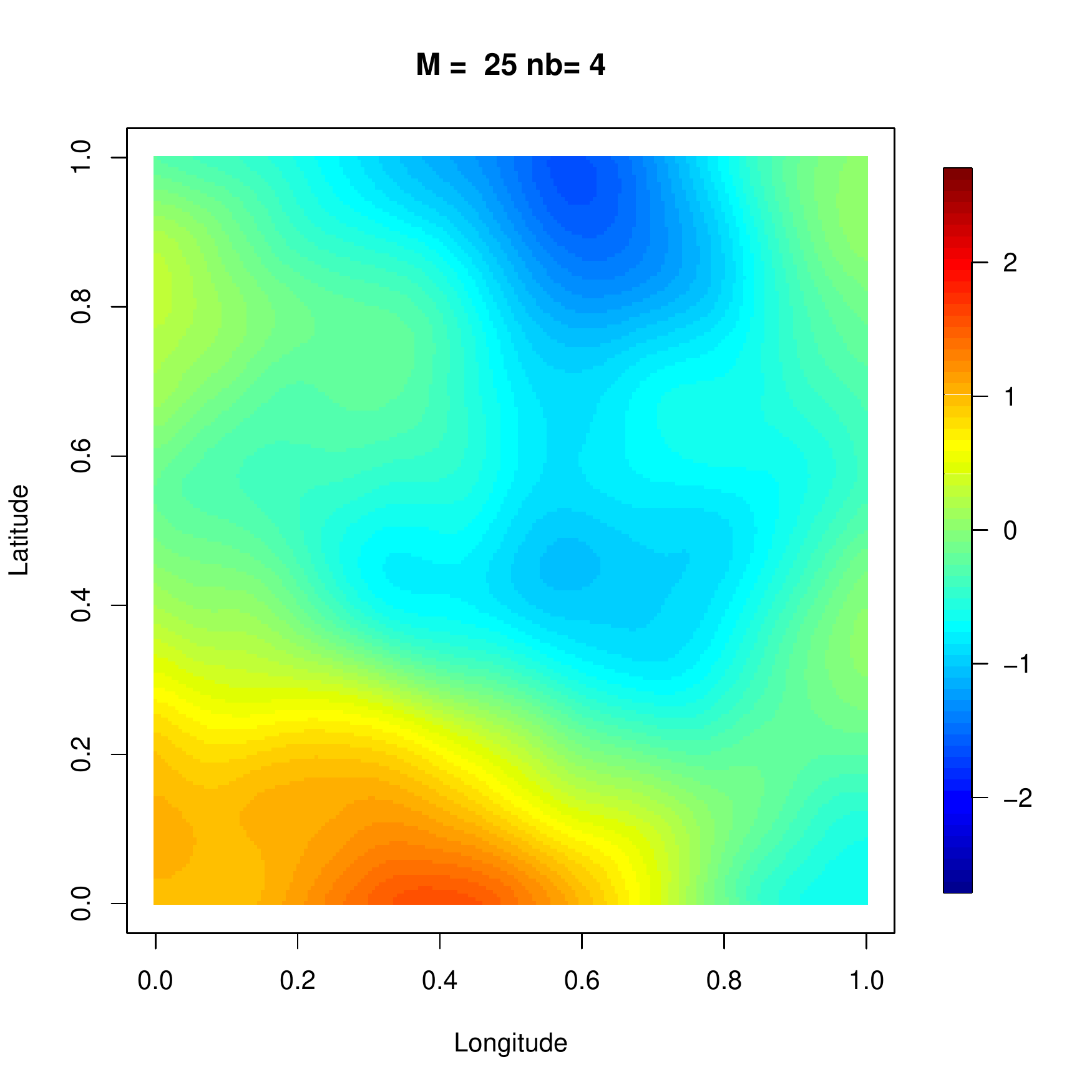}
\includegraphics[scale=0.22]{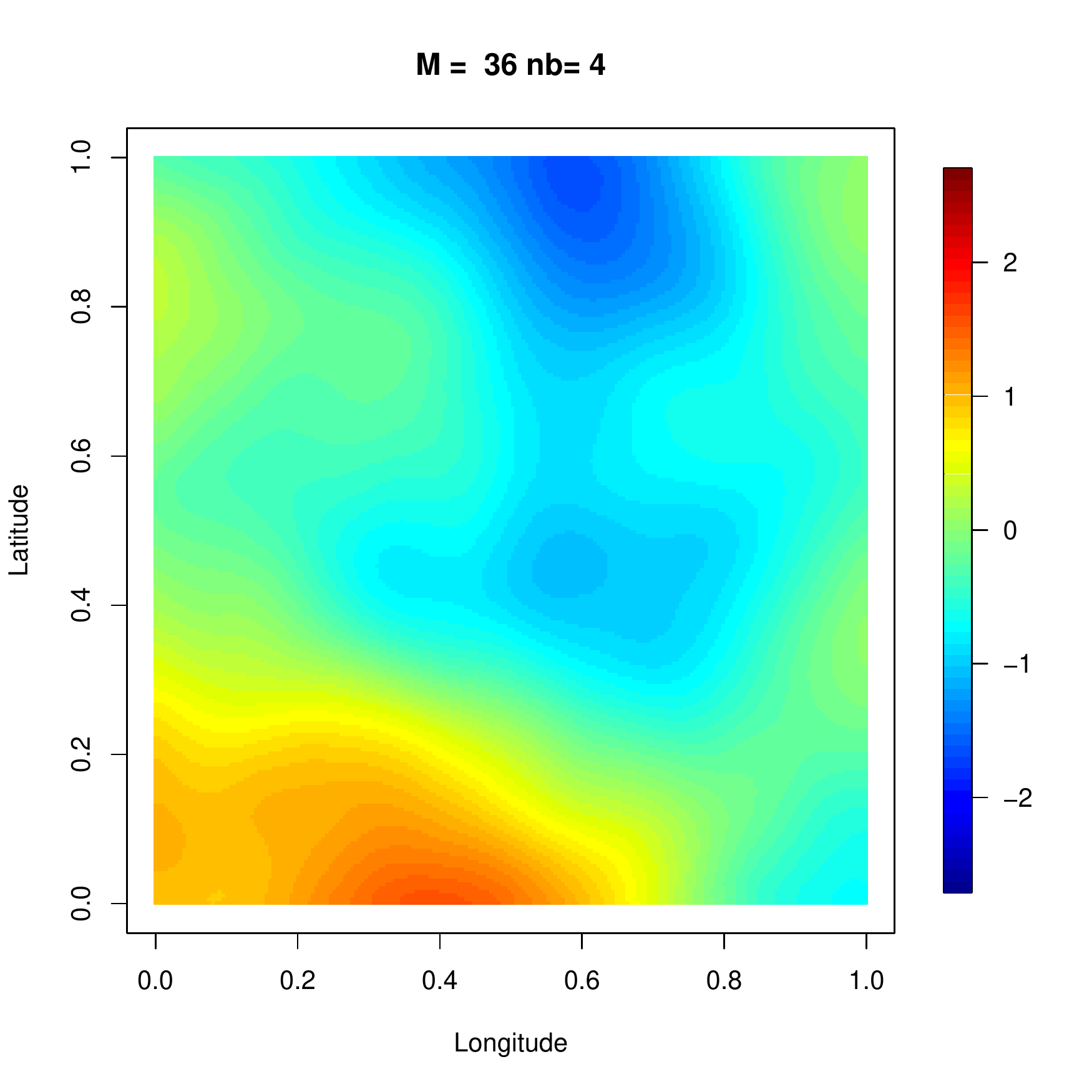}
\includegraphics[scale=0.22]{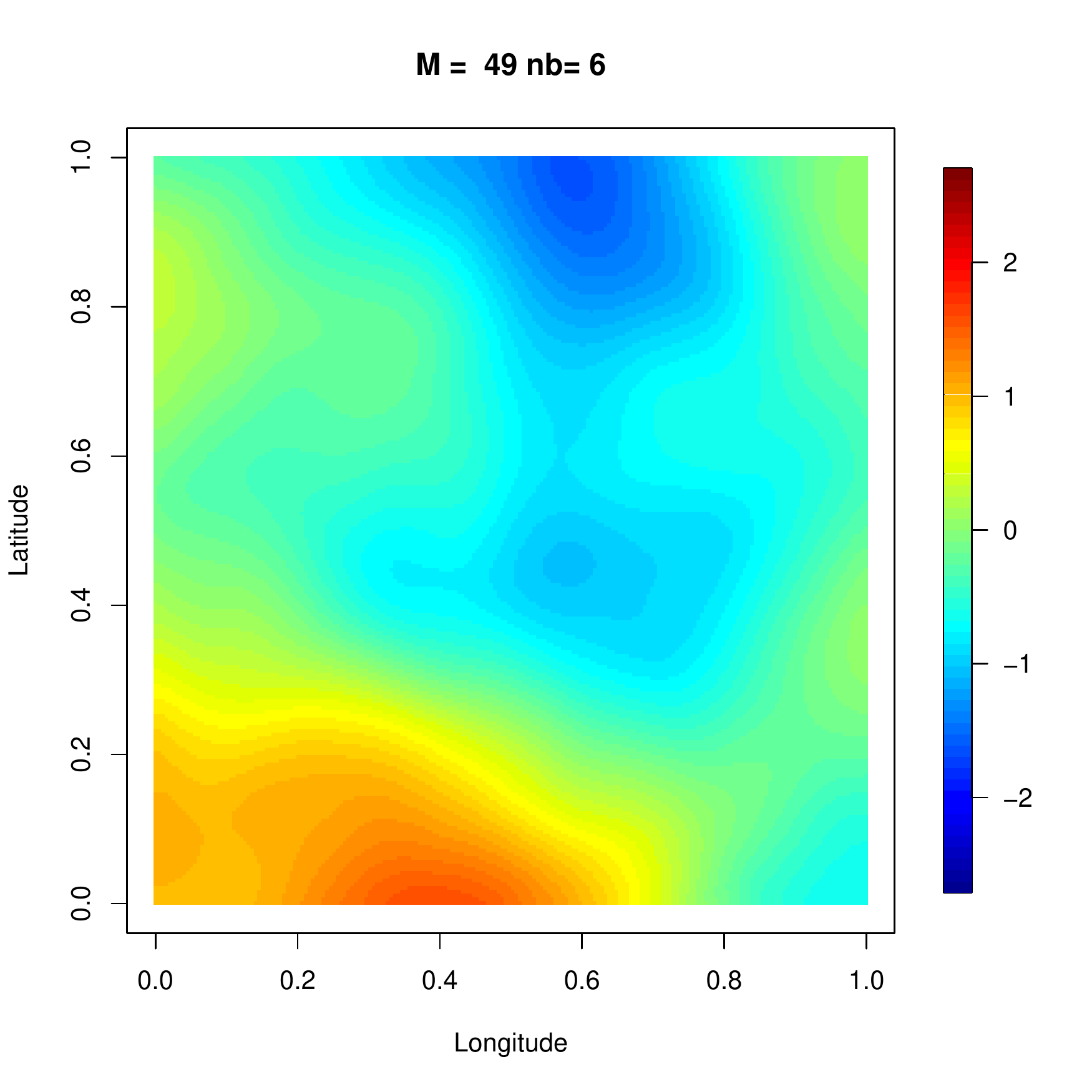}
\includegraphics[scale=0.22]{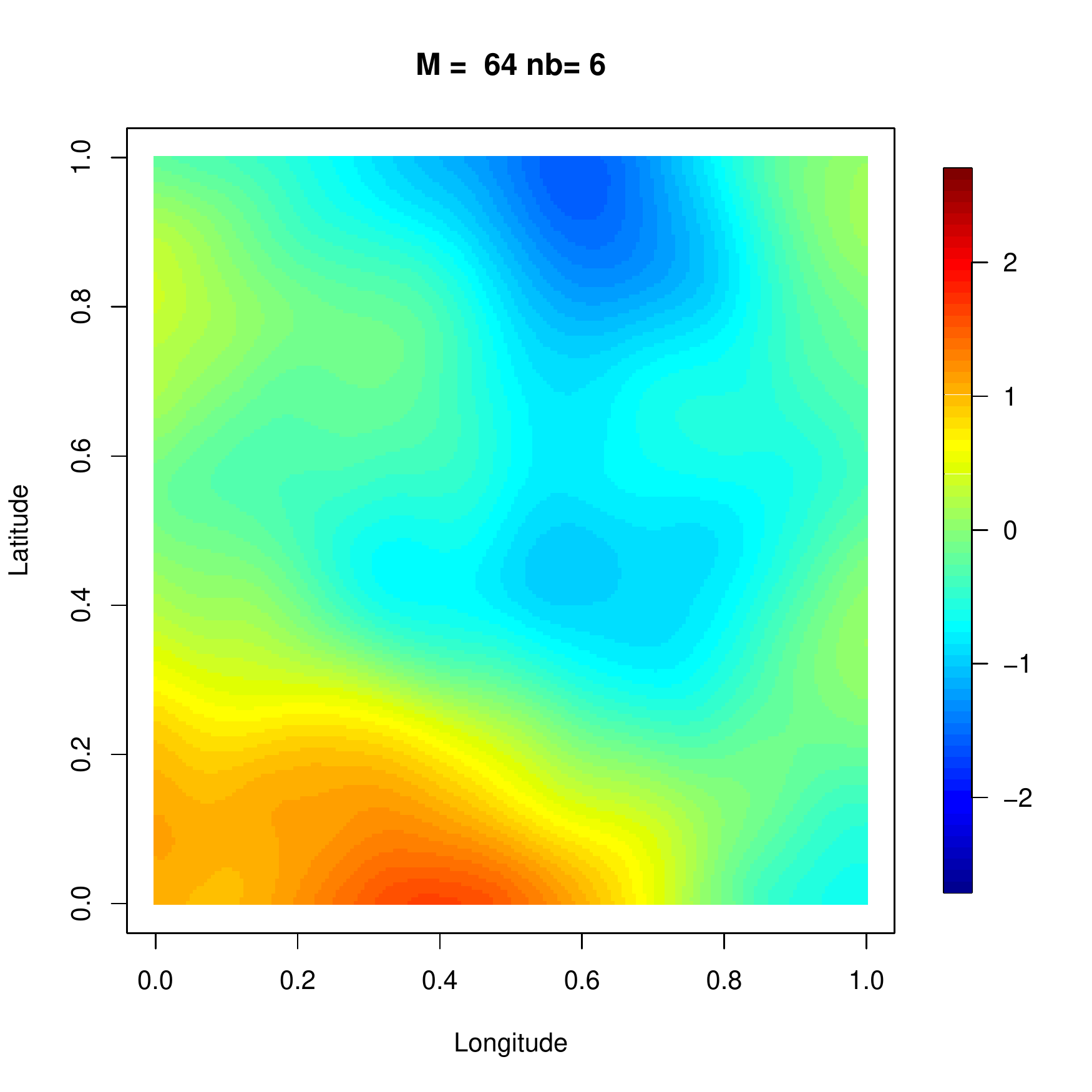}
\end{center}
\caption{\small  True spatial random effects $w$ with Mat\'{e}rn covariance function ($\nu=1.5$, $\phi=3.5$, $\sigma^2=1$ and $\tau^2=0.1$), and posterior mean estimates for NNGP models (upper panel) with $nb=10, 20, 30$  neighbors, and different block-NNGP models (lower panel) with regular blocks. }
  \label{fig:fig9}
  \end{figure}

The maps of the mean posterior estimates for block-NNGP models with regular blocks and  NNGP models (Figure~\ref{fig:fig9}) shows the inaccuracy of NNGP models when the number of neighbors  is small, they underestimate the spatial effects. 
In fact, a misspecified number of neighbors is more problematic for smoother processes. While the block-NNGP models show a good performance for different number of blocks. Indeed, the NNGP has proven to be successful in capturing local/small-scale variation of spatial processes, however, it might have one  disadvantage: inaccuracy in representing global/large scale dependencies. This might happen because the NNGP built the DAG based on observations, adversely, the block-NNGP built a chain graph based on blocks of observations, which captures both small and large dependence.

\section{Applications}
\label{s:app}

In this section, we illustrate the main advantages of the application of block-NNGP models throughout real data. 
First, the block-NNGP model is suitable for large data $(\ge10000 \mbox{ samples})$, this will be illustrated in analysing the mining  data, available by \citet{Eidsviketal:2014}. Second, the block-NNGP is suitable for data with large ranges and it can easily be combined with other model components within the general framework of latent Gaussian models and fitted efficiently using \texttt{R-INLA}. This is demonstrated in analysing the precipitation data in South America (available in \texttt{R} as \verb rnoaa::cpc_prcp ). 

\subsection{Mining data}
Here we study joint-frequency data acquired in an iron mine in the northern part of Norway \citep{Eidsviketal:2014}. 
We split the joint-frequency data in two subsets, a set composed by a random subset of $11000$  locations for estimation and the remaining $701$ observations were withheld to assess predictive performance. The joint-frequency data is modelled throughout a Gaussian geostatistical model, including only the effect of depth as covariate, along with an intercept, and a block-NNGP or NNGP using an exponential covariance function. The NNGP models are based on $10$ and $20$ neighbors, while the block-NNGP models are based on $280$ and $450$ regular blocks or $128$ and $289$ irregular blocks, and one neighbor block.  
Then we fit the models using \text{INLA} and our own implementation of the collapsed-MCMC \citep{Finleyetal:2019}. 

Table~\ref{tab:tab3} shows the parameter estimates and performance metrics, as expected, the performance of block-NNGP model is better for less blocks. We observe that the MCMC pre-processing takes approximately a month to run (based upon $10000$ iterations), while in INLA it takes a few hours or minutes. 
In particular, the time requirement for the block-NNGP model using irregular blocks was relatively small, although we use a small number of blocks, a possible explanation is that when using regular blocks, some blocks have a higher dimension. The goodness of fit of block-NNGP models is good, specifically for the root of mean square of error (RMSE), nevertheless, the WAIC and RSME shows a slightly better performance for NNGP models. 

\setlength{\tabcolsep}{0.3em}
\renewcommand{\arraystretch}{0.6}
\begin{table}[htb]
  \caption{\small  Mining data. Summary of mean parameter estimates. Criteria assessment.
}
  \centering
\scriptsize
  \begin{tabular}{lllllllll}
    \toprule 
& \textbf{INLA} & & & & & \textbf{MCMC}\\    
\textbf{Model}	&   \textbf{NNGP}  &  \textbf{NNGP}    & (R)\textbf{M=450} & (R)\textbf{M=280}  & (I)\textbf{M=128}  & (I)\textbf{M=289}   \\ 
		&    \textbf{(20)}  &  \textbf{(10)}    &  \textbf{nb=1}  &  \textbf{nb=1}   &  \textbf{nb=1}  &  \textbf{nb=1} \\ 
   \midrule
\textbf{$\beta_0$}	&-0.022	&-0.023	&0.014	&0.015	&0.014 & -0.02\\
\textbf{$\beta_1$}	&0.0004	&0.0004	&0.0004	&0.0004	&0.0004& 0.0\\
\textbf{$\tau^2$}	&0.098	&0.098	&0.099	&0.098	&0.098 & 0.1\\
\textbf{$\sigma^2$ }&0.159  &0.160	&0.156	&0.157	&0.157 & 0.16 \\
\textbf{$\phi$ }	&0.074	&0.072	&0.074	&0.075	&0.076 & 0.07\\
   \midrule
\textbf{LPML}	    &-4913.948    &	\textbf{-4910.236}	&-4983.081	&-4962.783	&-4957.957 &\\
\textbf{WAIC2}      &\textbf{9108.287}     &	9123.362	&9248.598	&9188.431	&9158.288 &\\
\textbf{RMSE}	    &0.254	      & 0.255	    &0.256	    &0.254	    &\textbf{0.253} &\\
\textbf{RMSP}	    &\textbf{1.096}	      & \textbf{1.096}	    &1.176	    &1.223	    &1.157  &\\
\textbf{time (sec)} &3030.121     &	2806.641	&7127.620	&7381.386	&7573.818 & 2912400\\
   \bottomrule    
    \end{tabular}
\label{tab:tab3}    
\end{table}
  
We also study the predictive performance across the models. This comparison is done over 701 randomized leave-out samples. We predict the joint frequency using the predictive distribution (see Section~\ref{s:inference}). Among the block-NNGP models, the RMSP shows a better predictive performance of the block-NNGP model with $M=128$ irregular blocks and $nb=1$ neighbor block.
Finally, maps of interpolated posterior mean estimates of joint-frequency data are displayed in the supplementary material D. We see little difference between these models. This emphasizes the small differences in the marginals for covariance parameters or the regression effects.

\subsection{Precipitation data}
\label{s:app2}

We also analyzed the precipitation data from the National Oceanic and Atmospheric Administration (NOAA) on January, 2017, over a region covering South America.   The data consist of $n=6000$ samples for estimation and $104$ to validate the models. 
Let $Y(s_i)$ be the observed precipitation in the location $s_i$, for $i=1, \dots, n$, we assume that $Y(s_i)\mid \bm{x}, \bm{\theta} \sim \mbox{Gamma}(\mu_i, \theta_0)$ where  $\theta_0$ is a precision parameter and the mean $\mu_i$ is defined as follows:
\begin{equation}
\log(\mu_i) = \bm{Z}^T(s_i) \bm{\beta} +  w(s_i),  \nonumber 
\end{equation}
where $\bm{\beta}$ is a vector of coefficient regression parameters, $\bm{Z}(s_i) = (1, z_1(s_i))^T$, $ z_1$ representing the latitude, $w(s_i)$ is a  spatial structured random effect, and  the precipitation field $\bm{w}=(w(s_1), \dots, w(s_n))^T$ follows a block-NNGP or a NNGP, with a  covariance function that depends on an exponential covariance function $\bm{{C}(\sigma^2, \phi)}$. 

We fit the block-NNGP models choosing a different number of irregular blocks $(M=64, 128)$ and $nb = 6$  neighbor blocks and NNGP with different number of neighbors $(nb =10, 20, 30, 50, 100)$. Full Bayesian inference was carried out using  INLA.
Based on the criteria assessment (see supplementary material D) the best NNGP model is the one with $nb=100$ neighbors, followed by the model with $nb=20$ neighbors.  While the best block-NNGP model is the one with $M=64$ and $nb=4$ neighbor blocks. Overall, the results are more {stable} for less blocks and more neighbor blocks. 
The RSME is lower for block-NNGP models. Note that all the models run in a reasonable time, 
for instance 
the  block-NNGP model with $M=64$ and $nb=4$ neighbor blocks (with approximately $376$ neighbors) runs in 13587.350 seconds.

\begin{figure}
\begin{center}
\includegraphics[scale=0.27]{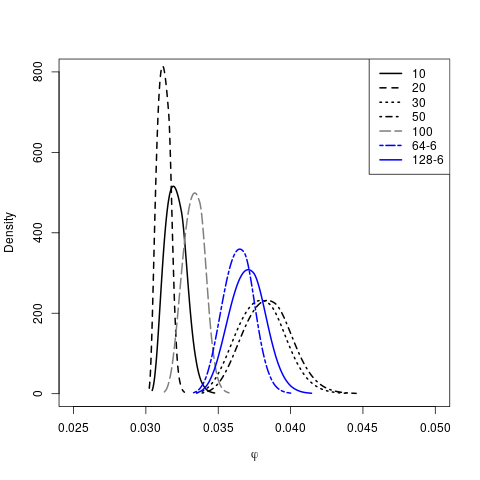}
\includegraphics[scale=0.27]{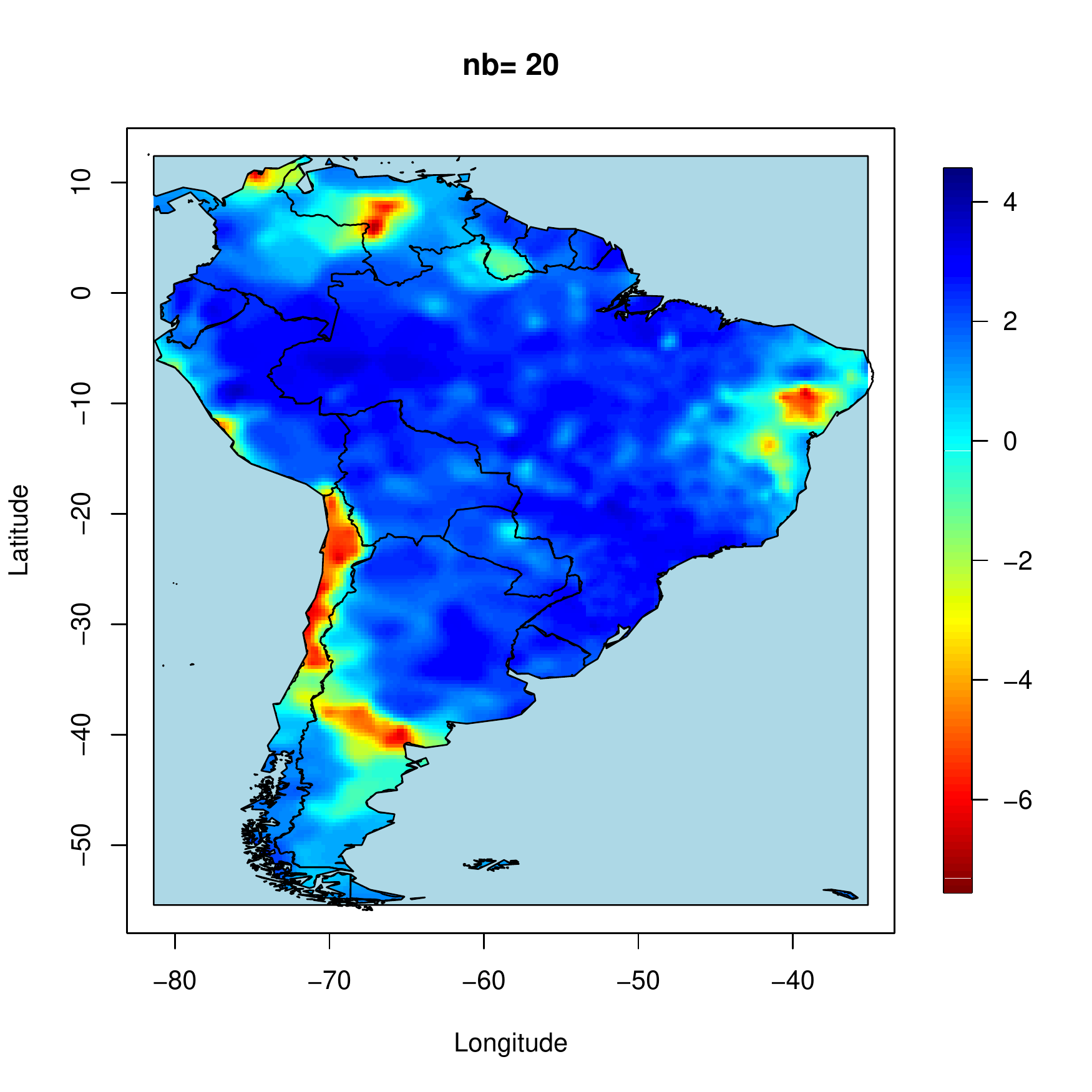}
\includegraphics[scale=0.27]{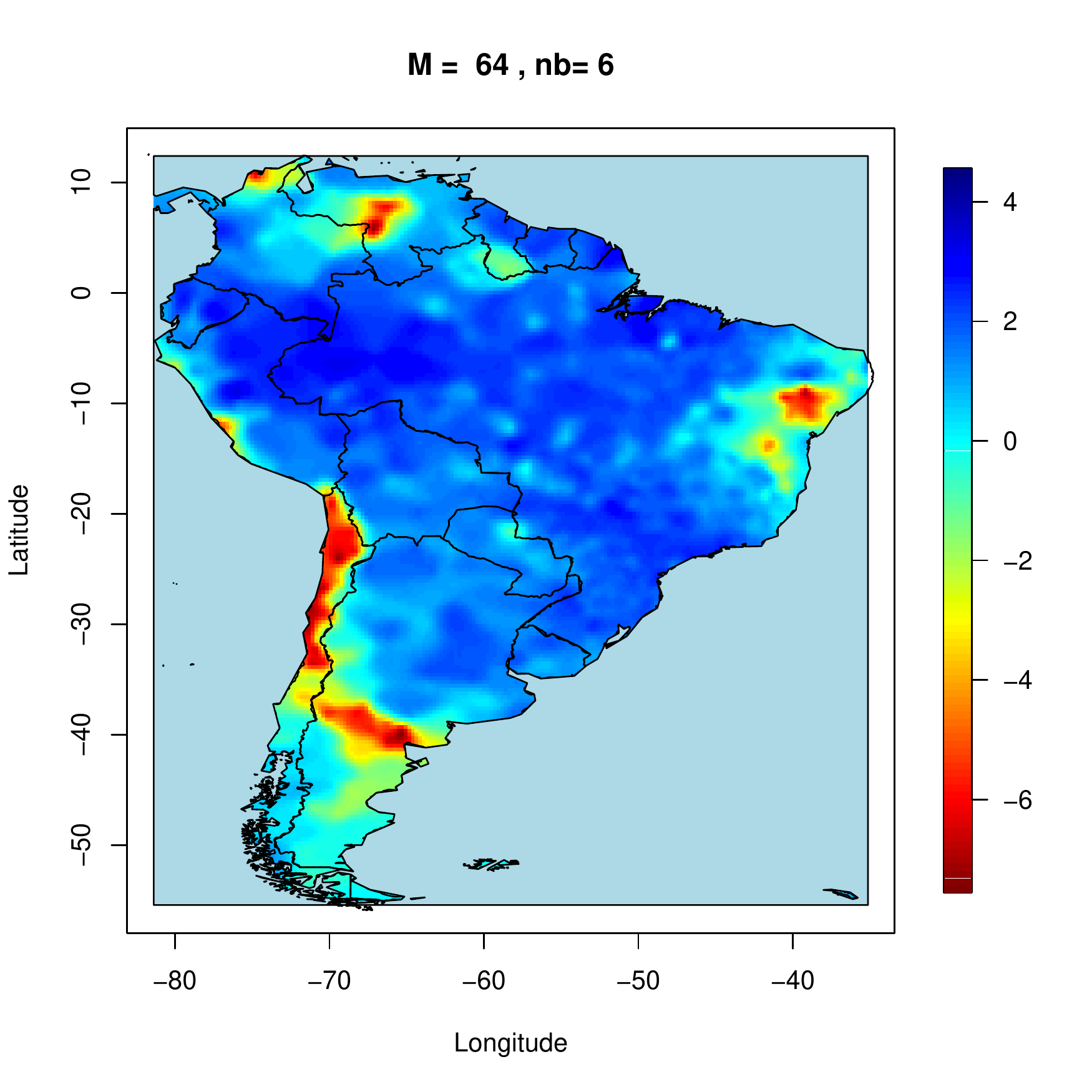}
\includegraphics[scale=0.27]{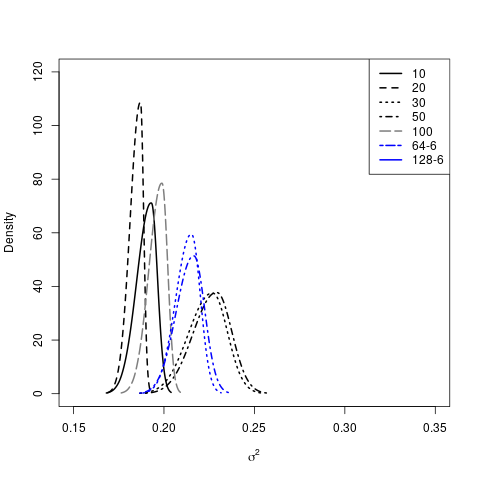}
\includegraphics[scale=0.27]{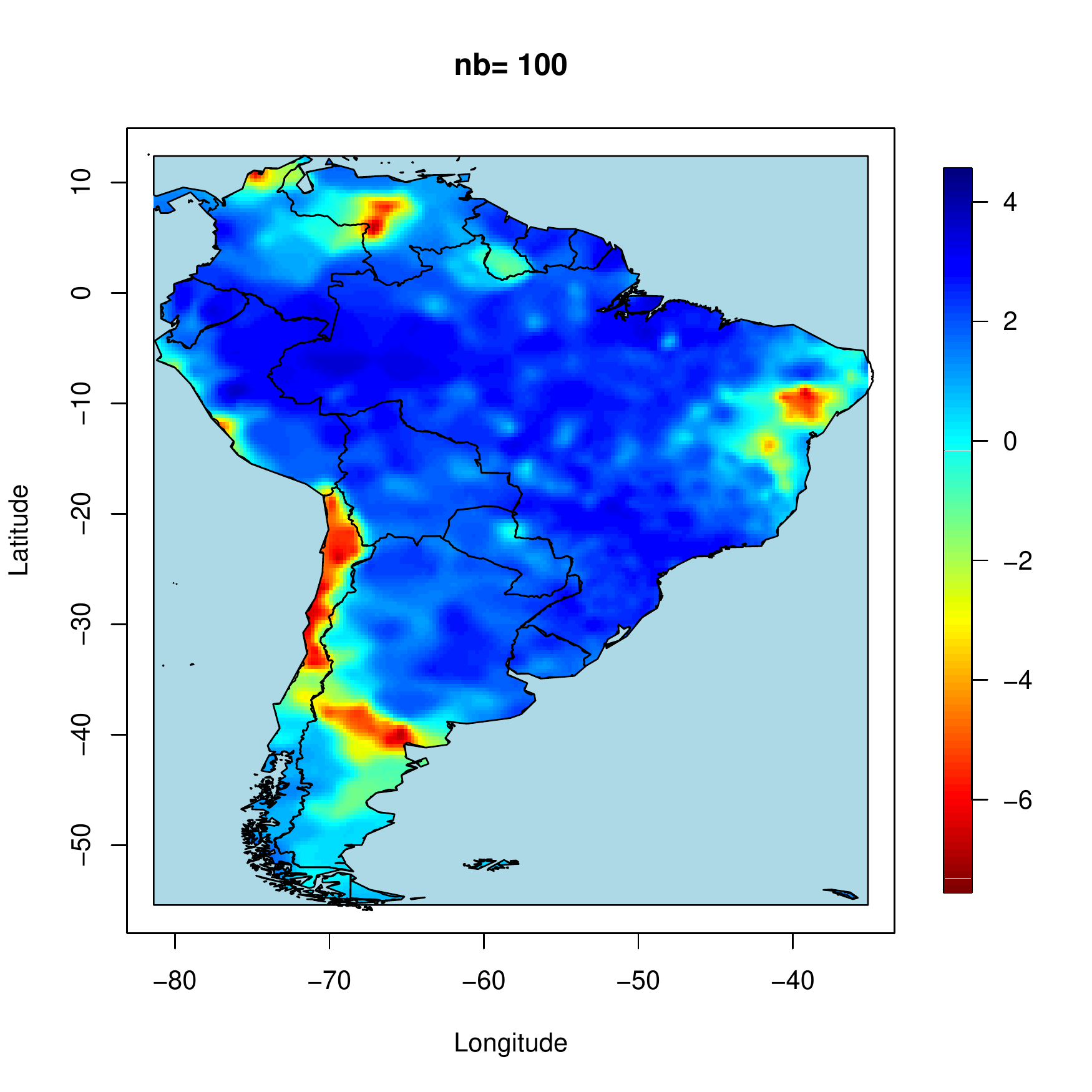}
\includegraphics[scale=0.27]{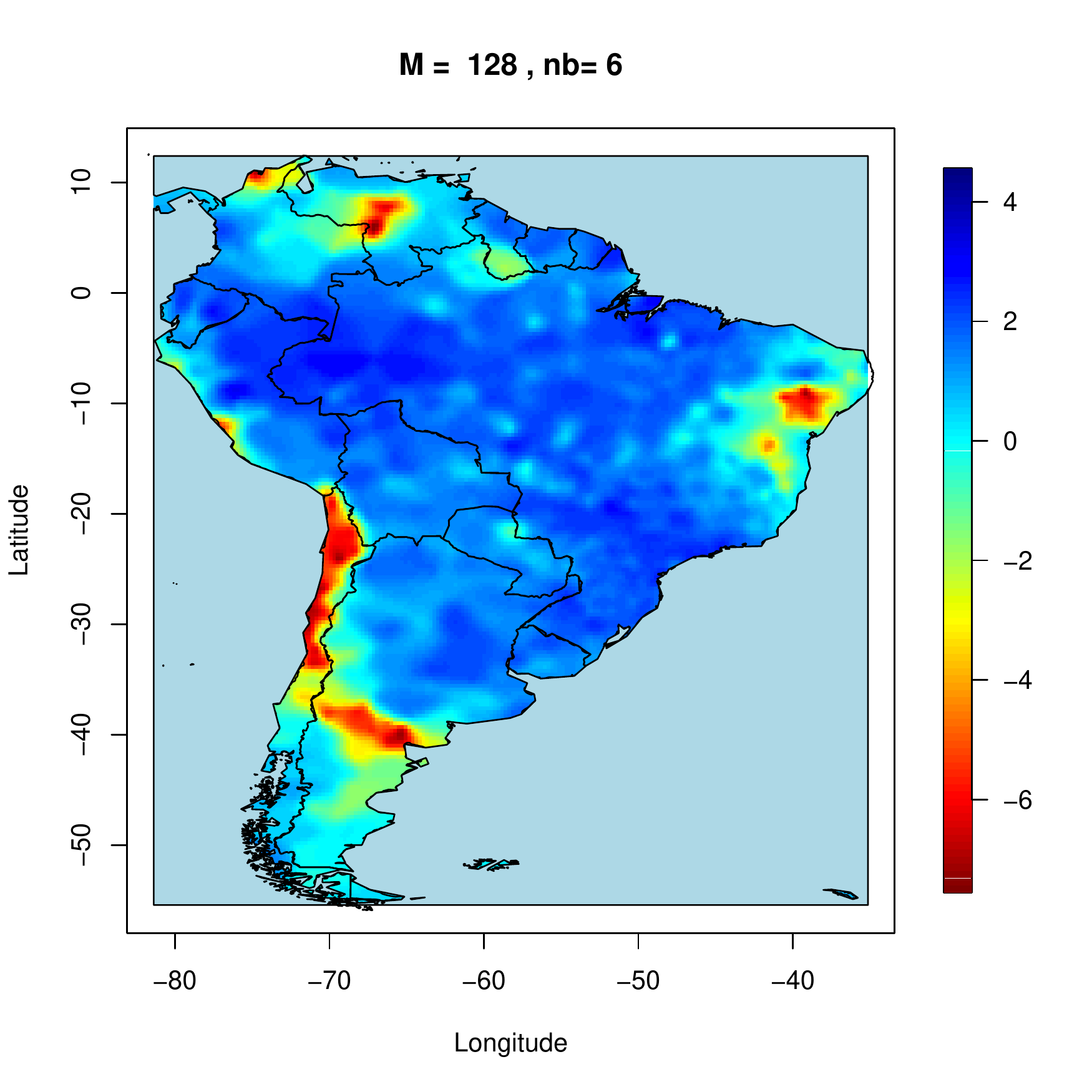}
\end{center}
\caption{\small First column: Posterior marginal densities of the spatial decay parameter $\phi$ and the marginal variance $\sigma^2$ for NNGP models with $nb = 10, 20, 30, 50, 100$ neighbors and block-NNGP models with $M = 64, 128$ irregular blocks and $nb=6$ neighbor blocks. Second column: Posterior mean estimates of spatial random effects for NNGP models. Third column: Posterior mean estimates of spatial random effects of  block-NNGP models. }
  \label{fig:fig8}
\end{figure}

Figure~(\ref{fig:fig8}) displays the marginal posterior densities for the $\phi$ and $\sigma^2$. We observe that there is not a clear pattern for NNGP models, for a small number of neighbors ($10$ and $20$), the mean posterior of $\phi$ hits the boundary of its prior uniform distribution, it increases for $30$ and $50$ neighbors, but it decreases again for $100$ neighbors.  A similar pattern is observed for $\sigma^2$. While, the block-NNGP models with a considerable number of neighbor blocks, ensures a quite similar estimation of $\phi$ and $\sigma^2$. Posterior means of the spatial fields under different models are also shown in Figure~(\ref{fig:fig8}).

\section{Conclusion}
\label{sec:conc}

This paper introduces the block-NNGP, a new GMRF for approximating Gaussian processes. An important feature is that this process is valid for any covariance function. Moreover, the precision matrix of the block-NNGP has a block-sparse structure, which allows scalable inference and distributed computations. Although there might not be an explicit number of blocks and neighbor blocks for optimal blocking, we are able to determine an adequate one by comprising the computational speed and statistical efficiency. Further, inference can be performed by the use of MCMC methods as well as the INLA approach \citep{Rue:2009}. However, we advocate the use of the INLA methodology.

INLA requires the latent field to be a GMRF as in our proposal, hence it provides a huge computational benefit, providing precise estimates in seconds or minutes. Furthermore, the combination of INLA and the block-NNGP directly allows for fast inference for non-Gaussian observations (e.g., Section~\ref{s:app2}). Scalable models for massive non-Gaussian data have been a challenge in the literature \citep{lee2019picar} with very few option available, we believe that the block-NNGP within INLA is an interesting alternative to perform such modeling in massive dataset preserving the characteristics of the data, e.g, avoiding transformation to achieve normality. 

Using theoretical results, large simulated data and real-data applications we have shown that the block-NNGP can provide a good approximation at low computational complexity and time. In particular, the block-NNGP performs better than the NNGP when the range is large or the spatial process is smoother. Extension of the block-NNGP process, e.g., multivariate and spatio-temporal extensions, will be investigated in future works. 

\bigskip
\begin{center}
{\large\bf SUPPLEMENTARY MATERIAL}
\end{center}

The supplementary material presents the formal proofs of the Theorems, Lemmas and Prepositions. Also, the full-MCMC algorithm is proposed. Finally, detailed simulation and application results are discussed.


\bibliographystyle{apalike}
\bibliography{biblio}

\newpage

\bigskip
\begin{center}
{\large\bf SUPPLEMENTARY MATERIAL}
\end{center}

\subsection*{A. Proofs of main results }

\textit{\textbf{Matrix Analysis Background}}
\\
\textit{{Theorem A1}: A matrix $\bm{B} \in \bm{\Re}^{m\times n}$ is  
 full column rank if and only if $\bm{B^{T}B}$ is invertible
}\\
\textit{{Theorem A2}: The determinant of an $n\times n$ matrix $\bm{B}$ is  
0 if and only if the matrix $\bm{B}$ is not invertible.
}\\
\textit{{Theorem A3}: Let $\bm{T_n}$ be a triangular matrix (either upper or
lower) of order n. 
Let $\det(\bm{T_n})$  be the determinant of $\bm{T_n}$. 
Then $\det(\bm{T_n})$ is equal to the product of all the diagonal elements of $\bm{T_n},$ 
that is, $\det(\bm{T_n}) = \prod_{k=1}^{n}(a_{kk}).$
}\\
\textit{{Proposition A1}: If $\bm{B}$ is  positive definite (p.d.), then 
 if $\bm{S}$ has full column rank, then $\bm{S^{T}BS}$ is positive definite. 
}\\
\textit{{Corollary A1}: If $\bm{B}$ is  positive definite, then 
 $\bm{B^{-1}}$ is positive definite. 
}

\begin{proof}[\textbf{Proof of Proposition 1}]

If $\pi(\bm{w_S})$ is a  valid multivariate joint density, $\pi(\bm{{w}_{b_k}} | \bm{{w}_{N(b_k)}})$ is also proper, and we have that $\int \pi(\bm{w_{b_k}}|\bm{w_{N(b_k)}}) d\bm{w_{b_k}} = 1, \forall k = 1, \dots, M. $ 
From the definitions of $\bm{G}$ and $\bm{G^b}$ there exists a set of nodes in block $\Delta(b_1)$,  $\bm{s_{\Delta(b_1)}}  \in \bm{G}$, such that ``the last node'' from a DAG $\bm{G^b}$ belongs to $\bm{s_{\Delta(b_1)}}$. Then the nodes in $\bm{s_{\Delta(b_1)}}$ do not have any directed edge originating from them. As a consequence, any node in block $\Delta(b_1)$ can not belong to the set of nodes of any other block. So the term in (2) where all locations of $\bm{\Delta(b_1)}$ appear is $\pi(\bm{w_{\Delta(b_1)}}|\bm{w_{N(\Delta(b_1))}})$. 
From Fubini's theorem, we can interchange the product and integral, thus 
\begin{eqnarray*}
\int {\pi}(\bm{w_S}) d \bm{w_S} &=& \int\dots\int \prod _{i=1}^M  \pi(\bm{w_{\Delta(b_i)}}|\bm{w_{N(\Delta(b_i))}}) d\bm{w_{\Delta(i)}} \\ 
 &=& \int\dots\int \prod _{i=2}^M  \pi(\bm{w_{\Delta(b_i)}}|\bm{w_{N(\Delta(b_i))}}) d\bm{w_{\Delta(i)}}. 
\end{eqnarray*}
Then, removing every node of $\bm{\Delta(b_1)}$ from $\bm{G}$ and $\bm{G^b}$, we have the chain graph $\bm{G'}$ and DAG $\bm{G'_b}$, respectively. There exists another set of nodes  $\bm{s_{\Delta(b_2)}}$ in $\bm{G'}$, such that ``the last node'' from a DAG $\bm{G'_b}$  belongs to $\bm{s_{\Delta(b_2)}}$. Then the nodes $\bm{s_{\Delta(b_2)}}$  do not have any directed edge originating from them. 
As consequence, any node in block $\bm{\Delta(b_2)}$ can not belong to the set of nodes of any other block. So the term in (2) where all locations of $\bm{\Delta(b_2)}$ appear is $\pi(\bm{w_{\Delta(b_2)}}|\bm{w_{N(\Delta(b_2))}})$.
Applying the Fubini's theorem again, 
\begin{eqnarray*}
\int \pi(\bm{w_S}) d \bm{w_S} 
 &=& \int\dots\int \prod _{i=3}^M  \pi(\bm{w_{\Delta(b_i)}}|\bm{w_{N(\Delta(b_i))}}) d\bm{w_{\Delta(i)}}.   
\end{eqnarray*}
In a similar way, we find $\bm{s_{\Delta(b_3)}}, \dots, \bm{s_{\Delta(M)}}$, such that, \\
$
\int \pi(\bm{w_S}) d \bm{w_S} = \int \prod _{i=1}^M  \pi(\bm{w_{\Delta(b_i)}}|\bm{w_{N(\Delta(b_i))}}) d\bm{w_{\Delta(i)}} = 1. 
$
\end{proof}

\begin{proof}[\textbf{Proof of Proposition 2}]
 Without loss of generality, assume that the data were reordered by blocks. 
From known properties of Gaussian distributions, 
$\bm{w_{b_k}}|\bm{w_{N(b_k)}} \sim  N(\bm{B_{b_k}} \bm{w_{N(b_k)}}, \bm{F_{b_k}})$,  
where $\bm{B_{b_k}} = \bm{C_{b_k, N(b_k)}} \bm{C^{-1}_{N(b_k)}}$ and 
$\bm{F_{b_k}} = \bm{C_{b_k}} - \bm{C_{b_k, N(b_k)}} \bm{C^{-1}_{N(b_k)}}\bm{C_{ N(b_k),b_k}}.$
Hence, 
\begin{eqnarray*}
\widetilde{\pi}(\bm{w_S}) &=& \prod_{k=1}^M p(\bm{w_{b_k}}|\bm{w_{N(b_k)}}) \\
&\propto& \prod_{k=1}^M  \frac{1}{|\bm{F_{b_k}}|^{1/2}} \exp\left\{ -\frac{1}{2} (\bm{w_{b_k}} - \bm{B_{b_k}} \bm{w_{N(b_k)}})^T \bm{F_{b_k}}^{-1} (\bm{w_{b_k}} - \bm{B_{b_k}} \bm{w_{N(b_k)}}) \right\} \\
 &\propto&   \frac{1}{ \prod_{k=1}^M |\bm{F_{b_k}}|^{1/2}} \exp\left\{ -\frac{1}{2} \sum_{k=1}^M  (\bm{w_{b_k}} - \bm{B_{b_k}} \bm{w_{N(b_k)}})^T \bm{F_{b_k}}^{-1} (\bm{w_{b_k}} - \bm{B_{b_k}} \bm{w_{N(b_k)}}) \right\}. 
\end{eqnarray*}
Let $\bm{w_{b_k}} - \bm{B_{b_k}} \bm{w_{N(b_k)}} = \bm{B^\star_{b_k}} \bm{w_S}$, and $j$ be the $j$-$th$ observation of block $\bm{b_k}$, then $\forall k = 1, \dots, M$, $i = 1, \dots, n$ and $ j = 1, \dots, n_{bk}$: 
\[ B^{\star}_{b_k} (j,i) =
  \begin{cases}
    1      & \quad \text{if } s_i \in \bm{s_{b_k}} \\
    B_{b_i}[j,l]  & \quad \text{if } s_i \in \bm{s_{b_k}}; s_i = s_{N(b_k)}[l]; l= 1, \dots, N_{bk} \\
    0  & \quad \text{otherwise,}  
  \end{cases}
\]
\\
and
\[  \bm{B^{\star}_{b_k}} =
\begin{bmatrix}
 \bm{B^{\star}_{b_k} (1)}  \\
  \vdots   \\
 \bm{B^{\star}_{b_k} (j)} \\
  \vdots \\
 \bm{B^{\star}_{b_k} (n_{b_k})}
 \end{bmatrix} _{n_{b_k} \times n}
,\]
\\
where $\bm{B^{\star}_{b_k} (j)}= (B^{\star}_{b_k} (j,1),B^{\star}_{b_k} (j,2),\dots, B^{\star}_{b_k} (j,n))$. 
From these definitions, $  \bm{B^{\star}_{b_k}}$ is a matrix with i-th column full of zeros if $s_i\notin \bm{s_{b_k}}$ or $ s_i\notin \bm{N(s_{b_k})}$. 
Since the data were reordered by blocks and the neighbor blocks are from the past, $\bm{B^{\star}_{b_k}}$ has also the next form: 
\[
 \bm{B^{\star}_{b_k}} =   \left[
    \begin{array}{ccccc}
     \bm{R_k}   &  \bm{A_k}   & 0   & \dots & 0 \\ \cline{2-2}
\end{array}\right]
,\]
where $\bm{A_k}$ is a $n_{bk}\times n_{bk}$ matrix and $\bm{R_k}$ is a $n_{bk}\times \sum_{r=1}^{k-1} n_{br}$ matrix with at least one column with none null-element if $nb \neq 0$. 
\\
Then,
\begin{eqnarray*}
\widetilde{\pi}(\bm{w_S}) &\propto&   \frac{1}{ \prod_{k=1}^M |\bm{F_{b_k}}|^{1/2}} \exp\left\{ -\frac{1}{2} \sum_{k=1}^M  (\bm{B^\star_{b_k}} \bm{w_S})^T \bm{F_{b_k}^{-1}} (\bm{B^\star_{b_k}} \bm{w_S}) \right\}\\
&\propto&   \frac{1}{ \prod_{k=1}^M |\bm{F_{b_k}}|^{1/2}} \exp\left\{ -\frac{1}{2} \sum_{k=1}^M  \bm{w_S^T} (\bm{B_{b_k}^\star})^T  \bm{F_{b_k}^{-1}} (\bm{B^\star_{b_k}} \bm{w_S}) \right\}\\
 &\propto&   \frac{1}{ \prod_{k=1}^M |\bm{F_{b_k}}|^{1/2}} \exp\left\{ -\frac{1}{2} \sum_{k=1}^M  \bm{w_S^T}  ( (\bm{B_{b_k}^\star})^T  \bm{F_{b_k}^{-1}} \bm{B_{b_k}}^{\star} ) \bm{w_S}  \right\}\\
 &\propto&   \frac{1}{ \prod_{k=1}^M |\bm{F_{b_k}}|^{1/2}} \exp\left\{ -\frac{1}{2}  \bm{w_S}^T  ( \sum_{k=1}^M (\bm{B_{b_k}}^\star)^T  \bm{F_{b_k}^{-1}} \bm{B_{b_k}}^{\star} ) \bm{w_S}  \right\}. 
\end{eqnarray*}
Let 
$\sum_{k=1}^M (\bm{B_{b_k}}^\star)^T  \bm{F^{-1}_{b_k}} \bm{B_{b_k}}^{\star} = (\bm{B^{\star}_s})^T \bm{F_s ^{-1}} \bm{B^{\star}_s} $, 
where 
$\bm{B_{s}} =   \left[
    \begin{array}{ccccc}
    \bm{B_{b1}^{\star}}   &  ...  & ...   & \dots & \bm{B_{bM}^{\star}} \\ 
\end{array}\right] 
$
and $\bm{F_s ^{-1}}  = \mbox{diag}( \bm{F_{b_k} ^{-1}})$.  
$ \bm{F_s ^{-1}} $  is a block diagonal matrix and (iii) is proved. 
And given that  $  \bm{B^{\star}_{b_k}}$ is a matrix with i-th column full of zeros for $i> \sum_{r=1}^{k} n_{br}$, then $\bm{B_s}$ is a block matrix and lower triangular, 
and (ii) is proved. 
\\
Finally, 
$\widetilde{p}(\bm{w_S}) \propto   \frac{1}{ \prod_{k=1}^M |\bm{F_{b_k}}|^{1/2}} \exp\left\{ -\frac{1}{2}  \bm{w_S^T}  ( \bm{B_s^T F_s ^{-1}B_s}) \bm{w_S}  \right\}$ and $ \bm{\widetilde C_s^{-1}} = \bm{\widetilde Q_s}= \bm{B_s^T F_s ^{-1}B_s}.$
\\
\\
\textbf{$\widetilde{C}_s$ is positive definite}
\\
From properties of the Normal distribution, the covariance of the conditional
distribution of $\bm{w_{b_k}}|\bm{w_{N(b_k)}}$ is also p.d. (by Schur complement conditions),
then
$\bm{F_{bk}} = \bm{C_{b_k}} - \bm{C_{b_k,N(b_k)}}\bm{C^{-1}_{N(b_k)}}\bm{C_{N(b_k),b_k}}$, 
 is p.d. 
 Moreover, $\bm{F_s} = diag(\bm{F_{bk}})$ and 
 a block diagonal matrix is p.d. iff each diagonal block is positive definite,
 so given that $\bm{F_{bk}}$ is p.d. and $\bm{F_s}$ is block diagonal with blocks
 $\bm{F_{bk}}$ p.d then $\bm{F_s}$ is p.d.  
By  Corollary A1,  $\bm{F_s}$ is p.d. then $\bm{F^{-1}_s}$ is p.d. 
By Theorem A1, $\bm{B_s}$ has full column rank if and only iff $\bm{R_s} = \bm{B^T_s B_s}$ is invertible. 
By Theorem A2, the inverse of $\bm{R_s}$ exists if and only if $\det(\bm{R_s})\neq 0$. 
Using the well-known matrix theorems, we can prove the following:  
$\det(\bm{R_s})= \det(\bm{B^T_s B_s}) = \det(\bm{B^T_s}) \det(\bm{B_s}) \neq 0$  
 if $\det(\bm{B^T_s}) =  \det(\bm{B_s}) \neq 0$. 
Given that $\bm{B_s}$ is a lower triangular matrix, by Theorem A3, 
$\det(\bm{B_s}) = \prod_{k=1}^{n}(b_{kk})$. And, $b_{kk} =1, \forall k$, then 
$\det(\bm{B_s}) \neq 0.$ So,  $\bm{R_s}$ is invertible and  $\bm{B_s}$ has full column rank. 
By Proposition A1, given that $\bm{B_s}$ has full column rank, and $\bm{F_s^{-1}}$ is p.d. then 
$\bm{\widetilde{Q}_s} = \bm{\widetilde{C}_s^{-1}} = \bm{B^T_s F_s^{-1} B_s}$ is p.d.
Also by corollary A1, $\bm{\widetilde{C}_s^{-1}}$ is p.d. then $\bm{\widetilde{C}_s}$ is p.d.  
\\
\\
Since $\widetilde p(\bm{w_S}) \propto   \frac{1}{ \prod_{k=1}^M |\bm{F_{b_k}}|^{1/2}} \exp\left\{ -\frac{1}{2}  \bm{w_S}^T  ( \bm{\widetilde C_s}^{-1}) \bm{w_S}  \right\}$, $ \bm{\widetilde C_s}^{-1} = \bm{B_s^T F_s ^{-1}B_s}$,  and $\bm{\widetilde{C}_s}$ is p.d., then $\widetilde p(\bm{w_S})$ is a pdf of a multivariate normal distribution.
\\
\\
If  $n_{bk} \ll n$ then $i> \sum_{r=1}^{k} n_{br}$ and $  \bm{B^{\star}_{b_k}}$ will be more sparse. Also, if $n_k$ is small, the block diagonal matrix $ \bm{F_s}^{-1} $  will be more sparse. As a result, 
$ \bm{\widetilde C_s^{-1}} = \bm{B_s^T F_s ^{-1}B_s}$, will still be sparse.
\end{proof}

\begin{proof}[\textbf{Proof of Lemma 1}]
We need to prove that the finite dimensional distributions in (6) are consistent with a stochastic process. The Kolmogorov consistency conditions are checked as follows: 

\textbf{Symmetry under permutation:} Let $\Delta_1, \dots, \Delta_n$ be any permutation of $1, \dots, n$, note that $\bm{S}$ is fixed, then it is clear that 
$\widetilde \pi(w{(v_1)}, \dots, w{(v_n)}) = \widetilde \pi(w{(v_{\Delta_1})}, \dots, w{(v_{\Delta_n}}))$ if and only if the same holds for the distribution of $u_i|\bm{N(u_i)}$. Since $\bm{w_U}|\bm{w_S}$ follows a l-multivariate normal distribution, then the symmetry condition is satisfied by $\pi(\bm{w_U}|\bm{w_S})$, and it holds that the next condition 
$\widetilde \pi({w{(u_1)}}, \dots, {w{(u_{l})}}|\bm{w_S}) = \widetilde \pi({w{(u_{\Delta_1})}}, \dots, {w{(u_{\Delta_{l}}}})|\bm{w_S})$ is necessary and sufficient to prove the symmetry condition of $\widetilde \pi(\bm{w_V})$. 
To prove this we define the next pdfs,
\begin{eqnarray*}
\widetilde \pi(w(u_1), \dots, w(u_l)|\bm{w_S} ) &=& |2\pi \bm{F_U}|^{-1/2}\exp \left\{-\frac{1}{2} (\bm{w_U} - \bm{B_U w_S})^T \bm{F_U} ^{-1} (\bm{w_U} - \bm{B_U w_S}) \right\}\\
 &=&  |2\pi \bm{F_U}|^{-1/2}\exp \left\{ \bm{Q(w_U)} \right\}, 
\end{eqnarray*}
 and
\begin{eqnarray*}
\widetilde p(w(u_{\Delta_1}), \dots, w(u_{\Delta_l})|\bm{w_S} ) &=& |2\pi \bm{\Sigma'}|^{-1/2}\exp \left\{-\frac{1}{2} (\bm{w_{U\Delta}} - \bm{m'})^T {\bm{\Sigma'}} ^{-1} (\bm{w_{U\Delta}} - \bm{m'}) \right\} \\
 &=& |2\pi \bm{\Sigma'}|^{-1/2}\exp \left\{ \bm{Q(w_{U\Delta}}) \right\}.  
\end{eqnarray*}
following \citet{Abrahamsen:1997}, we also define a permutation matrix $\bm{P}$ such that 
$(\Delta_1, \dots, \Delta_l)^T = \bm{P}(1, \dots, l)^T$. 
Then $\bm{P} w_U = \bm{P}(w(u_1), \dots, w(u_l))^T= (w(u_{\Delta_1}), \dots, w(u_{\Delta_l}))^T = \bm{w_{U\Delta}}$. 
And the mean and covariance matrix of $\bm{w_{U}{\Delta}}|\bm{w_S}$ are  $\bm{m'} = \bm{P B_{U} w_{S}} $ and $\bm{\Sigma'} = \bm{P F_{U} P'}$. 
Since $\bm{P}^{-1}=\bm{P}^T$  it follows that $|\bm{P}| =  \pm 1$ which implies that $|\bm{\Sigma'}| = |\bm{F_{U}}|$. 
Using this we have,\\
$$
\begin{aligned}
\bm{Q(w_{U\Delta})} &= (\bm{P w_U -m'})^T \bm{\Sigma'}^{-1}(\bm{P w_U -m'}) \\
 &= (\bm{P w_U  - P B_{U} w_{S}})^T  (\bm{P F_{U} P'})^{-1} (\bm{P w_U}  - \bm{P B_{U} w_{S}}) \\
&=  ( \bm{w_U}  - \bm{B_{U} w_{S}})^T \bm{P}^T (\bm{P}^T \bm{F_{U}}^{-1} \bm{P}^T) \bm{P} ( \bm{w_U}  - \bm{B_{U} w_{S}})\\
&=  
( \bm{w_U}  - \bm{B_{U} w_{S}})^T \bm{P}^T  \bm{\Sigma'}^{-1}  \bm{P} ( \bm{w_U}  - \bm{B_{U} w_{S}})  \\  
 &= ( \bm{w_U}  - \bm{B_{U} w_{S}})^T   \bm{F_{U}}^{-1}   ( \bm{w_U}  - \bm{B_{U} w_{S}}) = \bm{Q(w_U)}. 
\end{aligned}
$$ 
\\
Since both $|\bm{F_{U}}|$ and $\bm{Q(w_U)}$ are invariant under permutations, 
$\widetilde \pi(w{(u_1)}, \dots, w{(u_{l})}|\bm{w_S}) = \widetilde \pi(w{(u_{\Delta_1})}, \dots, w{(u_{\Delta_{l}}})|\bm{w_S})$ and 
hence the symmetry condition is satisfied.

\textbf{Dimensional consistency:} We also assume that $\bm{S}$ is fixed, so, this proof does not differ from the one found in \citep{Dattaetal:2016} although  $\widetilde \pi(\bm{w_S})$ has a different definition. 
\\
Let $\bm{V_1} = \bm{V} \cup \{v_0\}$ then $\bm{V_1} = \bm{S'}  \cup \{v_0\} \cup \bm{U}$. We need to verify $\widetilde \pi (\bm{w_V}) = \int \widetilde \pi (\bm{w_{V_1}}) d(w(v_0))$. 
So, we have two cases:\\
\underline{Case 1}: If $v_0\in \bm{S}$. By definition 
$\widetilde \pi (\bm{w_{V_1}}) = \int\widetilde \pi (\bm{w_{{V_1}|\bm{S}}}|\bm{w_S})\widetilde p(\bm{w_S}) \prod_{si \in \bm{S}|\bm{V_1}} d(w_{s_i})$, where ${V_1}|\bm{S}$ denotes ${V_1}$ without $\bm{S}$ and $\bm{S}|{V_1}$ denotes $\bm{S}$ without ${V_1}$. 
Then
\[ \int \widetilde \pi (\bm{w_{V_1}}) d(w(v_0)) = \int  \widetilde \pi (\bm{w_{{V_1}|\bm{S}}|\bm{w_S}})\widetilde \pi(\bm{w_S}) \prod_{si \in \bm{S}|\bm{V_1}}d(w(s_i)) d(w(v_0)).
\]  
If $v_0\in \bm{S}$, and $\bm{V} = \bm{S'} \cup \bm{U}$ then $v_0 \in (\bm{S'})^c$, and  $\prod_{si \in \bm{S}|\bm{V_1}}d(w(s_i) d(w(v_0)) = \prod_{si \in (\bm{S'})^c}d(w(s_i)$, and
\[
\int \widetilde \pi (w_{\bm{V_1}}) d(w(v_0))  = \int\widetilde \pi (\bm{w_{V_1|S}}|\bm{w_S})\widetilde \pi(\bm{w_S}) \prod_{si \in  (\bm{S'})^c} d(w_{s_i}). 
\]
Also, ${\bm{V_1}|\bm{S}} =  \bm{U}$ since $v_0\in \bm{S}$, then
\[\int \widetilde \pi (\bm{w_{V_1}}) d(w(v_0))  = \int\widetilde \pi (\bm{w_{U}}|\bm{w_S})\widetilde \pi(\bm{w_S}) \prod_{si \in  (\bm{S'})^c} d(w_{s_i}) = \widetilde \pi(\bm{w_V}).
\] 
\underline{Case 2}: If $v_0\notin \bm{S}$, then $\bm{V_1}|\bm{S} = \bm{U} \cup \{v_0\}$,  
$\widetilde \pi (\bm{w_{V_1|S}}|\bm{w_S}) = \widetilde \pi (\bm{w_{U|S}}|\bm{w_S}) \widetilde \pi (w(v_0)|\bm{w_S})$ and $\bm{S|V_1} =  (\bm{S'})^c$. Now,
\begin{eqnarray*}
\widetilde \pi (\bm{w_{V_1}}) &=& \int\widetilde \pi (\bm{w_{V_1|S}}|\bm{w_S})\widetilde \pi(\bm{w_S}) \prod_{si \in \bm{S}|\bm{V_1}} d(w_{s_i})\\
&=& \int\ \widetilde \pi (\bm{w_U}|\bm{w_S}) \widetilde \pi (w(v_0)|\bm{w_S}) \widetilde \pi(\bm{w_S}) \prod_{si \in  (\bm{S'})^c} d(w_{s_i}).
\end{eqnarray*}
Hence, 
\begin{eqnarray*}
\int \widetilde \pi (\bm{w_{V_1}}) d(w(v_0))  &=& \int\widetilde \pi (\bm{w_{U}}|\bm{w_S}) \widetilde \pi (w(v_0)|\bm{w_S}) \widetilde \pi(\bm{w_S}) \prod_{si \in  (\bm{S'})^c} d(w_{s_i}) d(w(v_0)) \\
  &=& \int  \widetilde \pi(\bm{w_S})\widetilde \pi (\bm{w_{U}}|\bm{w_S}) [\widetilde \pi (w(v_0)|\bm{w_S}) d(w(v_0))] \prod_{si \in  (\bm{S'})^c} d(w_{s_i}),  
\end{eqnarray*}
where $\int\widetilde \pi (w(v_0)|\bm{w_S}) d(w(v_0)) = 1$, since $w(v_0)$ does not appear in any other term.  Finally, 
\[
\int \widetilde \pi (\bm{w_{V_1}}) d(w(v_0))  = \int  \widetilde \pi(\bm{w_S})\widetilde \pi (\bm{w_{U}}|\bm{w_S}) \prod_{si \in  (\bm{S'})^c} d(w_{s_i})  = \widetilde \pi(\bm{w_V}). 
\]
\end{proof}

\begin{proof}[\textbf{Proof of Theorem 1}]
 To verify that $\widetilde \pi(\bm{w_V})$ is the pdf of the finite dimensional distribution of a Gaussian process, we only need to prove that  $\widetilde \pi(\bm{w_V})$ is the pdf of a multivariate normal distribution. Since  $\bm{w_U}|\bm{w_S}$ follows a l-multivariate normal distribution and  $\bm{w_S}$ follows a n-multivariate normal distribution, the product of these densities is also a multivariate normal distribution. 

Let $\widetilde C_{m,n}$ be an element of $\bm{\widetilde C_S}$. The cross-covariance is computed for the next possible cases:
\\
\underline{Case 1}: If $v_1\in \bm{S}$ and $v_2 \in \bm{S}$, that is, $v_1=s_i $ and $v_2 = s_j$, then $\mbox{cov}(w(v_1), w(v_2)|\theta)) = \widetilde C_{s_i, s_j}.$
\\
\underline{Case 2}: If $v_1\in U$ and $v_2 \in S$, we may suppose also that $v_2\in b_l$. 
Using the law of total covariance,
\[
\mbox{cov}(w(v_1), w(v_2)|\theta)) = \mbox{E}(\mbox{cov}(w(v_1), w(v_2)|w_S)|\theta) + \mbox{cov}(\mbox{E}(w(v_1)|w_S) , \mbox{E}(w(v_2)|w_S)|\theta).
\]
From our definition $ w(v_1)|\bm{w_S} \bot w(b_l)|\bm{w_S}$ and $v_2\in b_l$, then we have that 
$w(v_1)| \bm{w_S}  \bot w(v_2)|\bm{w_S}$ and  $\mbox{cov}(w(v_1)| \bm{w_S}  , w(v_2)|\bm{w_S} ) = 0$.
Further,  $\mbox{E}(w(v_1)|\bm{w_S}) = \bm{B_{v_1}} \bm{w_{N(v_1)}}$ and using the next property, $\mbox{E}(g(X)|X) = g(X)$, $\mbox{E}(w(v_2)|\bm{w_S}) = w(v_2)$. 
It follows that, 
\[
\mbox{cov}(w(v_1), w(v_2)|\theta)) = \mbox{E}(0|\theta) + \mbox{cov}(\bm{B_{v_1}} \bm{w_{N(v_1)}}, w(v_2)|\theta) = \bm{B_{v_1}} \bm{\widetilde C_{N(v_1), w(v_2)}}  = \bm{B_{v_1}} \bm{\widetilde C_{N(v_1), w(s_j)}}. 
\]
\underline{Case 3}: If $v_1\in U$ and $v_2 \in U$. This part of the proof is the same for the NNGP, found in \citep{Dattaetal:2016}. 
We have $\mbox{E}(w(v_1)|\bm{w_S}) = \bm{B_{v_1}} \bm{w_{N(v_1)}}$ and $\mbox{E}(w(v_2)|\bm{w_S}) = \bm{B_{v_2}} \bm{w_{N(v_2)}}$. Then, 
\begin{eqnarray*}
  \mbox{cov}(\mbox{E}(w(v_1)|\bm{w_S}) , \mbox{E}(w(v_2)|\bm{w_S})|\theta) &=& \mbox{cov}(\bm{B_{v_1}} \bm{w_{N(v_1)}}, \bm{B_{v_2}} \bm{w_{N(v_2)}}) \\
  &=& \bm{B_{v_1}} \mbox{cov}( \bm{w_{N(v_1)}},  \bm{w_{N(v_2)}})\bm{B^T_{v_2}}.
\end{eqnarray*}

Observe that if $v_1 \neq v_2$, then $w(v_1)|\bm{w_S} \bot w(v_2)|\bm{w_S}$ and  $\mbox{cov}(w(v_1), w(v_2)|\bm{w_S}) = 0$. Conversely, if 
$v_1 = v_2$ now $\mbox{cov}(w(v_1), w(v_2)|\bm{w_S}) = \mbox{var}(w(v_1)|\bm{w_S}) = F_{v_1}$. 
Then,\\
$\mbox{cov}(w(v_1), w(v_2)|\bm{w_S}) = \delta(v_1=v_2) F_{v_1},$ and $ \mbox{E}(\delta(v_1=v_2) F_{v_1}|\theta) = \delta(v_1=v_2) F_{v_1}$. \\
Hence, 
\[
\mbox{cov}(w(v_1), w(v_2)|\theta)) = \delta(v_1=v_2) F_{v_1} + \bm{B_{v_1}} \bm{\widetilde C_{N(v_1), N(v_2)}}\bm{B_{v_2}}^T.
\]
\end{proof}

\beginsupplementS

\subsection*{B. Bayesian inference through full-MCMC and collpased MCMC}

The LGM is also a hierarchical model, thus, in a Bayesian framework, inference can also be performed using simulation-based techniques, such as MCMC methods. Here we describe in detail how to achieve Bayesian inference for a Gaussian geostatistical model using the block-NNGP. This approach can also be extended for non-Gaussian families. 

Let $\bf{Y} = (Y(s_1), \dots, Y(s_n))^T$ be a realization of a spatial process defined for all $s_i \in \bm{D} \subset \Re^2$,  $i =1, \dots, n$. The basic geostatistical Gaussian model is of the form 
\begin{equation}
Y(s_i) = \bm{X}^T(s_i) \bm{\beta} +  w(s_i) + \epsilon(s_i),  
\label{eq:eqinf2} 
\end{equation}
where $\bm{\beta}$ is a coefficient vector (or regression parameter), $\bm{X}^T(s_i)$ is a  vector of covariates, $w(s_i)$ is a  spatial structured random effect, thus $\bm{w}=(w(s_1), \dots, w(s_n))^T$ captures the spatial association, and $\epsilon(s_i) \sim N(0, \tau^2)$ models the measurement error.

We assigned priors to $\beta, w, \tau$, and hyperparameters. The usual Gaussian process prior for $\bm{w} \sim N(0, \bm{C}(\theta_1))$, where $\bm{C}(.)$ is some specific covariance function which depends on $\theta_1=(\phi, \sigma^2)$. Instead of this prior we assume that $\bm{w} \sim \mbox{block-NNGP}(0,\bm{\widetilde{C}})$. We assumed $\bm{\beta} \sim N(\bm{\mu_{\beta}}, \bm{V_{\beta}})$ and $\bm{\theta} = ( \phi, \sigma^2, \tau^2) \sim \Delta(\bm{\theta})$.

The joint posterior distribution for the model in (\ref{eq:eqinf2}) is given by
\begin{equation}
\Delta(\bm{\theta},\bm{\beta},\bm{w} |\bm{y}) \propto \Delta(\bm{\theta}) \times \Delta_G(\bm{\beta}|\bm{\mu_{\beta},\Sigma_{\beta}}) 
\times  \Delta_G(\bm{w}|0,\bm{\widetilde{C}})\times \Delta_G(\bm{y}|\bm{X\beta} + \bm{w}, \bm{D}), 
\label{eq:eq5}  
\end{equation}
where  $\Delta_G(|.,.) $ denotes the Gaussian density, and $\bm{D}$ is a diagonal matrix with entries $\tau^2$. 

\noindent The parameters $\bm{\theta},\bm{\beta},\bm{w}$ are updated in a Gibbs sampler within Metropolis random-walk step  (full-MCMC) through the following algorithm: 
\begin{enumerate}
 \item[(i)] Block updating of $\bm{\theta}$ through Metropolis Random walk. The target log-density is 
\begin{equation}
\begin{aligned}
\log(\Delta(\bm{\theta}^{\star}|\bm{y}, \bm{w},\bm{\beta})) 
&\propto \log\Delta(\bm{\theta})  
-\frac{1}{2}\log|\bm{D}| - \frac{1}{2}\log|\bm{\widetilde{C}}|  -\\
& \frac{1}{2}(\bm{y}-\bm{X\beta} - \bm{w})^T \bm{D}^{-1} (\bm{y}-\bm{X\beta} - \bm{w})   -\frac{1}{2}\bm{w}^T \bm{\widetilde{Q}}\bm{w}, \nonumber 
\end{aligned} 
\end{equation}
where $\bm{\widetilde{Q}} = \bm{\widetilde{C}}^{-1}$;  \item[(ii)] Gibbs sampler that updates $\bm{\beta}$ from the full conditional 
$ \bm{\beta}|\bm{y}, \bm{w}, \bm{\theta} \sim N(\bm{Bb}, \bm{B})$, 
where
$\bm{B} = (\bm{\Sigma} _\bm{\beta}^{-1} + \bm{X}^T \bm{D}^{-1} \bm{X})^{-1}$
and
$\bm{b} = \bm{\Sigma _{\beta}}^{-1} \bm{\mu_{\beta}} + X^T \bm{D}^{-1} \bm{y} - X^T \bm{D}^{-1} \bm{w} $; 
 \item[(iii)]  Gibbs sampler that updates $\bm{w}$ from the full conditional 
$\bm{w}|\bm{y}, \bm{\beta}, \bm{\theta}^{\star} \sim N(\bm{Ff}, \bm{F})$, where 
$\bm{F} = ( \bm{\widetilde{Q}} + \bm{D}^{-1})^{-1}$ and $\bm{f} = \bm{D}^{-1}(\bm{y}- \bm{X\beta})$. Repeat step i). 
\end{enumerate}

In general it is fast to compute $\bm{\widetilde{Q}}$ and  $\log|\bm{\widetilde{C}}|$ using properties of block matrices, Cholesky decomposition and parallel algorithms for blocks.  Nevertheless, although $\bm{\widetilde{Q}}$ is an sparse precision matrix and $(\bm{\widetilde{Q}} + \bm{D}^{-1})$ has the same sparsity, the inverse of this last expression is not sparse, therefore subsequent computations are performed using this huge dense matrix, and the cost to sample $\bm{\theta},\bm{\beta},\bm{w}$ through this approach is still too {costly.}

An alternative method for simulation from the conditional posterior $\Delta(\bm{\theta},\bm{\beta},\bm{w} |\bm{y})$ is to use the collapsed MCMC sampling \citep{Liu:1994}. 
The collapsed MCMC for the Gaussian block-NNGP model follows the next steps:\\ 
(i) Block updating of $\bm{\theta}$ through a Metropolis random walk. The target log-density is 
\begin{equation}
\log(\Delta(\bm{\theta}|\bm{y})) 
\propto \log\Delta(\bm{\theta})  
 - \frac{1}{2}\log|\bm{\Sigma}_{y|\beta, \theta}|  -
 \frac{1}{2}(\bm{y}-\bm{X\beta} - \bm{w})^T \bm{\Sigma}_{y|\beta, \theta}^{-1} (\bm{y}-\bm{X\beta} - \bm{w}), \nonumber
\end{equation}
where $\bm{\Sigma}_{\bm{y}|\bm{\beta, \theta}} = \bm{\widetilde C} + \bm{D}$ and  $\bm{\Sigma}_{\bm{y}|\bm{\beta, \theta}}^{-1} = \bm{D}^{-1} - \bm{D}^{-1}(\bm{\widetilde Q} + \bm{D}^{-1})^{-1}\bm{D}^{-1}$.\\
(ii) Gibbs sampler that updates $\bm{\beta}$ from  
$ \bm{\beta}|\bm{y} \sim N(\bm{Bb}, \bm{B}),$
where
$\bm{B} = (\bm{\Sigma}_{\beta}^{-1} + \bm{X^T} \bm{\Sigma}_{y|\beta,\theta}\bm{X})^{-1}$
and
$\bm{b} = \bm{\Sigma}_{\beta}^{-1} \bm{\mu}_{\beta} + \bm{X}^T  \bm{\Sigma}_{y|\beta,\theta}^{-1} \bm{y}$.  Repeat step i) and ii) until convergence;\\
(iii) Post-MCMC sampling: Use all the posterior samples of $\bm{\theta}$ and $\bm{\beta} $ to estimate $\bm{w}$ from 
$\bm{w}|\bm{\beta}, \bm{\theta}, \bm{y} \sim N(\bm{Ff}, \bm{F})$, where
$\bm{F} = (\bm{\widetilde Q} + \bm{D}^{-1})^{-1}$ and $\bm{f} = \bm{D}^{-1}(\bm{y}- \bm{X\beta})$. 

We recall that this scheme also has the same dense matrix $\bm{F} = (\bm{\widetilde Q} + \bm{D}^{-1})^{-1}$, however the main advantage of the composite MCMC approach is to draw samples  of $\bm{\theta}$ and $\bm{\beta}$, and then use these samples to recover $\bm{w}$. Further, \cite{Finleyetal:2019} argued that the blocking and sampling schemes of the composite sampling result in good convergence properties. 

These MCMC algorithms are available to massive data, because the precision matrix of the block-NNGP is computed using fast parallel algorithms for blocks and it does not need to store $n\times n$ dense distance matrices at each iteration of the MCMC sampler. Moreover good parallel libraries are available, such as the multi-threaded BLAS/LAPACK libraries included in  Microsoft R Open and  parallel Packages in $R$ like the $doMC$ Package \citep{doMC:2017}.

\subsection*{C. Supplementary simulation results}

\begin{figure}
\begin{center}
\includegraphics[scale=0.3]{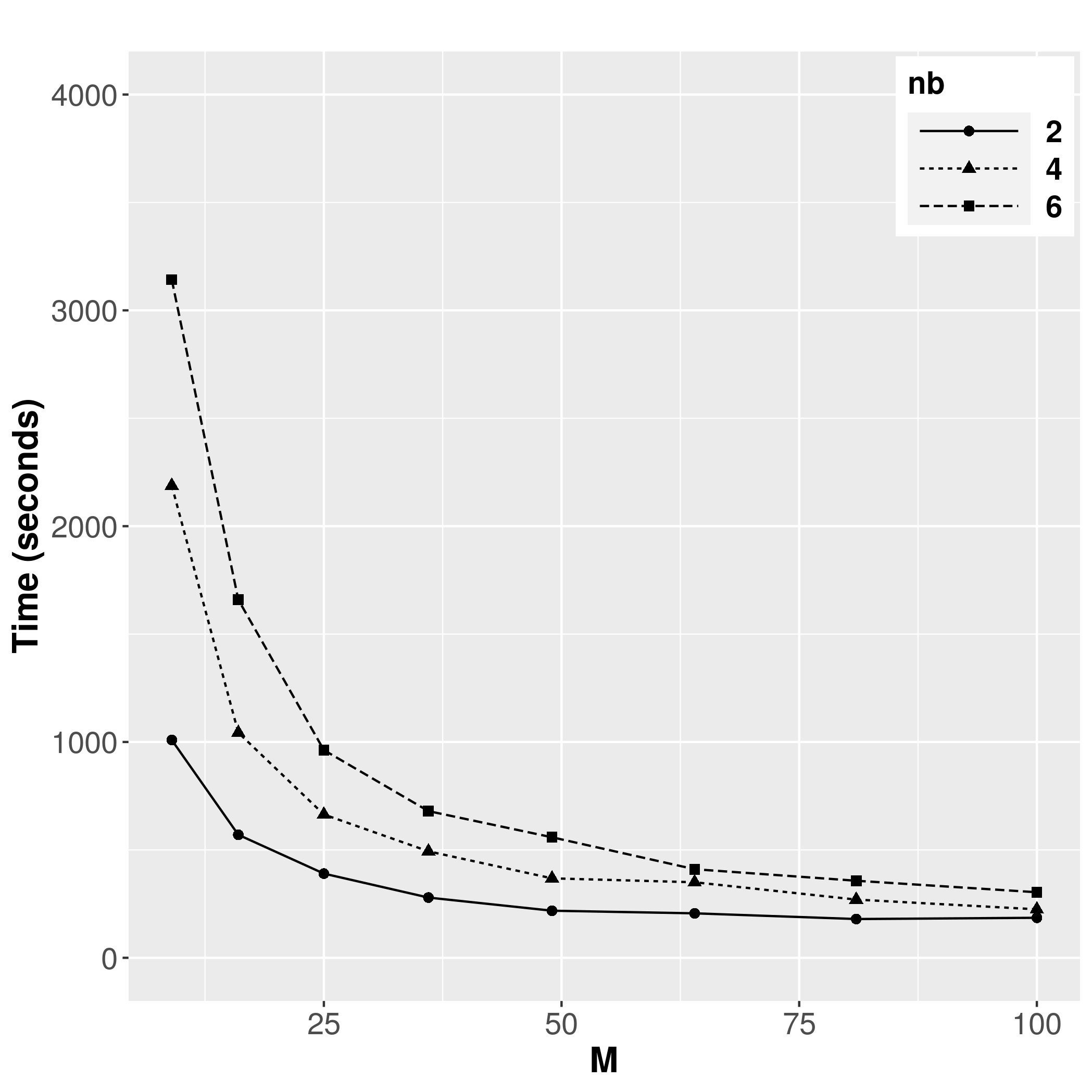} 
\includegraphics[scale=0.3]{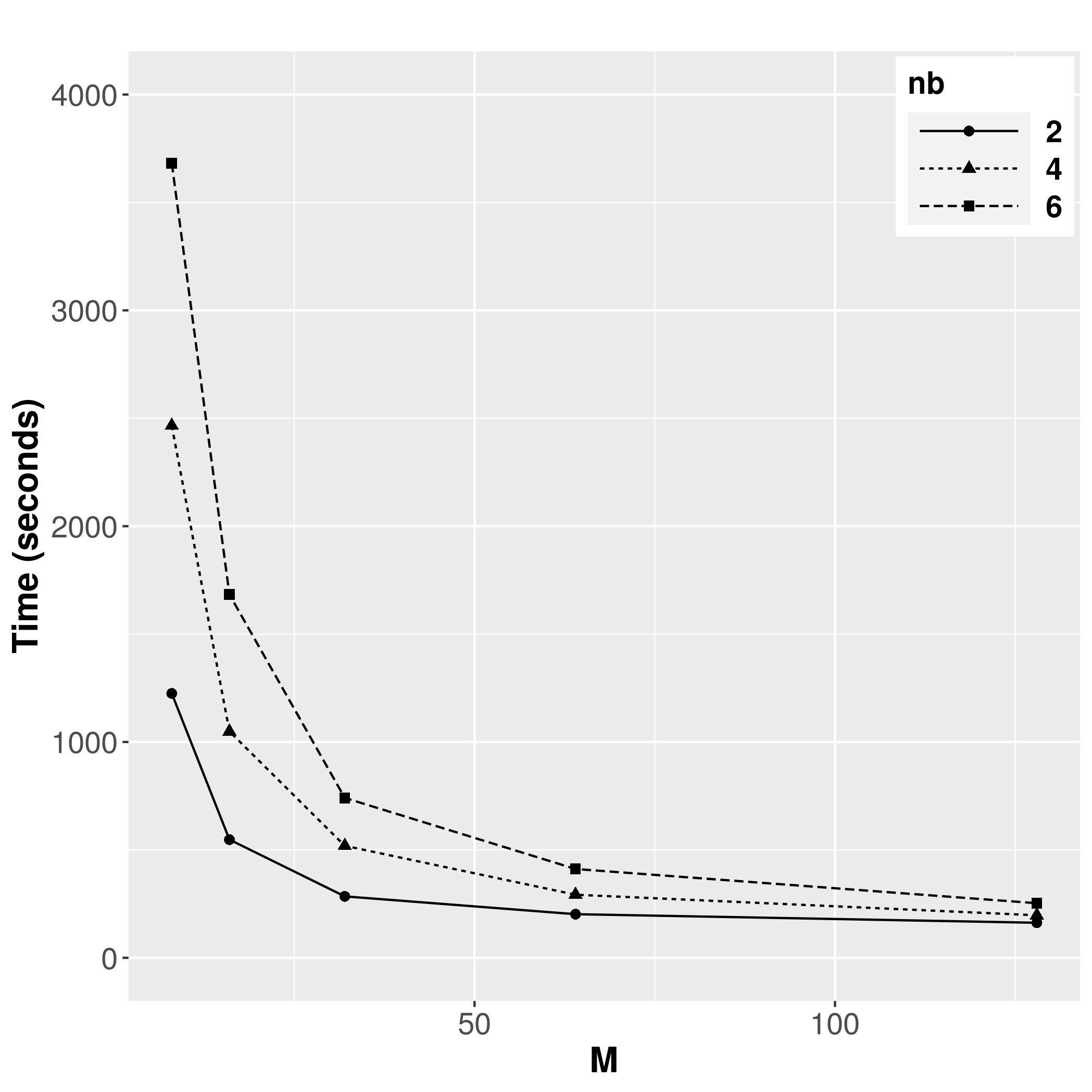} 
\includegraphics[scale=0.3]{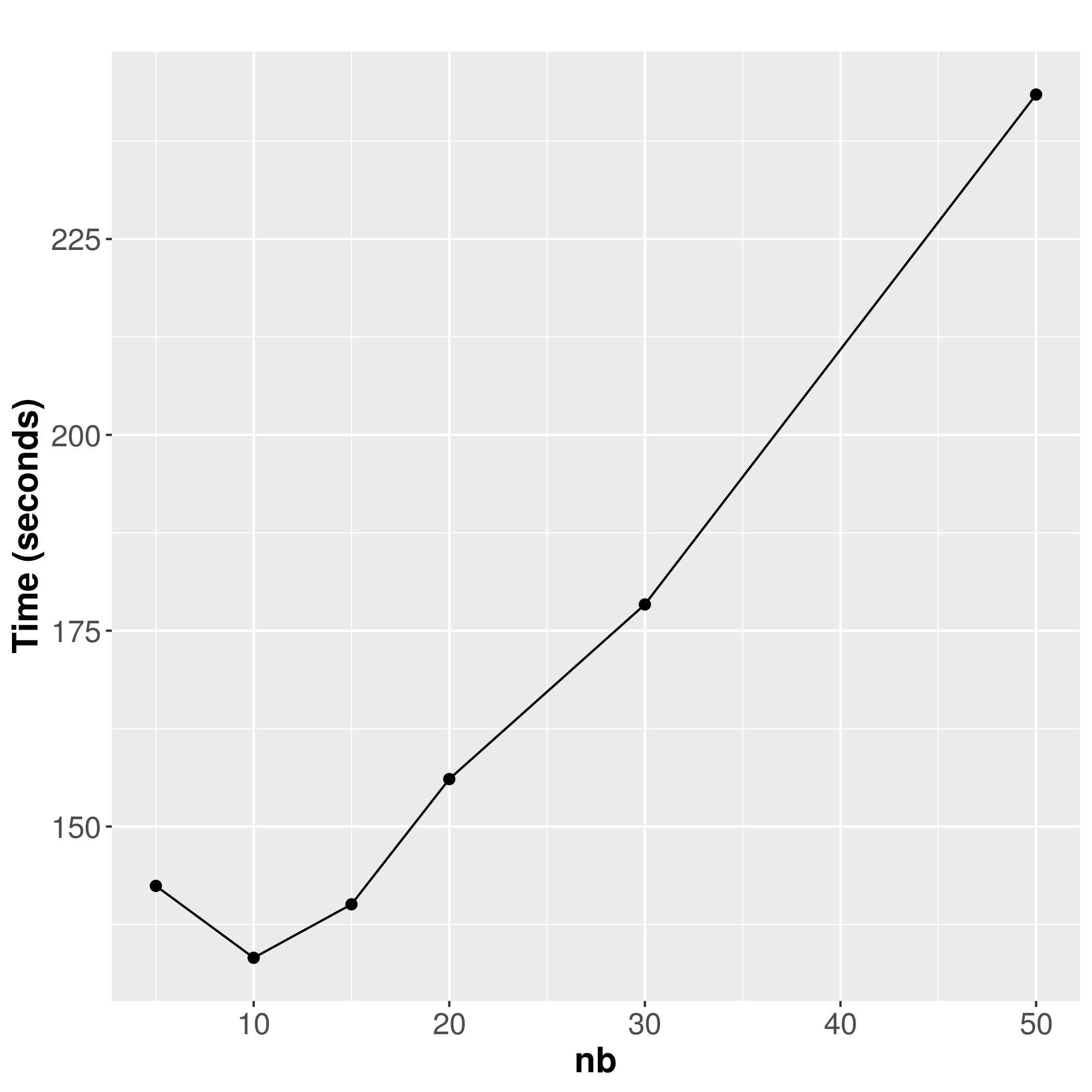} \\
\includegraphics[scale=0.3]{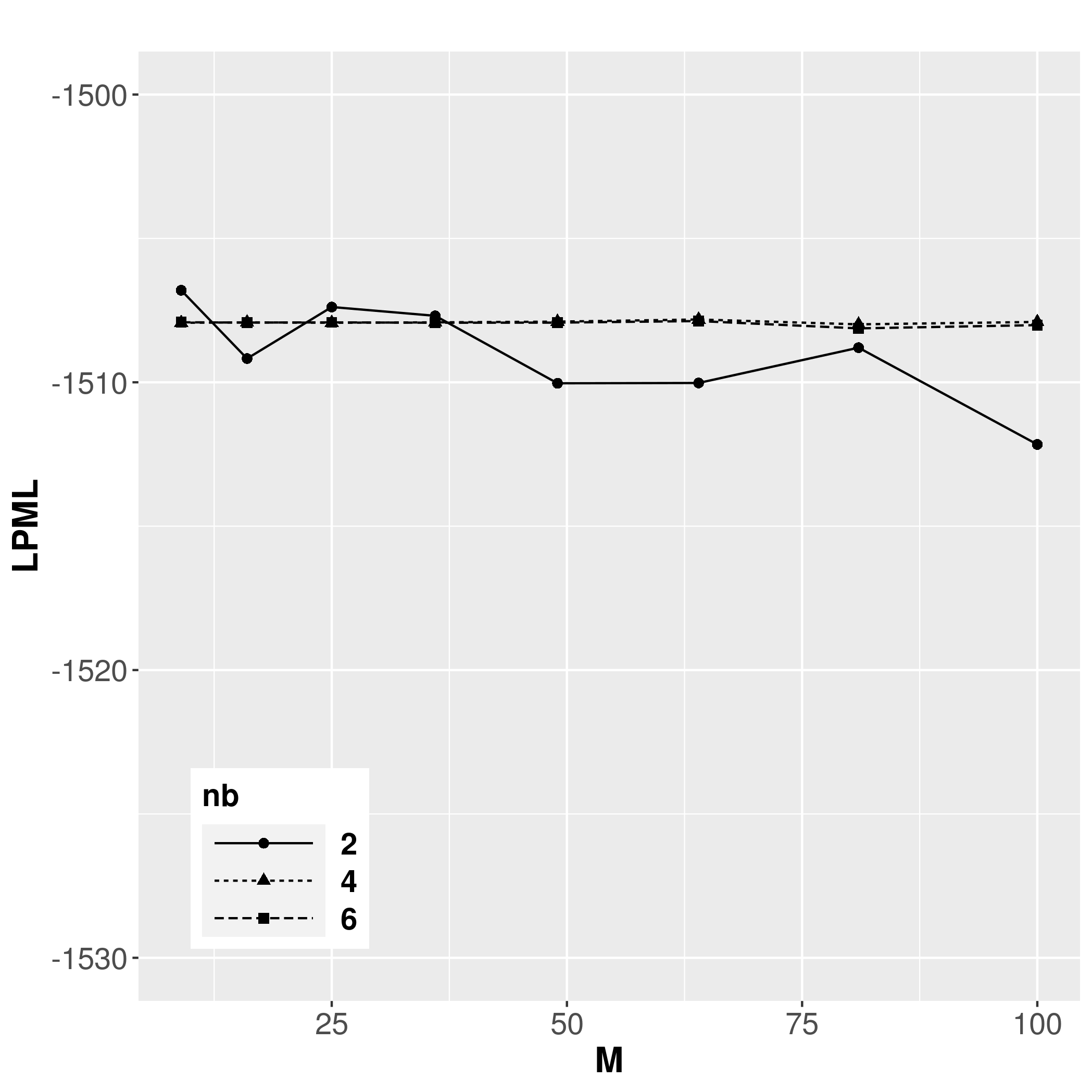} 
\includegraphics[scale=0.3]{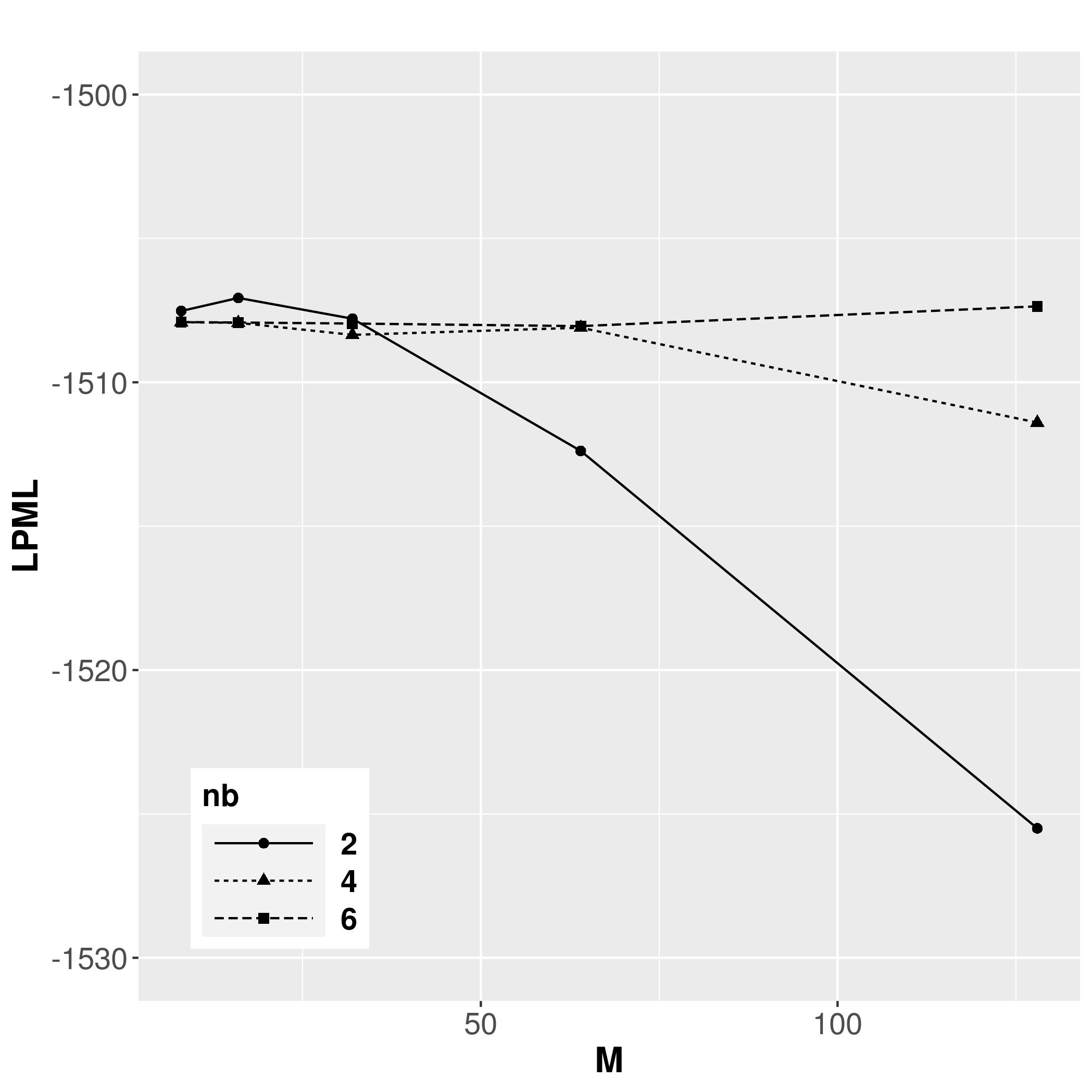} 
\includegraphics[scale=0.3]{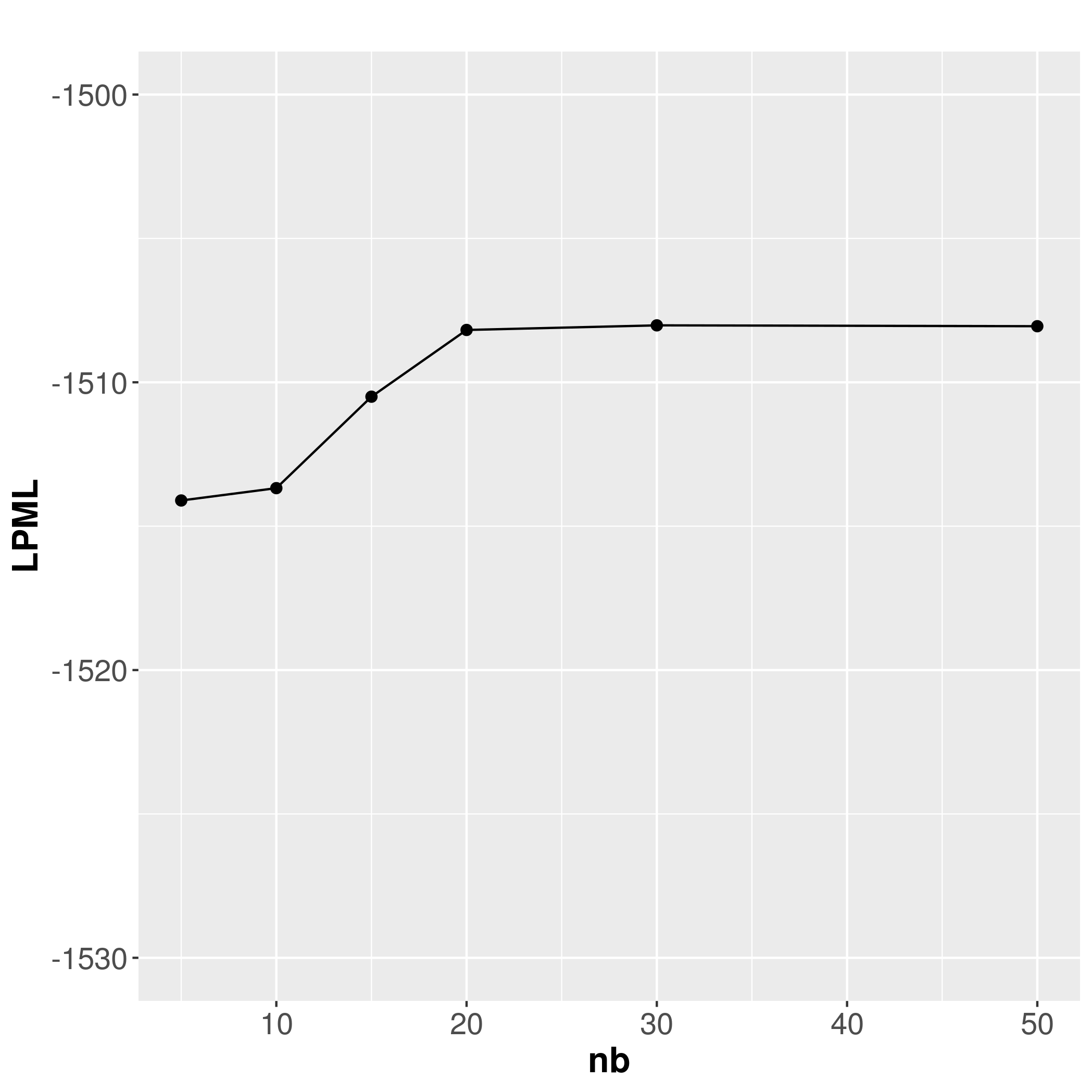} \\
\includegraphics[scale=0.3]{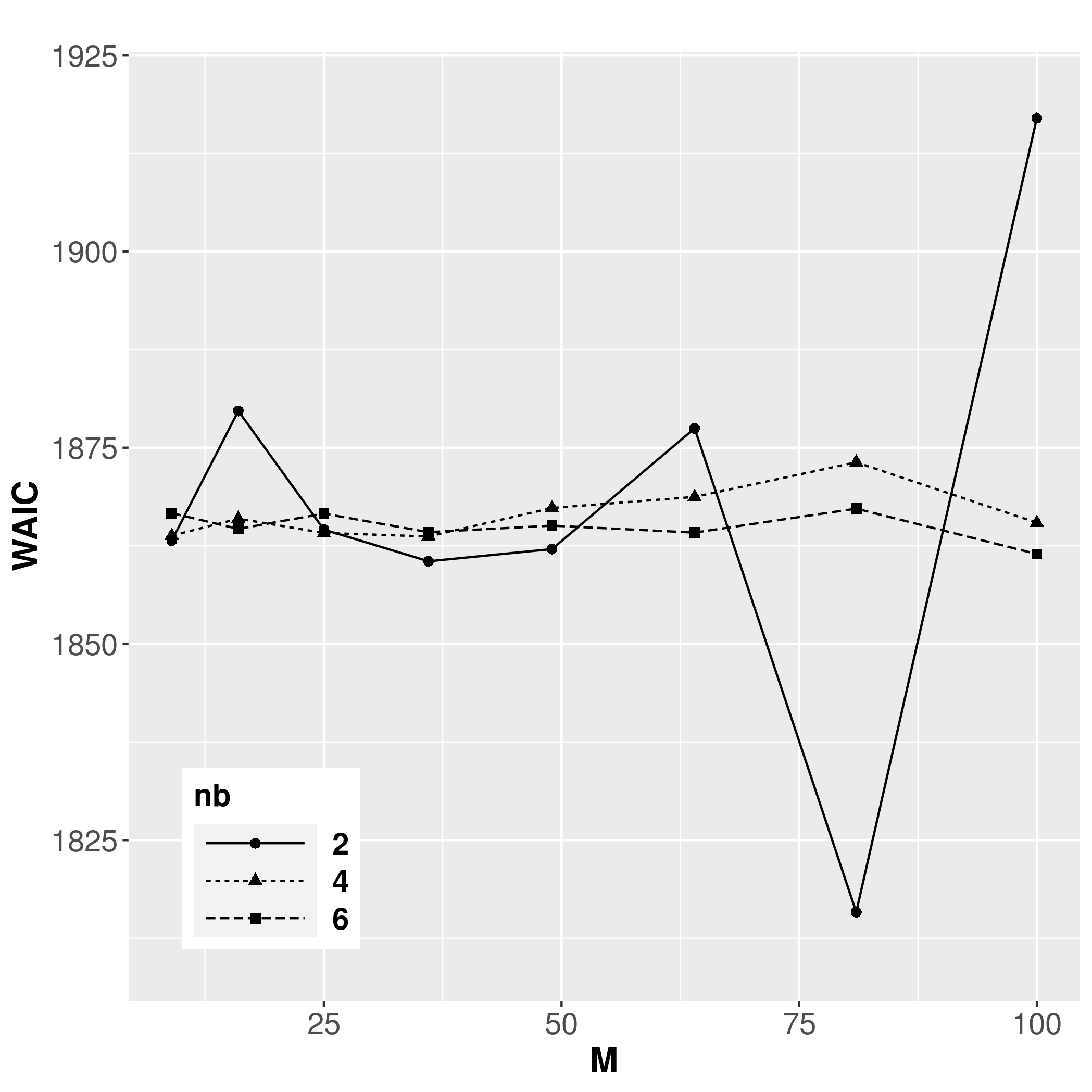} 
\includegraphics[scale=0.3]{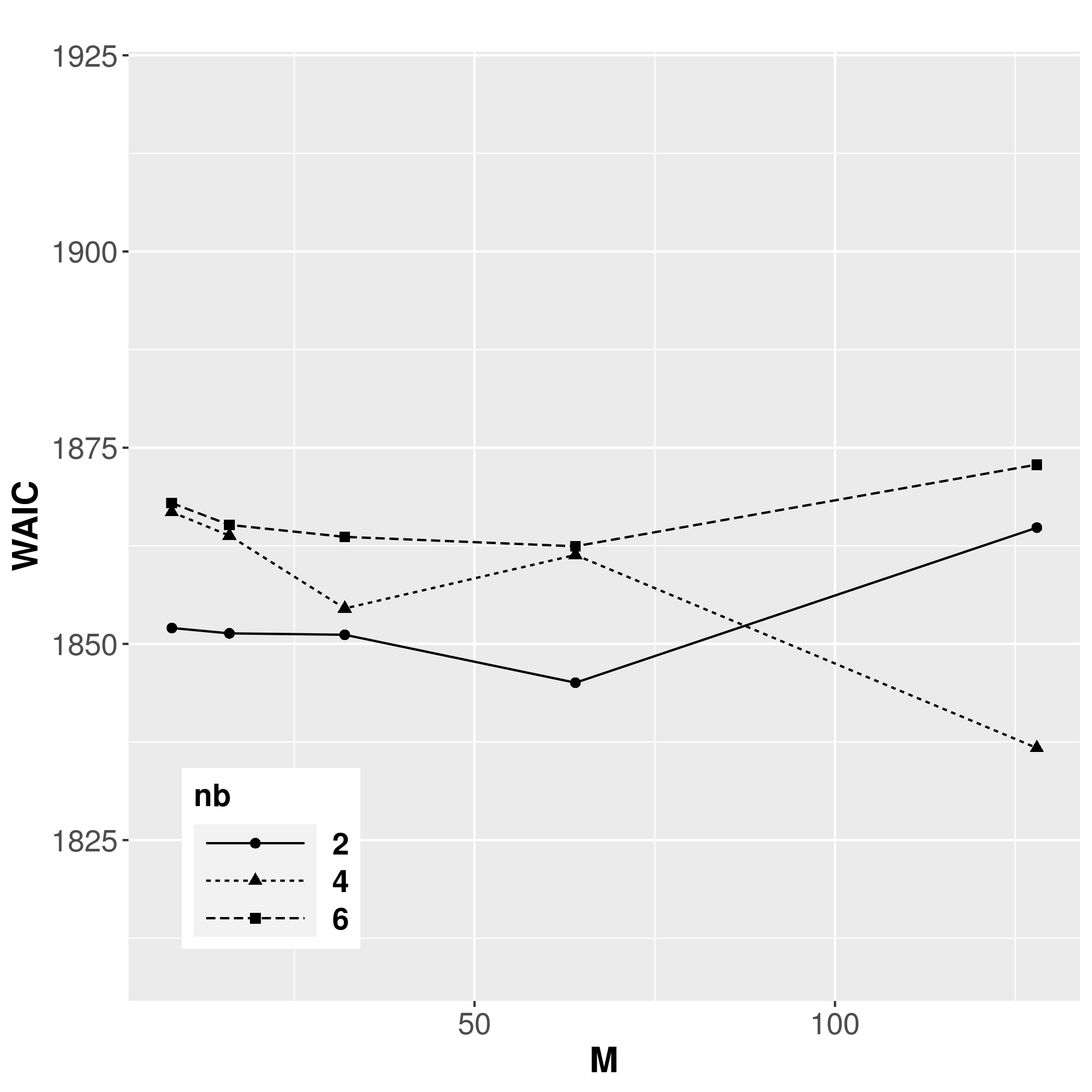} 
\includegraphics[scale=0.3]{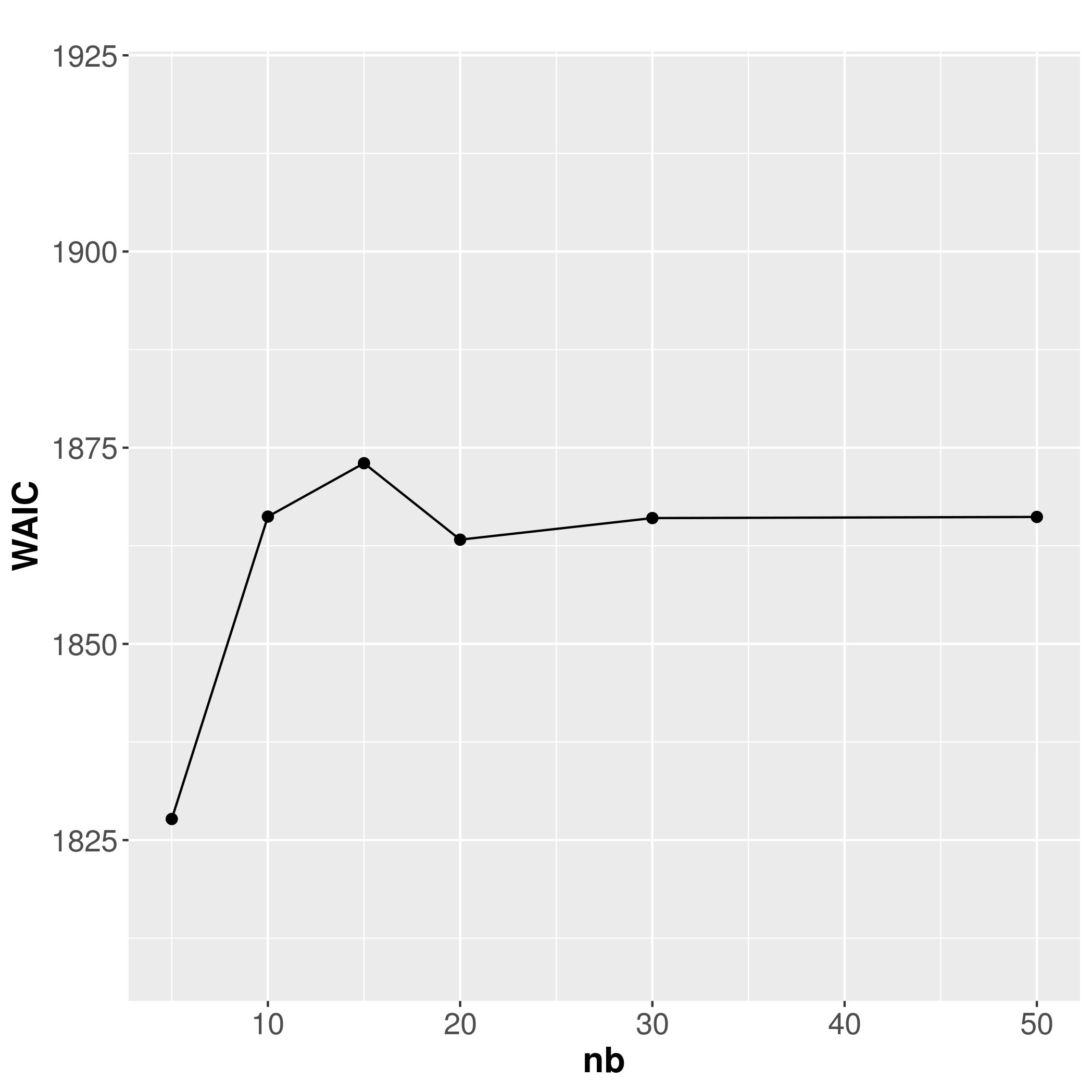} 
\end{center}
\caption{ SIM I ($\phi$ = 12). INLA results. Criteria assessment:  Running times (first row), LPML (second row)  and WAIC (third row), under block-NNGP models using regular blocks (left column), irregular blocks (middle column) and NNGP models (right column). }
  \label{fig:figs0}
\end{figure}

\begin{figure}
\begin{center}
\includegraphics[scale=0.3]{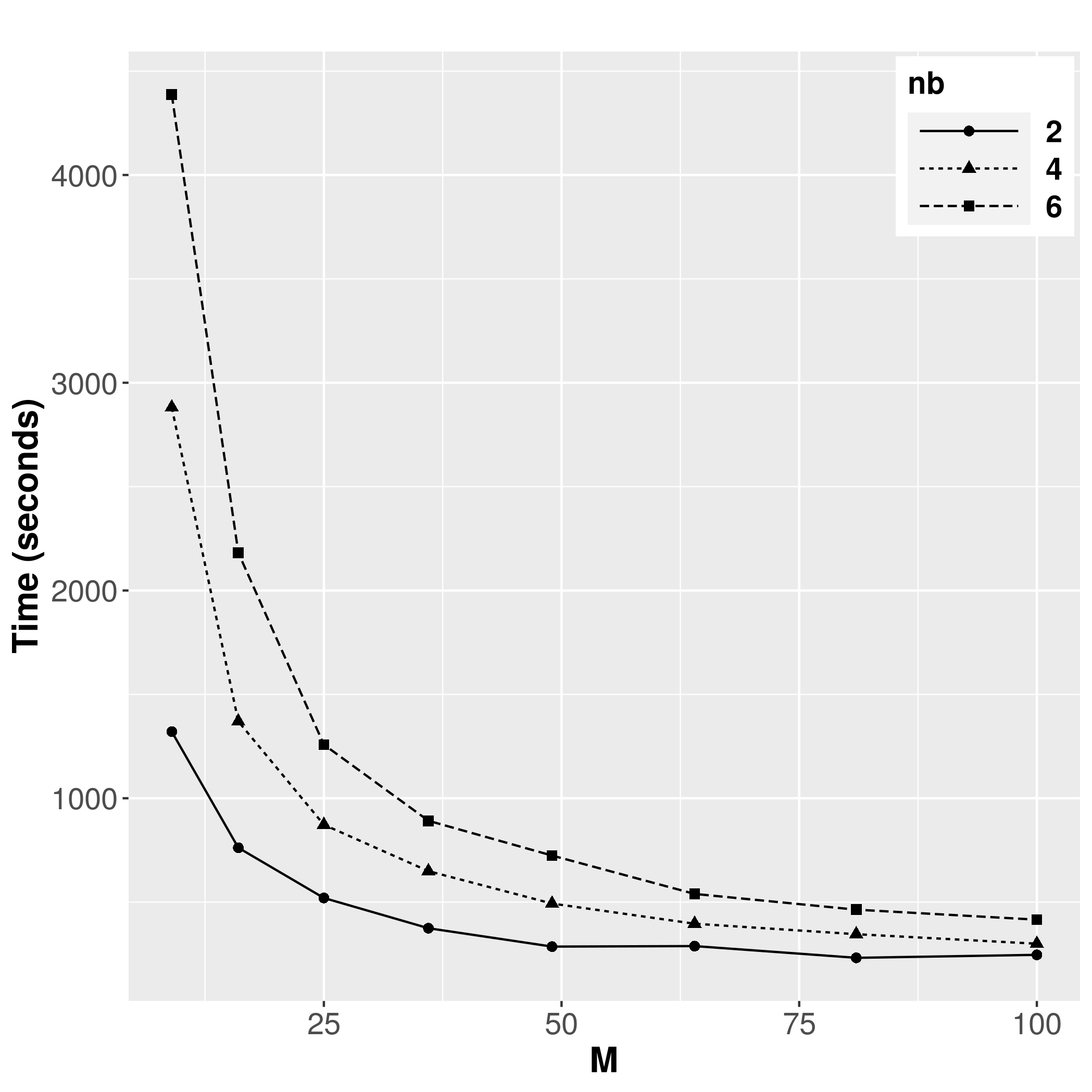} 
\includegraphics[scale=0.3]{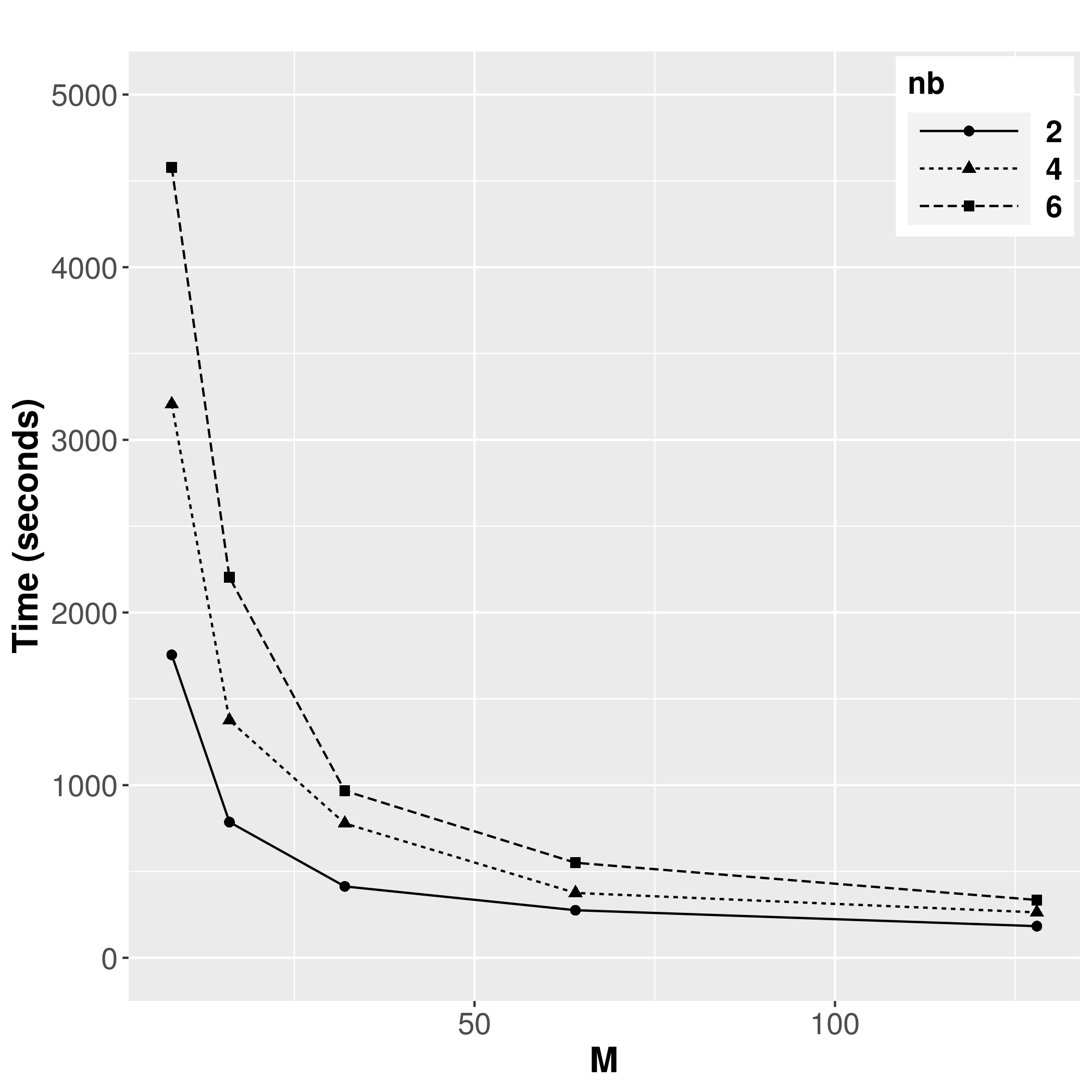} 
\includegraphics[scale=0.3]{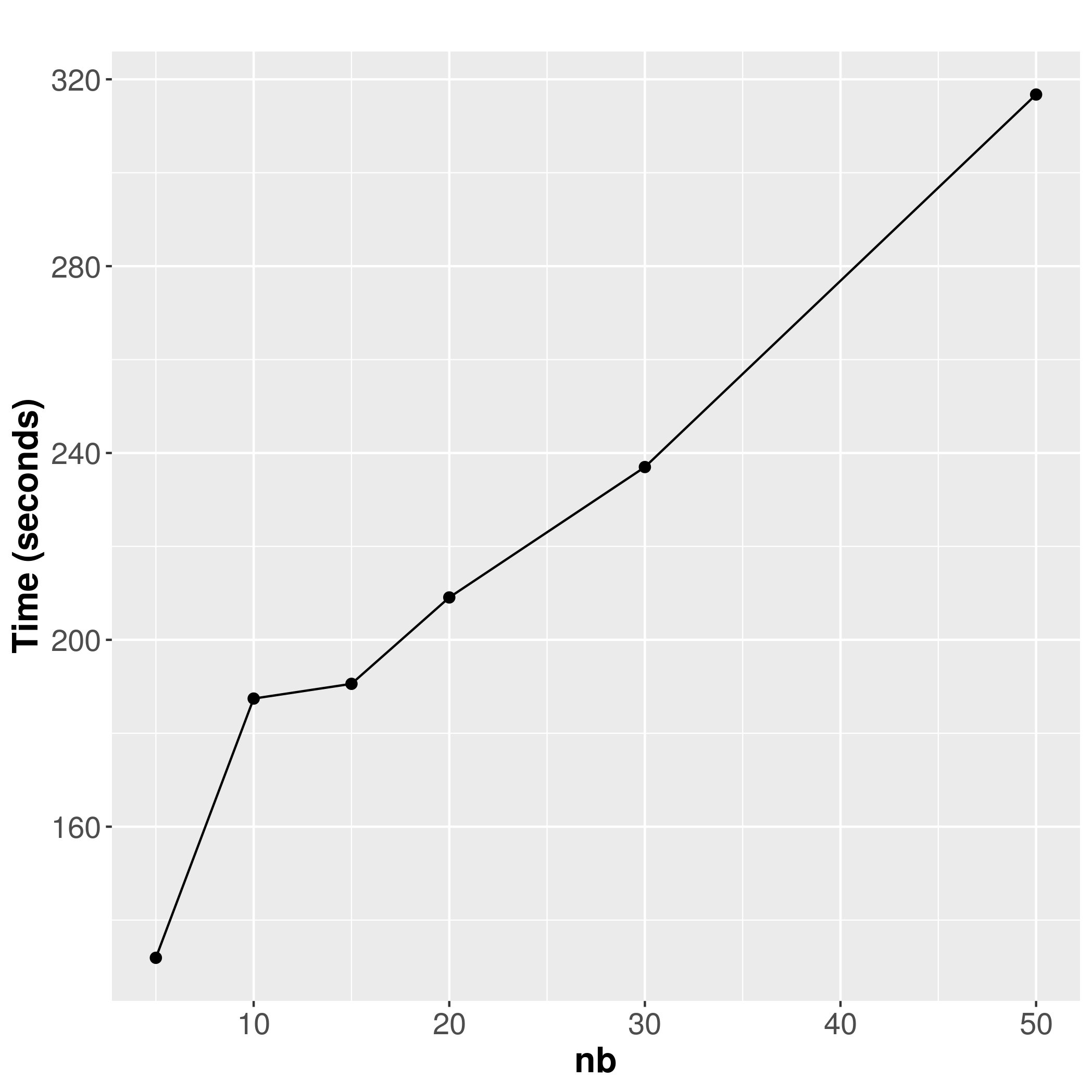} \\
\includegraphics[scale=0.3]{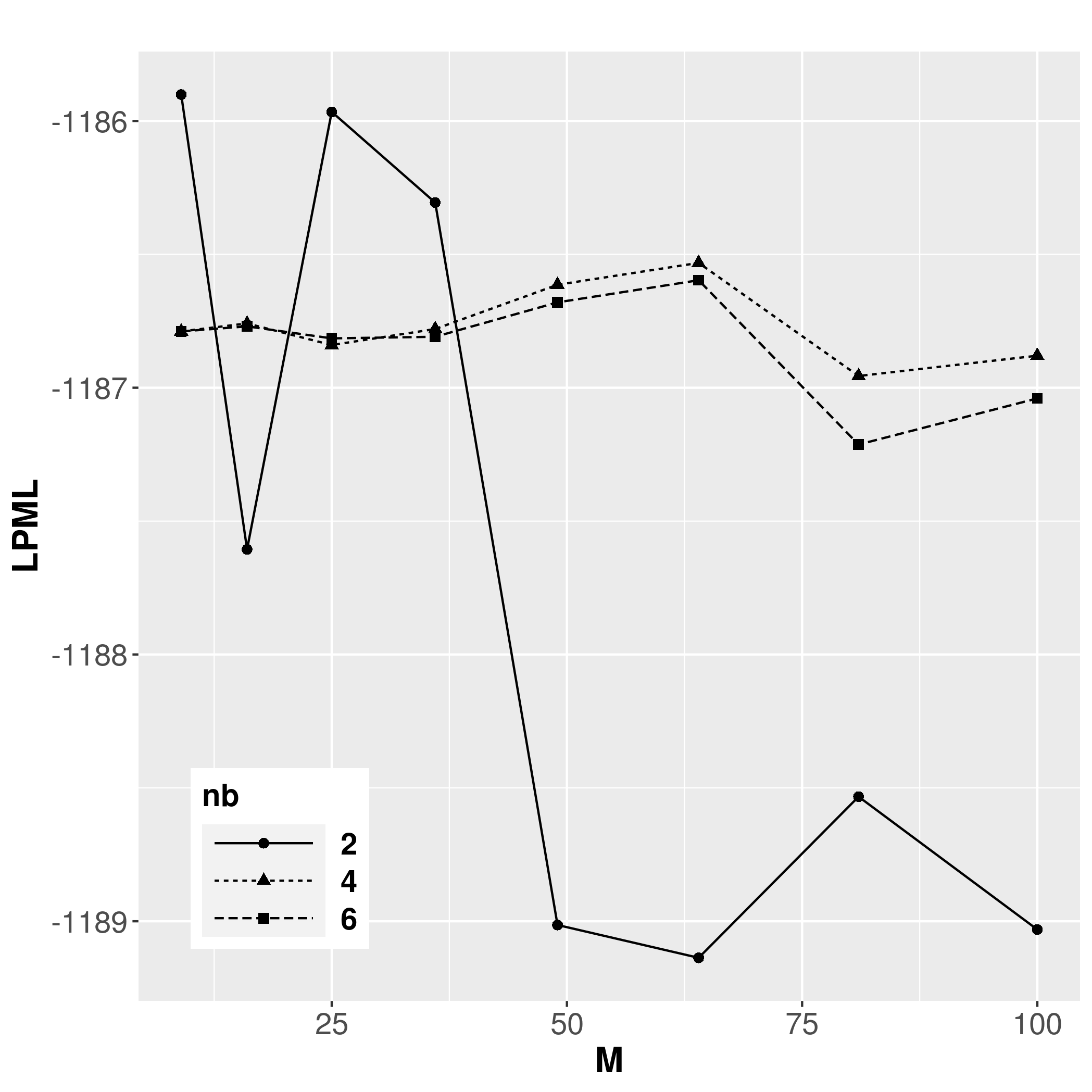} 
\includegraphics[scale=0.3]{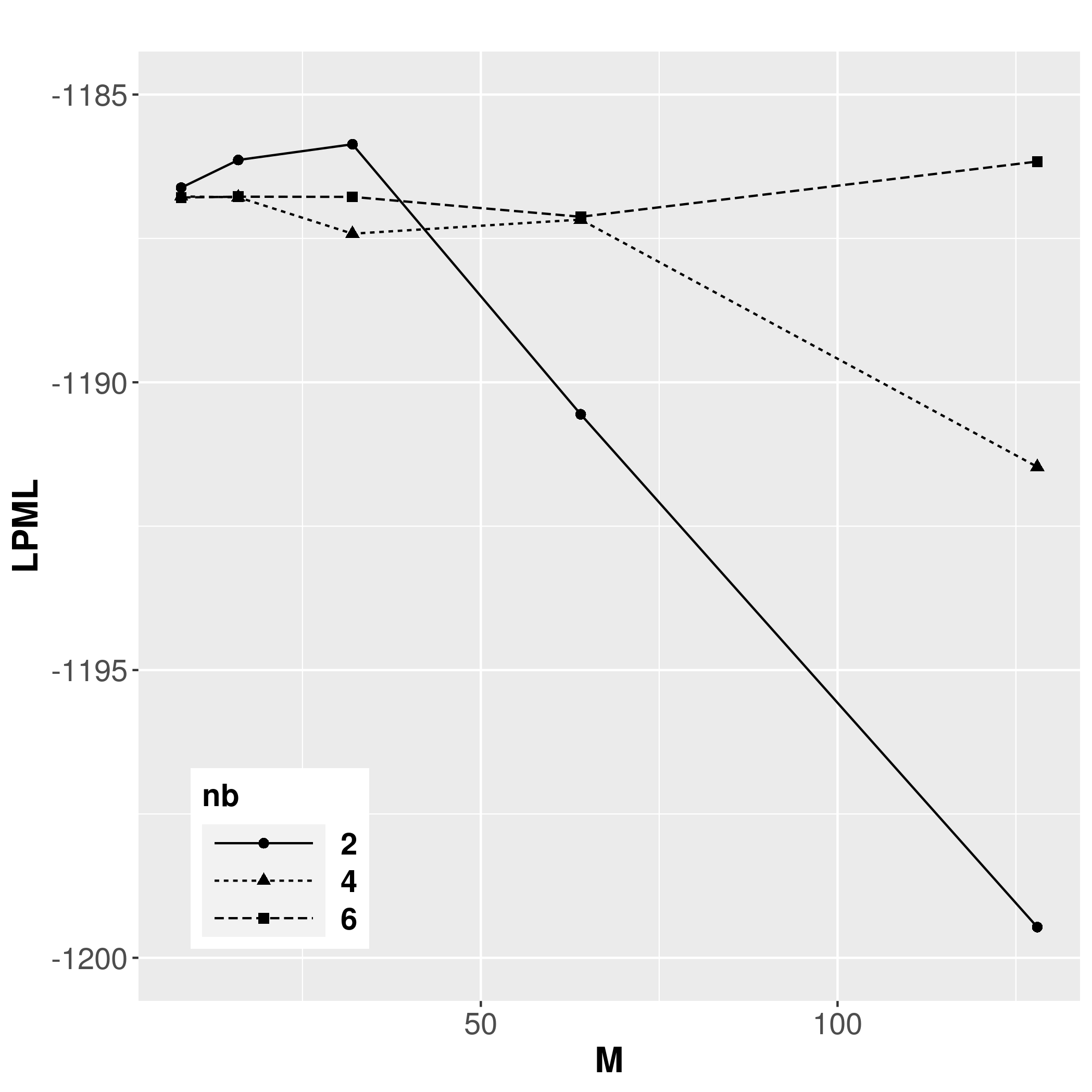} 
\includegraphics[scale=0.3]{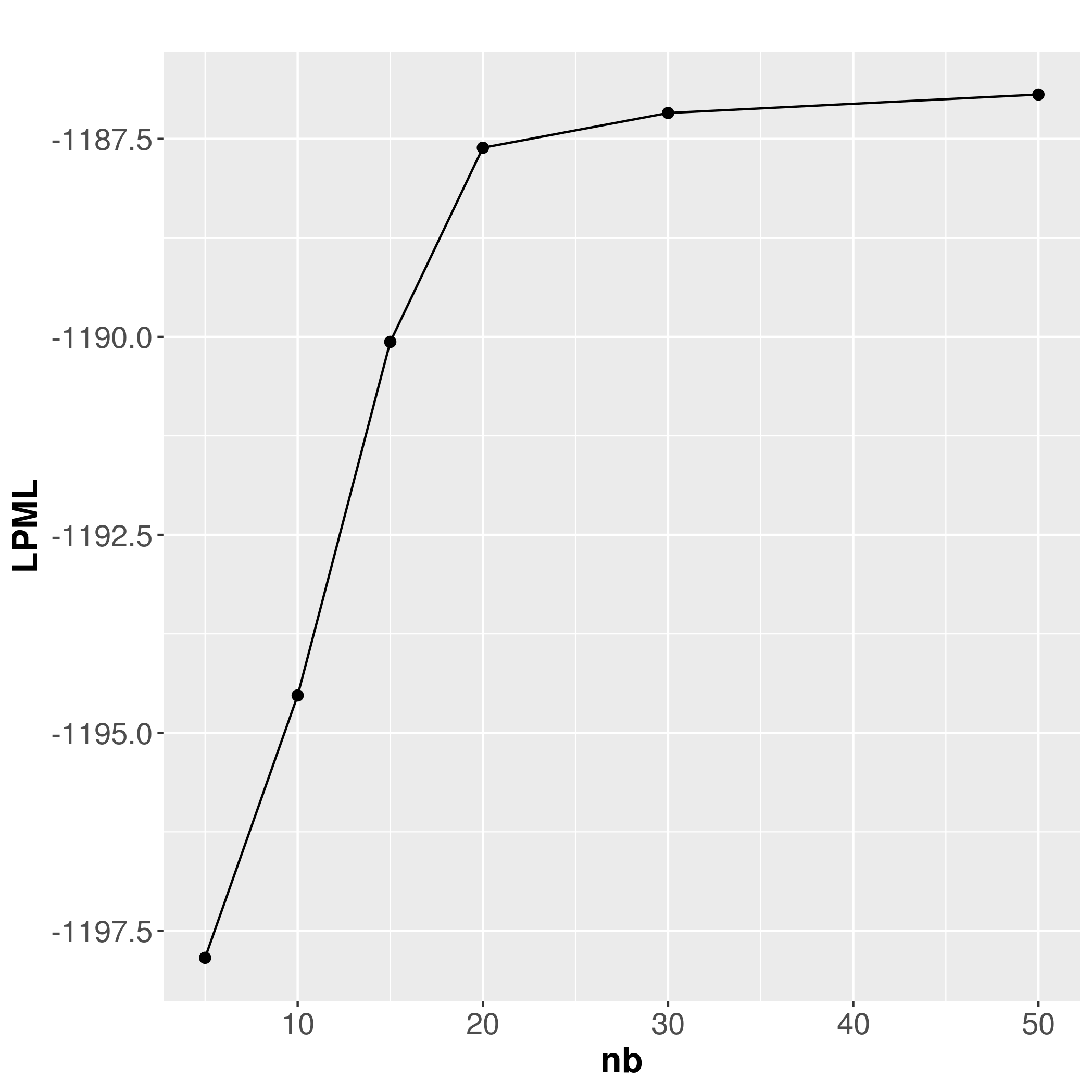} \\
\includegraphics[scale=0.3]{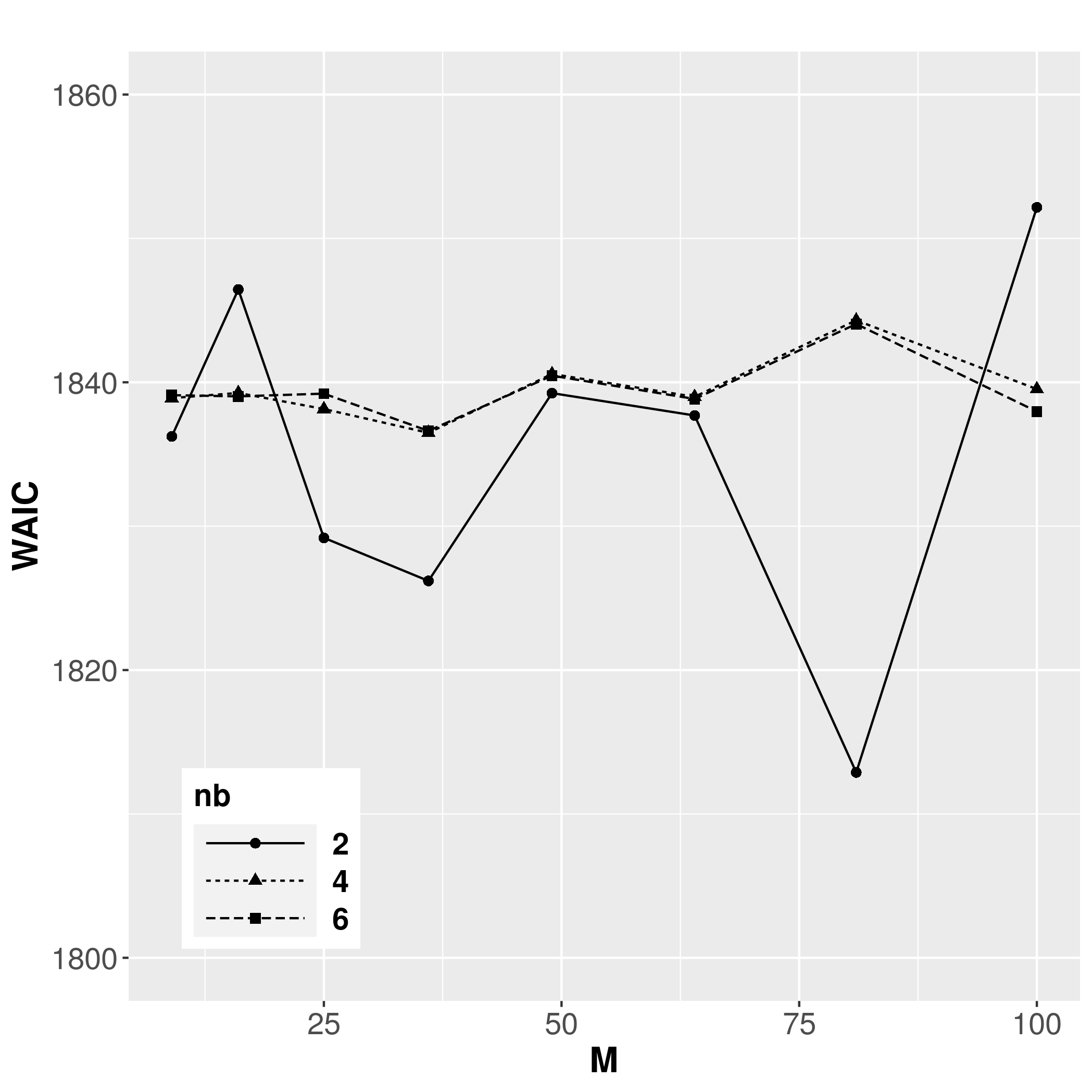} 
\includegraphics[scale=0.3]{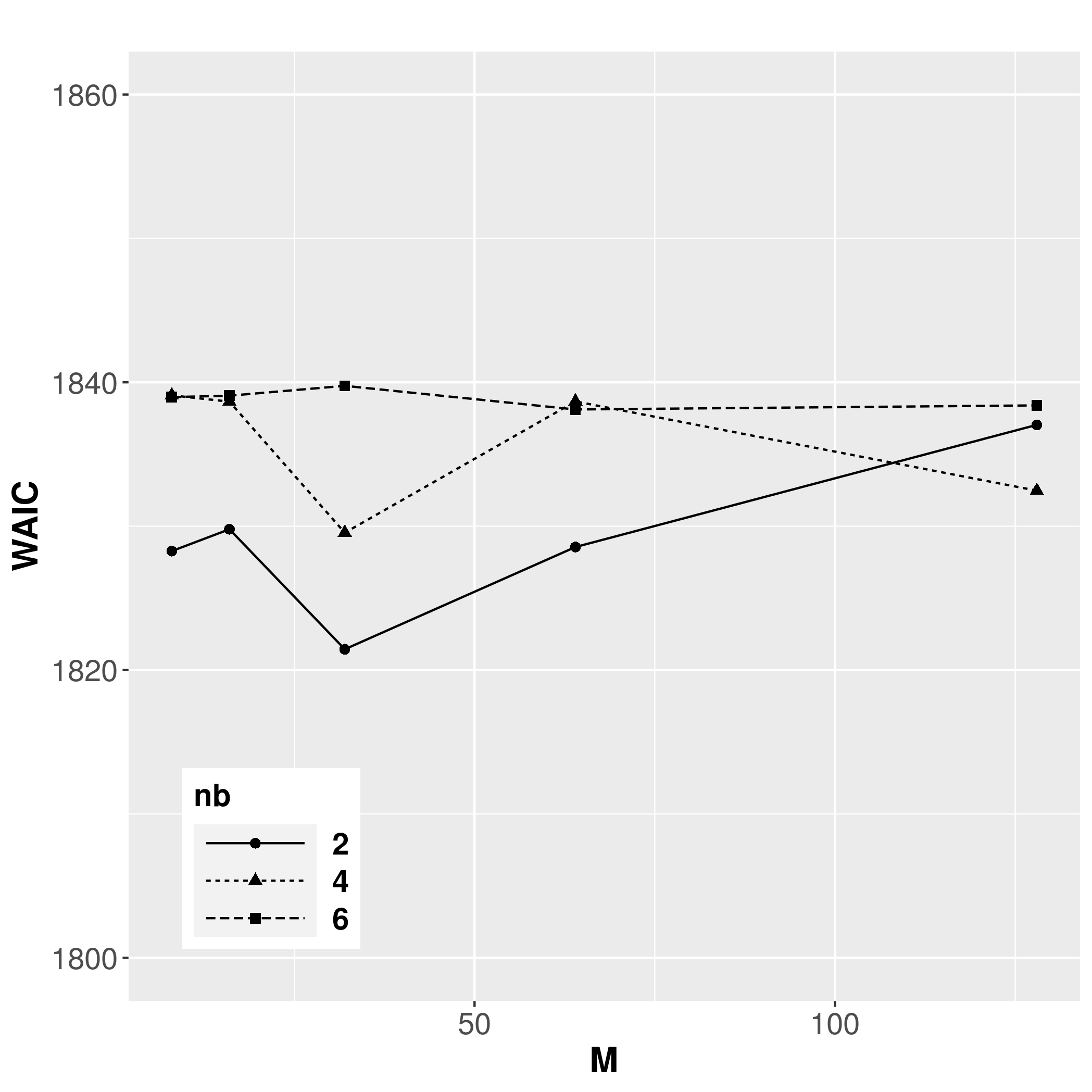} 
\includegraphics[scale=0.3]{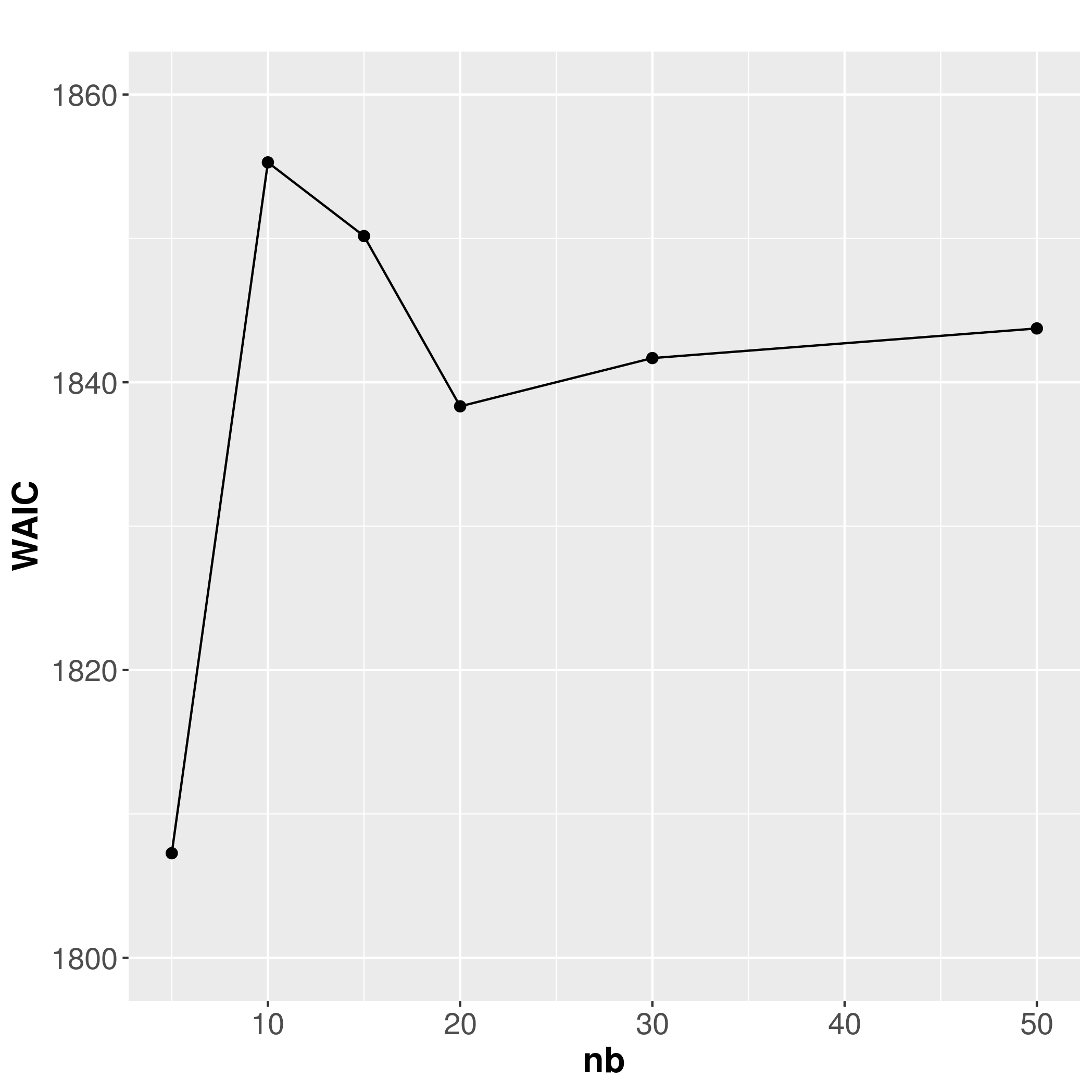} 
\end{center}
\caption{\small SIM II ($\phi$ = 6). INLA results. Criteria assessment:  Running times (first row), LPML (second row)  and WAIC (third row), under block-NNGP models using regular blocks (left column), irregular blocks (middle column) and NNGP models (right column). }
  \label{fig:figs2}
\end{figure}

\begin{figure}
\begin{center}
\includegraphics[scale=0.3]{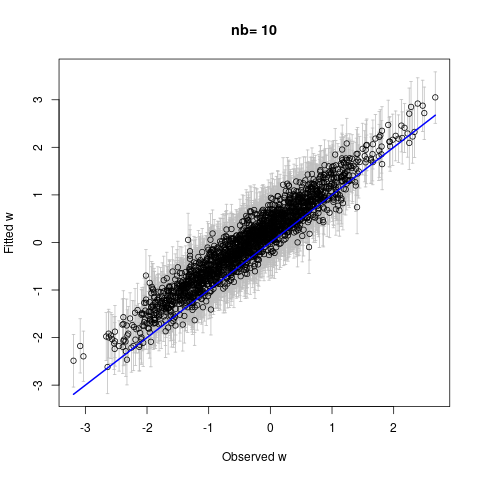}
\includegraphics[scale=0.3]{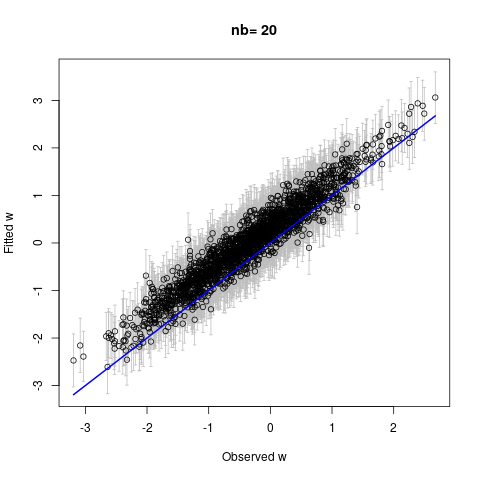}
\includegraphics[scale=0.3]{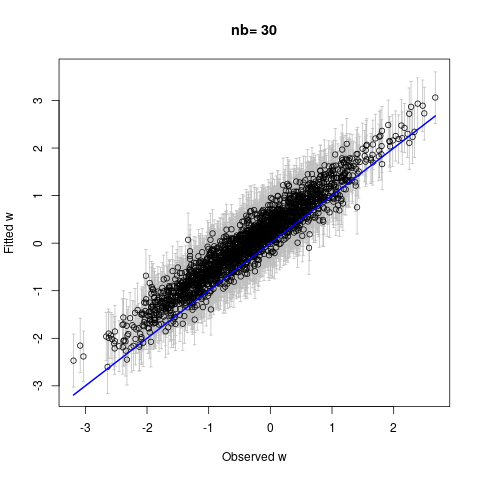}\\
\includegraphics[scale=0.3]{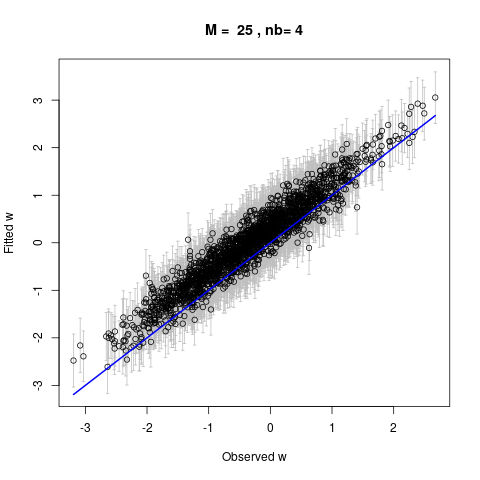}
\includegraphics[scale=0.3]{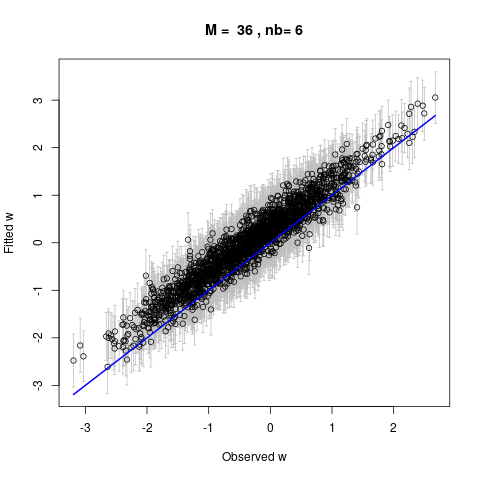}
\includegraphics[scale=0.3]{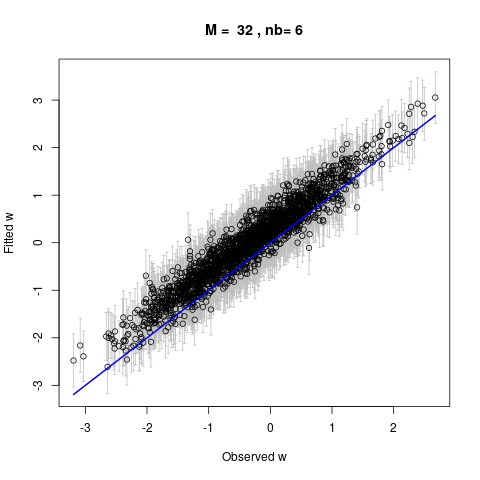}
\end{center}
\caption{ SIM I ($\phi = 12$). INLA results. Mean posterior estimates of spatial effects for different  NNGP (upper panel) and block-NNGP models (lower panel).}
  \label{fig:figs1}
\end{figure}

\begin{figure}
\begin{center}
\includegraphics[scale=0.3]{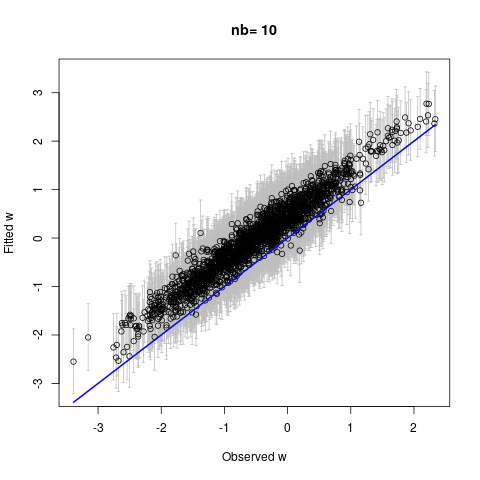}
\includegraphics[scale=0.3]{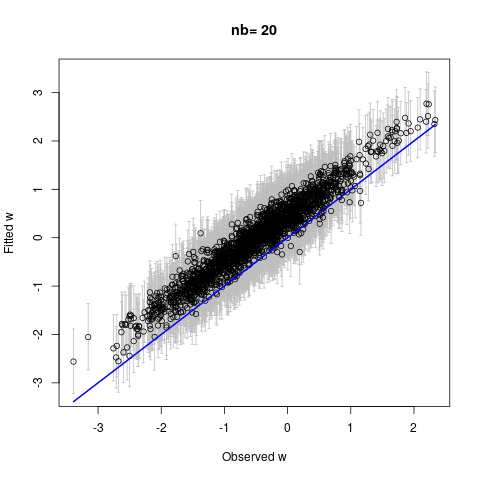}
\includegraphics[scale=0.3]{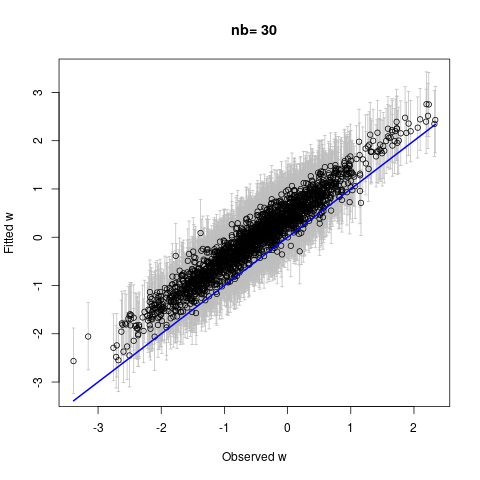}\\
\includegraphics[scale=0.3]{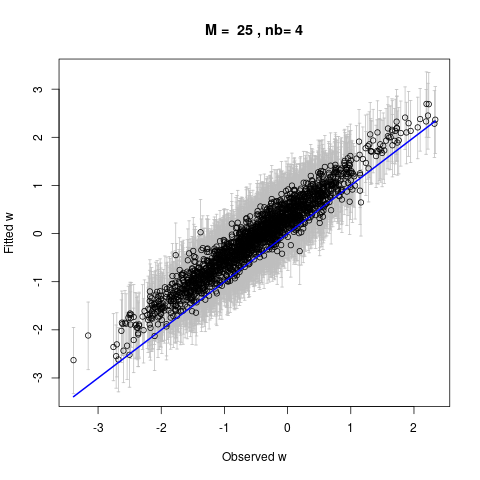}
\includegraphics[scale=0.3]{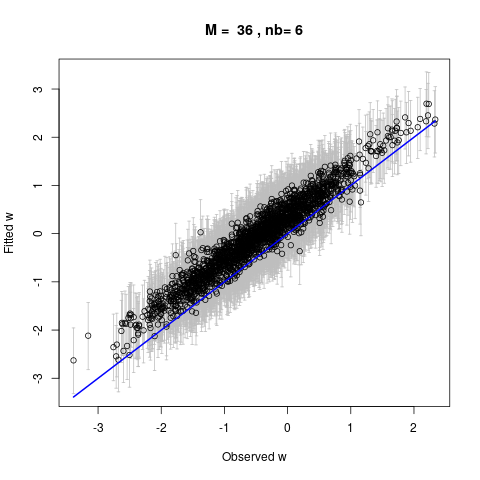}
\includegraphics[scale=0.3]{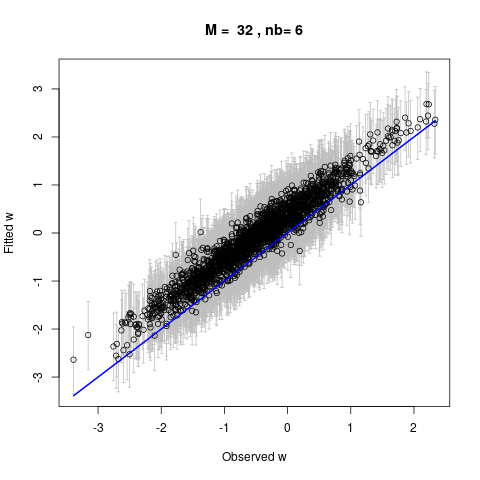}
\end{center}
\caption{ SIM II ($\phi = 6$). INLA results. Mean posterior estimates of spatial effects for different  NNGP (upper panel) and block-NNGP models (lower panel).}
  \label{fig:figs3}
\end{figure}

In this section more results on simulations are presented. 
Figure~\ref{fig:figs0} and Figure~\ref{fig:figs2} show  the criteria assessment and time requirements for scenarios SIM I ($\phi = 12$) and SIM II ($\phi = 6$).  
In general, NNGP models with $nb=4$ or $nb=6$ neighbor blocks show a better performance in terms of computational cost and goodness of fit. Almost all the models run in less than an hour, and in the best scenarios, they run in less than $1000$ seconds, showing the great advantage of running the block-NNGP models through INLA. It is observed that the computing times requirements for the block-NNGP models decreases as the number of neighbor blocks increases. While for the NNGP models, the computational cost increases as the number of neighbor increases. The LPML for all block-NNGP models with $nb=4$ or $nb=6$ neighbor blocks did not significantly change. A similar pattern is presented for NNGP-models with $nb=20$ or higher. The WAIC for block-NNGP models with regular blocks, $nb=4$ or $nb=6$ neighbor blocks are quite similar, while for irregular blocks, it tend to increase its value when the number of neighbor blocks increases. The posterior mean estimation of $w_S$  for scenarios SIM I ($\phi = 12$) and SIM II ($\phi = 6$) are displayed in Figure~\ref{fig:figs1} and Figure~\ref{fig:figs3}, respectively. These results confirm that the NNGP and block-NNGP models show a very good performance when the range is not too large and, as the range increases, the larger the credible intervals of the spatial effects. 

\begin{figure}
\begin{center}
\includegraphics[scale=0.22]{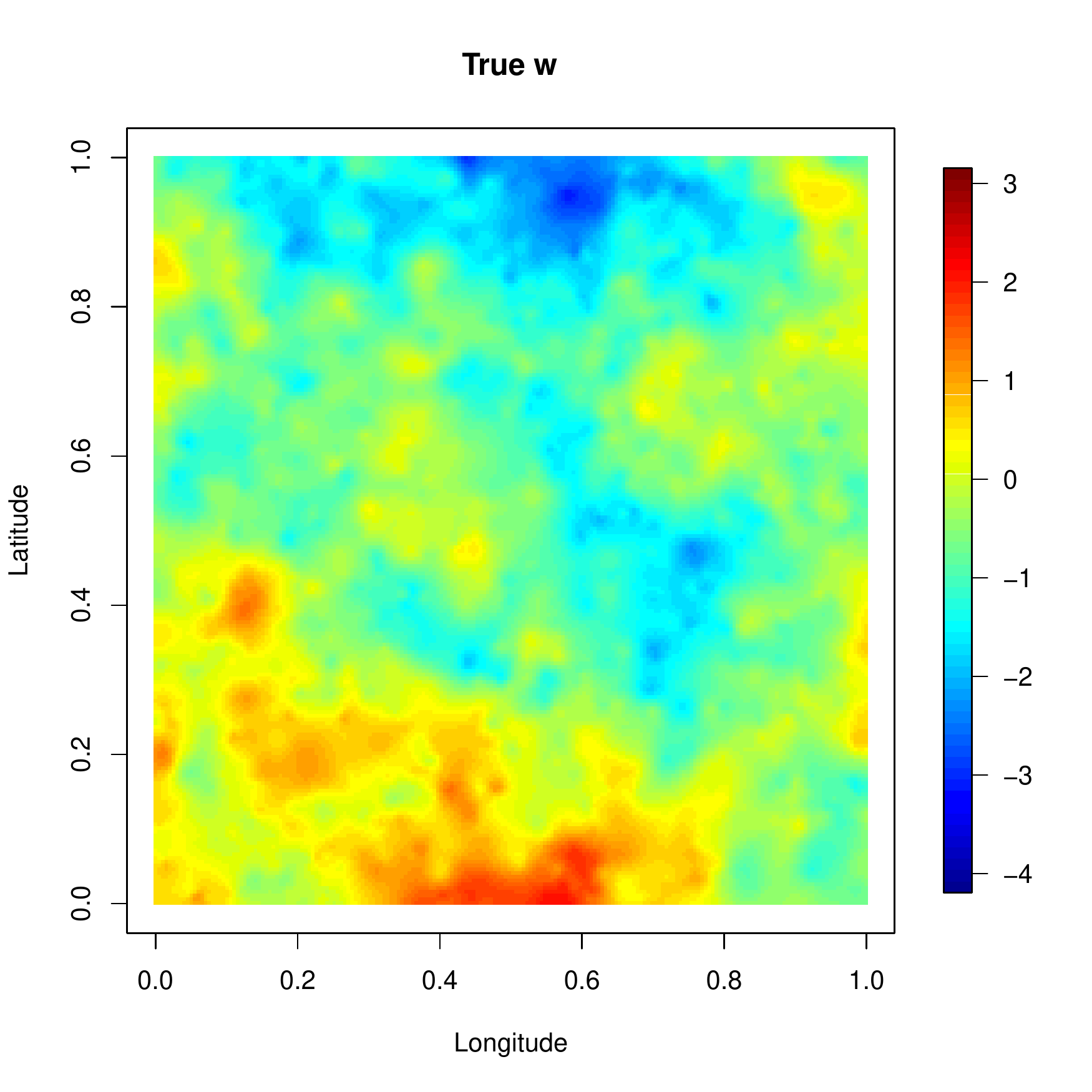}
\includegraphics[scale=0.22]{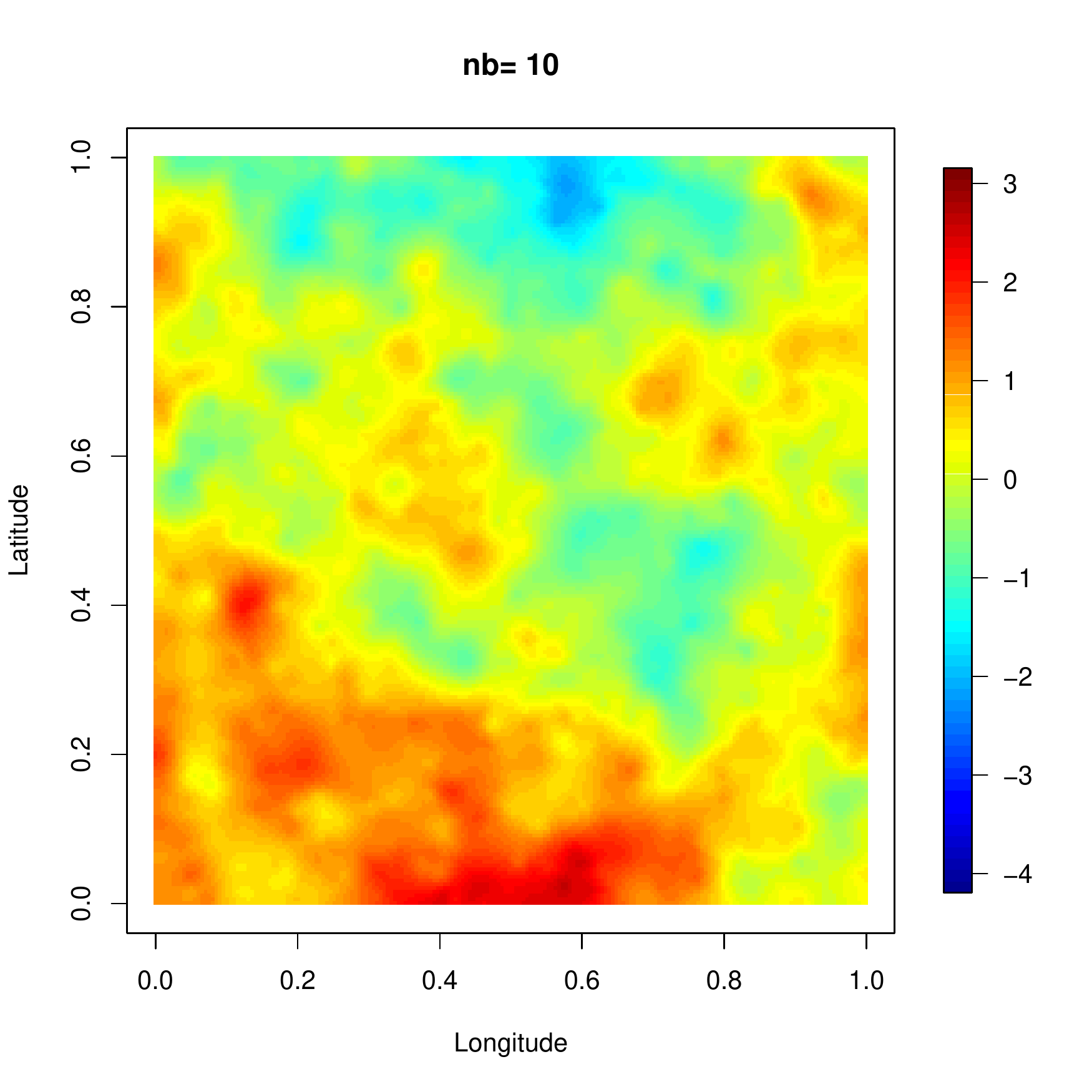}
\includegraphics[scale=0.22]{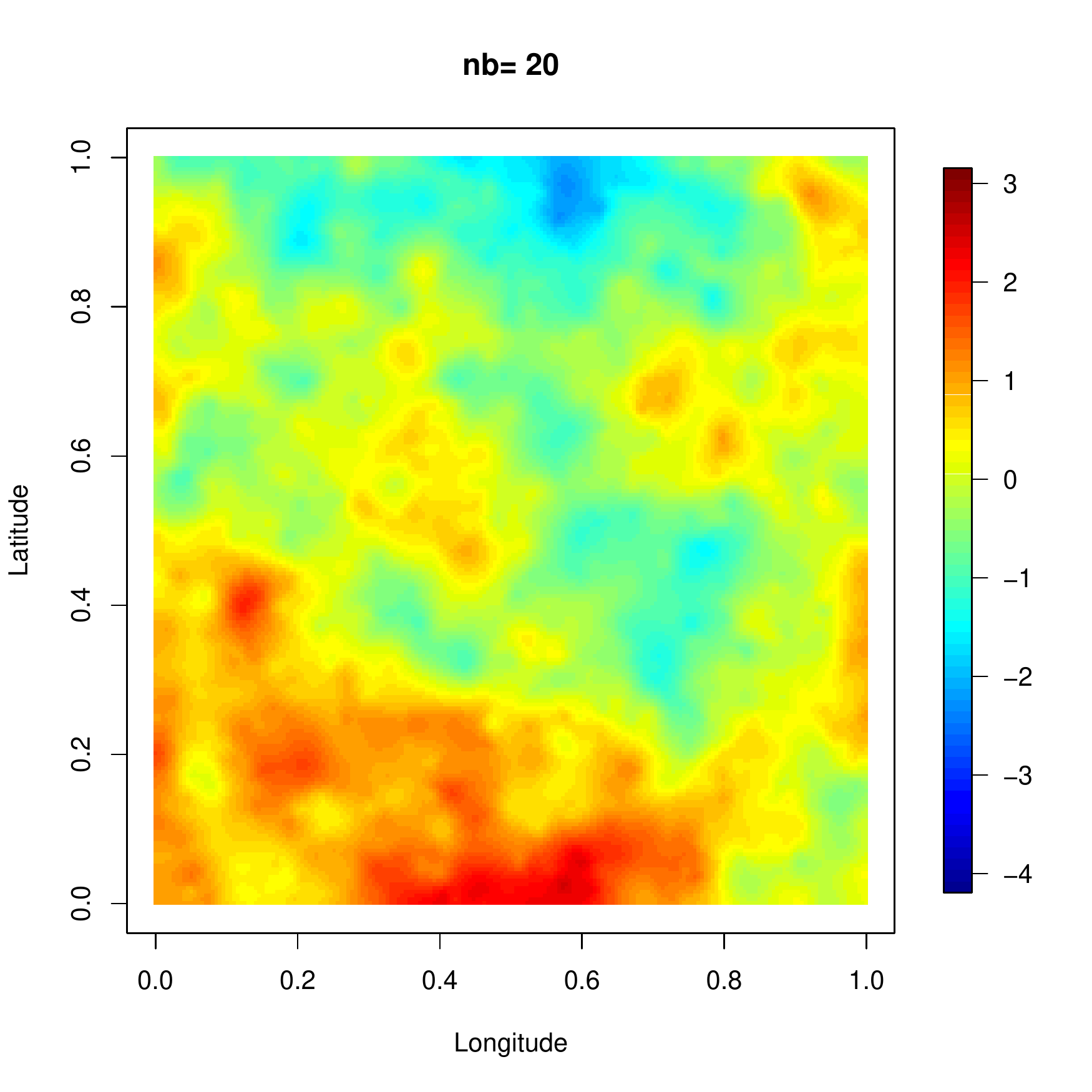}
\includegraphics[scale=0.22]{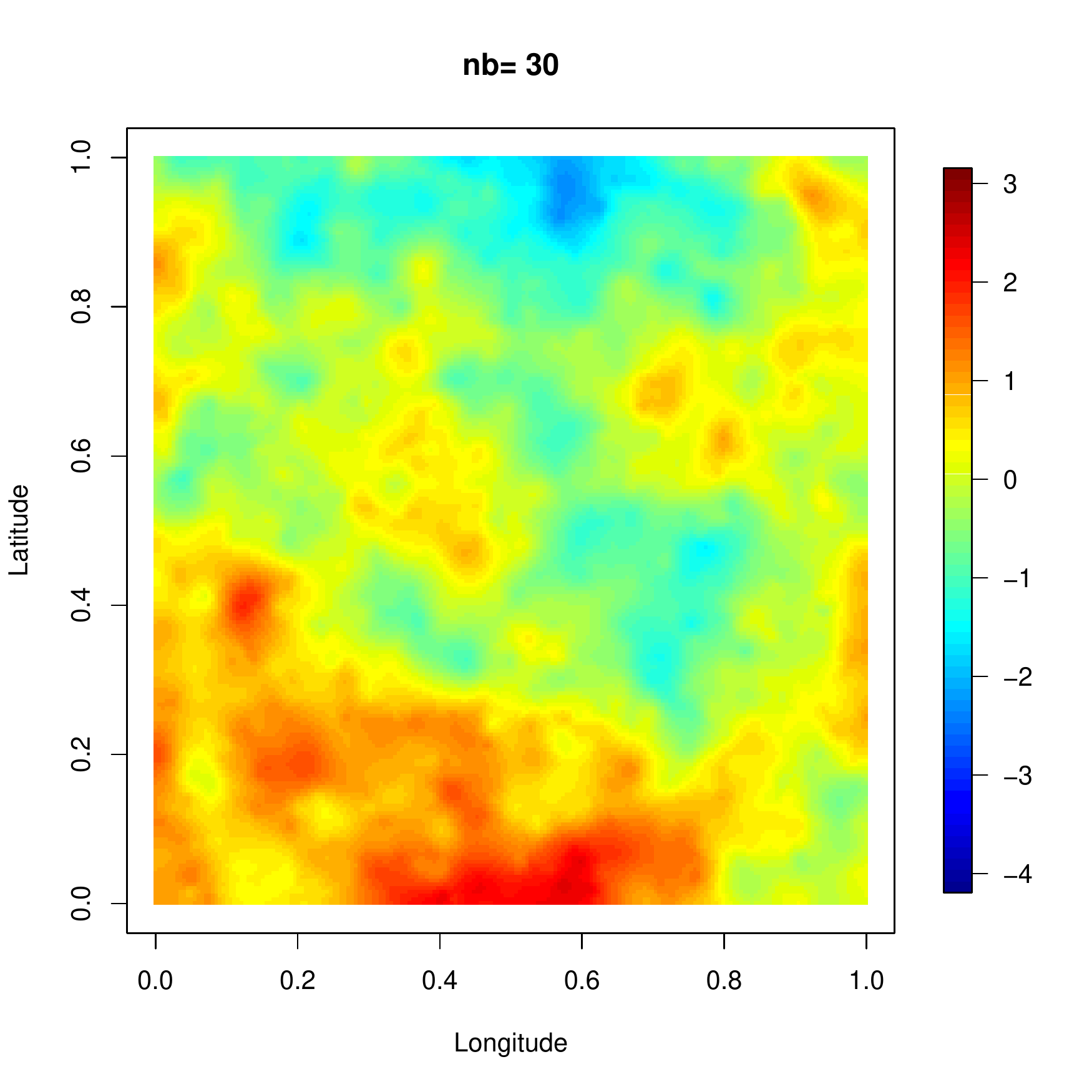} \\
\includegraphics[scale=0.22]{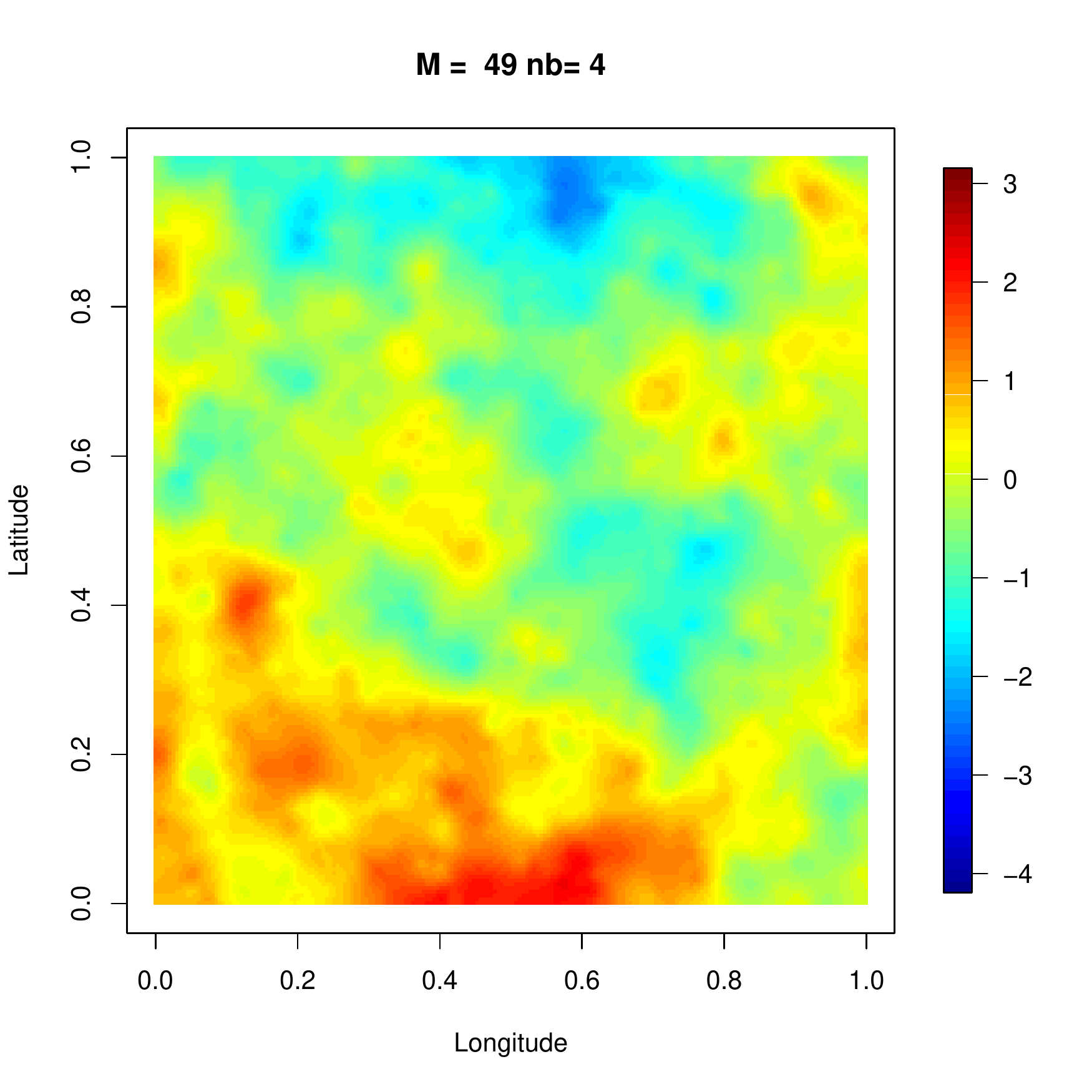}
\includegraphics[scale=0.22]{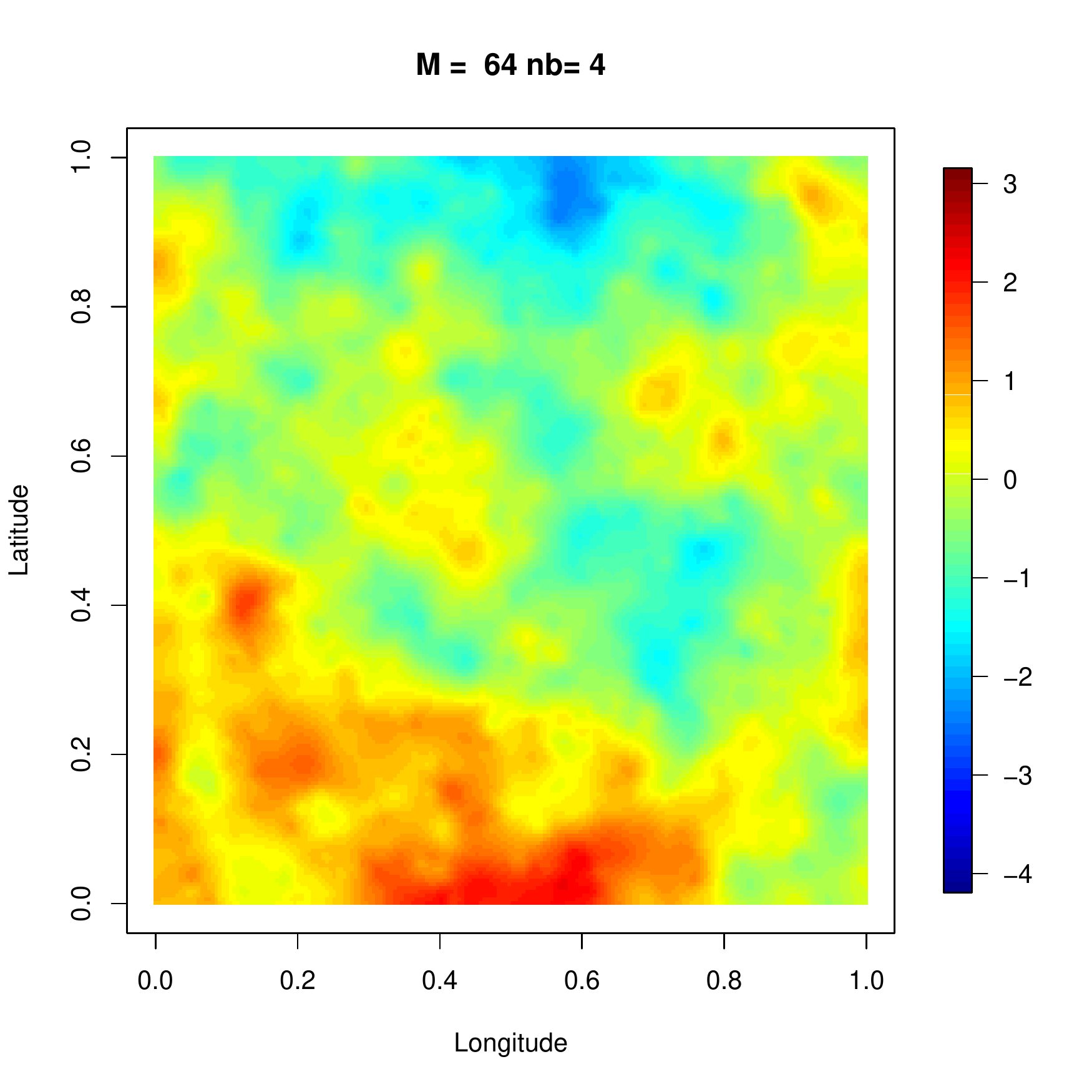}
\includegraphics[scale=0.22]{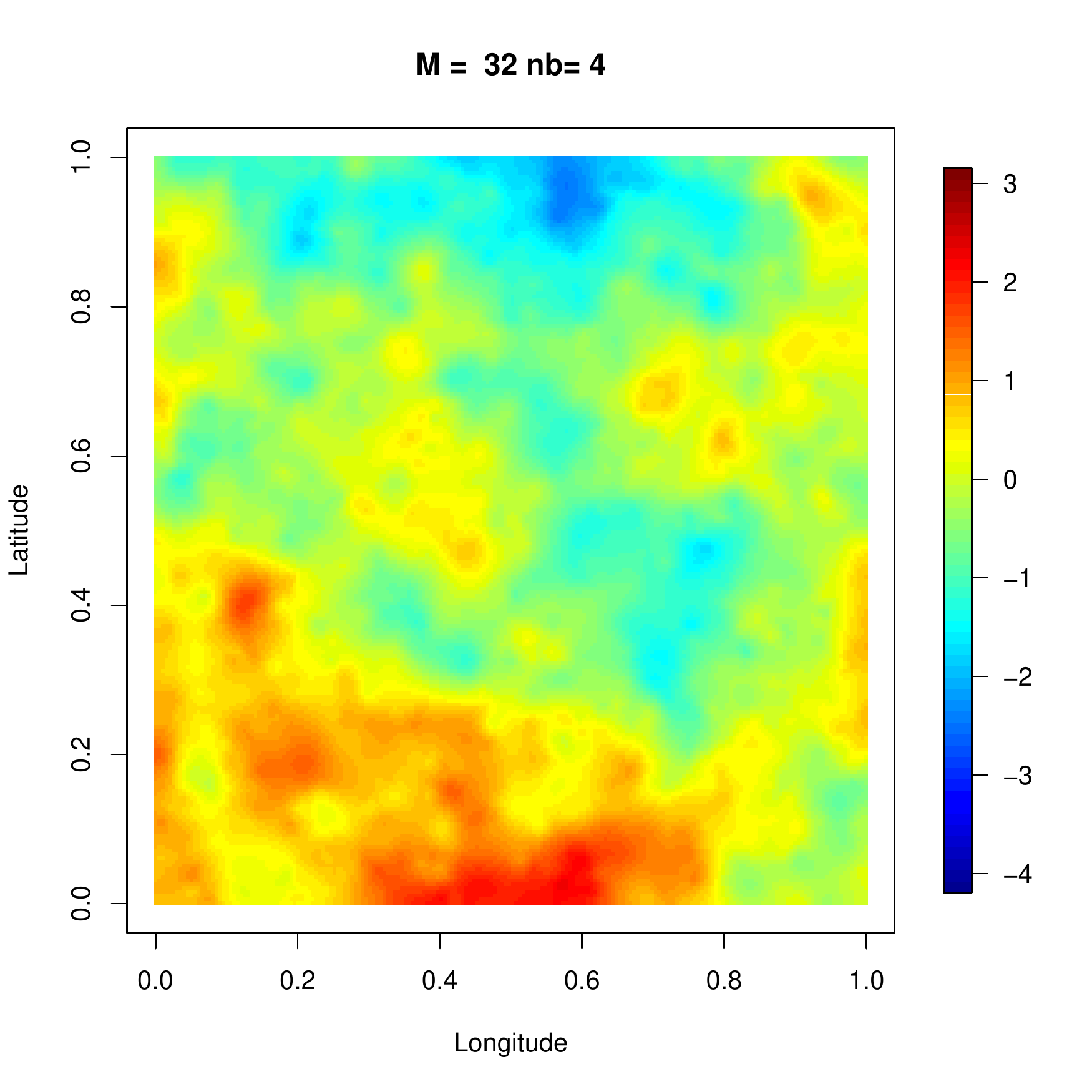}
\includegraphics[scale=0.22]{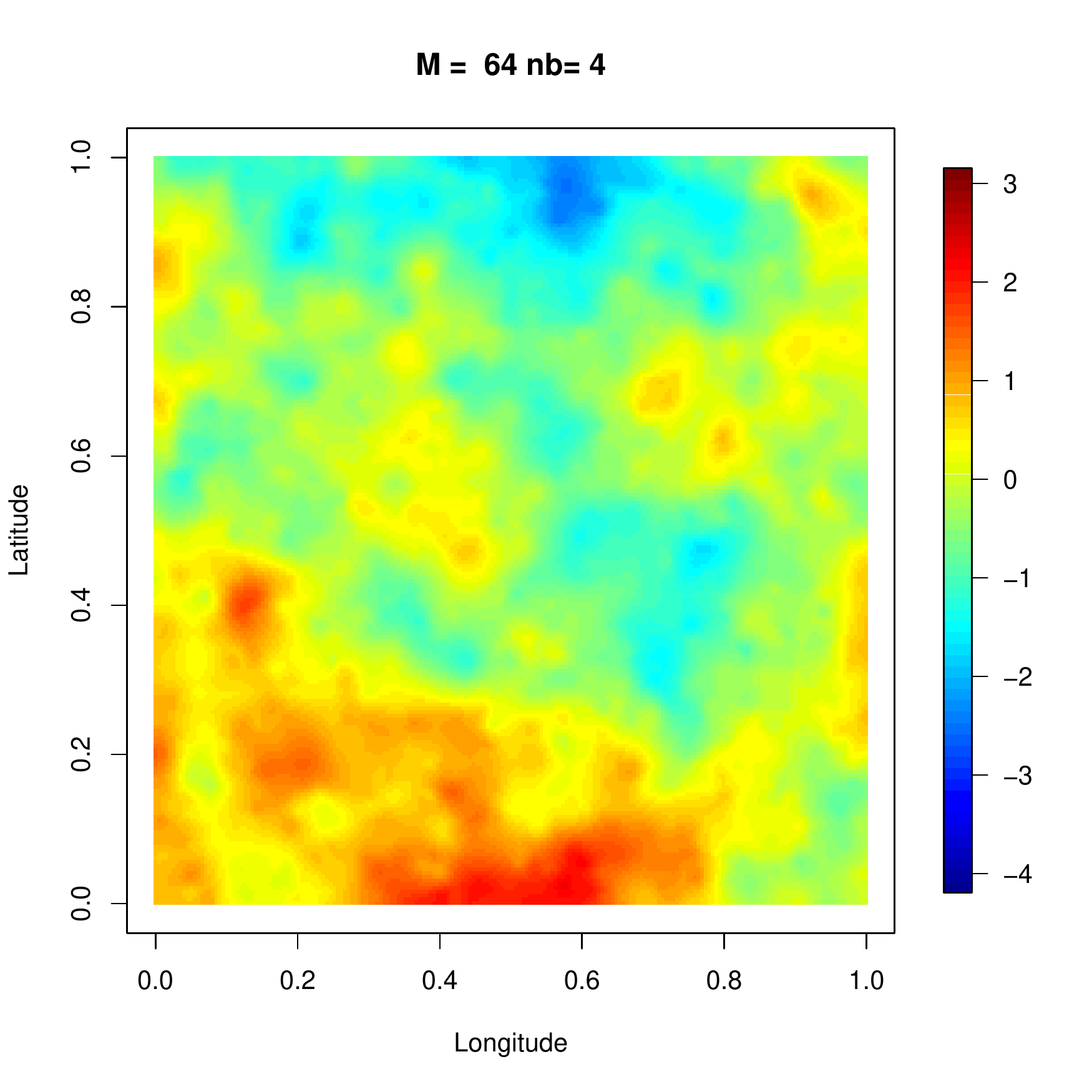}
\end{center}
\caption{ SIM III ($\phi$ = 3). True spatial random effects $w$, and their posterior mean estimates for NNGP models (upper panel) with $nb=10, 20, 30$  neighbors, and different block-NNGP models (lower panel) with regular blocks (R) and irregular blocks (I), using INLA. }
  \label{fig:fig5}
\end{figure}

Figure~\ref{fig:fig5} displays the posterior mean estimates of these spatial random effects interpolated over the domain for SIM III ($\phi=3$). 
The block-NNGP models show better approximations than the NNGP models. We can observe that the NNGP model with 10 neighbors did not approximate well the spatial process, it overestimate the spatial effects, specially in the south region. Indeed, the NNGP has proven to be successful in capturing local/small-scale variation of spatial processes, however, it might have one  disadvantage: inaccuracy in representing global/large scale dependencies. This might happen because the NNGP built the DAG based on observations, adversely, the block-NNGP built a chain graph based on blocks of observations, which captures both small and large dependence. 	

\begin{figure}
\begin{center}
\vspace{4cm}
\includegraphics[scale=0.3]{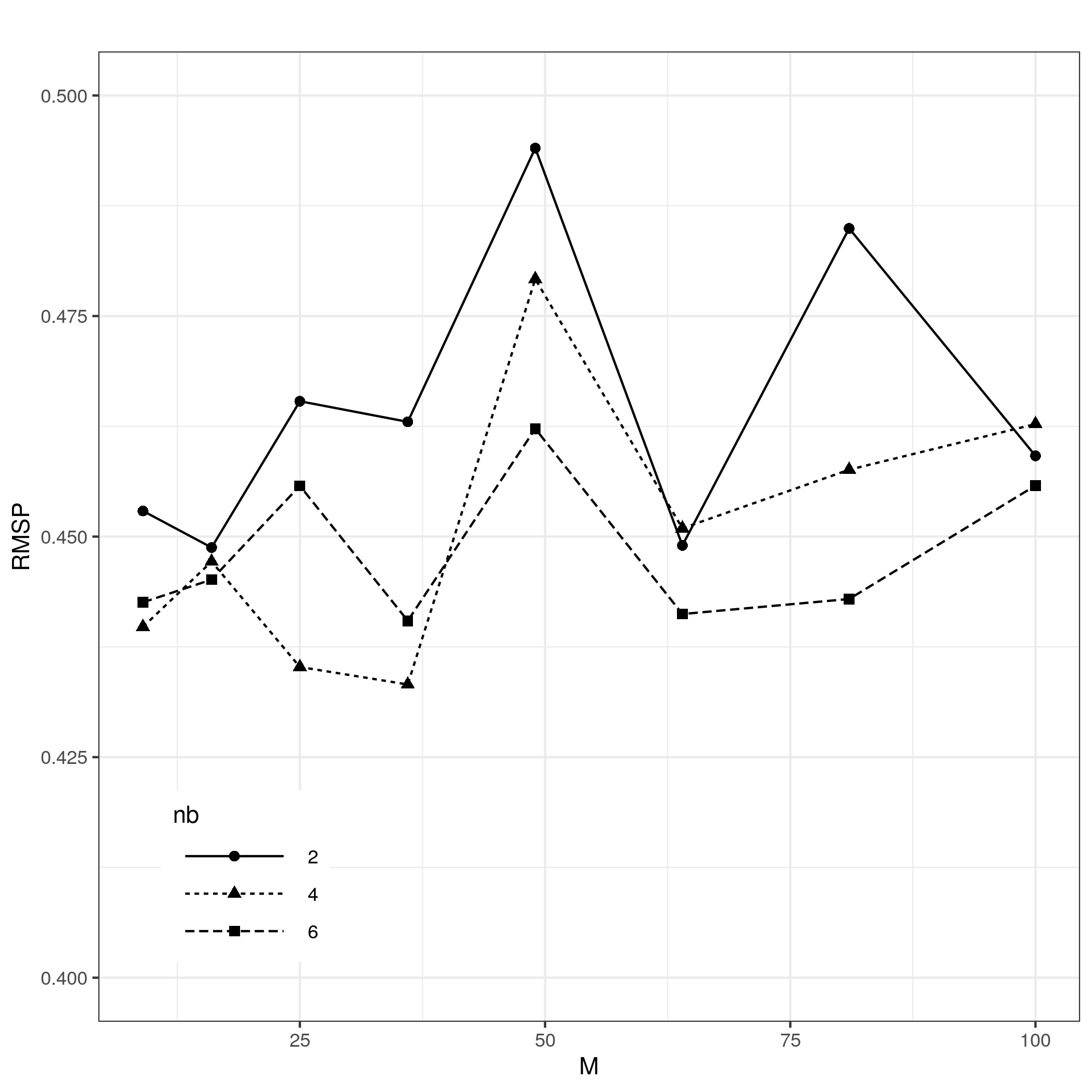} 
\includegraphics[scale=0.3]{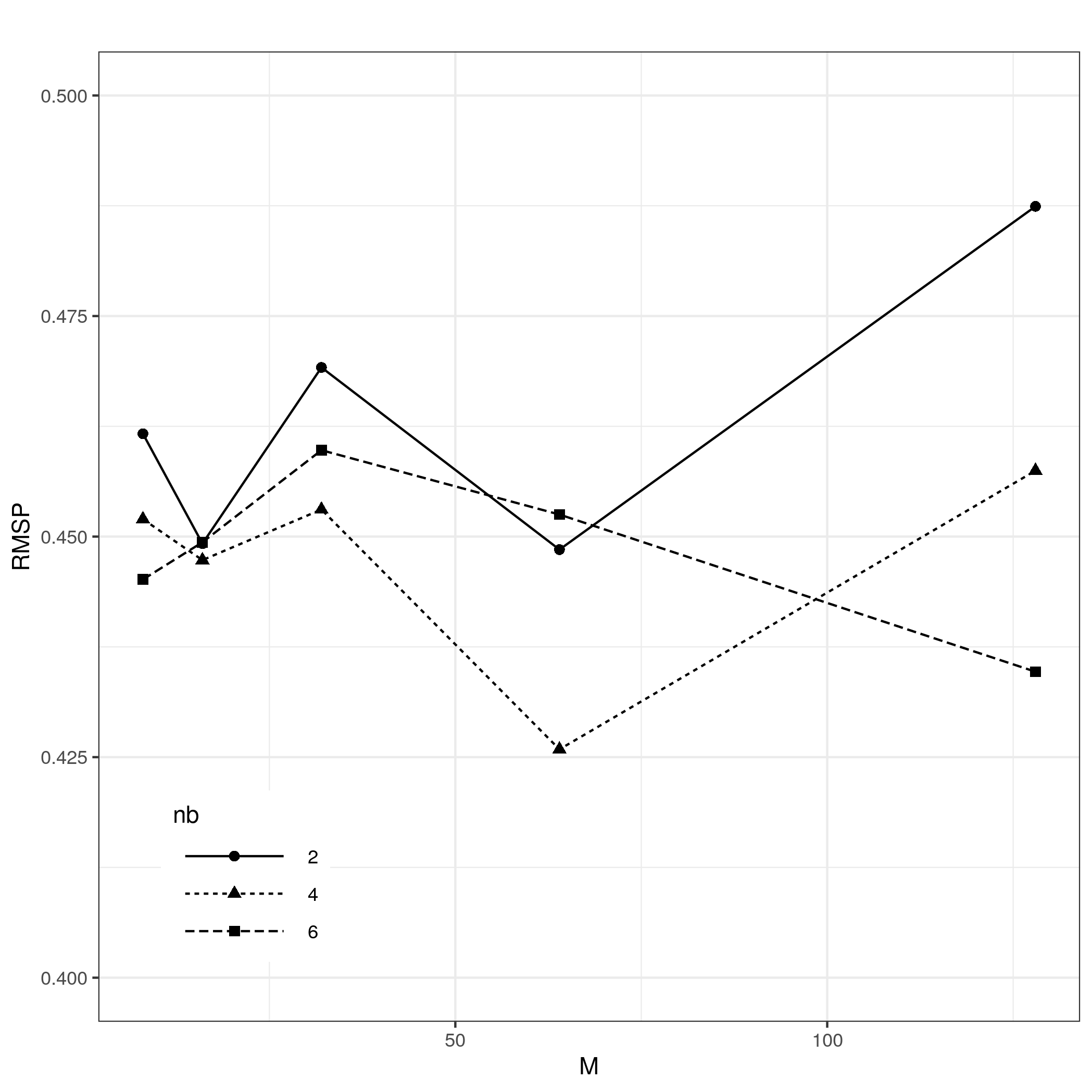} 
\includegraphics[scale=0.3]{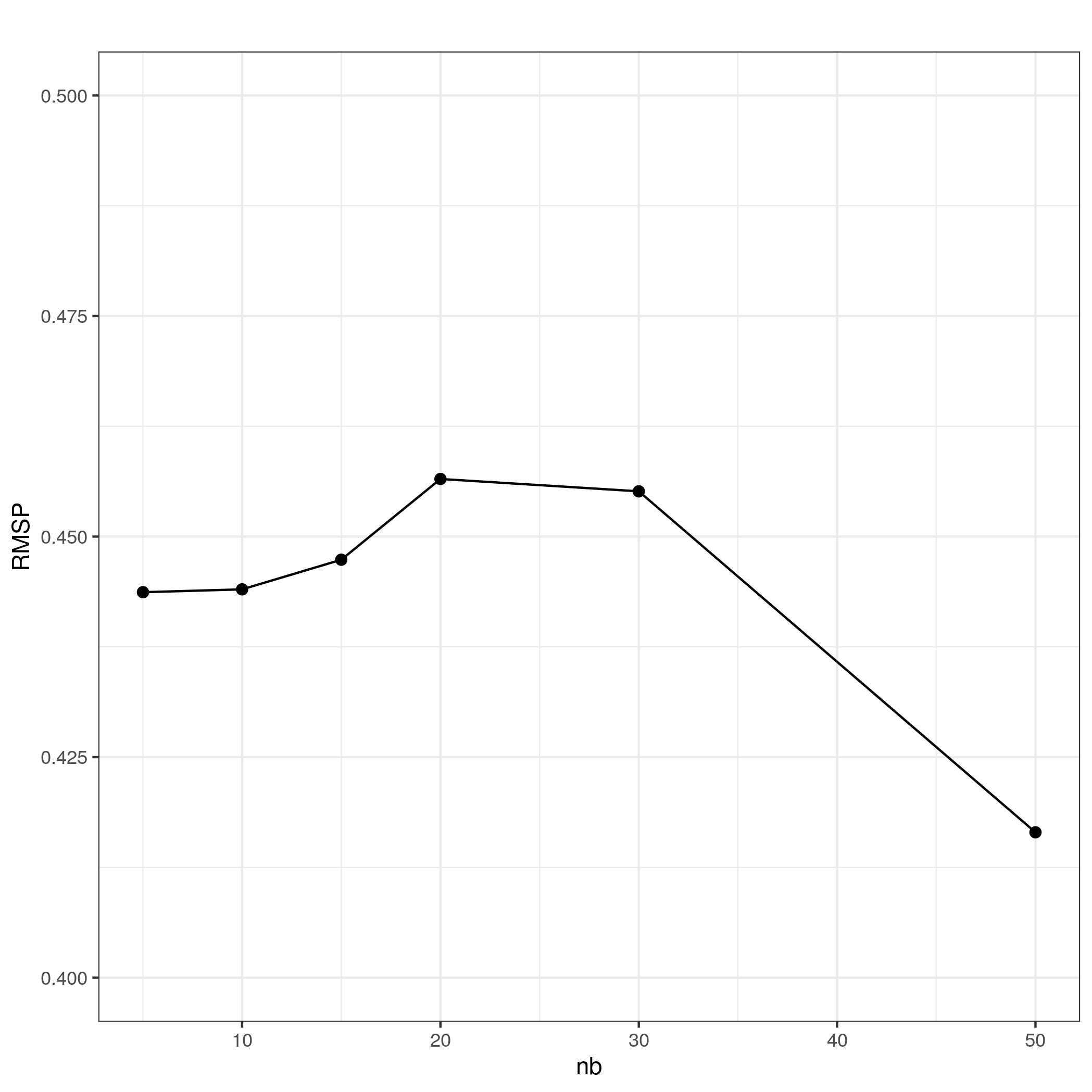} 
\end{center}
\caption{ SIM III ($\phi$ = 3). INLA results. Root mean square prediction error under block-NNGP models using regular blocks (left column), irregular blocks (middle column) and NNGP models (right column). }
  \label{fig:figs00}
\vspace{3cm}

\end{figure}

The root of mean square prediction error (RMSP) for all the fitted models, under SIM III, is shown in Figure~\ref{fig:figs00}. For block-NNGP models, as expected the prediction metric increases when the number of blocks is increased, being lower for a higher number of neighbor blocks. This result is a little bit more evident for irregular blocks. For NNGP models, the RMSP is higher when the number of neighbors is small.

\begin{figure}
\begin{center}
\includegraphics[scale=0.35]{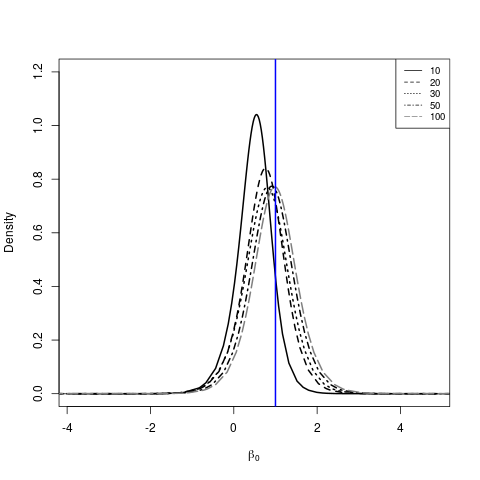}
\includegraphics[scale=0.35]{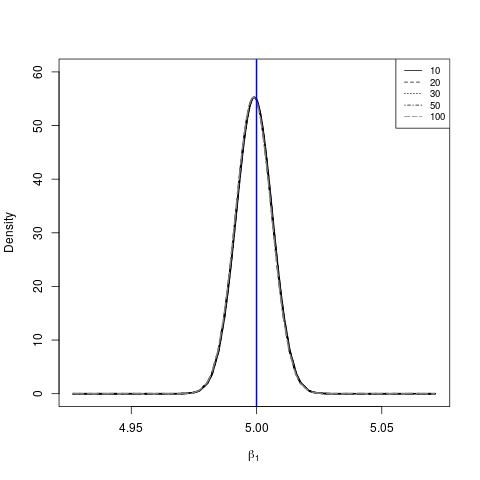}
\includegraphics[scale=0.35]{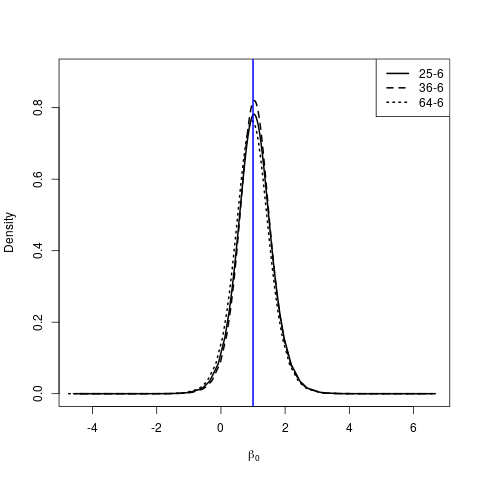}
\includegraphics[scale=0.35]{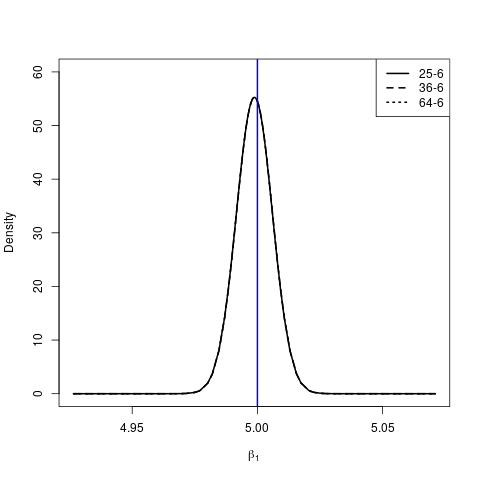}
\end{center}
\caption{INLA results for simulation of GP with Mat\'{e}rn covariance function ($\nu=1.5$, $\phi = 3.5$, $\sigma^2=1$ and $\tau^2=0.1$). Posterior marginal densities of regression coefficient effects for NNGP models (upper panel)with $nb = 10, 20, 30, 50, 100$ neighbors and block-NNGP models (lower panel) with $M = 25, 36, 64$ regular blocks and $nb=6$ neighbor blocks.  The solid vertical blue lines represent the true parameter values.
}
  \label{fig:figs8}
\end{figure}

\begin{figure}
\begin{center}
\includegraphics[scale=0.3]{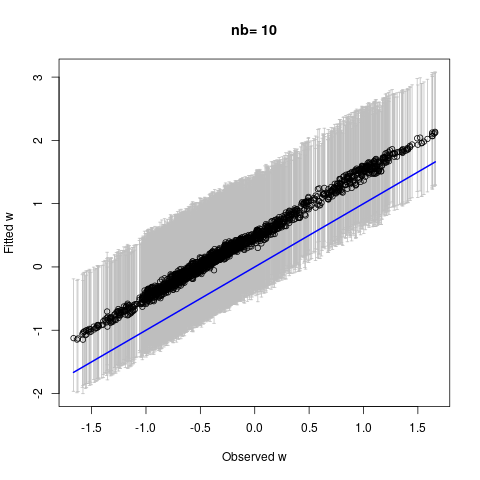}
\includegraphics[scale=0.3]{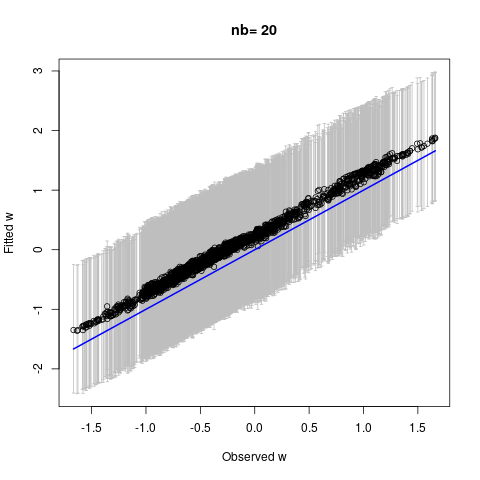}
\includegraphics[scale=0.3]{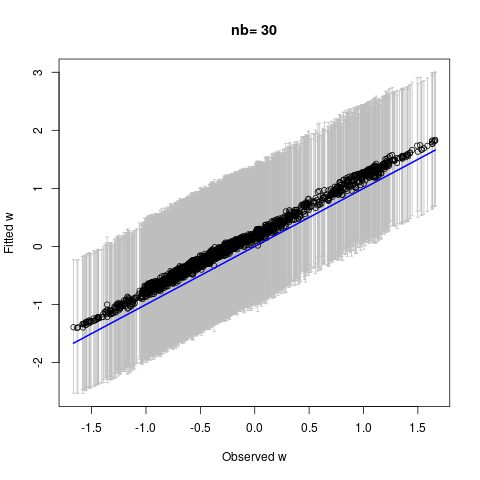}
\includegraphics[scale=0.3]{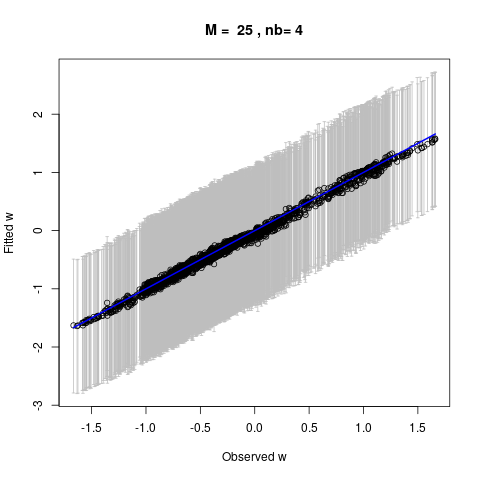}
\includegraphics[scale=0.3]{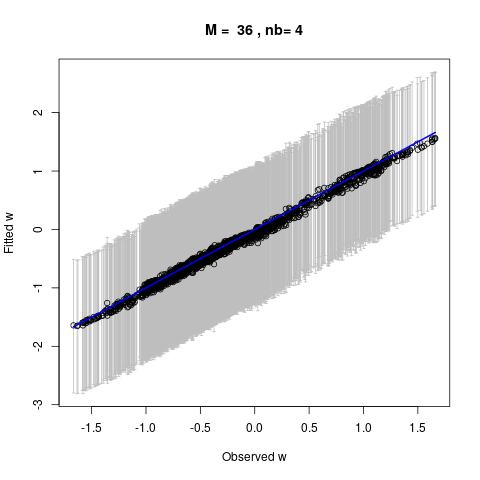}
\includegraphics[scale=0.3]{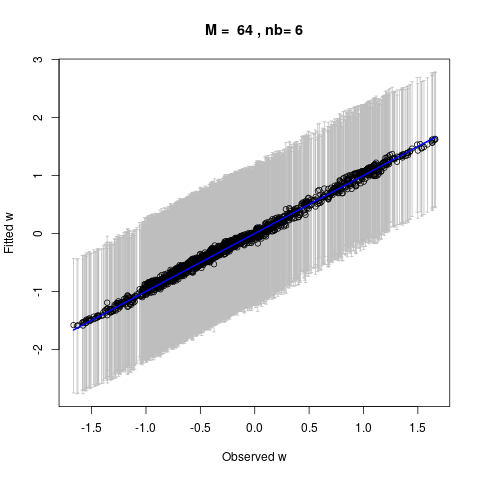}
\end{center}
\caption{ INLA results for simulation of GP with Mat\'{e}rn covariance function($\nu=1.5$, $\phi = 3.5$, $\sigma^2=1$ and $\tau^2=0.1$). Mean posterior estimates of spatial effects for different  NNGP (upper panel) and block-NNGP models using regular blocks (lower panel).}
  \label{fig:figs5}
\end{figure}

\begin{figure}
\begin{center}
\includegraphics[scale=0.3]{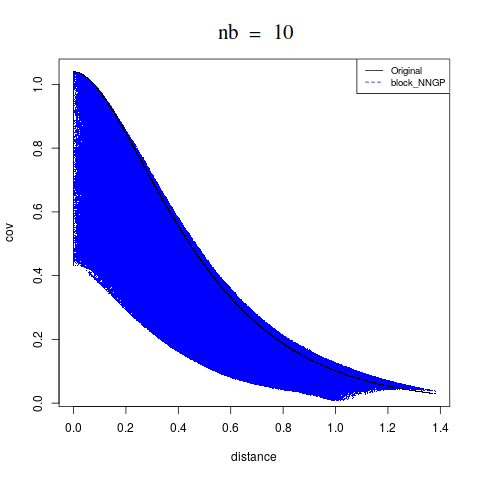}
\includegraphics[scale=0.3]{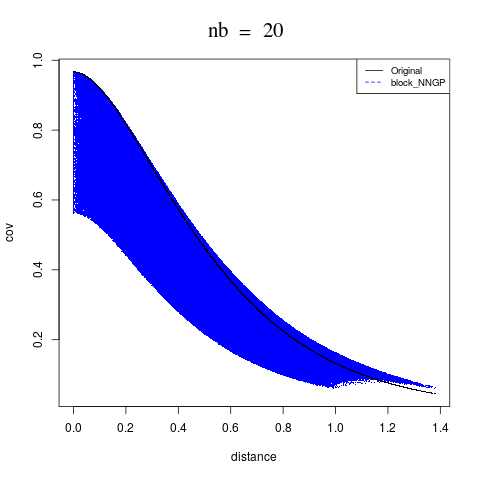}
\includegraphics[scale=0.3]{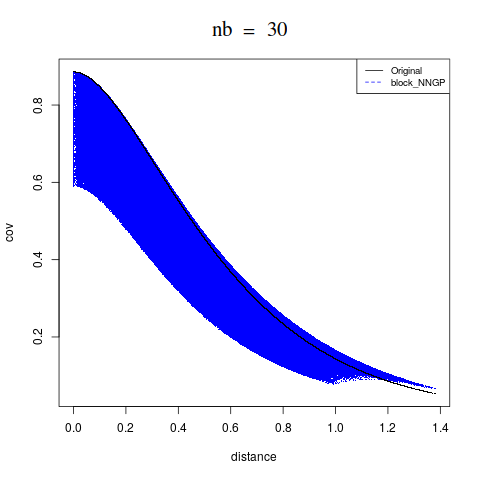}
\includegraphics[scale=0.3]{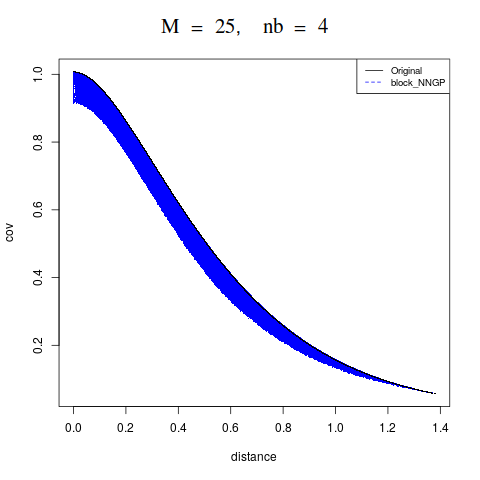}
\includegraphics[scale=0.3]{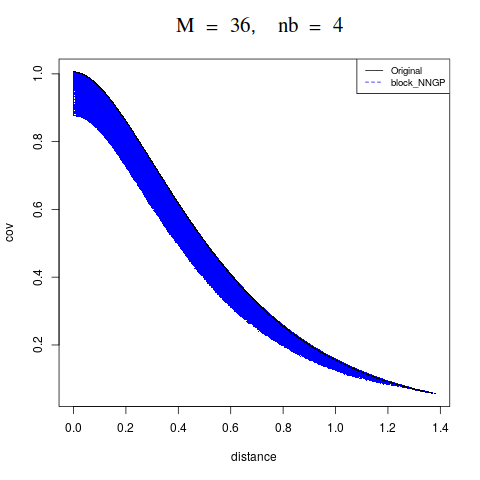}
\includegraphics[scale=0.3]{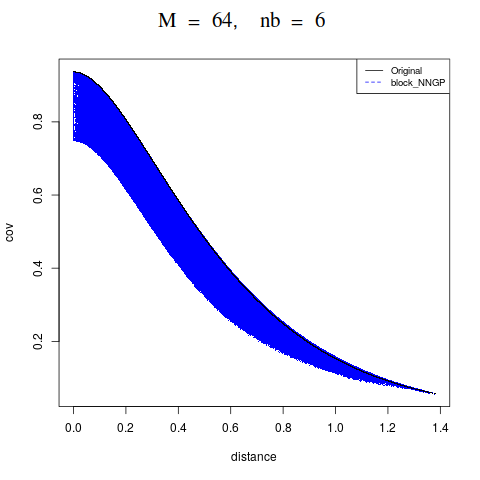}
\end{center}
\caption{ True Mat\'{e}rn covariance function ($\nu=1.5$, $\phi = 3.5$, $\sigma^2=1$ and $\tau^2=0.1$) of GP  against distance (black lines) and empirical approximated covariance of block-NNGP  (Regular blocks) against distance (blue dots) for different  NNGP (upper panel) and block-NNGP models (lower panel).
}
  \label{fig:figs6}
\end{figure}

We also simulate data from the same previous model 
but setting $\nu=1.5$. 
We computed the approximate posterior marginals for the fixed effects (Figure~\ref{fig:figs8}). The block-NNGP models give more accurate results than the NNGP models. For block-NNGP models, the posterior marginals for $\beta_0$ and $\beta_1$ with different number of blocks are overlaid. A different result is obtained for NNGP models, where the posterior marginals for $\beta_0$ with different number of neighbors dramatically change.
Figure~\ref{fig:figs5} displays scatter plots of the true spatial random effects $w(s_i)$ versus the mean posterior estimates. 
Finally, we also computed the theoretical Mat\'{e}rn covariance function (black line) and the empirical covariance function for the NNGP models and block-NNGP models (blue dots) { in Figure~\ref{fig:figs6}} for this scenario.  In general, the match between the theorical and empirical covariance is better for block-NNGP models, specially when the number of blocks is smaller. We further note that for NNGP models when the number of neighbors is small and the range is large, the theoretical covariance function  is far away from the true covariance function. 

\subsection*{D. Supplementary application results }

\begin{figure}[htb]
\begin{center}
\includegraphics[scale=0.45]{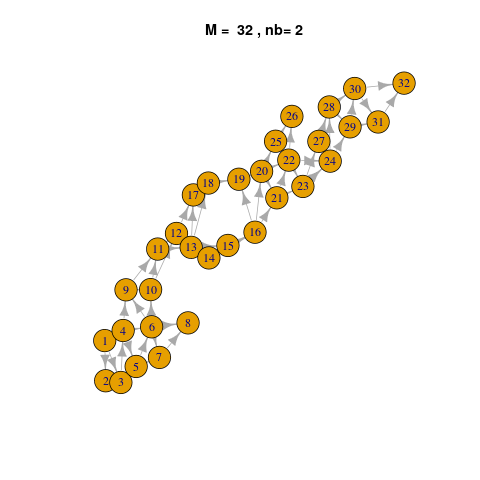}
\includegraphics[scale=0.45]{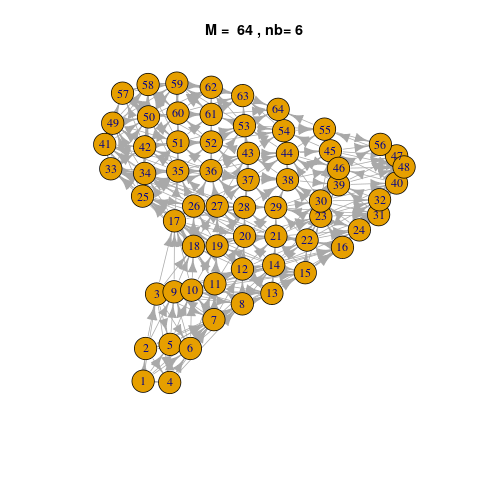}
\end{center}
\caption{ DAG of blocks for mining data (left) using 32 blocks and 2 neighbor blocks.  DAG of blocks for precipitation data (right) using 64 blocks and 6 neighbor blocks.}
  \label{fig:figs4}
  \end{figure}

\begin{figure}
 \centering
\includegraphics[scale=0.25]{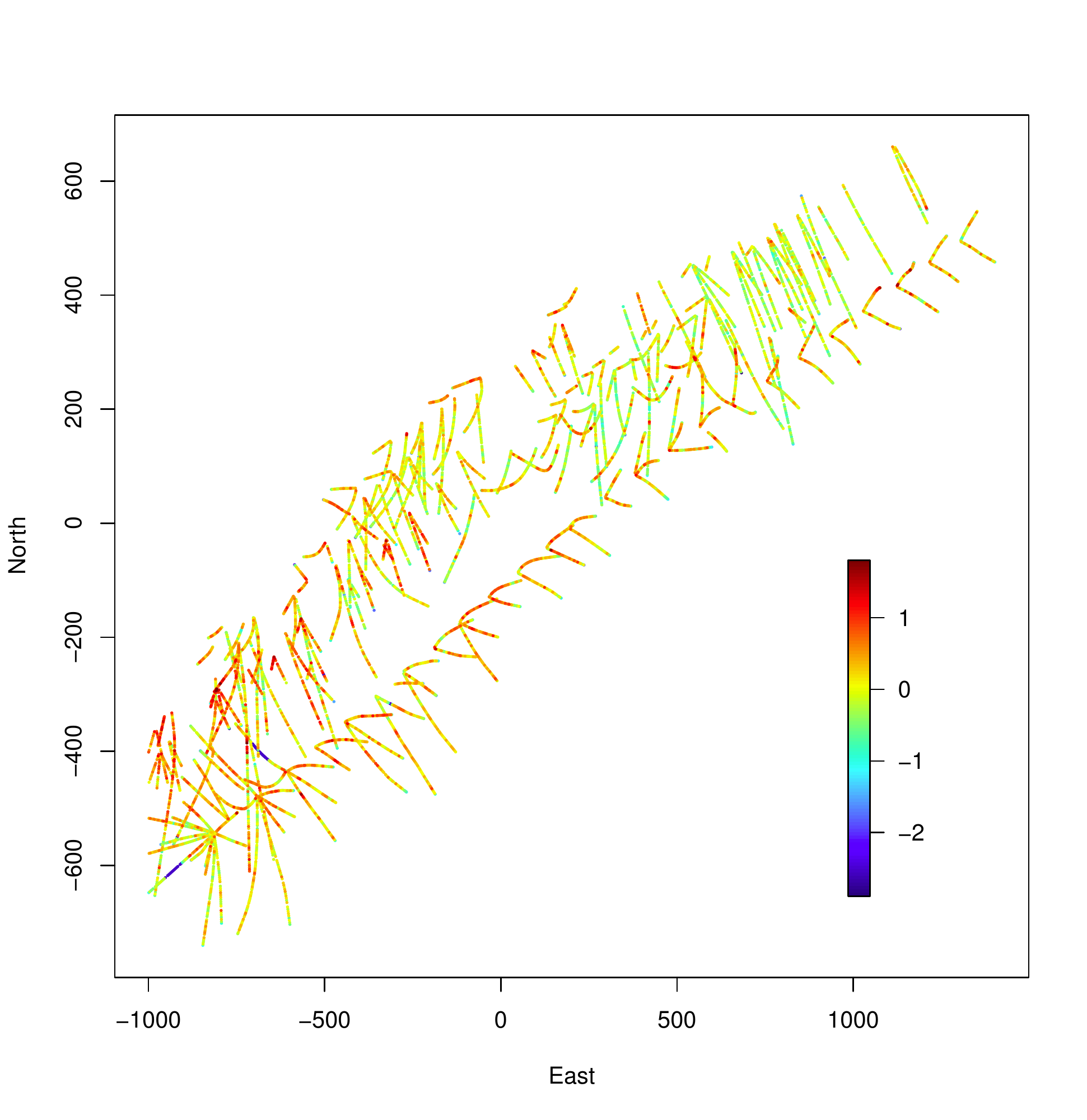}
\includegraphics[scale=0.3]{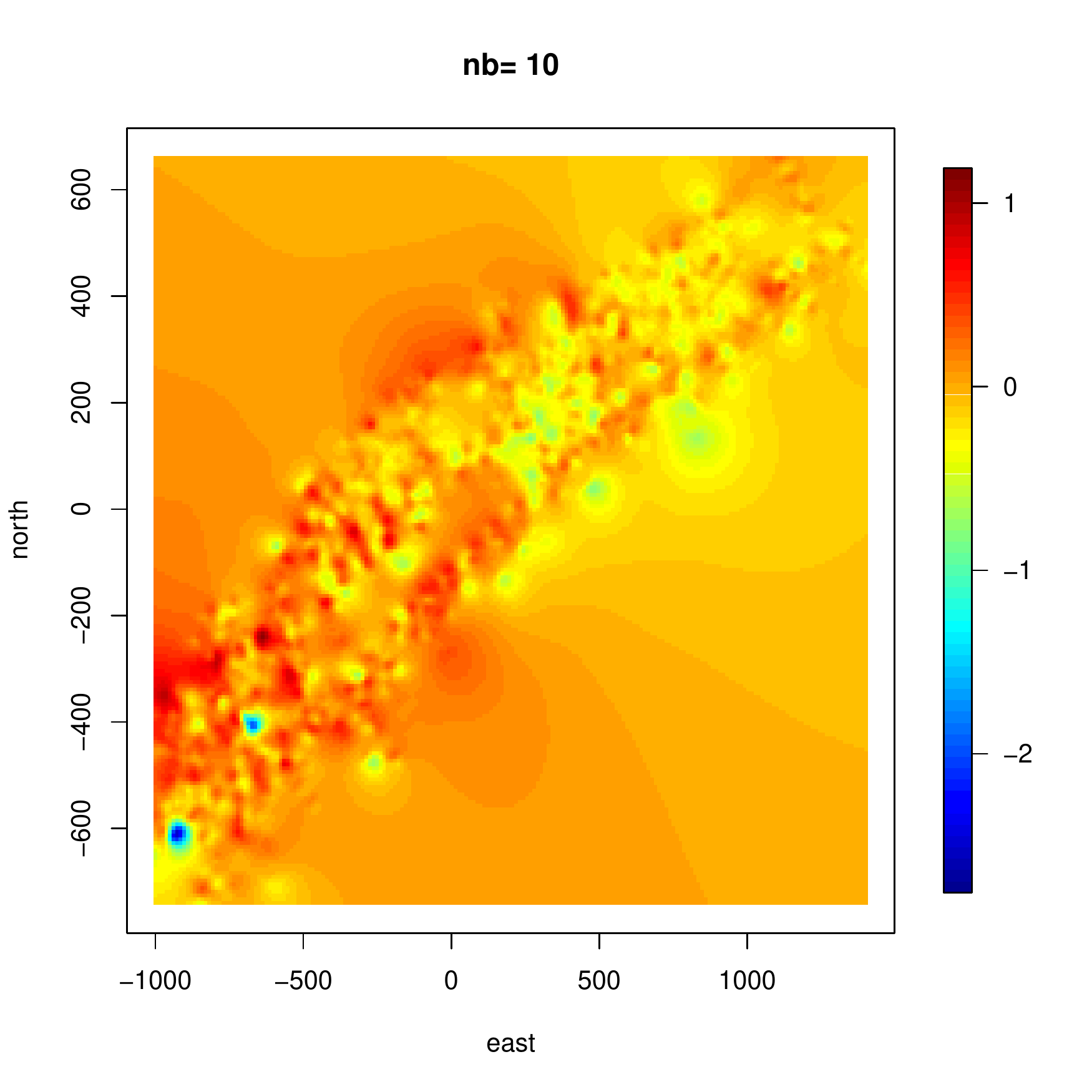}
\includegraphics[scale=0.3]{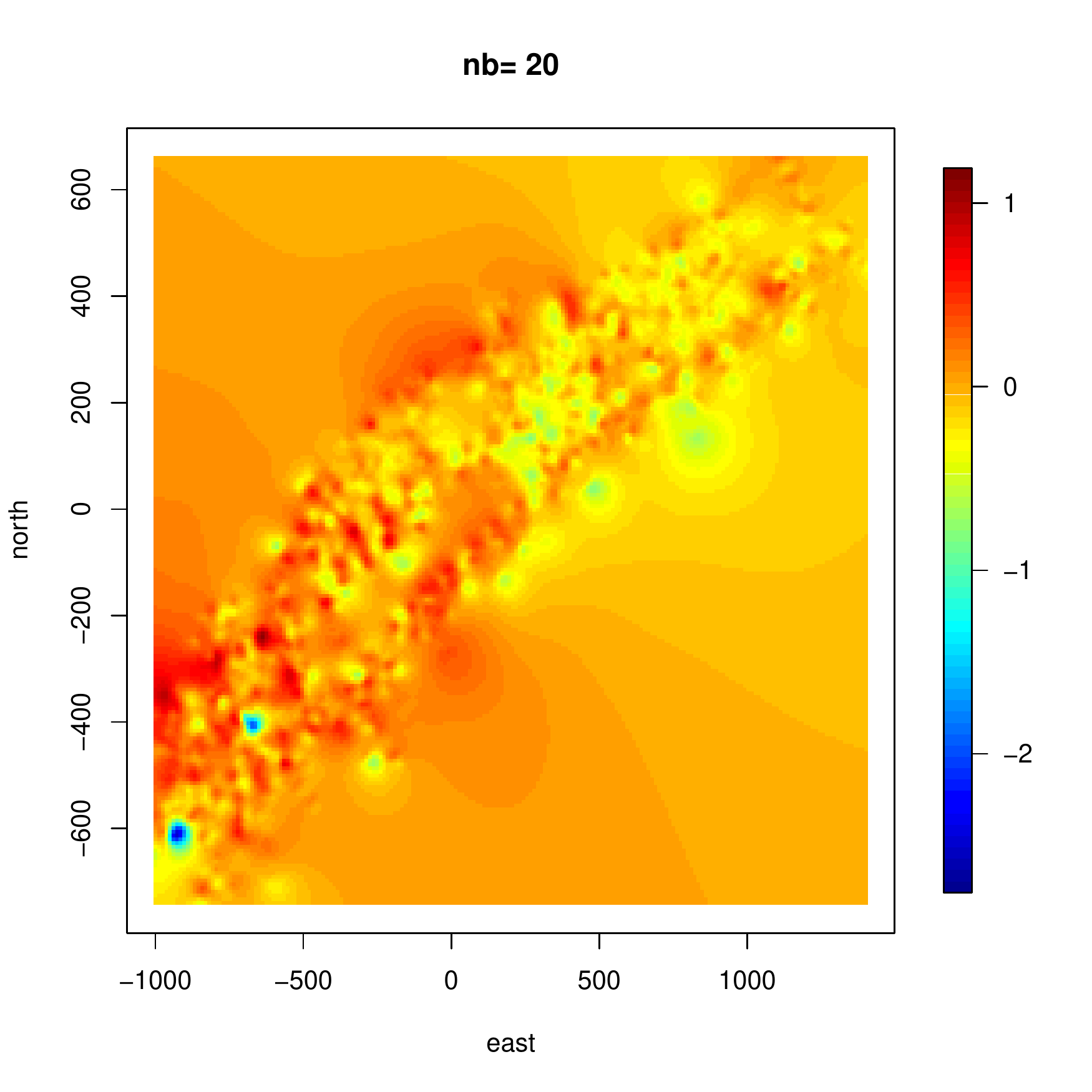}
\includegraphics[scale=0.3]{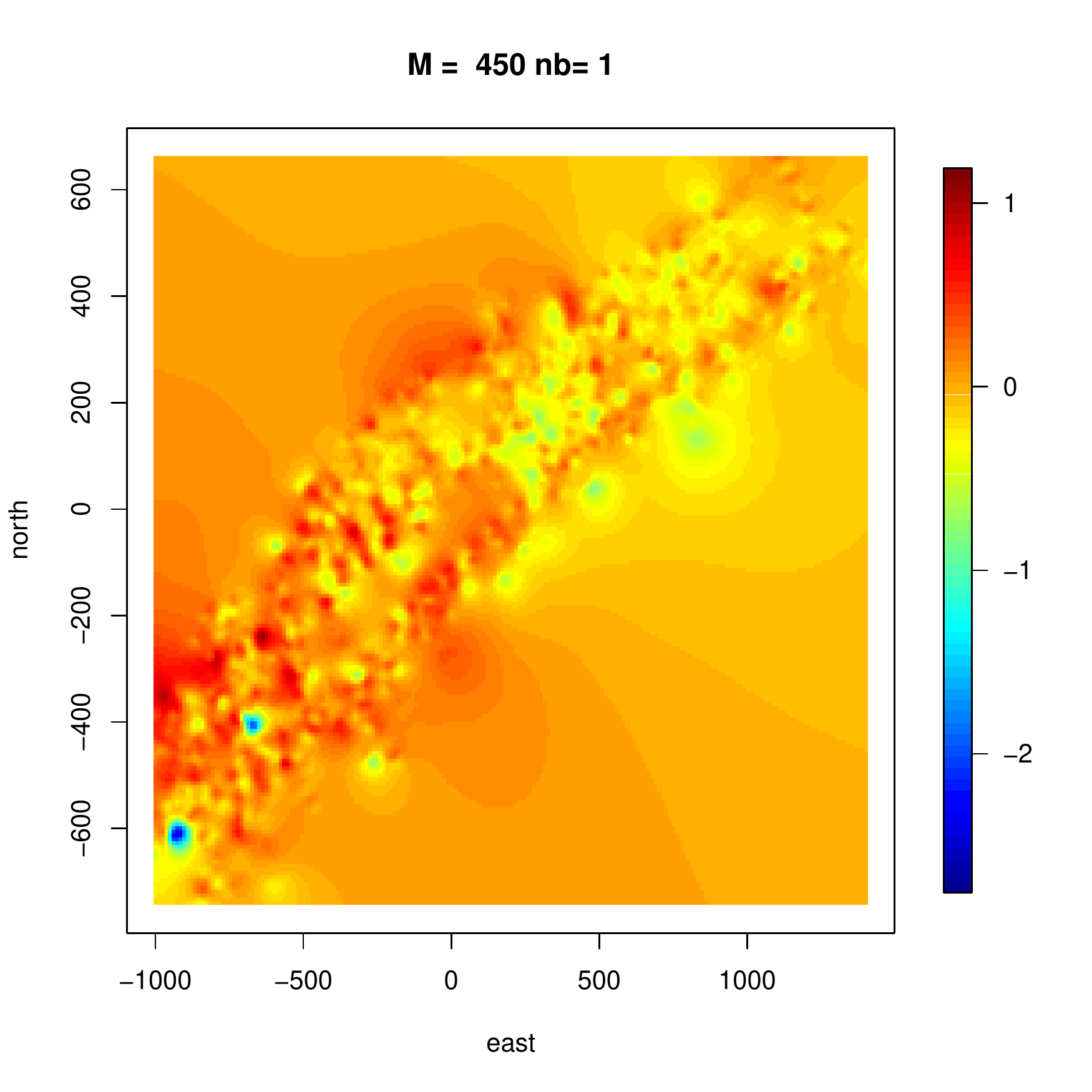}
\includegraphics[scale=0.3]{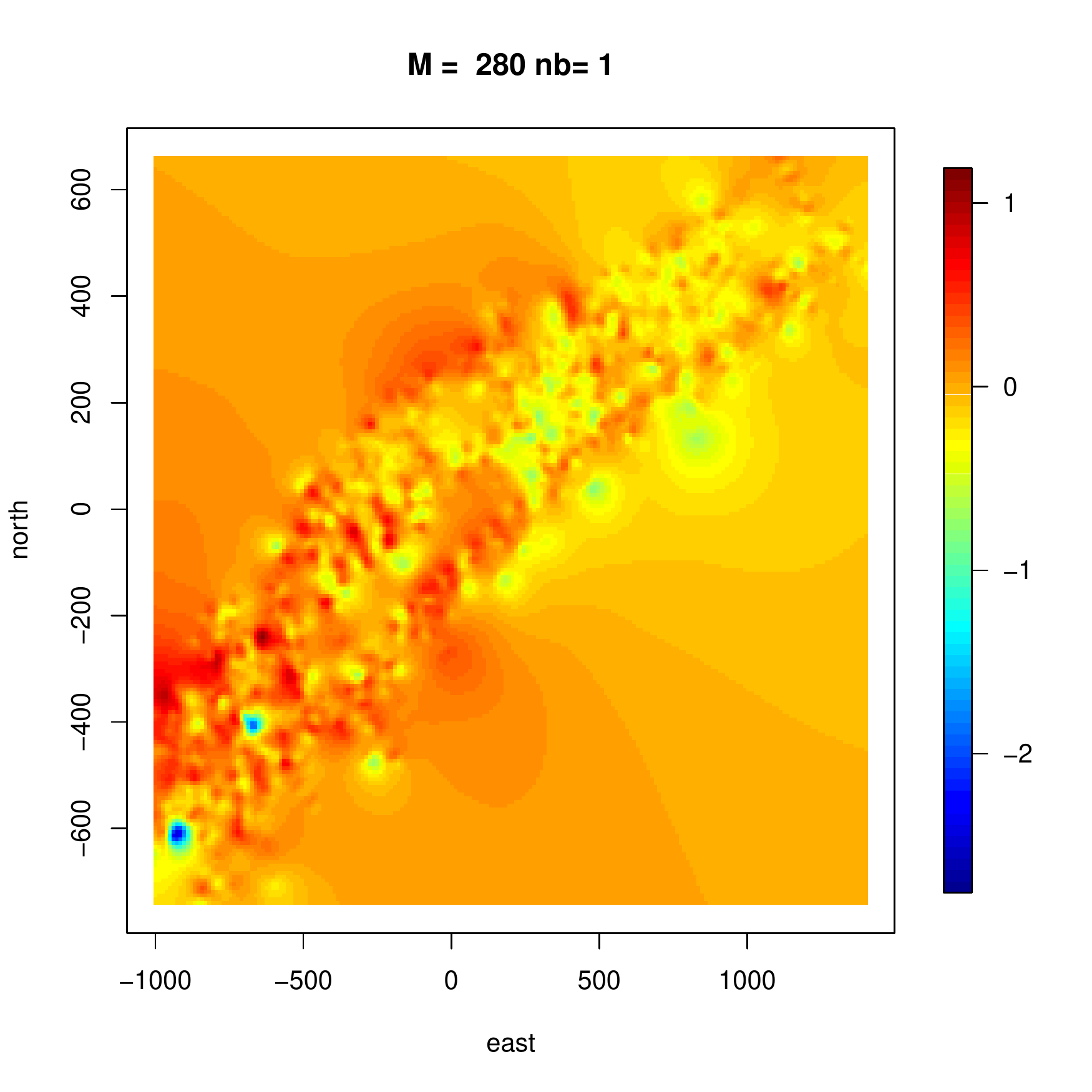}
\includegraphics[scale=0.3]{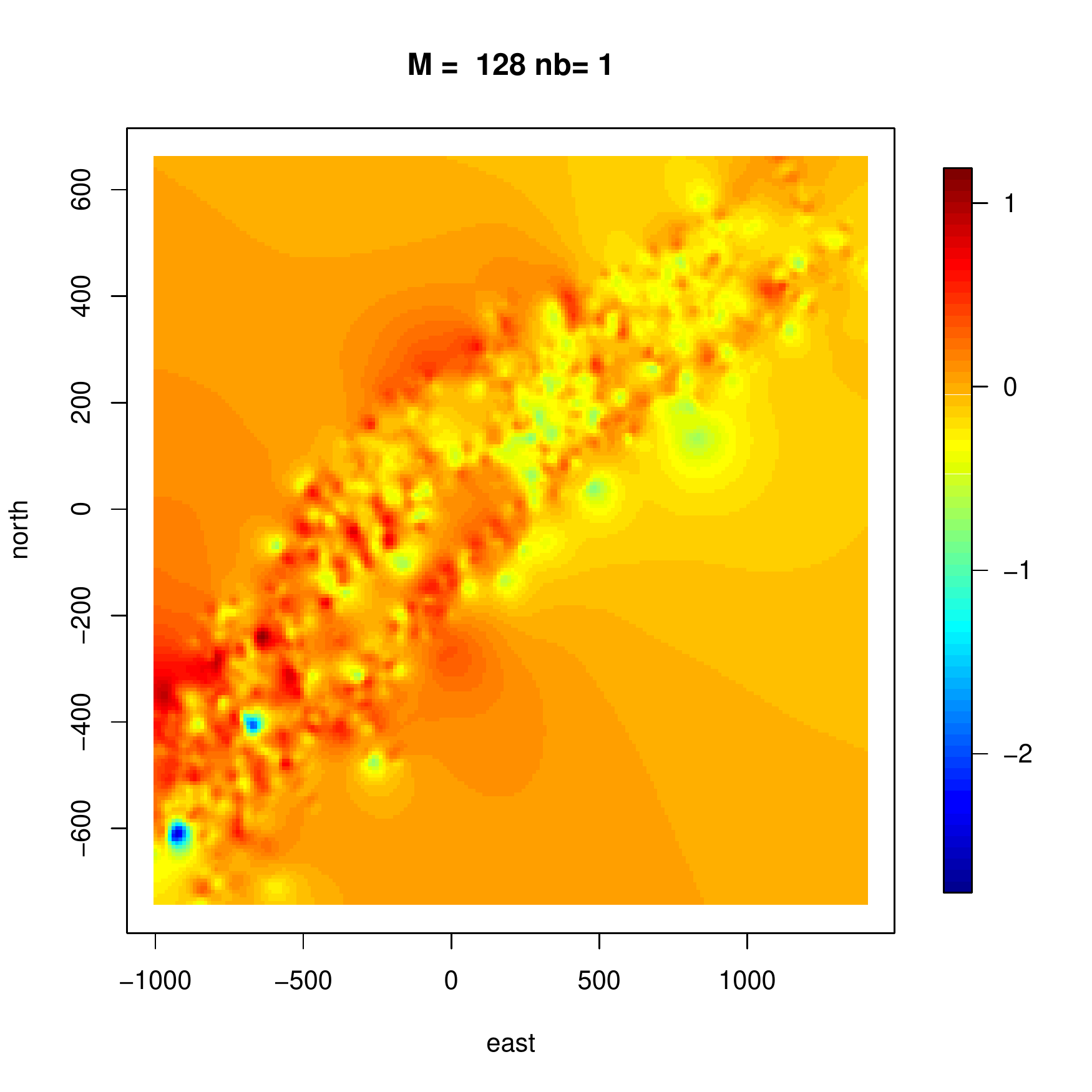}
\caption{ Original joint-frequency data (left upper plot). INLA results. Mean posterior of joint-frequency data using NNGP models with $nb=10, 20$ neighbors (upper panel). Mean posterior of joint-frequency data using block-NNGP models with regular blocks (M=450 and M=280) and irregular blocks (M=128), and nb=1 neighbor block (lower panel).}
  \label{fig:fig6}
\end{figure}

Fig.~\ref{fig:figs4} shows an example of DAGs, built using one location for each block, for the mining and precipitation data used in applications. Figure~\ref{fig:fig6} shows maps of interpolated posterior mean estimates of joint-frequency data. We see little difference between these models. 
Table~\ref{tab:tab4} presents the selection criteria of the fitted models. 
This result shows that it is quite difficult to choose the number of neighbors, because there is not a clear pattern that the more neighbors, the better the model. While the best block-NNGP model is the one with $M=64$ and $nb=4$ neighbor blocks, followed by the model with  $M=64$ and $nb=6$ neighbor blocks. Overall, the results are more {stable} for less blocks and more neighbor blocks. Table~\ref{tab:tab4} also presents the total running time for each model. 
\vspace{3cm}

\setlength{\tabcolsep}{0.3em}
\renewcommand{\arraystretch}{0.6}
\begin{table}[htb]
  \caption{ Precipitation data. INLA results. Criteria assessment and time requirements.
}
  \centering
\small
  \begin{tabular}{lllllllll}
    \toprule 
&M& nb& LPML& WAIC& RSME & RSMP & time (sec) \\ 
   \midrule
        &&10&	      -18407.420&	-11705.390&	0.0052 &0.523 &1535.664\\
        &&20&	      -18863.340&	-11750.800&	0.0051 &0.467 &2588.953\\
NNGP    &&30&   	  -19167.690&	-11666.300&	0.0051 &0.626 &1356.085\\
        &&50&	      -19228.520&	-11680.250&	0.0051 &0.561 &1745.064\\
        &&100&	      -18875.250&	-11783.170&	0.0050 &0.508 &5019.320\\
   \midrule
                &64&	2&	-19533.230&	-11642.630&	0.0050&0.598	& 7191.660\\
                &64&	4&	-19141.680&	-11736.000&	0.0050&0.680	& 13587.350\\
block-NNGP (I)  &64&	6&	-19436.940&	-11708.210&	0.0050&0.645	& 19992.250\\
                &128&	2&	-19661.130&	-11595.540&	0.0049&0.645	& 2740.382\\
                &128&	4&	-19483.990&	-11625.660&	0.0051&0.595	& 5434.684\\
                &128&	6&	-19381.560&	-11680.610&	0.0050&0.742	& 5092.628\\
   \bottomrule    
    \end{tabular}
\label{tab:tab4}    
\end{table}


\newpage 



\end{document}